\documentclass[letterpaper]{article}

\usepackage[T1]{fontenc}

\usepackage{geometry}
\usepackage{setspace}

\usepackage[backend=biber,style=chem-acs]{biblatex}
\usepackage{graphicx}
\usepackage{float}
\newfloat{scheme}{htbp}{los}
\floatname{scheme}{Scheme}
\floatname{chart}{Chart}
\newfloat{graph}{htbp}{loh}

\usepackage{chemformula} 
\usepackage[version = 4]{mhchem} 

\usepackage{hyperref}
\usepackage{authblk}
\author[1]{Ann S.~Y.~Lu}
\affil[1]{Department of Chemistry, University of Oxford, Inorganic Chemistry Laboratory, South Parks Road, Oxford, OX1 3QR, United Kingdom.}
\author[1,2]{Prajna Bhatt}
\affil[2]{CNR - Istituto Officina dei Materiali (IOM), Unità di Trieste, Strada Statale 14, km 163.5, 34149 Basovizza (TS), Italy}
\author[3]{Nathalie K.~Fernando}
\affil[3]{Department of Chemistry, University College London, 20 Gordon Street, London, WC1H~0AJ, United Kingdom.}
\author[4,5]{Laura E.~Ratcliff$\dagger$}
\affil[4]{Centre for Computational Chemistry, School of Chemistry, University of Bristol, Bristol BS8 1TS, United Kingdom}
\affil[5]{Hylleraas Centre for Quantum Molecular Sciences, Department of Chemistry, UiT The Arctic University of Norway, N-9037 Tromsø, Norway}
\author[1,3]{Anna Regoutz*}

\title{A comprehensive experimental and theoretical spectroscopic study of proteinogenic amino acids}
\date{Email: $\dagger$laura.ratcliff@bristol.ac.uk, *anna.regoutz@chem.ox.ac.uk}

\begin{document}

\maketitle

\begin{abstract}
  Amino acids are essential building blocks of life, yet our understanding of their chemistry and electronic structure in the solid state remains limited. This is particularly important because amino acids in the solid state are relevant to biological and pharmaceutical processes. X-ray photoelectron spectroscopy provides a powerful experimental probe of chemical states and occupied electronic structure; however, most spectroscopy studies of amino acids focus on gas-phase species or surface adsorbates, while crystalline amino acids remain underexplored, largely because of experimental challenges associated with radiation damage. Additionally, the spectra are often complex and difficult to interpret, motivating a combined experimental-theoretical approach. This study combines X-ray photoelectron spectroscopy and density functional theory calculations to systematically investigate the core, semi-core, and valence states of 20 proteinogenic amino acids as well as selenomethionine which can be incorporated during protein synthesis and deliver the essential nutrient, Se, required by humans. Calculated relative core binding energies show excellent agreement with experiment and enable reliable assignments. Projections of the density of states provide insight into the influence of local coordination and extended crystal structure, yielding a systematic understanding of the electronic structure and bonding in solid-state AAs. The insights gained from this study enhance the understanding of crystalline amino acids and validate the robustness of an integrated experiment--theory framework.
\end{abstract}

\section*{Keywords}

Density Functional Theory, X-ray Photoelectron Spectroscopy, Proteinogenic Amino Acids, Core States, Semi-Core States, Valence States

\section{Introduction}

Amino Acids (AAs) are fundamental building blocks of living organisms and play a crucial role in the synthesis of proteins in animals and plants.~\cite{Akram2011AminoArticle, Chandel2021AminoMetabolism, Dietzen2018AminoProteins} Beyond the fundamental importance of AAs for life, these compounds are of considerable scientific and technological interest, first and foremost due to their widespread applications in the food and pharmaceutical industries.~\cite{Essien2023PredictionElectra, Sathisaran2018EngineeringMedium} 
In the solid state, AAs adopt a zwitterionic structure, in which \ce{-NH2} is protonated, forming \ce{NH_3^+}, and \ce{-COOH} is deprotonated to a carboxylate group, \ce{COO^-}, and exist as molecular crystals. The crystal packing of AAs is strongly influenced by complex intermolecular interactions. To optimise the crystal packing and minimise the lattice energy, AA molecules adopt distinct orientations within the unit cell and, several AAs with long, bulky non-polar side chains, such as valine, leucine, and isoleucine, also exhibit multiple conformations (see Figures~\href{achemso-SI.pdf#Ali_crystal structure}{S1} to~\href{achemso-SI.pdf#S_Se_crystal structure}{S4} in the SI).

As such, the understanding of their chemical and electronic structure has been the subject of various photoelectron spectroscopy (PES) studies since the 1970s.~\cite{Clark1976AnPolypeptides, Cannington1979TheSurvey} However, most spectroscopy studies focus on gas-phase species,~\cite{Slaughter1988Core-photoelectronAcidity, Feyer2008CoreThreonine, Plekan2007InvestigationSpectroscopy, Powis2000PhotoelectronL-Alanine} or species adsorbed on metallic substrates, like Cu, Pd, and Au,~\cite{Ihs1990InfraredFl-Alanine, Eralp2011TheCu110, Feng2007ChemistryStudy, Uvdal1992L-CysteineStudy} leaving the crystalline AAs relatively underexplored.~\cite{Stevens2013QuantitativeXPS, Artemenko2021ReferenceAcids, Tzvetkov2010X-rayStudy, Chatterjee2008Core-levelHartreeFock} This is largely due to the experimental challenges associated with radiation damage,~\cite{Tzvetkov2010X-rayStudy, Zubavichus2004SoftStudy, Zubavichus2004SoftStudyb} resulting in a limited number of high-resolution X-ray photoelectron spectra reported in the literature. Furthermore, the complexity of the photoelectron spectra hinders the accurate interpretation of experimental features, thereby motivating a combined experimental and theoretical approach for precise spectral understanding. Density functional theory (DFT) is a widely used theoretical approach to simulate spectra due to its balance between accuracy and computational cost.~\cite{Hohenberg1964InhomogeneousGas, Kohn1965Self-consistentEffects} It has been successfully used to predict the core binding energies (BEs) for AAs in the gas-phase.~\cite{Tolbatov2014ComparativeGlycine, Tolbatov2017BenchmarkingAcids, Feyer2008CoreThreonine, Li2012First-principlesPhases} To date, studies employing combined XPS and DFT have remained limited in scope, primarily focusing on the core states of some simple crystalline AAs such as glycine, alanine, and serine,~\cite{Pi2020PredictingTheory} aromatic AAs, including phenylalanine, tyrosine, histidine, and tryptophan,~\cite{Regoutz2021AAcids} and alanine on adsorbate phase.~\cite{Gao2009Probing0}

This work provides a comprehensive experimental and theoretical spectroscopic study of the 20 proteinogenic AAs, as well as selenomethionine, the Se analogue of methionine, in the crystalline, solid state. The isolated 2D structures, including atom labels, are depicted in Figure~\href{achemso-SI.pdf#2D structure}{S5} in the SI. The X-ray photoelectron spectroscopy (XPS) measurements, including the core, semi-core, and valence states, form the experimental basis of this work with particular care taken to mitigate radiation damage during X-ray exposure. The theoretical spectra are calculated using DFT based on the known crystal structures of the AAs. Experimental and theoretical spectra are combined, providing a complete picture of the relationships between local coordination, extended crystal structure, and intra- and inter-molecular interactions.

\section{Methodology}

This work uses the three-letter symbol to refer to specific AAs throughout. Glycine (Gly) is the simplest AA, and is included throughout this manuscript to serve as a reference for discussing the chemical bonding and electronic structure. All other AAs are categorised into four groups, based on the properties of their side chain R groups as follows: aliphatic AAs, including alanine (Ala), valine (Val), leucine (Leu), isoleucine (Ile), and proline (Pro); aromatic AAs, such as phenylalanine (Phe), tyrosine (Tyr), histidine (His), and tryptophan (Trp); polar side chain-containing AAs, including asparagine (Asn), glutamine (Gln), arginine (Arg), lysine (Lys), aspartic acid (Asp), and glutamic acid (Glu); and S/Se-containing AAs, containing serine (Ser), threonine (Thr), cysteine (Cys), methionine (Met), and selenomethionine (SeMet). It should be noted that although Ser and Thr also contain polar side chains, they are included in the more specific S/Se group. This grouping is employed to enable systematic comparison of changes in side chain make-up of related AAs and the resulting effects on intermolecular interactions and crystalline structure. 

\begin{figure}[htp]
    \centering
    \includegraphics[width=1.0\textwidth]{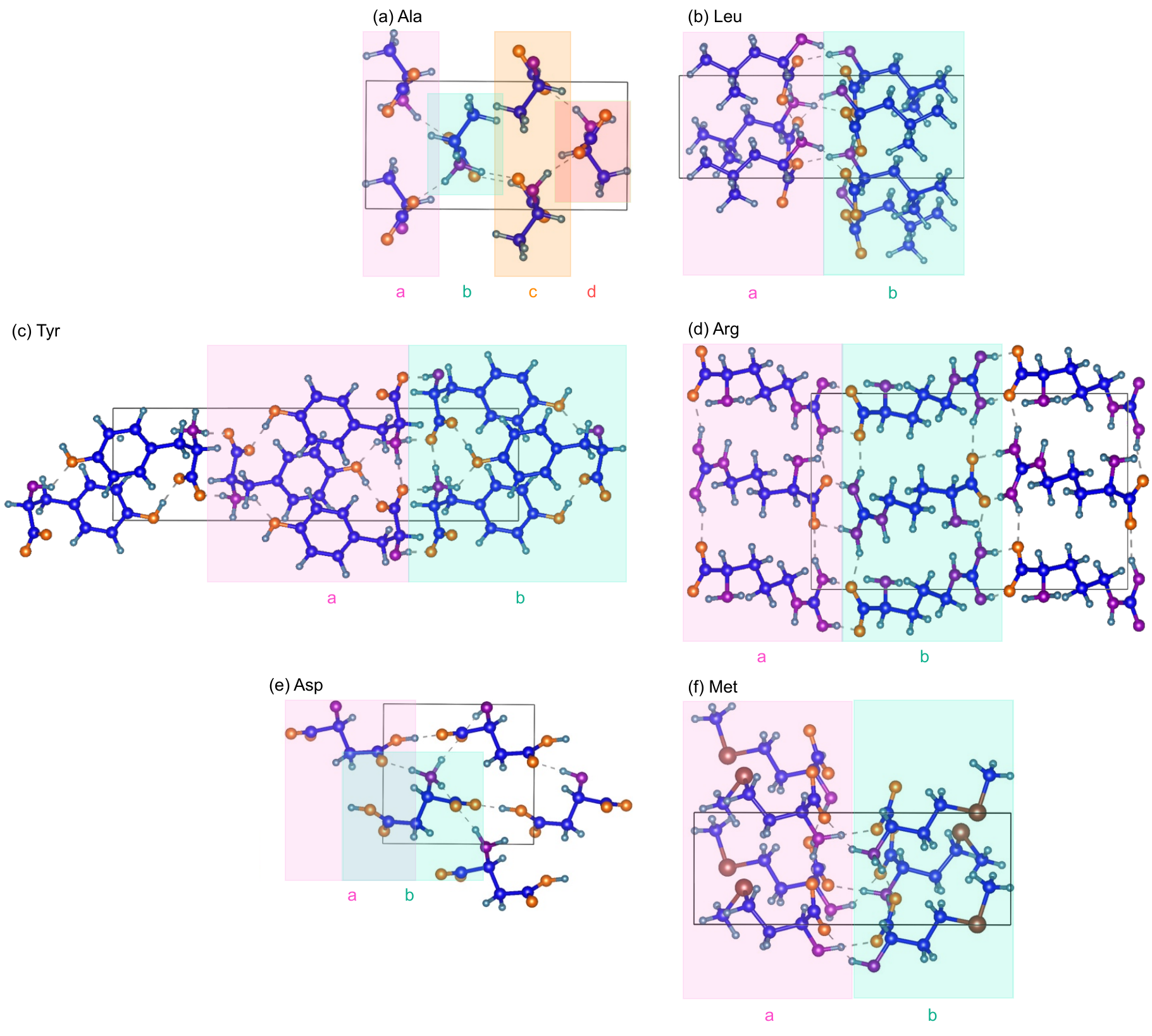}
    \caption{Visualisation of the unit cell crystal structures of the representative AAs with different columns and orientations highlighted by different colours for (a) Ala, (b) Leu, (c) Tyr, (d) Arg, (e) Asp, and (f) Met. Crystal structures of the representative AAs are chosen from the Cambridge Structural Database (CSD),~\cite{Waddell2024RODIN:Demonstration, Binns2016AccurateL-leucine, Courvoisier2012TheL-arginine, Bendeif2007TheAcid} and are visualised along the $a$ direction for all AAs except Arg, which is viewed along the $c$ direction. The first and second columns are coloured by pink and cyan for the representative AAs. The third and fourth columns are inked in orange and red for Ala. All H, C, N, O, and S atoms are inked in steel blue, deep blue, purple, orange, and brown, respectively. All structures were prepared in the VESTA software package.~\cite{Momma2011VESTA3Data}}
    \label{Crystal structure_MP}
\end{figure}

In the main manuscript data for a subset of representative AAs, selected based on their distinct crystal packing and molecular structure, is included alongside comprehensive analysis and discussion of all investigated AAs. The unit cell crystal structures for the representative AAs are visualised in Figure~\ref{Crystal structure_MP}. Complete datasets for all AAs are provided in the Supporting Information (SI). 

\subsection{Experimental Methods}

Powders of the L-stereoisomers of all investigated AAs were purchased from Sigma-Aldrich with a minimum purity of 98$\%$ and were used without further purification. XPS data were collected on Thermo Scientific K-\ce{Alpha^+} or NEXSA G2 XPS systems with a monochromated, microfocused Al K$\alpha$ X-ray source with a photon energy, h$\nu$, of 1486.7~eV (further referred to as 1.5~keV for simplicity). The X-ray source was operated at 6~mA emission current and 12~kV anode bias, and had a spot size of 400~$\mu$m. The base pressure was 2 $\times$ \ce{10^{-9}}~mbar for both systems. Survey spectra were obtained at a pass energy of 200~eV and dwell time of 25~ms; core level (CL) spectra were collected at a pass energy of 20~eV and dwell time of 50~ms; and valence band (VB) spectra were collected at a pass energy of 15~eV and dwell time of 75~ms. The samples were mounted on conducting carbon tape, and a dual Ar\textsuperscript{+} ion and electron source flood gun was used to compensate for any sample charging.~\cite{Baer2020XPSSamples} As AAs are prone to suffering from radiation damage, the samples were rastered, with data collection in each position limited to a duration where no significant damage could be observed. To determine the acceptable irradiation duration, iterative scans of all core state spectra were collected. The maximum data collection period was set based on the total measurement time accumulated before any changes in the spectra were observed. Data were collected at four points across each sample, which were then averaged to achieve the necessary signal statistics to allow peak analysis and comparison to theory. 

All data were analysed using the Thermo Scientific Avantage version 6.9.2 software package. For peak fit analysis, Shirley-type backgrounds and Voigt functions were used with both the full width at half maximum (FWHM) and Lorentzian/Gaussian (L/G) ratios refined within physically meaningful constraints. For each group of AAs, the average (avg.) value of binding energies (BEs) of C~1\textit{s} features, assigned to the \ce{C^$'$} atom, of each AA was taken, and the survey, CL, semi-core level (SCL), and VB spectra were aligned to this value. XPS data of Gly were included, and aligned to the avg.\ BE of the \ce{C^$'$} atom in each group for comparison. For each AA, the survey spectrum was normalised 0 to 1, all CL spectra were normalised to the area of the C~1\textit{s} feature assigned to the \ce{C^$'$} atom, and the VB spectra were normalised to their area, respectively. 

\subsection{Computational Methods}

DFT calculations were performed using the CASTEP code with a plane-wave (PW) basis set for both core and valence state calculations.~\cite{Clark2005FirstCASTEP, Segall2002First-principlesCode} Calculations employed norm-conserving pseudopotentials CASTEP 19.0 (NCP19) and the exchange-correlation (XC) functional, PBE.~\cite{Perdew1996GeneralizedSimple} Cut-off energy, ground-state $k$-point, and spectral $k$-point convergence tests, and geometry optimisations with no spin polarisation were performed. The Grimme's D2 correction scheme (G06) was applied to account for van der Waals interactions,~\cite{Grimme2006SemiempiricalCorrection} and a maximum force tolerance of 0.05~eV/\AA\ was applied when conducting the geometry optimisations. A cut-off energy of 1100~eV was employed for all AAs, except SeMet, which used a higher cut-off energy of 1300~eV due to the harder pseudopotential required by the heavier Se atom, maintaining the consistency of the convergence threshold of total energy (0.002~eV/atom) (see the cut-off convergence plot presented in Figure~\href{achemso-SI.pdf#cutoff}{S6} in the SI). Spectral $k$-points used for the smooth and accurate calculations of valence state spectra for all AAs are tabulated in Table~\href{achemso-SI.pdf#spectra kpoints}{S1} in the SI. The final relaxed lattice parameters after geometry optimisation were found to be within 4$\%$ of the initial selected experimental values (see Tables~\href{achemso-SI.pdf#structural table1}{S2} and~\href{achemso-SI.pdf#structural table2}{S3} in the SI). The differences in lattice parameters and unit cell angles for all AAs before and after geometry optimisation are tabulated in Tables~\href{achemso-SI.pdf#structural table1}{S2} and~\href{achemso-SI.pdf#structural table2}{S3} in the SI.

Projected density of states (PDOS) were calculated using a Mulliken-style population analysis and later post-processed using the OptaDOS code with 0.01~eV intervals to sample the DOS.~\cite{Morris2014OptaDOS:Codes} To account for the experimental broadening, a Gaussian broadening of 0.43~eV was applied, based on the total experimental resolution of the photoelectron spectrometers used in the experiment as determined from the 16/84\% width of the Fermi edge of a gold reference foil.~\cite{Wolstenholme2020ProcedureFrequently} To better compare with the experimental VB spectra, Scofield one-electron photoionisation cross sections for 1.48667~keV (Al~\ce{K_\alpha}) were applied to the calculated PDOS using the Galore package.~\cite{Jackson2018Galore:Spectroscopy} Theoretical spectra of the SCL and VB were normalised to the height and aligned to the BE of the following corresponding experimental features: the C~2\textit{s} feature with the highest BE for SCL and the lowest BE valence band feature for VB. These alignment strategies were chosen to allow the most robust comparison between theory and experiment to ensure ease of analysis and interpretation.\par

Core BE calculations were performed at the $\Delta$SCF level of theory,~\cite{Cavigliasso1999AccurateApproach, Pi2020PredictingTheory} which determines the BE as the total energy difference between the core-excited state and ground state, thereby allowing final state effects to be considered, using norm-conserving pseudopotentials (PSPs) generated by the on-the-fly PSP generator. The parameters used to generate the ground-state PSP for each element presented in AAs can be found in the SI. The same parameters were used for the core hole calculations with a core hole in the PSP. The core hole calculations have a net charge of $+$1. All calculated BEs have been aligned to the avg.\ of the BEs of C~1\textit{s} features assigned to the \ce{C^$'$} atom of each AA in each group. The remotemanager Python package was employed to submit all core BE and PDOS calculations and retrieve the results from a remote supercomputer.~\cite{Dawson2024ExploratoryCalculations}

\section{Results and Discussion}

The results and their discussion are presented first for the core, followed by the semi-core, and finally the valence states. In each Section the four AA groups, aliphatic, aromatic, polar side chain, and S/Se-containing, are discussed sequentially. 

\subsection{Core States}

The majority of AAs are composed solely of C, N, and O atoms, as expected based on the initial survey spectra shown in Figure~\href{achemso-SI.pdf#AAs_Survey}{S7} in the SI; accordingly, to characterise the core states, the C~1\textit{s}, N~1\textit{s}, and O~1\textit{s} CL spectra were measured. For the S/Se-containing AAs, including Cys, Met, and SeMet, additional S~2\textit{p} and Se~3\textit{d} CLs were measured. DFT-calculated core BEs are used to aid in the interpretation of the experimental core state features, and they generally describe the relative experimental BEs well. The following Section comprises, first, an overview of the core states of representative AA compounds. Then, discussions are organised into groups according to the character of the AAs, namely aliphatic, aromatic, polar side chain-containing, and S/Se-containing. The SI includes all experimental spectra, including peak fit analysis, and the corresponding relative theoretical core BEs for C, N, and O~1\textit{s} for the aliphatic, aromatic, polar side chain-containing, and S/Se-containing groups in Figures~\href{achemso-SI.pdf#Ali_CLs}{S8}-\href{achemso-SI.pdf#S_Se_CL_peakfitting}{S16}. Experimental and theoretical BE values of core states are summarised in Tables~\href{achemso-SI.pdf#Ali_CL_comp}{S4}-\href{achemso-SI.pdf#S_Se_CL_comp}{S7} in the SI.

\begin{figure}[htp]
    \centering
    \includegraphics[width=0.8\textwidth]{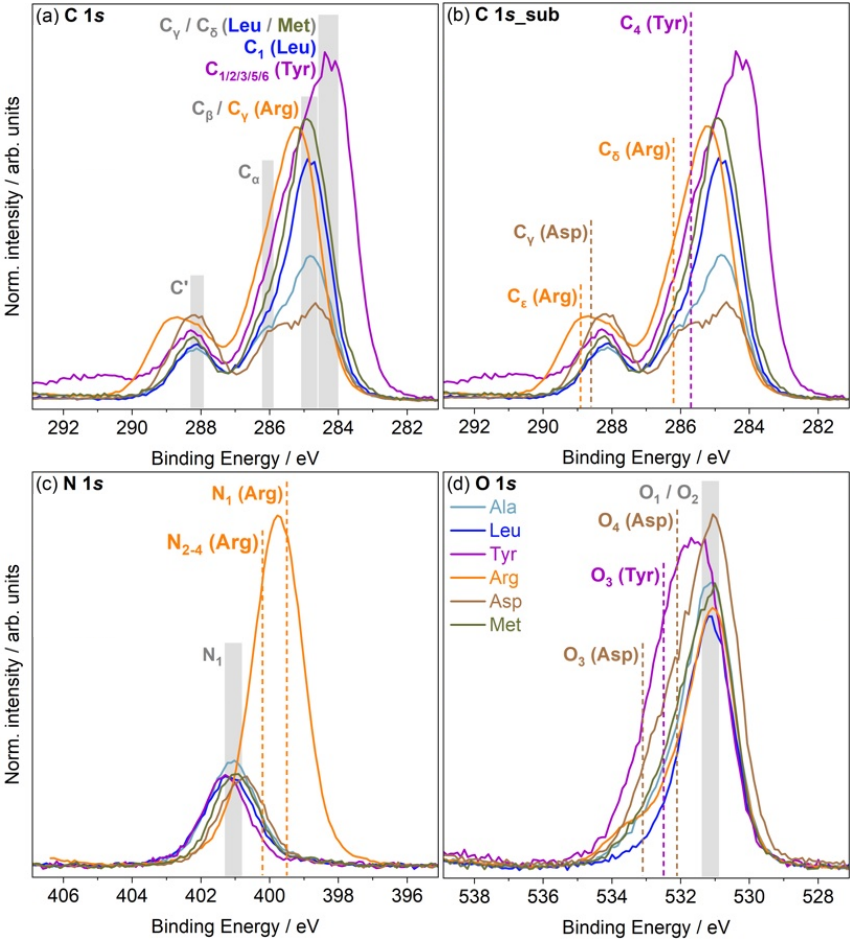}
    \caption{Combined CL spectra for representative AAs, including (a) and (b) C~1\textit{s} with (a) common features and (b) features to specific AAs highlighted, (c) N~1\textit{s} spectra, and (d) O~1\textit{s} spectra. Common features are labelled with grey solid lines of varying widths, representing the BE differences after alignment across chemical states for each CL. Features specific to individual AAs are labelled using coloured, dashed lines, where the colour of the dashed lines is the same as the label colour of the corresponding AA. Spectra are aligned to the average value of the BEs of C~1\textit{s} features, assigned to the \ce{C^$'$} atom, within each group of AAs.}
    \label{CLs_MP}
\end{figure}

The C, N, and O~1\textit{s} spectra for the representative AAs from each group are presented in Figures~\ref{CLs_MP}(a)/(c)/(d), respectively. Figure~\ref{CLs_MP}(b) highlights notable C~1\textit{s} features unique to specific AAs. The chemical state labels included in these Figures are based on theoretical core BEs, and shaded grey areas serve as illustrative guides of the observed experimental BE ranges of key chemical states. While these visual guides provide a qualitative overview of chemical shift trends, a rigorous statistical treatment and avg.\ BE values for each AA group are provided in the SI and are discussed in the following Sections.

In the C~1\textit{s} spectra (Figure~\ref{CLs_MP}(a)), two main features are consistently observed in all AAs: the carboxylic carbon (\ce{C$'$}) of the \ce{COO^-} group, and the $\alpha$-carbon (\ce{C_$\alpha$}) which is bound to the \ce{COO^-} and \ce{NH3^+} groups. \ce{C$'$} consistently has a higher BE, since it is bound to two highly electronegative O atoms, (\ce{\chi_O} $\approx3.44$),~\cite{Allred1961ElectronegativityData} that withdraw electron density from \ce{C$'$}. In contrast, \ce{C_$\alpha$} is bound to the less electronegative N atom (\ce{\chi_N} $\approx$ 3.04).~\cite{Allred1961ElectronegativityData} As a result of more electronegative bonded neighbours, the electron density around \ce{C$'$} is lower than that around \ce{C_$\alpha$}, due to the reduced electronic screening and a higher effective nuclear charge experienced by the core electrons. Consequently, more energy is required to remove a 1\textit{s} electron from \ce{C$'$}, resulting in a higher BE. Following this effect, the \ce{C_$\beta$} atom, located adjacent to \ce{C_$\alpha$}, is generally expected to exhibit the third-highest BE. However, this trend does not always hold; in some AAs, \ce{C_$\beta$} is indistinguishable from more distal C atoms in the side chain based on theoretical calculations. These variations are discussed in further detail in the subsequent group-specific Sections. 

Based on the zwitterionic structure of pristine AAs, the amino group attached to \ce{C_{$\alpha$}} is protonated with the positive charge localised on the $\alpha$-N atom, labelled as \ce{N1}. In the N~1\textit{s} spectra (Figure~\ref{CLs_MP}(c)), \ce{N1} consistently appears at a BE of approximately 401~eV, in agreement with previous studies.~\cite{Regoutz2021AAcids, Pi2020PredictingTheory, Stevens2013QuantitativeXPS, Zubavichus2004SoftStudyb, Tzvetkov2010X-rayStudy, Chatterjee2008Core-levelHartreeFock, Peeling1976EvaluationMeals} Arg represents a notable exception, where the entire N~1\textit{s} CL spectrum is globally shifted to lower BE, commensurate with previous solid-state studies.~\cite{Stevens2013QuantitativeXPS, Artemenko2021ReferenceAcids} This is because of the location of the positive charge. In Arg, the positive charge is delocalised over the \ce{N2}, \ce{N3}, and \ce{N4} atoms of the side-chain guanidinium group, and these environments appear at a higher BE than the neutral \ce{N1}. Moreover, the --\ce{COO^-} group attached to \ce{C_{$\alpha$}} is deprotonated with the excess negative charge delocalised on \ce{O1} and \ce{O2}. In the O~1\textit{s} spectra (Figure~\ref{CLs_MP}(d)), electron densities around \ce{O1} and \ce{O2} are nearly equivalent due to the delocalisation, giving rise to one single peak, commensurate with the literature.~\cite{Pi2020PredictingTheory, Regoutz2021AAcids, Stevens2013QuantitativeXPS, Artemenko2021ReferenceAcids, Zubavichus2004SoftStudyb} For AAs without side-chain O (e.g., Ala, Leu, Arg, and Met), the shoulder observed on the higher BE side of the \ce{O1} and \ce{O2} feature results from surface hydroxylation and adsorption of e.g., \ce{H2O} and \ce{CO2}, due to exposure to ambient conditions. For AAs containing hydroxyl or carboxyl groups in their side chains (e.g., Tyr and Asp), this feature is more pronounced in intensity due to contributions from the --OH or --COOH groups. 

\subsubsection{Aliphatic Group}~\label{Ali CLs}

The combined experimental C, N, and O~1\textit{s} CL spectra for the aliphatic group (Ala, Val, Leu, Ile, and Pro) are shown in Figure~\href{achemso-SI.pdf#Ali_CLs}{S8} in the SI. Common features among the aliphatic AAs are labelled with grey solid lines of varying widths, representing BE ranges for different chemical states. Features which are assigned to Pro specifically are presented using coloured, dashed lines in Figure~\href{achemso-SI.pdf#Ali_CLs}{S8}(b) in the SI. The assignment of experimental features is based on results from the theoretical calculations and is applied consistently to all other groups of AAs.

For the C~1\textit{s} spectra (Figure~\href{achemso-SI.pdf#Ali_CLs}{S8}(a) in the SI), three main features, \ce{C$'$}, \ce{C_$\alpha$}, and \ce{C_$\beta$}, are observed across all aliphatic AAs based on the theoretical calculations, except Leu. Their avg.\ experimental BEs are 288.2~eV, 286.3 $\pm$ 0.3~eV, and 285.1 $\pm$ 0.9~eV, respectively. The BE differences of these assignments are in good agreement with existing literature, particularly for Ala in both gas and solid states,~\cite{Feyer2008CoreThreonine, Pi2020PredictingTheory, Regoutz2021AAcids, Zubavichus2004SoftStudyb, Powis2003InvestigationSpectroscopy} and Val in aqueous solution.~\cite{Mocellin2017SurfaceXPS, Kumar2000AmphoterizationMolecules} It is important to note that BE positions reported in the literature for gas and solution studies are often higher than in the solid state because of different BE alignment strategies, i.e., BE of C~1\textit{s} of \ce{CO2}, 297.6~eV, in gas studies;~\cite{Feyer2008CoreThreonine} and energy of the highest occupied molecular orbital (1\ce{b1}) of liquid water in solution studies.~\cite{Mocellin2017SurfaceXPS, Winter2006PhotoemissionSolutions} The lowest BE feature in all aliphatic AAs has some overlap with adventitious C from handling in ambient air. In Pro, the feature near 286.4~eV is attributed to both \ce{C_$\alpha$} and \ce{C_5} (adjacent to the ring N), reflecting comparable C--N bonding environments in the pyrrolidine ring. The lowest BE feature is from \ce{C_3} and \ce{C_4}, which share similar chemical environments at 285.4~eV, consistent with previous gas-phase assignments.~\cite{Plekan2007InvestigationSpectroscopy} In Val, Leu, and Ile, the more distal side-chain carbons (\ce{C_$\gamma$}/\ce{C1} and \ce{C_$\beta$} for Leu) contribute to the lowest BE feature at 284.8 $\pm$ 0.1~eV). Notably, in Leu, \ce{C_$\beta$} is indistinguishable from more distal C atoms in the side chain, including \ce{C_$\gamma$}, \ce{C_$\delta$}, and \ce{C1}, based on calculations. This is explained in terms of the local chemical environment of \ce{C_$\beta$} in Leu and its structural isomer, Ile. \ce{C_$\beta$} in Leu is a secondary carbon, whereas in Ile it is a tertiary carbon, bound to three C atoms. Owing to the higher electronegativity of carbon (\ce{\chi_C} $\approx2.55$) relative to hydrogen (\ce{\chi_H} $\approx2.04$),~\cite{Allred1961ElectronegativityData} the C--H bond is polarised toward carbon, in contrast to the essentially nonpolar C--C bond. Since \ce{C_$\beta$} in Leu is bound to H, it experiences a higher local electron density, enhancing screening and reducing the BE in Leu compared to Ile. 

Finally, N and O~1\textit{s} spectra presented in Figures~\href{achemso-SI.pdf#Ali_CLs}{S8}(c)/(d) in the SI, possess core state features consistent with the literature.~\cite{Regoutz2021AAcids, Pi2020PredictingTheory, Stevens2013QuantitativeXPS, Tzvetkov2010X-rayStudy, Chatterjee2008Core-levelHartreeFock, Peeling1976EvaluationMeals, Artemenko2021ReferenceAcids} N~1\textit{s} has a single symmetric feature at an avg.\ experimental BE of 401.2 $\pm$ 0.3~eV with features visible at lower BE arising from radiation damage during the experiment. This results in deprotonation of the protonated amino group, forming the neutral --\ce{NH2} group. In addition, O~1\textit{s} spectra also present a single peak with an avg.\ BE of 531.1 $\pm$ 0.2~eV, arising from \ce{O1} and \ce{O2}.

\subsubsection{Aromatic Group}~\label{Aro CLs}

The combined C, N, and O~1\textit{s} CL spectra for the aromatic group (Phe, Tyr, His, and Trp) are presented in Figure~\href{achemso-SI.pdf#Aro_CLs}{S10} in the SI. For the C~1\textit{s} spectra (Figure~\href{achemso-SI.pdf#Aro_CLs}{S10}(a) in the SI), the BE positions of \ce{C$'$}, \ce{C_$\alpha$}, and \ce{C_$\beta$} are in good agreement with previous solid-state studies,~\cite{Regoutz2021AAcids, Stevens2013QuantitativeXPS} having avg.\ BEs of 288.3~eV, 286.3 $\pm$ 0.3~eV, and 285.4 $\pm$ 0.6~eV, respectively. The PBE-calculated relative BEs describe most experimental BE deviations well, however, larger discrepancies between the calculated and experimental BEs are observed for Trp. These can be attributed to the conjugated indole ring and strong final-state effects, consistent with a previous study.~\cite{Regoutz2021AAcids} C atoms in the aromatic ring, including \ce{C2} to \ce{C6} for Phe, \ce{C1}/\ce{C2}/\ce{C3}/\ce{C5}/\ce{C6} for Tyr, and \ce{C3}/\ce{C_{3a}}/\ce{C4}/\ce{C5}/\ce{C6}/\ce{C7} for Trp, are attributed to the feature at a BE around 284.1 $\pm$ 0.8~eV. Regarding the peaks assigned to specific AAs (Figure~\href{achemso-SI.pdf#Aro_CLs}{S10}(b) in the SI), \ce{C1} in Phe appears at a slightly higher BE around 285.0~eV than other C atoms in the ring, consistent with inductive withdrawal by --\ce{NH3^+} and --\ce{COO-} substituents bound to the \ce{C_$\alpha$}. In Tyr, the phenolic \ce{C4} lies at 285.7~eV, reflecting the C--OH environment. In Trp, the bridgehead \ce{C_{7a}} sits at 285.5~eV, comparable to \ce{C_$\beta$}, due to proximity to \ce{N2}. \ce{C2}, also adjacent to \ce{N2}, is observed at a lower BE of 284.6~eV.~\cite{Regoutz2021AAcids} In His, the three imidazole carbons (\ce{C2}, \ce{C4}, and \ce{C5}) do not contribute to the feature at 284.1~eV. In contrast, they show a clear distinction in BEs: \ce{C2} is flanked by two ring N atoms, \ce{N2} and \ce{N3}, and attributed to the feature at 285.8~eV. \ce{C4} and \ce{C5} are adjacent to \ce{N2} and \ce{N3}, respectively, appearing at 285.3~eV, consistent with the literature.~\cite{Regoutz2021AAcids, Stevens2013QuantitativeXPS}
In addition to the features discussed above, all C~1\textit{s} spectra exhibit \ce{\pi-\pi^*} shake-up satellites at BEs between 291-293~eV with relative intensities $\leq$ 4$\%$ of the lowest BE C~1\textit{s} feature, in good agreement with previous observations for conjugated systems, including aromatic AAs.~\cite{Clark1976AnPolypeptides} 

The symmetric feature denoted \ce{N1} has an avg.\ BE of 401.2~eV $\pm$ 0.7~eV (Figure~\href{achemso-SI.pdf#Aro_CLs}{S10}(c) in the SI). In His and Trp, N atoms from the non-protonated amino groups in the side chain lead to additional peaks at the lower BE side. BE positions of two N atoms, \ce{N2} (400.5~eV) and \ce{N3} (398.8~eV), from the imidazole ring in His, and \ce{N2} (399.4~eV) in Trp show good agreement with literature.~\cite{Regoutz2021AAcids, Stevens2013QuantitativeXPS, Artemenko2021ReferenceAcids} In O~1\textit{s} (Figure~\href{achemso-SI.pdf#Aro_CLs}{S10}(d) in the SI), \ce{O1} and \ce{O2} consistently give rise to the asymmetric peak at an avg.\ BE of 531.3 $\pm$ 0.3~eV. The phenolic O in Tyr contributes to the extra feature on the higher BE side at 532.5~eV.

\subsubsection{Polar Side Chain-Containing Group}~\label{other CLs}

The combined C, N, and O~1\textit{s} CL spectra for the polar side chain-containing AAs (Asn, Gln, Arg, Lys, Asp, and Glu) can be found in Figure~\href{achemso-SI.pdf#other_CLs}{S12} in the SI. For the C~1\textit{s} spectra (Figure~\href{achemso-SI.pdf#other_CLs}{S12}(a) in the SI), \ce{C$'$} and \ce{C_$\alpha$} have avg.\ BEs of 287.9~eV, and 286.2~eV $\pm$ 0.3~eV, respectively. The lowest BE feature at 284.8 $\pm$0.4~eV is attributed to the more distal side-chain carbons, which are assigned to \ce{C_$\beta$} in Gln/Arg/Glu/Lys, \ce{C_$\gamma$} in Gln/Arg/Lys/Glu, and \ce{C_$\delta$} in Lys, as well as adventitious C. Moreover, various functional groups in the side chains introduce additional, chemically specific C~1\textit{s} features, shown in Figure~\href{achemso-SI.pdf#other_CLs}{S12}(b) in the SI. In Asn and Gln, carbon atoms from the carbamido group (\ce{C_$\gamma$} in Asn and \ce{C_$\delta$} in Gln) appear at nearly the same BE as \ce{C$'$}. This can be explained by the fact that \ce{C_$\gamma$} and \ce{C_$\delta$} are bound to two electronegative atoms, N and O, making the electron density around \ce{C_$\gamma$} and \ce{C_$\delta$} become comparable to that around \ce{C$'$}. This matches previous solid-state studies for Asn and Gln.~\cite{Zubavichus2004SoftStudyb, Artemenko2021ReferenceAcids} In Arg, \ce{C_$\epsilon$} is bound to the guanidino group and strongly electron-deficient, as it attaches with three more electronegative N atoms. Therefore, it appears at the highest BE of 288.9~eV. The adjacent C atom, \ce{C_$\delta$}, is inductively influenced by the guanidino group and appears at 286.2~eV.~\cite{Stevens2013QuantitativeXPS} In Lys, the terminal C atom, \ce{C_$\epsilon$}, is bound to a neutral amino group, \ce{NH2}, giving a shift to the higher BE side relative to other C atoms in the side chain, appearing at 285.8~eV. For Asp and Glu, the carboxylic group, --COOH, is attached to \ce{C_$\gamma$} and \ce{C_$\delta$}, respectively. In the solid-state, the --COOH group in the side chain remains neutral, and the BEs of \ce{C_$\gamma$} and \ce{C_$\delta$} are higher than that of \ce{C$'$} in the \ce{COO^-} group, observed at an avg.\ BE of 288.7~eV $\pm$ 0.1~eV, in accordance with the higher electron density at \ce{C$'$} in the \ce{COO^-} group.~\cite{Briggs2003AnalysisToFSIMS, Alexander2001InteractionPseudoboehmite} 

The N~1\textit{s} spectra for the polar side chain-containing AAs are illustrated in Figure~\href{achemso-SI.pdf#other_CLs}{S12}(c) in the SI. The \ce{N1} gives rise to the feature at an avg.\ BE of 400.9 $\pm$ 0.4~eV, in line with previous data for solid-state AAs.~\cite{Stevens2013QuantitativeXPS, Clark1976AnPolypeptides, Artemenko2021ReferenceAcids, Zubavichus2004SoftStudyb, Peeling1976EvaluationMeals} The single asymmetric peak denoted to \ce{O1} and \ce{O2} shows an avg.\ BE of 531.0 $\pm$ 0.6~eV in Figure~\href{achemso-SI.pdf#other_CLs}{S12}(d) in the SI. In Asn/Gln, the side-chain carbonyl O from the amide group (\ce{O3}), is observed at a higher BE of 532.2 $\pm $ 0.2~eV. The carbonyl O (\ce{O4}), and hydroxyl O (\ce{O3}) from the side chain --COOH group in Asp and Glu contribute to additional features at 532.4 $\pm$ 0.6~eV, and 533.5 $\pm$ 0.7~eV, respectively.~\cite{Stevens2013QuantitativeXPS}

\subsubsection{S/Se-containing Group} ~\label{S Se CLs}

Lastly, for the S/Se-containing AAs (Ser, Thr, Cys, Met, and SeMet), the C, N, and O~1\textit{s} CL spectra are presented in Figure~\href{achemso-SI.pdf#S_Se_CLs}{S14} in the SI. In C~1\textit{s} (Figure~\href{achemso-SI.pdf#S_Se_CLs}{S14}(a) in the SI), two main features, \ce{C$'$}, and \ce{C_$\alpha$}, are consistently observed, at 288.3~eV, and 286.3 $\pm$ 0.3~eV. The lowest BE peak at 284.9~eV $\pm$ 0.1~eV arises from distal C (\ce{C_$\beta$} in Cys/Met/SeMet, \ce{C_$\gamma$} in Thr/Met/SeMet, and \ce{C_$\delta$} in Met/SeMet) and adventitious C. Though heteroatoms (i.e., S and Se) are present in the side chain in Cys, Met, and SeMet, the electronegativities of S (\ce{\chi_S} $\approx$ 2.58) and Se (\ce{\chi_Se} $\approx$ 2.55) are fairly similar to that of C.~\cite{Allred1961ElectronegativityData} Consequently, the local electron density around C bound to them remains nearly unchanged relative to C in the side chain which is not directly bound to (\ce{C_$\beta$} in Met/SeMet). As such, no large differences in BE are observed for S/Se--C bonding environments in C~1\textit{s}. In Ser and Thr, the electron-withdrawing --OH group, is bound to \ce{C_$\beta$}, increasing electron density resulting in a higher BE value, (286.6 $\pm$ 0.5~eV), as depicted in Figure~\href{achemso-SI.pdf#S_Se_CLs}{S14}(b) in the SI, which has also been noted previously.~\cite{Feyer2008CoreThreonine, Pi2020PredictingTheory} 

The N and O~1\textit{s} CL spectra are shown in Figure~\href{achemso-SI.pdf#S_Se_CLs}{S14}(c) and (d) in the SI, respectively. The feature attributed to \ce{N1} at an avg.\ BE of 401.2 $\pm$ 0.3~eV is in agreement with the literature.~\cite{Pi2020PredictingTheory, Brizzolara1996CysteineSpectroscopy, Artemenko2021ReferenceAcids} Small features on the lower BE side, arising from the radiation damage incurred during the experiment, are also labelled with purple dashed lines. The chemical equivalent O atoms, \ce{O1} and \ce{O2}, contribute a single peak at 531.2 $\pm$ 0.7~eV. In Ser and Thr, the \ce{O3} atom from the --OH group is observed at 532.8 $\pm$ 0.8~eV. Additional S~2\textit{p} and Se~3\textit{d} CL spectra for Cys, Met, and SeMet are illustrated in Figure~\href{achemso-SI.pdf#S_Se_CLs}{S15} in the SI to provide a complete picture of chemical states for the S/Se-containing group of AAs. S~2\textit{p} CL spectra in Figure~\href{achemso-SI.pdf#S_Se_CLs}{S15}(a) in the SI exhibit two peaks observed at avg.\ experimental BEs of 164.5 $\pm$ 0.5~eV and 163.3 $\pm$ 0.4~eV, correspond to S~2\ce{\textit{p}_{1/2}} and 2\ce{\textit{p}_{3/2}}, respectively. The Se~3\textit{d} spectrum in Figure~\href{achemso-SI.pdf#S_Se_CLs}{S15}(b) in the SI shows Se~3\ce{\textit{d}_{3/2}} and 3\ce{\textit{p}_{5/2}} peaks appeared at 56.3 and 55.4~eV.

Overall, the discussion of the C, N and O core states across the different subgroups of AAs reveals the intricacies in the local chemical environments. These are supported by DFT, which is able to provide relative BE values for specific chemical environments that are difficult to deconvolve with XPS alone, given limitations of resolution. Across all groups of AAs, the relative BE positions of two main features: \ce{C$'$}, and \ce{C_$\alpha$} are highly comparable, reflecting the similar chemical environment and providing a comprehensive understanding of chemical states and environment. More distal C atoms appear at lower BEs and provide a rich understanding of the differences in various chemical motifs in the side chain and electron density effects as a result of covalent bonding. N and O, in turn, support these observations, providing a complete picture of the main chemical environments in 20 proteinogenic AAs, as well as SeMet.

\subsection{Semi-Core States}~\label{SCL_discussion}

Following the discussion of the core states, the semi-core states can provide further insights into the detailed nature of the chemical bonding in AAs. Different from the core states, semi-core states exhibit strong hybridisation among different 2\textit{s} states, making them particularly sensitive to the local bonding environment. 

\begin{figure}[htp]
    \centering
    \includegraphics[width=0.9\textwidth]{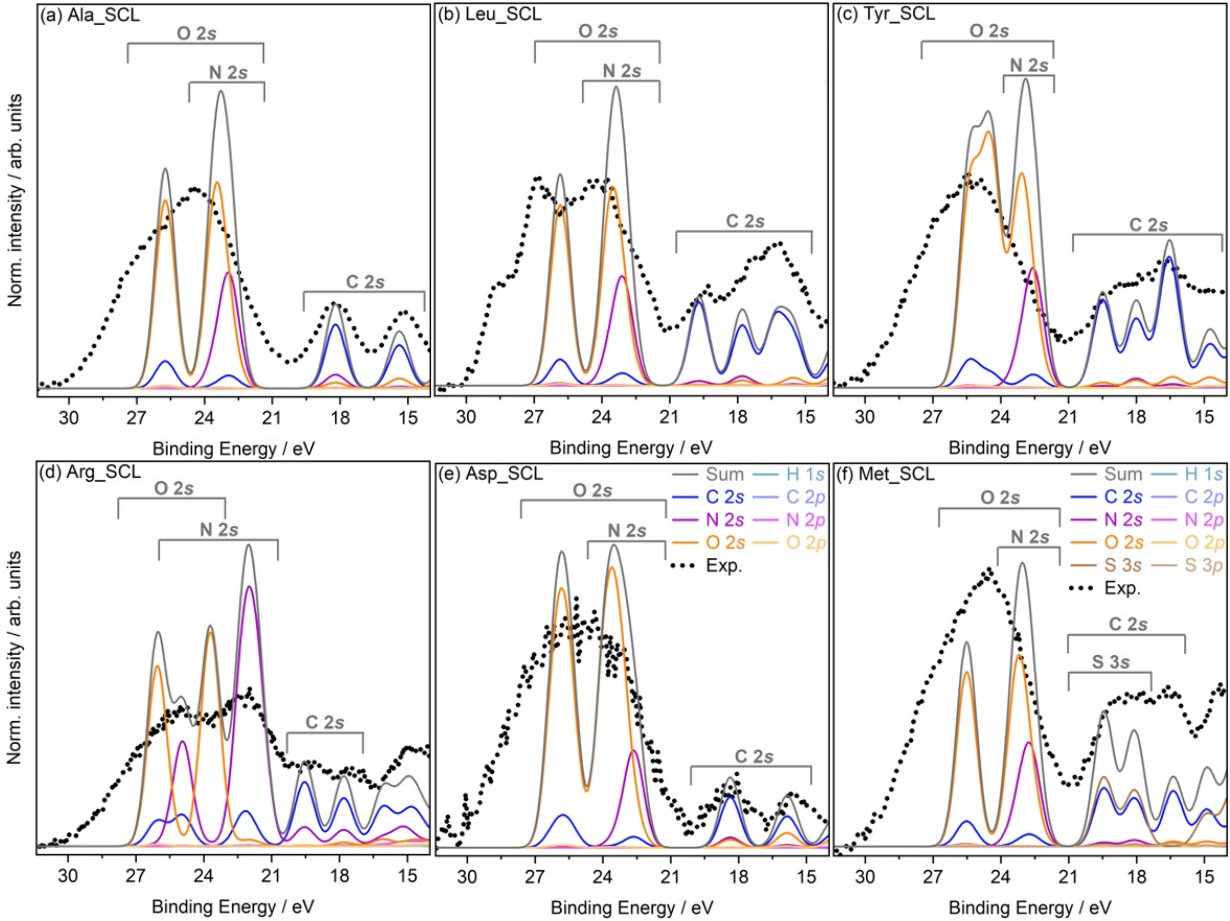}
    \caption{SCL spectra of the representative AAs, including (a) Ala, (b) Leu, (c) Tyr, (d) Arg, (e) Asp, and (f) Met. All plots display the one-electron photoionisation cross-section weighted PDOS, as well as the sum of all PDOS, from PBE-based DFT calculations and the experimental XP spectra. The labels in dark grey indicate the majority orbital contribution to the spectral features determined from DFT. The weighted PDOS have been aligned and normalised to the experimental BE peak of the highest theoretical C~2\textit{s} feature. The legend shown in (e) applies to subfigures (a)-(d), and Met has its own legend shown in (f).}
    \label{SCL_MP}
\end{figure}

\begin{figure}[htp]
    \centering
    \includegraphics[width=0.9\textwidth]{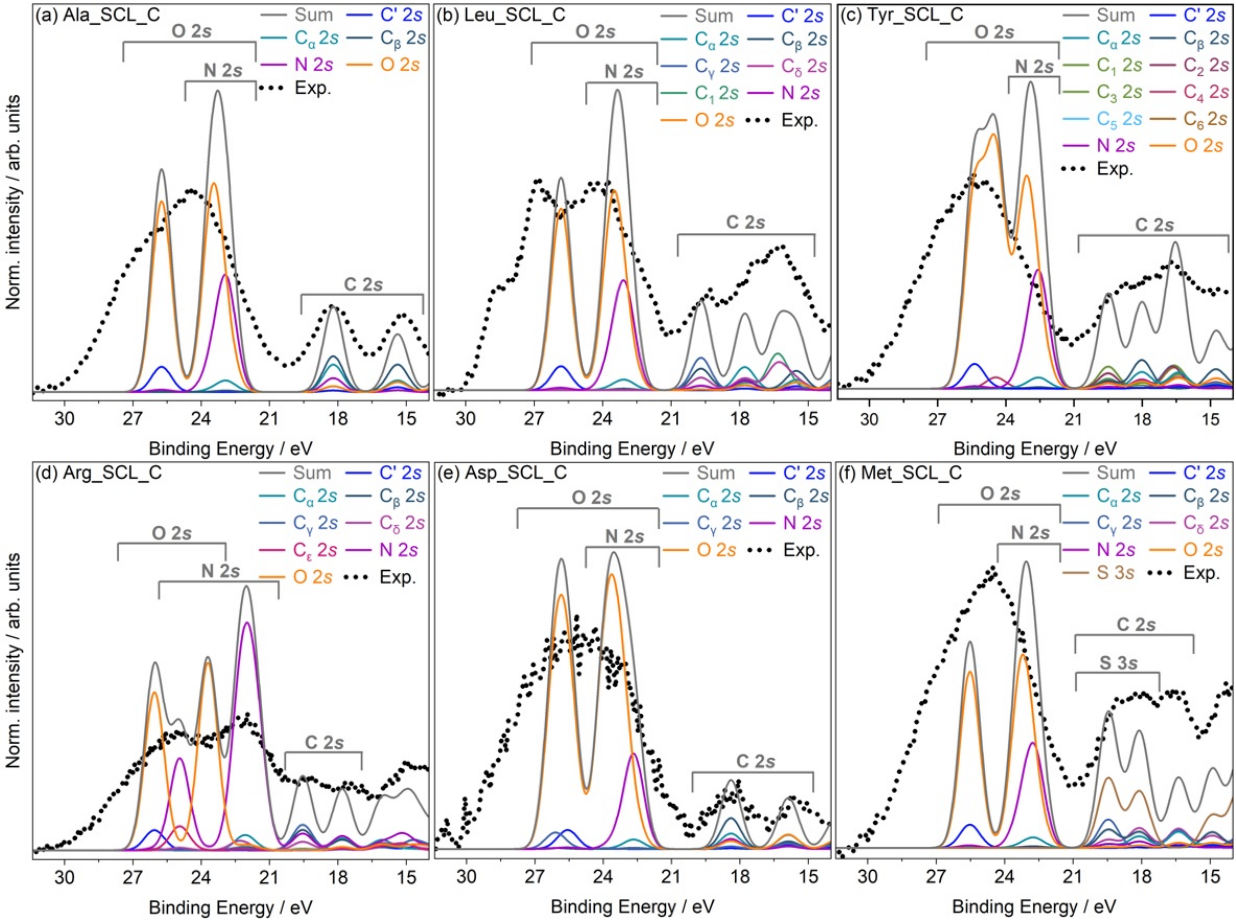}
    \caption{SCL spectra of the representative AAs with contributions projected onto specific C atoms, including (a) Ala, (b) Leu, (c) Tyr, (d) Arg, (e) Asp, and (f) Met. All plots display the one-electron photoionisation cross-section weighted PDOS, as well as the sum of all PDOS, from PBE-based DFT calculations and the experimental XP spectra. To better visualise the contributions from specific C atoms, only 2\textit{s} states are shown as 2\textit{p} states do not present contributions in the SCL region. The labels in dark grey indicate the majority orbital contribution to the spectral features determined from DFT. The weighted PDOS have been aligned and normalised to the experimental BE peak of the highest theoretical C~2\textit{s} feature.}
    \label{SCL_C_MP}
\end{figure}

\begin{figure}[htp]
    \centering
    \includegraphics[width=0.9\textwidth]{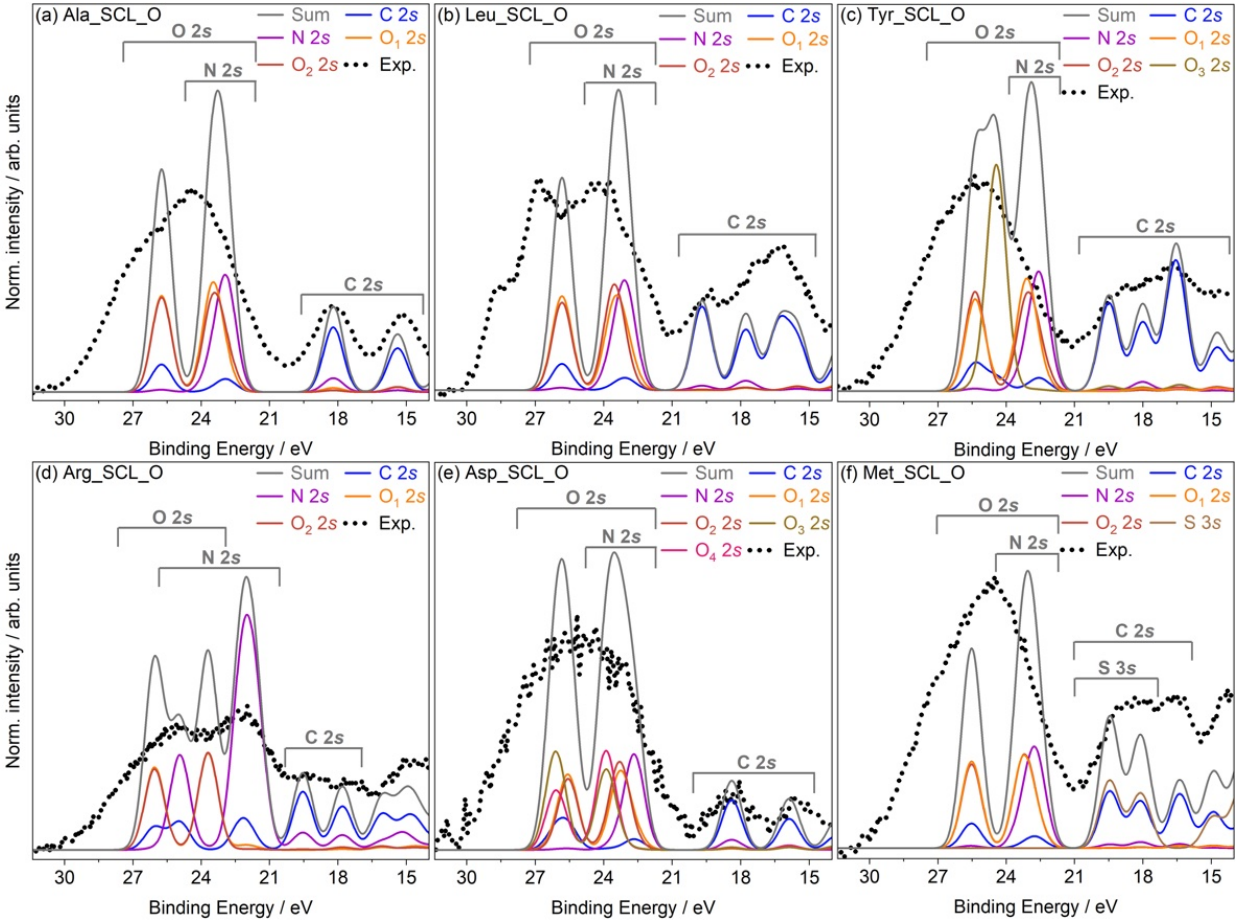}
    \caption{SCL spectra of the representative AAs with contributions projected onto specific O atoms, including (a) Ala, (b) Leu, (c) Tyr, (d) Arg, (e) Asp, and (f) Met. All plots display the one-electron photoionisation cross-section weighted PDOS, as well as the sum of all PDOS, from PBE-based DFT calculations and the experimental XP spectra. To better visualise the contributions from specific C atoms, only 2\textit{s} states are shown as 2\textit{p} states do not present contributions in the SCL region. The labels in dark grey indicate the majority orbital contribution to the spectral features determined from DFT. The weighted PDOS have been aligned and normalised to the experimental BE peak of the highest theoretical C~2\textit{s} feature.}
    \label{SCL_O_MP}
\end{figure}

The SCL spectra of the representative AAs are shown in Figure~\ref{SCL_MP}. Three different 2\textit{s}-dominated features, associated with O, N, and C~2\textit{s}, appear subsequently in the SCL region from higher to lower BE. Due to the distinct photoionisation cross section for each atom, the relative intensities of O, N, and C~2\textit{s} contributions calculated in theory can vary. Across these AAs, two main O~2\textit{s}-dominated features attributed to the backbone \ce{COO^-} group are consistently observed between 21-30~eV in both experiment and theory, except for Leu and Tyr, with the aid of DOS projected onto specific C and O atoms, presented in Figures~\ref{SCL_C_MP} and~\ref{SCL_O_MP}. In Leu (Figure~\ref{SCL_MP}(b)), additional experimental features are from surface contamination. In Tyr (Figure~\ref{SCL_MP}(c)), there are three O~2\textit{s}-dominated features noticeable in experiment, with the higher BE peak splitting into two features due to the additional --OH group in the side chain, assigned based on theory. It can be seen that the two O~2\textit{s} peaks in Asp are broader than the others (Figure~\ref{SCL_MP}). This broadening arises from the side chain --COOH group. Based on the DOS projected onto specific C and O atoms (Figures~\ref{SCL_C_MP}(e) and ~\ref{SCL_O_MP}(e)), each O~2\textit{s} feature can be further resolved into two components, with the higher BE side representing the --COOH group and the lower BE corresponding to the \ce{COO^-} group. As expected from the zwitterionic structure of AAs, one N~2\textit{s} feature from the backbone \ce{NH3^+} group is consistently notable, being mixed with the lower BE O~2\textit{s} feature. Arg is a notable exception. In Arg, the positive charge is delocalised around the guanidino group in the side chain, and it significantly affects the chemical bonding, resulting in a different SCL spectral pattern. Consequently, the N~2\textit{s} states are no longer mixed with O~2\textit{s} states. Instead, there are two distinct N~2\textit{s}-dominated features, where the higher BE peak lies between the two O~2\textit{s} features, and the lower BE feature is located at an even lower BE (Figure~\ref{SCL_MP}(d)). A different number of C~2\textit{s}-dominated features observed in the BE region of 21-15~eV depend highly on the chemical environment and the number of side chain C atoms, being mixed with some contributions from S~3\textit{s} or Se~4\textit{s} states in the case of S/Se-containing AAs, for example, Met (Figure~\ref{SCL_MP}(f)). 

The relative theoretical BE positions and overall bandwidths match well with the experimental features for all AAs. While the discrepancies in BE position and height of the observed features between theory and experiment for all AAs can be attributed to limitations of the projection scheme applied to create the PDOS, which is Mulliken population analysis. In addition, it is well-known that SCL BEs are underestimated using the PBE functional in DFT due to the lack of the intrinsic property of derivative discontinuity and resulting problem of delocalisation error, which can also cause discrepancies.~\cite{Perdew1997CommentEigenvalue, Hait2018DelocalizationNumber} The influence of surface species not accounted for in the DFT calculations can also cause these mismatches. Moreover, the theoretical peak widths are considerably narrower compared to the experimental data due to the absence of robust approaches of applying Lorentzian broadening to SCLs to describe the lifetime broadening coupled with the influence of deviation from a highly-ordered crystalline structure of the sample.~\cite{Muller1998ConnectionsAlloys} The Gaussian smearing applied in the PDOS calculations only reflects the experimental broadening, i.e., the energy resolution of the spectrometer.~\cite{Farahani2014Valence-bandFilms}

\subsubsection{Aliphatic Group}~\label{Ali SCL discussion}

The SCL of the aliphatic AAs (Figure~\href{achemso-SI.pdf#Ali_SCL}{S17} in the SI) show the two aforementioned O~2\textit{s}-dominated features at avg.\ experimental BEs of 23.8 $\pm$0.2~eV and 26.8 $\pm$0.2~eV, supported by theory, with additional features at higher BE from surface contamination. The theoretical BE positions of the two O~2\textit{s} features are comparable with BE differences of 0.1-0.2~eV. From the PDOS it can be seen that the lower BE O~2\textit{s} feature is hybridised with N~2\textit{s} states. For all aliphatic AAs except Pro, the N~2\textit{s} feature appears at a lower BE relative to the O~2\textit{s} peak. This is due to the \ce{NH2} character of N in the pyrrolidine ring in Pro. As can be expected from the difference in molecular structure, the main difference in SCL features of this group is observed in the C~2\textit{s}-dominated features, due to distinct numbers of side chain C atoms and arrangements. To aid a better understanding of the origin of specific spectral features, and therefore, the nature of the chemical bonding, PDOS with contributions projected onto specific C and O atoms were also performed for all AAs, shown in Figures~\href{achemso-SI.pdf#Ali_SCL_C}{S18}-\href{achemso-SI.pdf#Ali_SCL_O}{S19} in the SI.\par

To establish how the local chemical bonding and intermolecular hydrogen-bonding environments can be reflected from the SCL spectra, the orbital contributions and their degree of hybridisation are systematically analysed across all groups of AAs. Based on the DOS projected onto specific C atoms (Figure~\href{achemso-SI.pdf#Ali_SCL_C}{S18} in the SI), the contributions from C atoms to the two O~2\textit{s}-dominated features are highly consistent across the group, with the higher BE O~2\textit{s} peak overlapping with \ce{C$'$}~2\textit{s} and the lower BE O~2\textit{s} feature being mixed with \ce{C_$\alpha$}~2\textit{s} state. The overlapping contributions between O and \ce{C$'$}~2\textit{s} states at the higher BE and between N and \ce{C_$\alpha$}~2\textit{s} states at the lower BE match well with the structure of AAs in the solid state, suggestive of the \ce{COO^-} group and \ce{C_\alpha}--N bond. To the best of our knowledge, this direct atom-resolved assignment of the SCL features has not previously been reported for AAs in the solid-state. To further investigate the level of hybridisation between different states and probe the relative strength of intermolecular hydrogen bonds formed between the \ce{COO^-} and \ce{NH3^+} groups of AAs, four 2\textit{s} peak area ratios based on PDOS calculations projected onto specific C and O atoms were extracted: \ce{C$'$}:\ce{O1}; \ce{C$'$}:\ce{O2}; \ce{N1}:O; and \ce{C_$\alpha$}:\ce{N1}. These ratios were determined by selecting appropriate BE ranges and using the integration function in the Origin software package, and are tabulated in Table~\href{achemso-SI.pdf#area ratio table_Ali}{S8} in the SI. Note that the peak area ratio, \ce{N1}:O, was calculated using the area under the N~2\textit{s} state at the lower BE and the area under the O~2\textit{s} state at the higher BE, as the lower BE O~2\textit{s} state represents the antibonding molecular orbital (MO) based on the visualisation of the electron density distribution calculated from general PDOS calculations. This is also applied to other groups of AAs, which will be discussed in detail in later Sections. 

The \ce{C$'$}:\ce{O1} and \ce{C$'$}:\ce{O2} ratios are high across all the aliphatic AAs , with values between 1:3.48 and 1:3.74 for \ce{C$'$}:\ce{O1} ratios, and 1:3.08 and 1:3.45 for \ce{C$'$}:\ce{O2} ratios. This reflects the fact that the electron density around the \ce{COO^-} group is strongly pulled away from \ce{C$'$} by the highly electronegative O atoms, resulting in covalent bonds with highly polar character, thus making the O atoms strong H-bond acceptors. It is worth mentioning that the \ce{C$'$}:\ce{O1} ratios are slightly higher than \ce{C$'$}:\ce{O2} ratios, reflected by their corresponding bond lengths, where \ce{C$'$}--\ce{O1} (1.26665-1.27951\(\text{\AA}\)) is longer than \ce{C$'$}--\ce{O2} (1.25753-1.26721\(\text{\AA}\)) (listed in Table~\href{achemso-SI.pdf#area ratio table_Ali}{S8} in the SI). This is because of the asymmetric delocalisation of the negative charge around \ce{COO^-}, due to the influence of intermolecular hydrogen bonds and crystal packing effects, consistent with the findings in the literature.~\cite{Dixon1976ElectronicAtoms., Lin2017ShortProtons} This observation contrasts with the primitive understanding, in which the \ce{COO^-} group are found to have identical C--O bond lengths due to resonance.~\cite{Engh1991AccurateRefinement} This asymmetric charge distribution can be further supported by the unequal electron density distribution around \ce{O1} and \ce{O2} in the \ce{COO^-} group (Figure~\href{achemso-SI.pdf#Ali_electron density distribution}{S20} in the SI). Furthermore, different Mulliken bond populations of \ce{C$'$}--\ce{O1}, and \ce{C$'$}--\ce{O2}, alongside the Mulliken charges on each O atom in the \ce{COO^-} group and the corresponding bond lengths ($r_{C'-O_1}$, and $r_{C'-O_2}$), which are tabulated in Table~\href{achemso-SI.pdf#bond population_Ali}{S10} in the SI, support this observation. Bond population, also referred to as the overlap population, serves as a quantitative measure of the covalent or ionic nature of a bond.~\cite{Segall1996PopulationMaterials} A strong inverse correlation between the bond population and bond length is observed, as demonstrated in Figure~\href{achemso-SI.pdf#Ali_population correlation}{S21} in the SI. An increase in the bond length corresponds to a decrease in the bond population, indicating a reduced shared electron density and a greater ionic contribution. Therefore, the longer \ce{C$'$}--\ce{O1} bonds are consistent with the higher \ce{C$'$}:\ce{O1} ratios and suggest that the \ce{C$'$}--\ce{O1} bond is more polarised than the \ce{C$'$}--\ce{O2} bond. In contrast, the Mulliken charges exhibit a much weaker correlation with bond length. This discrepancy suggests that the local charge on each O atom is influenced not only by the \ce{C$'$}--O bond length, but also by the specific geometry of the intermolecular hydrogen-bonding network. This phenomenon is also observed for other groups of AAs, as discussed later.

The \ce{N1}:O ratios are comparable across the aliphatic AAs, except in Pro, which exhibits a higher ratio of 1:1.53. Similarly, the \ce{C_$\alpha$}:\ce{N1} ratios are consistently high across the group, ranging from 1:10.61 for Ala to 1:12.11 for Pro. The higher \ce{C_$\alpha$}:\ce{N1} ratio corresponds to a longer \ce{C_$\alpha$}--\ce{N1} bond length, resulting in a lower level of hybridisation. As Pro has the highest ratio and longest bond length, the level of hybridisation between the \ce{C_$\alpha$} and \ce{N1} is the lowest, making \ce{N1} more polarised and a stronger H-bond donor within the group.

As the intermolecular hydrogen bonds for crystalline AAs are formed between the \ce{COO^-} and \ce{NH3^+} groups, different \ce{C$'$}:O ratios and the corresponding \ce{C$'$}--O bond lengths influence the relative ability of \ce{O1} and \ce{O2} to act as hydrogen bond acceptors, and thus the geometry and relative strength of the intermolecular hydrogen bonds. The relevant bond lengths, including $r_{C'-O_1}$, $r_{C'-O_2}$, and $r_{H-N}$, together with the intermolecular hydrogen-bond distances, $r_{O_1-H}$ and $r_{O_2-H}$, for the aliphatic group are tabulated in Table~\href{achemso-SI.pdf#bond length study_Ali}{S10} in the SI, including different conformations of Val, Leu, and Ile. The presence of different conformations for Val, Leu, and Ile in the crystal structure can be attributed to their bulky, non-polar side-chain R groups, which can form van der Waal's interactions, leading to steric hindrance. Therefore, to minimise the strain, molecules adopt different conformations to form the most stable crystal structure, influencing the packing of molecules in the crystal, and consequently, bond lengths. Such a situation is also found in other AAs with bulk and non-polar side chain groups, such as Phe, Trp, Arg, Lys, Met, and SeMet, addressed in the following Sections. Note that for Trp, Arg, and Lys, although more electronegative N atoms are included in the side chain, this situation persists due to the bulk, aromatic ring in Trp and longer aliphatic side chains in Arg and Lys. The torsion angle, $\psi$, defined along the \ce{O1}--\ce{C^'}--\ce{C_\alpha}--N direction,~\cite{Brunner2018TheCCOcis} is used to distinguish the different conformations of Val, Leu, and Ile. Similar intermolecular hydrogen bond motifs and distances can be seen throughout the group, except for Pro, where the pyrrolidine ring in the side chain strongly influences the hydrogen bonding pattern. In Val, Leu, and Ile, one crystallographically distinct conformation exhibits an opposite intermolecular hydrogen-bonding arrangement, reflecting the influence of crystal-packing effects, consistent with the literature.~\cite{KazuoTorii1971TheL-isoleucine}
 
Turning to C~2\textit{s}-dominated features, an increasing number of features are observed for Ala, Val, Leu, and Ile. This is attributed to the increasing complexity and number of chemically distinct carbon environments in the side chains, with two features in Ala and Val, and four features in Leu and Ile. In Ala, the two observable features are dominated by contributions from \ce{C_$\beta$}~2\textit{s}. The higher BE feature also shows significant contribution from \ce{C_$\alpha$}~2\textit{s}, while the lower BE feature shows diminished contribution from \ce{C_$\alpha$}, along with \ce{C$'$}, and O~2\textit{s} states. For Val, the highest BE C~2\textit{s} feature is also dominated by contributions from \ce{C_$\beta$}~2\textit{s}, the central branching carbon of the isopropyl group, overlapping with contributions from two terminal C atoms (\ce{C_$\gamma$} and \ce{C1}~2\textit{s} states), indicative of their identical chemical environment. Similarly, the contributions from \ce{C_$\alpha$} and N~2\textit{s} states overlap, making them difficult to decipher in the Figure and reflecting their connected chemical environment. Interestingly, the lower BE C~2\textit{s} feature shows a clear distinction between Ala and Val. In Ala, it is a peak primarily from \ce{C_$\beta$}~2\textit{s}. In Val, however, this feature is asymmetric. This asymmetry arises from distinct sub-components: the lower BE side is dominated by the terminal \ce{C1} atom, while the higher BE shoulder within the same peak has significant contributions from \ce{C_$\alpha$}~2\textit{s}, mixed with states from C$'$, N, and O~2\textit{s} states. The broader intensity distribution for \ce{C_$\gamma$}~2\textit{s} is consistent with their environment in the terminal methyl groups. The isomers Leu and Ile display similar components of C~2\textit{s} features. In Leu, a specific feature is dominated by \ce{C_$\gamma$}~2\textit{s}, which acts as the branch point to the two terminal methyl groups (\ce{C_$\delta$}~2\textit{s} and \ce{C1}~2\textit{s}), whose contributions overlap. Conversely, in Ile, it is \ce{C_$\beta$}~2\textit{s} that dominates the analogous feature, consistent with its role as the branch point for the secondary-butyl chain. This assignment illustrates how the PDOS not only fingerprints the side chain but also sensitively distinguishes between isomeric structures.

\subsubsection{Aromatic Group} 

The experimental SCL spectra and the DFT-simulated spectra for the aromatic group are presented in Figure~\href{achemso-SI.pdf#Aro_SCL}{S22} in the SI, and are increasingly complex due to the heteroatoms, i.e., N and O atoms, introduced in the side chain, especially for Tyr, His, and Trp.

Two common O~2\textit{s}-dominated features are consistently observed with avg.\ experimental BEs of 23.9$\pm$1.6 and 27.5$\pm$1.2~eV. From DFT-calculations, the BE positions of these two features are observed with BE differences of 1.9 and 1.8~eV. It is obvious that the BE ranges for the aromatic group are comparably larger than those for the aliphatic group due to the complicated aromatic side chains. An additional O~2\textit{s} feature is observed in Tyr, indicative of the --OH group attached to \ce{C4} in the side chain based on PDOS calculations projected onto specific C and O atoms in Figure~\href{achemso-SI.pdf#Aro_SCL_C}{S23} and~\href{achemso-SI.pdf#Aro_SCL_O}{S25} in the SI, respectively. Multiple N~2\textit{s}-dominated peaks are expected for His and Trp as both include several N atoms, and these N atoms are observed to have different BEs in the N~1\textit{s} CL spectra (Figure~\href{achemso-SI.pdf#Aro_CLs}{S10}(c) in the SI). In His, three N~2\textit{s} features are observed with different N atoms dominating, in agreement with three features identified in its N~1\textit{s} CL spectrum. Based on the PDOS projected onto specific N atoms in Figure~\href{achemso-SI.pdf#Aro_SCL_N}{S24}(a) in the SI, both \ce{N2} and \ce{N3} are attributed to the highest and lowest BE N~2\textit{s} features, with \ce{N2} showing the highest intensity in the highest BE feature and both atoms showing comparable intensity in the lowest BE feature, respectively. The second-highest BE N~2\textit{s} feature is not surprisingly dominated by \ce{N1}, same as the findings in other AAs. Trp only shows a broad N~2\textit{s} predominant feature, overlapping with the lower O~2\textit{s} feature. While in the N~1\textit{s} CL spectrum, two distinct peaks are present. 

Multiple 2\textit{s} peak area ratios for the aromatic AAs are listed in Table~\href{achemso-SI.pdf#area ratio table_Aro}{S11} in the SI. Note that for the $\alpha$-polymorph of Trp which contains bulk non-polar side chain: an indole ring, the bond lengths $r_{C'-O_1}$, $r_{C'-O_2}$, and $r_{C_\alpha-N_1}$ show more than two values, and additional bond lengths $r_{C_2-N_2}$ and $r_{C_{7a}-N_2}$. Therefore, these values are not tabulated in Table~\href{achemso-SI.pdf#area ratio table_Aro}{S11} in the SI but presented in Table~\href{achemso-SI.pdf#bond length study2_aro}{S13} in the SI for the study of the relative strength of intermolecular hydrogen bonds. Four common 2\textit{s} peak area ratios, including \ce{C$'$}:\ce{O1}; \ce{C$'$}:\ce{O2}; \ce{N1}:O; and \ce{C_$\alpha$}:\ce{N1}, are highly comparable with those for aliphatic AAs. In Tyr, an additional 2\textit{s} peak area ratio, \ce{C4}:\ce{O3}, is determined and displays a significantly higher value than \ce{C$'$}:\ce{O1} and \ce{C$'$}:\ce{O2} ratios. Correspondingly, the length of \ce{C4}--\ce{O3} bond is longer than that of \ce{C$'$}--\ce{O1} and \ce{C$'$}--\ce{O2} bonds. This results in a lower level of hybridisation between the \ce{C4} and \ce{O3}~2\textit{s} states, making \ce{O3} a stronger hydrogen bond acceptor. This is further reinforced by the shorter \ce{O3}$\cdots$H\ce{O2} distance shown in Table~\href{achemso-SI.pdf#bond length study2_aro}{S13} in the SI.

Turning to C~2\textit{s}-dominated features, Phe and Tyr present four C~2\textit{s} features with similar components and patterns, with differences arising from varying contributions from \ce{C4}. For the highest BE feature, \ce{C1} contributes the most for both AAs; while two pairs of C atoms; \ce{C2} and \ce{C6}, and \ce{C3} and \ce{C5} have the same weight of contribution, suggestive of the same chemical environment. This is consistent with the position rule for organic molecules, where \ce{C2} and \ce{C6} are in the ortho position, and \ce{C3} and \ce{C5} are in the meta position. For the other three features, different C atoms contribute distinctly, with different extent of N~2\textit{s} and O~2\textit{s} state contributions.

\subsubsection{Polar Side Chain-Containing Group}~\label{SCL other discussion}

Two O~2\textit{s}-dominated features from the \ce{COO^-} group are consistently observed for the polar side chain-containing group of AAs, with avg.\ experimental BEs of 26.9$\pm$1.3 and 24.2$\pm$1.2~eV. The theory reproduces these two features with slightly smaller deviations (0.9 and 0.6~eV, respectively). The SCL spectra for this group can be found in Figure~\href{achemso-SI.pdf#other_SCL}{S27} in the SI. Based on the PDOS, the lower BE O~2\textit{s} feature is consistently hybridised with N~2\textit{s} states, with the N~2\textit{s} feature appearing at the lower BE, except for Arg. In Arg, the guanidino group in the side chain significantly affects the chemical bonding, resulting in a different SCL spectral pattern. Due to the presence of functionally diverse side chains such as the carbamido group (--\ce{CONH2}) in Asn and Gln, and the carboxylic group (--COOH) in Asp and Glu, additional features are also observed in the SCL spectra, leading to significantly more complexities than those of other groups. Therefore, PDOS with contributions projected onto specific C, N, and O atoms are also performed to help identify each experimental feature and provide a better understanding of the nature of the chemical bonding, presented in Figures~\href{achemso-SI.pdf#other_SCL_C}{S28}-\href{achemso-SI.pdf#other_SCL_O}{S30} in the SI.

Except for Arg, two O~2\textit{s}-dominated features, with avg.\ experimental BEs of 26.9 $\pm$ 1.3 and 24.2 $\pm$ 1.2~eV, respectively, are consistently observed. In Asp and Glu, each of the two O~2\textit{s} features splits into two closely spaced peaks, matching the presence of chemically inequivalent oxygen atoms. According to PDOS projected specific C and O atoms (see Figures~\href{achemso-SI.pdf#other_SCL_C}{S28}(c)-(e) and~\href{achemso-SI.pdf#other_SCL_O}{S30}(c)-(e) in the SI), the higher BE component within each feature originates predominantly from \ce{O3} and \ce{O4} of the neutral --COOH group; while the lower BE component is dominated by \ce{O1} and \ce{O2} from the deprotonated \ce{COO^-} group. The higher BE O~2\textit{s}-dominated features, are hybridised with the \ce{C_\gamma} and \ce{C_\delta} atoms for Asp and Glu, respectively, consistent with the labelling of the molecular structure of the atoms, while the lower BE components are mixed with \ce{C$'$} atoms. Moreover, the BE positions of the \ce{C_\gamma}, \ce{C_\delta} and \ce{C$'$} atoms match well with the corresponding BE positions in the C~1\textit{s} spectra, as illustrated in Figure~\href{achemso-SI.pdf#other_CLs}{S12}(b) in the SI.

Due to the presence of carbamido groups in the side chains in Asn and Gln, the higher BE O~2\textit{s}-dominated features also split into two closely packed peaks, similar to those for Asp and Glu. For this feature, the higher BE component represents the \ce{COO^-} group, supported by contributions from the \ce{O1}, \ce{O2}, and \ce{C$'$}~2\textit{s} states. The lower BE component is attributed to the \ce{O3}~2\textit{s} states, overlapping with contributions from the \ce{N2}, \ce{C_\gamma} and \ce{C_\delta}~2\textit{s} states for Asn and Gln, respectively, indicative of the carbamido groups, as shown in Figures~\href{achemso-SI.pdf#other_SCL_C}{S28}(a)/(b) and~\href{achemso-SI.pdf#other_SCL_N}{S29}(a)/(b) in the SI. The BE positions of the \ce{C_\gamma}, \ce{C_\delta}, and \ce{C$'$}~2\textit{s} states are consistent with those in the C~1\textit{s} spectra (see Figure~\href{achemso-SI.pdf#other_CLs}{S12}(b) in the SI). An additional N~2\textit{s}-dominated feature appears at the lower BE of 23.1 $\pm$ 0.2~eV. This feature includes minor O~2\textit{s} contributions and arises from \ce{N2} and \ce{O3} within the carbamido functional group, as indicated in Figures~\href{achemso-SI.pdf#other_SCL_N}{S29}(a)/(b) and~\href{achemso-SI.pdf#other_SCL_O}{S30}(a)/(b) in the SI, representing the antibonding molecular orbitals. 

In Lys, the hybridisation between the lower BE O~2\textit{s} feature and the N~2\textit{s} state is significantly stronger than in other AAs of the group, due to the overlap contributions from both \ce{N1} and \ce{N2} atoms (Figure~\href{achemso-SI.pdf#other_SCL_N}{S29}(d)). A less intense, N~2\textit{s}-dominated peak is observed at a BE of 21.7~eV, attributed primarily to \ce{N2}. The spectral characteristics of Arg differ substantially from other AAs. The lower BE O~2\textit{s} feature no longer overlaps with the N~2\textit{s} state; instead, the O~2\textit{s} and N~2\textit{s} components appear clearly separated, giving rise to two distinct N~2\textit{s}-dominated peaks at avg.\ experimental BEs of 25.8 and 21.9~eV, respectively. The higher BE N~2\textit{s} peak arises mainly from \ce{N2} to \ce{N4} atoms within the guanidino group, where the positive charge is delocalised on the guanidino group. The lower BE N~2\textit{s} peak exhibits split contributions from all four N atoms, with \ce{N1} providing the dominant component near the peak centre. \ce{N2} contributes to the higher BE side of this feature, while \ce{N3} and \ce{N4}~2\textit{s} states overlap on the lower BE side, consistent with a similarity in local chemical environments. 

The hybridisation between different states is notably complex, and the intermolecular hydrogen bonds are not only formed between the \ce{COO^-} and \ce{NH3^+} groups. Hence, six AAs in this group are further classified into three subgroups based on the properties of functional groups in the side chains: AAs including the --\ce{CONH2} group (Asn and Gln); --COOH group (Asp and Glu); and exceptions (Arg and Lys). This classification allows a clearer comparison of the degree of hybridisation and the relative strength of intermolecular hydrogen bonds in systems with different side-chain hydrogen-bonding capabilities. In Asn and Gln, eight 2\textit{s} peak area ratios, including \ce{C$'$}:\ce{O1}; \ce{C$'$}:\ce{O2}; \ce{N1}:\ce{O}; \ce{C_\alpha}:\ce{N1}; \ce{C_x}:\ce{O3}; \ce{N2}:\ce{O2}; \ce{N2}:\ce{O3}; \ce{C_x}:\ce{N2}, where \ce{C_x} represents the C atom attached to the --\ce{CONH2} group, are presented in Table~\href{achemso-SI.pdf#area ratio table1_other}{S15} in the SI. The high \ce{C$'$}:\ce{O1} and \ce{C$'$}:\ce{O2} ratios observed for Asn and Gln, and also for other AAs in the group as discussed later, are consistent with the nature of the \ce{COO^-} group, as described in the ``Aliphatic Group'' Section. The slightly higher \ce{C$'$}:\ce{O1} ratios reinforce the presence of an asymmetric electron density distribution within the \ce{COO^-} group, which is further supported by the Mulliken bond populations, Mulliken charges, and corresponding bond lengths (Table~\href{achemso-SI.pdf#bond populaion_other}{S20} in the SI). The correlation between Mulliken bond population and bond length follows the same trend as the aliphatic group, and a strong correlation is expected and illustrated in Figure~\href{achemso-SI.pdf#other_population correlation}{S31} in the SI. Given that intermolecular hydrogen bonding is significantly more complex in this section, the Mulliken charges are inherently more difficult to interpret. 

Unlike the aliphatic and aromatic groups, where multiple values of bond lengths exist for $r_{C'-O_1}$ and $r_{C'-O_2}$, single values of bond lengths are observed for Asn and Gln. This is because the side chain --\ce{CONH2} group participates in the formation of additional, directional hydrogen bonds that constrain the orientation of molecules in the unit cell, which is visualised in Figures~\href{achemso-SI.pdf#other_crystal structure}{S4}(a) and (b) in the SI. This is also the case for Asp and Glu, as discussed later. In contrast, the \ce{O3} atoms display higher \ce{C_x}:\ce{O3} ratios than \ce{C$'$}:\ce{O1} and \ce{C$'$}:\ce{O2} ratios and longer bond lengths than $r_{C'-O_2}$, together with the longer intermolecular hydrogen bond distances and the corresponding shorter bond length of H--N, indicated in Table~\href{achemso-SI.pdf#bond length study1_other}{S16} in the SI. This suggests a lower level of hybridisation at \ce{O3} and a weaker intermolecular hydrogen bond accepting ability. The \ce{C_\alpha}:\ce{N1} and \ce{N1}:O ratios for both Asn and Gln are comparable with those observed for other groups of AAs, which are also the case for the corresponding bond lengths of \ce{C_\alpha}--N and H--N tabulated in Tables~\href{achemso-SI.pdf#area ratio table1_other}{S15} and~\href{achemso-SI.pdf#bond length study1_other}{S16} in the SI, respectively. The \ce{N2} atom in the side chain also participates in the hydrogen bond framework. The \ce{N2}:\ce{O3} ratios presented in Table~\href{achemso-SI.pdf#area ratio table1_other}{S15} in the SI are significantly higher than \ce{N2}:\ce{O2}, indicating a lower hybridisation between \ce{N2} and \ce{O3}~2\textit{s} states. This suggests that the relative hydrogen bond strength formed between \ce{N2} and \ce{O3} is weaker. This is further supported by the shorter bond lengths of \ce{O_2}--H and longer bond lengths of H--\ce{N_2}, listed in Table~\href{achemso-SI.pdf#bond length study1_other}{S16} in the SI. 

Eight 2\textit{s} peak area ratios for Asp and Glu can be found in  Table~\href{achemso-SI.pdf#area ratio table2_other}{S17} in the SI, together with the intermolecular hydrogen bond length study shown in Table~\href{achemso-SI.pdf#bond length study2_other}{S18} in the SI. The high and comparable \ce{C$'$}:\ce{O1} and \ce{C$'$}:\ce{O2} peak area ratios are consistently found for Asp (1:3.64) and Glu (1:3.68). Similar to Asn and Gln, Asp and Glu display single values of bond lengths for $r_{C'-O_1}$ and $r_{C'-O_2}$, presented in Table~\href{achemso-SI.pdf#bond length study2_other}{S18} in the SI, due to the polar side chain --COOH group, as seen from the crystal structure in Figures~\href{achemso-SI.pdf#other_crystal structure}{S4}(e)/(f) in the SI. While the two peak area ratios of \ce{C_x}:\ce{O3} and \ce{C_x}:\ce{O4} exhibit clear discrepancies, with \ce{C_x}:\ce{O3} having higher values. This indicates that the \ce{O3} and \ce{O4} atoms are chemically different, where the \ce{O3} forms a single covalent bond with \ce{C$'$}, while \ce{O4} forms a double bond instead. In addition, the higher \ce{C_x}:\ce{O3} ratios mean a lower level of hybridisation, corresponding to a higher bond length of $r_{C_\delta-O_3}$. This makes the \ce{O3} atom a better hydrogen bond donor instead, forming an intermolecular hydrogen bond with the \ce{O1} atom. The \ce{C_x}:\ce{O3} ratios are comparable to those of the \ce{C$'$}:\ce{O1} and \ce{C$'$}:\ce{O2} peak area ratios. Therefore, the \ce{O4} is consistently a hydrogen bond acceptor. The \ce{C_\alpha}:\ce{N1} peak area ratios, with 1:10.41 for Asp and 1:11.10 for Glu, and their corresponding bond lengths are comparable with those observed for other groups of AAs tabulated in Table~\href{achemso-SI.pdf#area ratio table2_other}{S17} in the SI. The \ce{N1} atom is not only involved in the formation of intermolecular hydrogen bonds with the \ce{O1} and \ce{O2} atoms from the \ce{COO^-} group, but also the \ce{O4} atoms in the --COOH group (see Table~\href{achemso-SI.pdf#bond length study2_other}{S18} in the SI). For Asp, these hydrogen bonds have similar \ce{C^$'$-O2}$\cdots$H distances, leading to comparable \ce{N1}:\ce{O2} and \ce{N1}:\ce{O4} peak area ratios. In contrast, for Glu, the hydrogen bond involving the \ce{N1} and \ce{O4} atoms has a weaker relative strength, reflected by the longer \ce{C_\delta-O4}$\cdots$H distance, than hydrogen bonds formed between the \ce{N1} and \ce{O2} atoms. This is further indicated by the lower \ce{N1}:\ce{O4} ratios and higher \ce{N1}:\ce{O2} ratios. The \ce{C_x}:\ce{O3} ratios are significantly higher than the \ce{C_x}:\ce{O4} ratios. This suggests that the relative hydrogen bond strength formed involving \ce{O3} is stronger than that involving \ce{O4}. This is further supported by the shorter bond lengths of $r_{O_3-H}$, and the corresponding longer bond lengths of $r_{H-N}$ compared to those involving the \ce{O4} atoms, indicating a more preferred acceptor site for the intermolecular hydrogen bonds.

The various 2\textit{s} peak area ratios and the intermolecular hydrogen bond length study for Arg and Lys are presented in Tables~\href{achemso-SI.pdf#area ratio study3_other}{S19} and~\href{achemso-SI.pdf#bond length study3_other}{S20} in the SI, respectively. As observed in other AAs in the group, the \ce{C$'$}:\ce{O1} and \ce{C$'$}:\ce{O2} ratios are consistently high. The bond lengths, $r_{C'-O_1}$ and $r_{C'-O_2}$, however, display multiple values due to the two distinct conformations distinguished by the torsion angle $\psi$. Similar to Val/Leu/Ile from the aliphatic group, Arg and Lys have longer aliphatic chains before the guanidino and neutral amino groups bound to the end, leading to steric hindrance. The molecules, therefore, tend to adopt different conformations to form the most stable crystal structure, resulting in various values of bond lengths. Different from other AAs, the neutral --\ce{NH2} group attached to the \ce{C_\alpha} does not participate in the formation of main intermolecular hydrogen bonds. Instead, the guanidino group with a delocalised positive charge is involved in the intermolecular hydrogen bond scheme, which is depicted in Figure~\href{achemso-SI.pdf#other_crystal structure}{S4}(c) in the SI. The \ce{N4}:O ratio (1:5.89) is higher than the \ce{N3}:O (1:5.46) as tabulated in Table~\href{achemso-SI.pdf#area ratio3_other}{S19} in the SI. This illustrates that the level of hybridisation between \ce{N4} and O is lower, suggesting that the intermolecular hydrogen bond formed between the \ce{N3} and O is relatively weaker than the bond formed between the \ce{N4} and O. This is further supported by the longer \ce{N3}H--\ce{O2} distances shown in Table~\href{achemso-SI.pdf#bond length study3_other}{S20} in the SI. For Lys, the \ce{N1} from the \ce{NH3^+} group is the main hydrogen bond donor, similar to the aliphatic AAs, and the \ce{N2} from the side chain is not involved in the main intermolecular hydrogen bond scheme. The \ce{N1}:O and \ce{C_\alpha}:\ce{N1} ratios are comparable to those for the aliphatic AAs, which is also the case for the corresponding bond length.

Moving to the C~2\textit{s}-dominated region, all AAs in this group, except for Lys, which displays three less-defined peaks, exhibit two well-resolved features in both experiment and theory, with avg.\ experimental BEs of 18.8~eV and 16.5~eV and corresponding BE ranges of 0.8~eV and 1.7~eV, respectively. The higher BE C~2\textit{s} feature is consistently dominated by the \ce{C_$\beta$}~2\textit{s} state, with \ce{C_$\gamma$}~2\textit{s} contributions of comparable intensity becoming more notable in Gln and Glu. C atoms directly bound to strongly electronegative functional groups, such as \ce{C_$\gamma$} and \ce{C_$\delta$} adjacent to carbamido and carboxylic groups, contribute to the O~2\textit{s}-dominated feature with higher BE, showing stronger hybridisation with O~2\textit{s} orbitals. This feature also exhibits pronounced mixing, with the \ce{N1}~2\textit{s} state overlapping the \ce{C_$\alpha$}~2\textit{s} contribution, and minor O~2\textit{s} contributions arising from \ce{O1} and \ce{O3} in Asn and Gln and \ce{O3} and \ce{O4} in Asp and Glu. In Arg, the higher BE feature is instead dominated by the \ce{C_$\gamma$}~2\textit{s} state, with descending intensities of \ce{C_$\beta$} and \ce{C_$\delta$}. The N~2\textit{s} state also contributes to this feature to a lesser degree, mainly from \ce{N1} and \ce{N2}, with less intense contribution from \ce{N3} and \ce{N4}~2\textit{s} at the bottom. The \ce{C_$\epsilon$}~\textit{s} state, directly bonded to the guanidino group, appears at the higher BE N~2\textit{s} feature. The lower BE C~2\textit{s} feature is primarily dominated by \ce{C_$\alpha$}~2\textit{s}, with comparable \ce{C_$\beta$}~2\textit{s} intensity in Asn and Asp, and \ce{C_$\gamma$}~2\textit{s} intensity in Gln and Glus. Minor contributions from \ce{C_\gamma} and \ce{C$'$} in Asn and Asp and from \ce{C_\delta} and \ce{C$'$} in Gln and Glu are observed near the bottom of this feature, accompanied by minor contributions from the N~2\textit{s} and O~2\textit{s} states. For Lys, contributions from \ce{C_\beta}, \ce{C_\gamma}, and \ce{C_\delta} hybridise across three less-defined features, overlapping with some contributions from \ce{C_\alpha} and \ce{C_\epsilon} appearing at the third peak. \ce{N1} and \ce{N2}~2\textit{s} states present comparable but weaker contributions. \ce{C$'$} and O~2\textit{s} states contribute only minimally at the bottom.

\subsubsection{S/Se-Containing Group}

The SCL spectra for the last group, the S/Se-containing AAs, as well as the PDOS projected onto specific C and O atoms, are presented in Figures~\href{achemso-SI.pdf#S_Se_SCL}{S32} to~\href{achemso-SI.pdf#S_Se_SCL_O}{S34} in the SI. Two main O~2\textit{s}-dominated features, which exhibit avg.\ experimental BEs of 26.9 $\pm$ 1.2 and 23.8 $\pm$ 1.2~eV, are observed across all AAs in the group. Theory predicts BE ranges of 0.7~eV for both features. Negative contributions are present for S~3\textit{s} in Cys due to the limitations of the projection scheme applied (see full explanation in ``Aliphatic Group'' Section). For Ser and Thr, an additional O~2\textit{s}-dominated feature is noticeable between these two features in both experiment and theory, with an avg.\ experimental BE of 25.4~eV $\pm$ 0.1~eV. This peak is attributed to the --OH group in the side chain, similar to Tyr. The contributions from C atoms to the two main O~2\textit{s}-dominated features are consistent with the groups of AAs discussed so far. For Ser and Thr, the additional O~2\textit{s} peaks overlap with the contributions from the \ce{C_\beta}~2\textit{s} states, in agreement with the structure, where the --OH group is attached to the \ce{C_\beta} atom. Five 2\textit{s} peak area ratios are given in Table~\href{achemso-SI.pdf#area ratio table_S_Se}{S22} in the SI. 

The high \ce{C$'$}:\ce{O1} and \ce{C$'$}:\ce{O2} ratios are comparable with other groups of AAs. For Ser and Thr, the \ce{C$'$}:\ce{O1} ratios are lower than the \ce{C$'$}:\ce{O2} ratios. This might be attributed to the extra --OH group in the side chain, acting as an additional H-bond acceptor, resulting in the electron density around \ce{O1} and \ce{O2} being more distorted. The \ce{C$'$}:\ce{O1} and \ce{C$'$}:\ce{O2} ratios are comparable to those of the aliphatic group for Cys, Met, and SeMet. The unequal electron density distribution around the \ce{COO^-} group for the S/Se-containing AAs is presented in Figure~\href{achemso-SI.pdf#S_Se_electron density}{S35} in the SI. Consistent with observations in the aliphatic group, the higher \ce{C$'$}:\ce{O1} ratios stem from the asymmetric delocalisation of negative charge across the carboxylate group. This phenomenon is further quantified by the Mulliken bond populations and the corresponding bond lengths (see Table~\href{achemso-SI.pdf#bond populaion_S_Se}{S24} in the SI). As previously noted for the aliphatic group, a robust and inverse correlation exists (Figure~\href{achemso-SI.pdf#S_Se_population correlation}{S36} in the SI): longer bond lengths correspond to lower Mulliken bond populations. Once again, Mulliken charges do not strictly adhere to this trend, likely due to the inherent complexity of the intermolecular hydrogen-bonding networks.

Met and SeMet, again, exhibit two values for $r_{C'-O_1}$, and $r_{C'-O_2}$, respectively. The various bond lengths of $r_{C'-O_1}$, $r_{C'-O_2}$, and $r_{C_\beta-O_3}$ and the corresponding bond distances of $r_{O_1-H}$, $r_{O_2-H}$, and $r_{N-H}$ are presented in Table~\href{achemso-SI.pdf#bond length study_S_Se}{S23} in the SI, enabling better understanding of the patterns of intermolecular interactions.

For the \ce{N1}:O ratios, all AAs exhibit comparable values, with Ser showing a slightly lower ratio (1:1.28), indicative of similar N--H hydrogen bond donation strengths. Though Cys, Met, and SeMet contain S and Se in the side chain, S and Se do not contribute to the formation of intermolecular H-bonds due to the aforementioned similarity in electronegativity values of S, Se, and C. Therefore, the polarisability of N--H bonds is weakly affected by S and Se atoms, leading to similar \ce{N1}:O ratios to \ce{Ser_\gamma}, \ce{Ser_\alpha}, and Thr. The \ce{C_$\alpha$}:\ce{N1} ratios are comparable to those of the aliphatic AAs. Notably, \ce{Ser_\gamma}, \ce{Ser_\alpha}, and Thr display higher ratios than Cys, Met, and SeMet, with \ce{Ser_\alpha} having the highest ratio of 1:13.78. This can be understood as the electron-withdrawing group, --OH group attached in the side chain, distorts the electron density around the \ce{C_$\alpha$}--\ce{N1} bond, weakening the strength of the \ce{C_$\alpha$}--\ce{N1} bond and lengthening the bond. Correspondingly, higher \ce{C_$\alpha$}:\ce{N1} ratios and resulting lower levels of hybridisation are observed. Considering the additional --OH group in Ser and Thr, significantly higher ratios of \ce{C_$\beta$}:\ce{O3} are observed, with values of 1:15.62 for Thr and 1:16.46 for Ser. This reflects \ce{O3} as a stronger H-bond acceptor, forming stronger intermolecular H-bonds.

The C~2\textit{s}-dominated features, appearing in the BE region of around 15–20~eV, exhibit two distinct peaks for Ser, Thr, and Cys, and three features for Met and SeMet, where additional S~3\textit{s} and Se~4\textit{s} contributions emerge in Cys, Met, and SeMet, respectively. The theoretical predictions of the peak positions match the experimental peak positions reasonably well, with subtle discrepancies in intensities. The highest BE C~2\textit{s}-dominated feature shows an avg.\ experimental BE of 18.9~eV with a range of 1.7~eV, spanning from 17.9~eV in Ser (the lowest) to 19.6~eV in Met (the highest). This feature is primarily dominated by the \ce{C_\alpha}~2\textit{s} state in Ser, while in the other AAs, it is dominated by the C directly bonded to the side-chain heteroatoms: \ce{C_\beta} in Thr and Cys, and \ce{C_\gamma} in Met and SeMet. In SeMet, the \ce{C_\beta}~2\textit{s} contribution exhibits comparable intensity to \ce{C_\gamma}~2\textit{s} state and overlaps, making these components difficult to decipher in the Figure. The presence of an additional --OH group in the side chain of Ser and Thr increases the O~2\textit{s} contribution, resulting in a higher intensity peak inside this feature. Noticeable contributions from \ce{C_\alpha} and \ce{C1}~2\textit{s} states are also evident in Thr. For Cys, Met, and SeMet, the \ce{C_\alpha} and N~2\textit{s} states contribute jointly to this feature, with similar, overlapping intensities, consistent with the local bonding environment and molecular structure of AAs. For the lower BE C~2\textit{s} features, they are dominated by the \ce{C_\beta}~2\textit{s} state and more distal C atoms, with the \ce{C_\alpha}~2\textit{s} state dominated in Cys. Ser and Thr display single features with experimental BEs of 14.9 and 16.7~eV, respectively, with minor contributions from \ce{C_\alpha}, \ce{C$'$}, and O~2\textit{s} states, and a subtle contribution from N~2\textit{s} state in Thr. Moving to Met and SeMet, which exhibit two features at the lower BE side, the second highest BE C~2\textit{s} feature is mainly attributed to \ce{C_\alpha}, \ce{C_\beta}, and \ce{C_\delta}~2\textit{s} states, with comparable intensities in Met and \ce{C_\delta} having the largest intensity in SeMet, being mixed with small contributions from N~2\textit{s} and O~2\textit{s} states at the bottom, as well as Se~4\textit{p} state. Components of the third C~2\textit{s} feature are the same as the second-highest BE C~2\textit{s} feature.

Overall, the detailed discussion of the SCL spectra presented provides a systematic and comprehensive understanding of bonding behaviour and establishes direct relationships between spectral features, orbital hybridisation, local bonding, intermolecular hydrogen bonding patterns, and crystal packing effects across the four AA groups. To further understand the electronic structure of AAs, probing the occupied valence electronic structure with both XPS and DFT is required.

\subsection{Valence States}~\label{VB_discussion}

\begin{figure}[htp]
    \centering
    \includegraphics[width=0.9\textwidth]{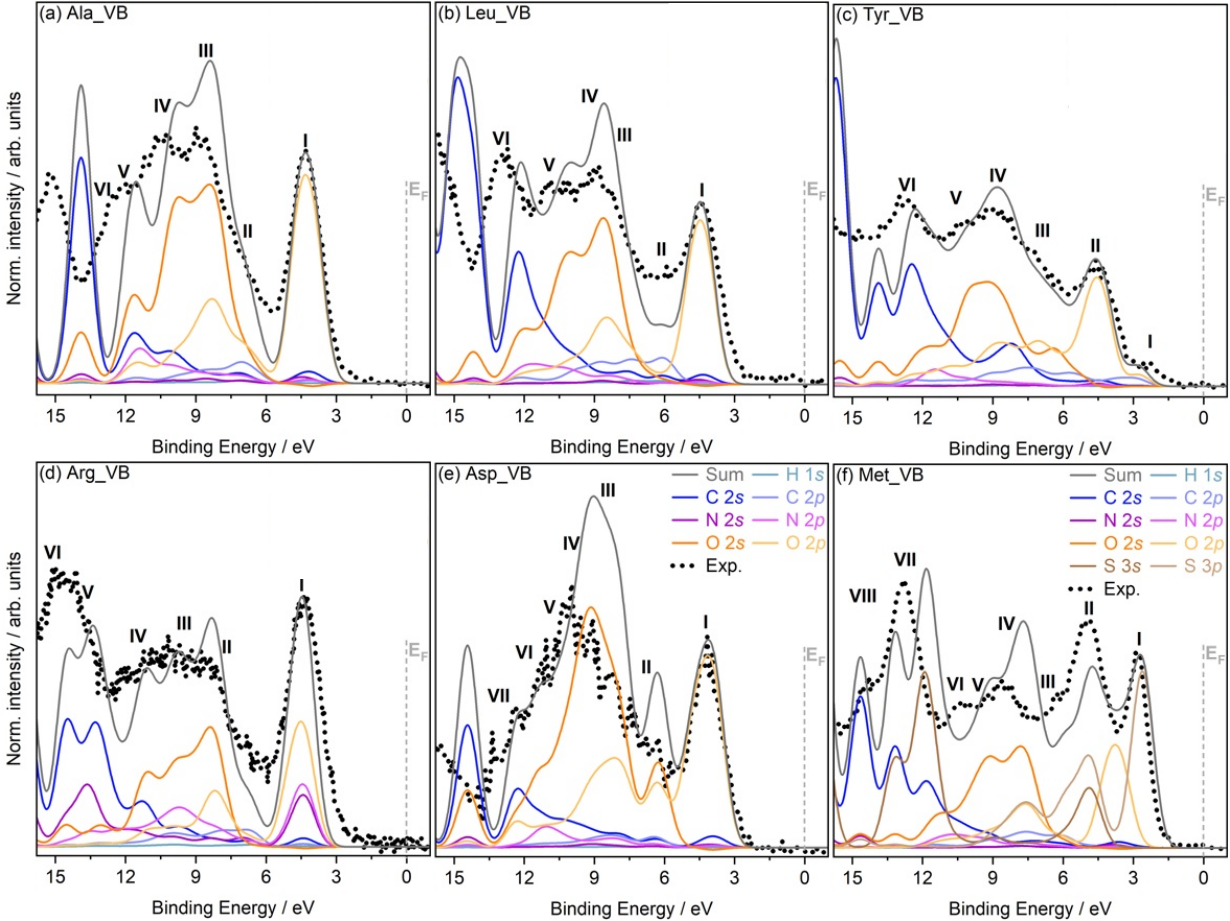}
    \caption{VB spectra of the representative AAs, including (a) Ala, (b) Leu, (c) Tyr, (d) Arg, (e) Asp, and (f) Met. All plots display the one-electron photoionisation cross-section weighted PDOS, as well as the sum of all PDOS, from PBE-based DFT calculations and the experimental XP spectra. The Roman letters in black indicate the main spectral features with the Fermi level labelled as \ce{E_F} and marked by a light grey dashed line. The weighted PDOS have been aligned and normalised to the lowest BE feature in the experiment. The legend shown in (e) applies to subfigures (a)-(d), and Met has its own legend shown in (f).}
    \label{VB_MP}
\end{figure}

The comparison between the calculated PDOS and the experimental VB spectra for the representative AAs is presented in Figure~\ref{VB_MP}. It is noticeable that the complexity of the VB spectra across these representative AAs has increased relative to Ala, both in the component of feature I and the number of features observed in the BE region of 6-15~eV. This increased complexity reflects the chemical diversity of the side chains, as longer carbon chains and additional functional groups introduce distinct valence orbital contributions and more convoluted orbital hybridisation. This, therefore, necessitates the detailed group-by-group electronic structure analysis presented in the following Sections.

\subsubsection{Aliphatic Group} 

The comparison between the calculated PDOS and the experimental VB spectra for the aliphatic AAs is shown in Figure~\href{achemso-SI.pdf#Ali_VB}{S37} in the SI. Feature I, the VB maximum or highest occupied molecular orbital (HOMO), appears at an avg.\ experimental BE of 4.4 $\pm$ 0.1~eV. In the BE region of 6-15~eV, an increased number of features is observed across the aliphatic group, with five distinct features identifiable for each aliphatic AA. This region is not purely dominated by O~2\textit{s}; instead, calculations exhibit a significant increase in the weight of C~2\textit{s} states, particularly in the BE region of 11-13~eV. Features III and IV remain dominated by O~2\textit{s} states, being mixed with C~2\textit{p} and O~2\textit{p}. In Ala, feature VI is still dominated by O~2\textit{s} states but with an increase in C~2\textit{s} intensity compared with Gly. In Val, Leu, and Ile, the less intense feature II is primarily dominated by C~2\textit{p}. Feature VI is dominated by C~2\textit{s} states, mixed with O~2\textit{s} and minor C~2\textit{p}, N~2\textit{p}, and O~2\textit{p} contributions. Pro, owing to its pyrrolidine ring, exhibits distinctive VB behaviour. Rather than being dominated by a particular state, feature II shows the hybridisation of C~2\textit{p} with O~2\textit{s} and 2\textit{p}. The dominant states of features III-V align with the other aliphatic AAs, whereas features VI–VIII are predominantly C~2\textit{s} states with hybridisation from N~2\textit{s}, N~2\textit{p}, and O~2\textit{s} states. In addition, features VII and VIII are observed in the BE region of 13-15~eV, which are mostly contributed to by C~2\textit{s} states, showing some hybridisation with N~2\textit{s} and O~2\textit{s} at the bottom.

PDOS with contributions projected onto specific C atoms are shown in Figure~\href{achemso-SI.pdf#Ali_VB_C}{S38} in the SI. Feature I consistently presents some contributions from \ce{C$'$} and \ce{C_\alpha}~2\textit{s} states, hybridising with N~2\textit{s} and 2\textit{p} states, at the bottom. This indicates the bonding of the protonated amino and carboxylate groups, in line with the structure of crystalline AAs. The lower BE features, II-IV in Ala/Val, and II-V in Leu/Ile/Pro, are also dominated by \ce{C$'$} and \ce{C_\alpha}~2\textit{s}. These are likely from more distal C atoms in the side chain. In Ala and Val, \ce{C_\beta}~2\textit{s} present significant contributions to feature V, as well as to feature VI in Val, with contributions from the \ce{C_\gamma} and \ce{C1}~2\textit{s} overlapping in feature VI, having comparable intensities and line shapes due to the same chemical environment. Features VI and VII are dominated by \ce{C_\gamma}~2\textit{s} in Leu and Ile. In particular, in Leu, contributions from the \ce{C_\delta} and \ce{C1}~2\textit{s} states have comparable intensities in feature VII because of the same chemical bonding and environment. Whereas in Ile, \ce{C_\beta}, \ce{C_\delta}, and \ce{C1}~2\textit{s} states show descending intensities. Turning to Pro, the major contribution in feature VI is from \ce{C3}~2\textit{s} states, followed by \ce{C$'$}~2\textit{s}. Features VII and VIII arise primarily from C atoms in the pyrrolidine ring, in which feature VII is dominated by \ce{C5}~2\textit{s} states. Feature VIII is dominated by the \ce{C3} and \ce{C4}~2\textit{s} states with comparable intensities, consistent with the similar environments imposed by the pyrrolidine ring. Note that the \ce{C3} and \ce{C4} atoms exhibit a different chemical environment compared to \ce{C5} as \ce{C5} is directly bound to \ce{N1}, resulting in discrepancies in the contributions to each experimental feature. Additionally, \ce{C$'$} and \ce{C_\alpha}~2\textit{s} show notable contributions to this feature. 

\subsubsection{Aromatic Group}

The calculated PDOS with the experimental spectra for the aromatic AAs is shown in Figure~\href{achemso-SI.pdf#Aro_VB}{S39} in the SI. Features I (the VBM) and II (for Tyr and Trp only) are observed at an avg.\ BE of 4.5~eV with a spread of 0.4~eV. Tyr and Trp exhibit pronounced shoulders near the Fermi edge in both theory and experiment, and have experimental BEs of 2.32 and 1.56~eV, respectively. In Tyr, the lone pair electrons on the O atom of the --OH group are highly delocalised into the aromatic $\pi$ system via conjugation, pushing the electron density into the aromatic $\pi$ system. This raises the energy of the HOMO, which in turn lowers the BE, resulting in the distinct shoulder, having a BE of 2.32~eV, which is dominated by the \ce{O3} and C~2\textit{p} states according to the PDOS projected onto specific O atoms (see Figure~\href{achemso-SI.pdf#Aro_VB_O}{S42}(b) in the SI). In Trp, the HOMO energy is further pushed to a higher value due to the extended $\pi$-conjugation provided by the indole group in the side chain. The indole group is a fused system and composed of a benzene ring and a pyrrole ring, resulting in a total of 10 $\pi$ electrons that are highly delocalised across the indole group. This extensive conjugation significantly raises the energy of the HOMO relative to the simple aromatic systems, for example, Phe. This therefore lowers the BE to the exceptionally low value and exhibits a shoulder at the BE of 1.56~eV, which is dominated by the \ce{N2} and C~2\textit{p} states based on the PDOS projected onto specific C and N atoms presented in Figures~\href{achemso-SI.pdf#Aro_VB_C}{S40}(e)/(f) and~\href{achemso-SI.pdf#Aro_VB_N}{S41}(c)/(d) in the SI, respectively. 

Across AAs in the group, Features I and II are consistently dominated by O~2\textit{p} states. In His, which contains additional N atoms in the side chain, feature I shows strong hybridisation with N~2\textit{s} and 2\textit{p}. Within the BE region of 6-13~eV, five identifiable features are observed in Phe and His, while Tyr and Trp with only four notable features. Relative to Gly, the dominant contributions to feature II in Phe and to feature III in Trp are C~2\textit{p} states rather than O~2\textit{s} states. In His, feature II shows hybridisation from all valence states, excluding H~1\textit{s} state. Features III-V in Phe, IV-V in Tyr and Trp, and III-IV in His are primarily dominated by O~2\textit{s}, with mixing from C~2\textit{s}, C~2\textit{p}, and O~2\textit{p} states; in His and Trp, increased contributions from N~2\textit{p} states emerge at the bottom due to the additional N atoms contained. As in the aliphatic group, the highest BE feature within this region is attributed to C~2\textit{s} states, with hybridisation from O~2\textit{s} in Phe and Tyr, and from N~2\textit{s} in Trp. However, in His, the highest BE feature VI is dominated by N~2\textit{s} from \ce{N2} and \ce{N3} atoms in the imidazole ring based on N-specific PDOS. The second-highest BE feature V is consistently attributed to C~2\textit{s} states. 

PDOS with contributions projected onto specific C atoms for each aromatic AA, and specific N atoms for His and Trp, are presented in Figures~\href{achemso-SI.pdf#Aro_VB_C}{S40}, and~\href{achemso-SI.pdf#Aro_VB_N}{S41} in the SI, respectively. In Phe, all C atoms are attributed to features III-V with \ce{C1}~2\textit{s} state showing a slightly higher intensity. Feature VI, indicating the benzene ring attached in the side chain, is dominated on its higher BE side by the chemically equivalent \ce{C3} and \ce{C5}~2\textit{s} with comparable intensities, followed by the \ce{C2} and \ce{C6}~2\textit{s}. The lower BE side is mainly attributed to the \ce{C1} and \ce{C4}~2\textit{s} states, with the \ce{C1}~2\textit{s} state exhibiting a slightly higher intensity due to its proximity to the \ce{NH3^+} and \ce{COO^-} groups, which leads to a different local chemical environment than \ce{C4}. Similar to Phe, all C atoms contribute to features III-V in Tyr. In feature VI, the higher BE side is dominated by the \ce{C2}~2\textit{s}, followed by the \ce{C3} and \ce{C6}~2\textit{s} states. The lower BE side shows a broad \ce{C5}~2\textit{s} state contribution, followed by the \ce{C1}~2\textit{s} state. Notably, the \ce{O3}~2\textit{p} state displays a comparable line shape to the \ce{C4}~2\textit{s} state at the bottom, which strongly suggests the \ce{C4}--\ce{O3} chemical bond, matching the phenolic structure. Considering His, feature IV is dominated by the \ce{C$'$}~2\textit{s}. Feature V shows major contributions from the \ce{C4} and \ce{C5}~2\textit{s} states of the imidazole ring, with minor contributions from the \ce{C_\beta} and \ce{C$'$}~2\textit{s} at the bottom. Feature VI is dominated by the \ce{C2}~2\textit{s} state, which is mixed with the \ce{N2} and \ce{N3}~2\textit{s} states, consistent with the \ce{N3}=\ce{C2}--\ce{N2} moiety. Due to the larger indole system, features IV-VI are complex, and all C atoms contribute to different extents.

\subsubsection{Polar Side Chain-Containing Group}

The comparison of calculated PDOS with experimental VB spectra for the polar side chain-containing group of AAs is depicted in Figure~\href{achemso-SI.pdf#other_VB}{S43} in the SI. The VBM (feature I) appears at an avg.\ BE of 4.3~eV with a spread of $\approx$ 0.5~eV. As in the aliphatic and aromatic groups, feature I is dominated by O~2\textit{p} states, with notable contributions from N~2\textit{s} and 2\textit{p} states, especially in Asn, Gln, Arg, and Lys, which is consistent with their side-chain chemistry. Furthermore, in Asn, Gln, Asp, and Glu, feature I carries weak C~2\textit{s} contributions at the bottom. Across the BE region of 6–13~eV, multiple features are observed, with three features in Arg, four features in Lys, five features in Asn, Gln, and Glu, and six features in Asp, respectively. Features II–V in Asn, III-V in Gln, II-IV in Arg, III-IV in Lys, II-VI in Asp, and IV-V in Glu are consistently dominated by O~2\textit{s} states and mixed with C~2\textit{s}, C~2\textit{p}, N~2\textit{s}, and O~2\textit{p} states. Relative to the aliphatic and aromatic groups discussed previously, the highest BE features (features VI/VII) in the BE region of 13-15~eV, observed for Arg, Lys, and Glu, are coherently dominated by C~2\textit{s} states due to the side chain C atoms with major hybridisation from O~2\textit{s} states and minor contributions from N~2\textit{p}, and O~2\textit{p} states at the bottom. The weighting from N~2\textit{s} states is notably increased for AAs with extra N atoms introduced, such as Asn, Gln, Arg, and Lys. 

PDOS with contributions projected onto specific C, N, and O atoms are presented in Figures~\href{achemso-SI.pdf#other_VB_C}{S44} to~\href{achemso-SI.pdf#other_VB_O}{S46} in the SI, respectively. O~2\textit{p} states from all O atoms present in all AAs contribute to the VBM, feature I. Features II and III show contributions from C~2\textit{p} states with hybridisation of N~2\textit{p} states in all other groups of AAs, except Arg, where the C~2\textit{p} state is hybridised with other states at the bottom in feature III. For Asn, the \ce{C^$'$}~2\textit{s} state shows the largest contribution in features IV and V compared to other C atoms. Feature VI shows major contributions from \ce{C_\gamma}~2\textit{s}, followed by \ce{C_\beta}, \ce{O3}, and \ce{N2}~2\textit{s}, consistent with the molecular structure. Gln shows similar C, N, and O~2\textit{s} components, with feature VII showing contributions from \ce{C_\beta}~2\textit{s}, followed by \ce{C_\delta}~2\textit{s}, overlapping with \ce{C_\alpha}, \ce{O3}, \ce{N2}, and \ce{C^$'$}~2\textit{s} states. Components of features VI and VII for Asp and features V-VII for Glu are similar to those in Asn and Gln. Features V in Asp and Glu and VI in Asp are mainly attributed to the \ce{C^$'$}~2\textit{s} states. The \ce{C_\beta}~2\textit{s} state contributes the most to feature VII for Asp, followed by the \ce{C_\gamma}, \ce{O3}, and \ce{O4}~2\textit{s} states, matching well with the structure of the side chain --COOH group. The \ce{C_\alpha} and \ce{C^$'$}~2\textit{s} states are also attributed to this feature with little contribution from the \ce{O1} and \ce{O2}~2\textit{s} states at the bottom. In Glu, feature VI is dominated by the \ce{C_\gamma}~2\textit{s}, followed by \ce{C_\beta}, \ce{C^$'$}, and \ce{C_\alpha}~2\textit{s}, as well as 2\textit{s} states from all four atoms. While the \ce{C_\beta}~2\textit{s} state contributes the most to feature VII, with decreasing intensities of \ce{C_\delta}, \ce{C_\alpha}, and \ce{C^$'$}~2\textit{s} states, and similar O components with feature VI. The presence of contributions from different C and O~2\textit{s} states matches the structure of the \ce{COO^-} and --COOH groups. Contributions from specific C atoms for Lys are more complicated than for Asn, Gln, Asp, and Glu due to an increased number of side chain C atoms. According to Figures~\href{achemso-SI.pdf#other_VB_C}{S44}(d) in the SI, feature V is dominated by contributions from the \ce{C_\beta} and \ce{C_\gamma}~2\textit{s} states with comparable intensity but appearing at different BE positions with the \ce{C_\gamma}~2\textit{s} at the lower BE side and the \ce{C_\beta}~2\textit{s} at the higher BE side. The \ce{C_\delta} and \ce{C_\epsilon}~2\textit{s} states are also attributed to features V with noticeable intensity. The \ce{C_\epsilon}~2\textit{s} contributes the most to feature VI, followed by the \ce{C_\alpha} and \ce{C_\gamma}~2\textit{s}, together with the \ce{N1} and \ce{N2}~2\textit{s} states, matching the covalent bonds formed: the \ce{C_\alpha}--\ce{N1} and the \ce{C_\gamma}--\ce{N2}. Considering Arg with a number of N atoms included in the structure, the weights of N atoms to the VB electronic structure are significantly increased. 2\textit{p} states from \ce{N1} to \ce{N4} atoms contribute to features III and IV, respectively, with the \ce{N3} and \ce{N4}~2\textit{p} having similar line shapes and \ce{N1} and \ce{N2}~2\textit{p} having distinct line shapes. This can be attributed to the different chemical bonds in which they are formed, where the \ce{N3} and \ce{N4} atoms participate in the intermolecular hydrogen bonds, while the \ce{N1} and \ce{N2} atoms do not. Regarding features V and VI, the \ce{C_\beta} and \ce{C_\gamma}~2\textit{s} states show the largest contribution to feature V, with a minor contribution attributed to the \ce{C^$'$}~2\textit{s} state at the bottom. Feature VI is dominated by the \ce{C_\alpha}~2\textit{s} state, followed by the \ce{C_\gamma} and \ce{C_\delta}~2\textit{s} states. An obvious contribution from the N~2\textit{s} state is spread between features V and VI (see Figures~\href{achemso-SI.pdf#other_VB}{S43} in the SI), showing some hybridisation with \ce{C_\delta} and \ce{C_\epsilon}~2\textit{s}, suggestive of the bonding between the \ce{C_\delta} and guanidino group.

\subsubsection{S/Se-Containing Group}

The comparison of the calculated PDOS with the experimental spectra for the S/Se-containing group of AAs, as well as PDOS with contributions projected onto specific C and O atoms, are shown in Figures~\href{achemso-SI.pdf#S_Se_VB}{S47} to~\href{achemso-SI.pdf#S_Se_VB_O}{S49}in the SI, respectively.

For Ser and Thr, feature I is observed with a value comparable to other groups of AAs with an avg.\ BE of 4.4~eV and a BE range of 0.3~eV, and is consistently dominated by O~2\textit{p} states, in line with Gly and other groups of AAs. In contrast, in Cys, Met, and SeMet, the VBM appears at a much lower BE, ranging from 3.8~eV for Cys to 2.6~eV for SeMet, and has a large spread $\approx$ 1.2~eV. In addition, feature I is dominated by S~3\textit{p} states in Cys, which are mixed with O~2\textit{p} state and Met, and Se~4\textit{p} state in SeMet, instead, indicative of the lone pairs of electrons localised on the S and Se atoms, respectively. Within the region of 6-15~eV, an ascending number of features is observed, with five distinct features identifiable in Ser and Thr, six features in Cys, and seven features in Met and SeMet. For Ser and Thr, features III-V are dominated by O~2\textit{s} states, showing some hybridisation from the \ce{C^$'$}~2\textit{s}, N~2\textit{p}, and O~2\textit{p} states at the bottom. Feature VI in Ser is found to be dominated by O~2\textit{s} states, which is the same as Ala, probably due to the same number of side chain C atoms. Instead, the highest BE feature VI in Thr is dominated by C~2\textit{s} with notable contributions from O~2\textit{s}, consistent with other AAs. This feature is dominated by the \ce{C_\beta}~2\textit{s}, followed by the \ce{C1}~2\textit{s} (see Figure~\href{achemso-SI.pdf#S_Se_VB_C}{S48}(c) in the SI). The \ce{C^$'$} and \ce{C_\alpha}~2\textit{s} also show little contribution at the bottom, along with contributions from both \ce{O1} and \ce{O2}~2\textit{s} and~2\textit{p}. 

Turning to the S/Se-containing AAs, as SeMet is a selenium analogue of Met, the VB spectra of Met and SeMet show comparable line shapes, in both experiment and theory. Features III-VI in Cys, and IV-VI in Met and SeMet are dominated by O~2\textit{s} states, being mixed with O~2\textit{p}, S~3\textit{p}, Se~4\textit{p} states, consistent with the components for other AAs. Compared with other groups of AAs, the highest BE features VII in Cys, and VII-VII in Met and SeMet are again dominated by \textit{s} states of the S and Se atoms, with notable hybridisation from C~2\textit{s}. The lower BE features II in Cys and II-III in Met show comparable contributions from S~3\textit{s} and 3\textit{p}; features II-III in SeMet are dominated by Se~4\textit{p}, hybridised with the Se~4\textit{s} and O~2\textit{p}. Contributions from specific C atoms for Cys, Met, and SeMet are presented in Figures~\href{achemso-SI.pdf#S_Se_VB_C}{S48}(d)-(f) in the SI. For Cys, the components of contributions from specific C atoms are similar to those in Ser and Ala due to the same number of side chain C atoms. The \ce{C^$'$}~2\textit{s} shows minor contributions to features V and VI, and feature VII is dominated by the \ce{C_\beta}~2\textit{s}, being mixed with the \ce{C^$'$} and \ce{C_\alpha}~2\textit{s} at the bottom. Though Met and SeMet show comparable line shapes of VB spectra, contributions from specific C atoms exhibit some discrepancies. In both Met and SeMet, features V and VI are dominated by the \ce{C^$'$} and \ce{C_\alpha}~2\textit{s} states, and VII by the \ce{C_\alpha}~2\textit{s} states. For feature VIII, the \ce{C_\beta}~2\textit{s} shows the largest contribution, followed by the \ce{C_\delta}~2\textit{s} in Met. While in SeMet, the \ce{C_\delta}~2\textit{s} contributes the most instead, and \ce{C_\beta}~2\textit{s} exhibits a less intense peak. 

\subsubsection{Correlation of Valence Band Maximum and Optical Band Gap}

Having assigned the orbital characters of the VB features and analysed the electronic structure of the occupied states, the electronic and optical properties of AAs, which are commonly described by their electronic and optical band gaps, are further investigated. These properties are of great scientific interest because of their profound importance in the widespread use of AA films in biosensors and optoelectronic devices, as well as in the adhesion of AAs on inorganic surfaces including insulators and semiconductors.~\cite{Flores2008OpticalCrystals, Willett2005DifferentialSurfaces} From a solid-state chemistry perspective, the electronic band gap, \ce{E_g}, is defined as the energy separation between the valence band maximum (VBM) and the conduction band minimum (CBM). The positions of the VBM for each AA group are determined from the as-collected VB spectra as the intersection of linear fits to the final drop in intensity and the background lines. The optical band gap for each AA in the solid state, \ce{E^{solid}_g}, can be experimentally determined using a spectroscopic technique, UV-Vis spectroscopy, by measuring the reflectance data in the 200–1000~nm wavelength range and as the intersection of the linear fits of the Tauc plots made for the reflectance data and the x-axis.~\cite{Tauc1968OpticalSi}

The UV-Vis reflectance data, the Tauc plot, and the correlation graph between the values of VBM and optical \ce{E^{solid}_g} for all groups of AAs are presented in Figures~\href{achemso-SI.pdf#Ali_UV}{S50},~\href{achemso-SI.pdf#other_UV}{S52},~\href{achemso-SI.pdf#other_UV}{S54}, and~\href{achemso-SI.pdf#S_Se_UV}{S56} in the SI, respectively. Values of determined VBM and the optical \ce{E^{solid}_g} for four groups of AAs are tabulated successively in Tables~\href{achemso-SI.pdf#Ali_VBM}{S25} to~\href{achemso-SI.pdf#S_Se_VBM}{S28} in the SI. It should be aware that the PBE functional is well-known to underestimate \ce{E_g}.~\cite{Perdew1983PhysicalDiscontinuities, Sham1983Density-functionalGap} Employing a better functional, such as hybrid functional PBE0, would be substantially more computationally costly, particularly for AAs with larger unit cell sizes, for example Trp. Therefore, only qualitative theoretical results are included here, alongside the discussion of the experimental results.

The nature of the VBM for all aliphatic AAs is consistent, which arises from the negative charge on the \ce{COO^-} group located on an antibonding molecular orbital, as visualised in Figure~\href{achemso-SI.pdf#Ali_HOMOs}{S51} in the SI. This results in comparable VBM values, except for Pro, which has a lower VBM value of 2.36~eV. Its HOMO is attributed to both the negative charge delocalised over the two O atoms and the lone pair of electrons from N, as shown in Figure~\href{achemso-SI.pdf#Ali_HOMOs}{S51}(e) in the SI. This increased delocalisation reduces the localisation of the electron density on the O atoms in the \ce{COO^-} group, and further stabilises the HOMO, lowering the VBM. Compared with the optical \ce{E_g} values, the VBM does not correlate directly with the \ce{E^{solid}_g} values. In particular, \ce{E^{solid}_g} values does not correspond to twice the VBM BEs, which would be expected if the \ce{E_F} perfectly lies in the middle between the VBM and CBM. This indicates that the energy separation between the VBM and \ce{E_F} is not exactly equal to that between \ce{E_F} and the CBM, suggesting some fluctuations in the position of the CBM between different AAs, which can be attributed to complex intermolecular interactions and crystal packing effects.

Compared with the aliphatic group, the aromatic AAs exhibit generally lower VBMs, especially for Tyr and Trp with values of 1.27 and 1.03~eV, respectively. This reflects the influence of the aromatic rings, in which the delocalised $\pi$-system provides additional stabilisation to the HOMO, further lowering the VBM. This observation is consistent with a past study,~\cite{Bachtold2001FullyInfrared} and further supported by the visualisation of HOMOs in Figure~\href{achemso-SI.pdf#Aro_HOMOs}{S53} in the SI. In Tyr and Trp, the presence of heteroatoms within the side chains allows their lone pairs to delocalise into the aromatic $\pi$-system, leading to additional stabilisation of the HOMO and a consequent lowering of the VBM positions. However, His has a value of 2.87~eV comparable to the aliphatic group due to the partial delocalisation of the lone pair from \ce{N2} in the imidazole ring (see Figure~\href{achemso-SI.pdf#Aro_HOMOs}{S53}(c) in the SI). The lone pair from \ce{N1} remains localised in an \ce{$sp$^2} orbital, which reduces the overall delocalisation and therefore destabilises the HOMO, resulting in a higher VBM. Distinct from the aliphatic group, a higher VBM value corresponds to a higher optical band gap, and a strong positive correlation is observed in the aromatic group, which is presented in Figure~\href{achemso-SI.pdf#Aro_UV}{S52}(d) in the SI. 

For the polar side chain-containing group of AAs, the VBM values are comparable to those of the aliphatic group due to the similar nature of their HOMOs, as visualised in Figure~\href{achemso-SI.pdf#other_HOMOs}{S55} in the SI. Here, additional polar side chains cause the HOMO to also localise on the lone pair electrons of the O and N atoms. In particular, Arg and Lys exhibit slightly lower VBM values than the other AAs in the group. In Arg (Figure~\href{achemso-SI.pdf#other_HOMOs}{S55}(c) in the SI), the positive charge within the guanidinium group is delocalised across the C and N atoms, forming a resonance-stabilised structure that lowers the energy of the HOMO, resulting in a correspondingly lower VBM. In Lys (Figure~\href{achemso-SI.pdf#other_HOMOs}{S55}(d) in the SI), the HOMO is instead located on the side chain $\epsilon$-\ce{NH2} group, dictating the stabilisation of the HOMO. The overall correlation between VBM and the optical \ce{E^{solid}_g} is notably poor and negative (see Figure~\href{achemso-SI.pdf#other_UV}{S54} in the SI). This lack of a strong linear relationship may be attributed to surface degradation of the samples during prolonged atmospheric exposure, which can introduce surface states or impurities that disproportionately affect the measured \ce{E^{solid}_g} values. 

Turning to the S/Se-containing AAs, the VBM of Thr is 2.92~eV, similar to those of the aliphatic group, consistent with the origin of HOMOs. Ser shows a slightly higher VBM value of 3.42~eV. Though both the negative charge and lone pairs of electrons on O in the side chain --OH group contribute to the HOMOs for Ser and Thr, Ser has a denser crystal packing, leading to stronger overlap of the MOs and greater band dispersion, ultimately pushing the VBM to a higher energy level. Cys, Met, and SeMet exhibit significantly lower VBM values due to the presence of S and Se atoms in the side chains, particularly SeMet with the lowest value of 1.61~eV. In these systems, the lowest BE feature is dominated by S~3\textit{p} and Se~4\textit{p} states instead of the O~2\textit{p} state. As the 3\textit{p} and 4\textit{p} orbitals are more diffuse than the O~2\textit{p} orbital, their overlap with neighbouring orbitals is reduced, resulting in weaker bonding interactions and more stabilised valence electronic states. Consequently, the VBM values are smaller. Generally, a higher VBM value correlates with a larger optical \ce{E^{solid}_g}. Met, however, serves as an exception: despite having the second-lowest VBM, it shows a relatively high optical \ce{E^{solid}_g} of approximately 4.56~eV. 

The combined interpretation of experimental VB spectra and DFT-calculated PDOS enables the complex spectral features to be assigned to specific atomic and orbital contributions, providing a systematic understanding of how side-chain chemistry, local bonding, and crystal packing govern the occupied electronic structure across the four groups of AAs.

\section{Conclusion} 

This work presents the first comprehensive experimental and theoretical investigation of the core, semi-core, and valence states of the 20 proteinogenic AAs and SeMet in the solid state. The combination of XPS measurements with DFT calculations enables reliable assignment of complex spectral features that cannot be resolved from experiment alone. The calculated relative core BEs reproduce the experimental chemical shifts well and reveal the complexities of local chemical environments.

The SCL spectra provide further insight into chemical bonding through the strong hybridisation of the C, N, and O~2\textit{s} states. Two O~2\textit{s}-dominated features associated with the \ce{COO^-} group are consistently identified, with their orbital mixing reflecting the \ce{C$'$}--\ce{O1}, \ce{C$'$}--\ce{O2} and \ce{C_\alpha}--N bonding environments. Contributions from additional side-chain functional groups are also clearly resolved in the PDOS. Analysis of PDOS with projections onto specific atoms, integrated peak area ratios, electron density distributions, and Mulliken bond populations demonstrates an asymmetric distribution of negative charge within the \ce{COO^-} group. This asymmetry is governed by intermolecular hydrogen bonding and crystal-packing effects and is reflected in the inequivalent C--O bond lengths and bonding characteristics. Furthermore, the C~2\textit{s}-dominated region sensitively distinguishes side-chain bonding motifs, molecular conformations, and closely related structures, including structural isomers.

The VB spectra become increasingly informative with increasing side-chain complexity, with principal features being consistently reproduced and assigned using the calculated PDOS. The VBM is predominantly derived from O~2\textit{p} states for most AAs, whereas aromatic conjugation and the presence of S or Se atoms introduce substantial side-chain contributions, markedly modifying its position. The generally weak correlations between the measured VBM, calculated molecular HOMO energy, and optical band gap further demonstrate that the occupied electronic structure of crystalline AAs is governed not only by the intrinsic molecular structure but also by intermolecular interactions and extended crystal packing. Overall, this study helps address the difficulty of interpreting radiation-sensitive systems with inherently complex characteristics, and establishes an integrated XPS-DFT framework for connecting spectral features with local chemical bonding and extended solid-state structure, providing a reference for interpreting the electronic structure of amino acids and related molecular crystals.

\section*{Supporting information}

The Supporting Information is available free of charge.
\begin{itemize}
  \item Additional computational details, molecular and crystal structures of investigated AAs, additional results, including XPS survey, combined CL spectra, full set of the CL spectra with peak fitting and calculated BEs, PDOS comparisons with XPS semi-core states and valence band for all AAs categorised by groups, visualisations of electron density distribution, correlation of Mulliken bond population with bond length, tables of the extracted peak area ratios based on PDOS calculations projected onto specific C and O atoms, various bond lengths and intermolecular hydrogen bond distances, UV-Vis data, Tauc plots, tables of VBM and optical band gap, and visualisation of HOMOs, and additional references.
\end{itemize}

\section*{Author Information}

\subsection*{\textbf{Corresponding Author}}

\textbf{Anna Regoutz} - Department of Chemistry, University of Oxford, Inorganic Chemistry Laboratory, South Parks Road, Oxford, OX1 3QR, United Kingdom; orcid.org/0000-0002-3747-3763; 
Email: anna.regoutz@chem.ox.ac.uk 

\textbf{Laura E.~Ratcliff} - Centre for Computational Chemistry, School of Chemistry, University of Bristol, Bristol BS8 1TS, United Kingdom; Hylleraas Centre for Quantum Molecular Sciences, Department of Chemistry, UiT The Arctic University of Norway, N-9037 Tromsø, Norway; orcid.org/0000-0002-9760-5465; Email: laura.ratcliff@bristol.ac.uk

\subsection*{\textbf{Authors}}

\textbf{Ann S.~Y.~Lu} - Department of Chemistry, University of Oxford, Inorganic Chemistry Laboratory, South Parks Road, Oxford, OX1 3QR, United Kingdom; orcid.org/0009-0000-5555-6776

\textbf{Prajna Bhatt} - Department of Chemistry, University College London, 20 Gordon Street, London, WC1H~0AJ, United Kingdom; CNR - Istituto Officina dei Materiali (IOM), Unità di Trieste, Strada Statale 14, km 163.5, 34149 Basovizza (TS), Italy; orcid.org/0000-0002-5699-2841

\textbf{Nathalie K.~Fernando} - Department of Chemistry, University College London, 20 Gordon Street, London, WC1H~0AJ, United Kingdom; orcid.org/0000-0001-7814-1741

\subsection*{\textbf{Notes}}

The authors declare no competing financial interest.

\section*{Data Availability}

The data underlying this study are available in the published article and its Supporting Information (SI). Relaxed crystal structures used for DFT calculations, together with the PDOS and XPS data for main figures, including the core, semi-core, and valence states, are available at Zenodo as .zip archive and Origin file, respectively, at https://doi.org/10.5281/zenodo.21360191.

\section*{Acknowledgements}

N.~K.~F.~acknowledges support from the Engineering and Physical Sciences Research Council (EP/L015277/1). Calculations were performed on the University of Oxford Advanced Research Computing (ARC) facility, https://doi.org/10.5281/zenodo.22558. XPS data collection was performed at the EPSRC UK National XPS Service, grant number EP/Y023587/1.

\printbibliography

@article{Boldyreva2006AA,
    title = {{A comparative study of the anisotropy of lattice strain induced in the crystals of DL-serine by cooling down to 100 K, or by increasing pressure up to 8.6 GPa. A}},
    year = {2006},
    journal = {Zeitschrift f{\"{u}}r Kristallographie-Crystalline Materials},
    author = {Boldyreva, Elena V. and Kolesnik, Evgenia N. and Drebushchak, Tatyana N. and Sowa, Heidrun and Ahsbahs, Hans and Seryotkin, Yuri V.},
    pages = {150--161},
    volume = {221},
    url = {https://www.degruyterbrill.com/document/doi/10.1524/zkri.2006.221.2.150/html},
    issn = {00442968}
}

@article{Novelli2021AccurateK,
    title = {{Accurate H-atom parameters for the two polymorphs of l-histidine at 5, 105 and 295 K}},
    year = {2021},
    journal = {Structural Science},
    author = {Novelli, Giulia and McMonagle, Charles J. and Kleemiss, Florian and Probert, Michael and Puschmann, Horst and Grabowsky, Simon and Maynard-Casely, Helen E. and McIntyre, Garry J. and Parsons, Simon},
    month = {10},
    pages = {785--800},
    volume = {77},
    publisher = {International Union of Crystallography},
    url = {https://journals.iucr.org/paper?S205252062100740X},
    issn = {20525206}
}

@article{Binns2016AccurateL-leucine,
    title = {{Accurate hydrogen parameters for the amino acid l-leucine}},
    year = {2016},
    journal = {Structural Science},
    author = {Binns, Jack and Parsons, Simon and McIntyre, Garry J.},
    month = {12},
    pages = {885--892},
    volume = {72},
    publisher = {International Union of Crystallography},
    url = {https://journals.iucr.org/paper?fx5005},
    issn = {20525206},
    pmid = {27910839}
}

@article{Sosa-Rivadeneyra2023CrystallizationPolytypism,
    title = {{Crystallization of the Third Polymorphic Modification of L-Tryptophan: A Case of Non-order-disorder Polytypism}},
    year = {2023},
    journal = {Crystal Growth and Design},
    author = {Sosa-Rivadeneyra, Martha and Zavala, Aranzazu and Rivas-Silva, Juan Francisco and Uriza-Prias, Diana and Bern{\`{e}}s, Sylvain},
    month = {10},
    pages = {7031--7036},
    volume = {23},
    publisher = {American Chemical Society},
    url = {https://pubs.acs.org/doi/full/10.1021/acs.cgd.3c00704},
    issn = {15287505}
}

@article{Ruggiero2016ExaminationSpectroscopy,
    title = {{Examination of l -Glutamic Acid Polymorphs by Solid-State Density Functional Theory and Terahertz Spectroscopy}},
    year = {2016},
    journal = {Journal of Physical Chemistry A},
    author = {Ruggiero, Michael T. and Sibik, Juraj and Zeitler, J. Axel and Korter, Timothy M.},
    month = {9},
    pages = {7490--7495},
    volume = {120},
    publisher = {American Chemical Society},
    url = {https://pubs.acs.org/doi/full/10.1021/acs.jpca.6b05702},
    issn = {15205215},
    pmid = {27588684}
}

@article{Yamada2007L-Asparagine,
    title = {{L-Asparagine}},
    year = {2007},
    journal = {Structure Reports},
    author = {Yamada, Kazuhiko and Hashizume, Daisuke and Shimizu, Tadashi and Yokoyama, Shigeyuki},
    month = {8},
    pages = {o3802-o3803},
    volume = {63},
    publisher = {International Union of Crystallography},
    url = {https://journals.iucr.org/paper?zl2057 https://journals.iucr.org/e/issues/2007/09/00/zl2057/},
    issn = {16005368}
}

@article{Williams2015L-Lysine:Acids,
    title = {{L-Lysine: Exploiting Powder X-ray Diffraction to Complete the Set of Crystal Structures of the 20 Directly Encoded Proteinogenic Amino Acids}},
    year = {2015},
    journal = {Angewandte Chemie International Edition},
    author = {Williams, P. Andrew and Hughes, Colan E. and Harris, Kenneth D.M.},
    month = {3},
    pages = {3973--3977},
    volume = {54},
    publisher = {John Wiley {\&} Sons, Ltd},
    url = {https://onlinelibrary.wiley.com/doi/full/10.1002/anie.201411520 https://onlinelibrary.wiley.com/doi/abs/10.1002/anie.201411520 https://onlinelibrary.wiley.com/doi/10.1002/anie.201411520},
    issn = {1521-3773},
    pmid = {25651303}
}

@article{Jiang2015Non-topotactic-glycine,
    title = {{Non-topotactic phase transformations in single crystals of {$\beta$}-glycine}},
    year = {2015},
    journal = {Crystal Growth and Design},
    author = {Jiang, Qi and Shtukenberg, Alexander G. and Ward, Michael D. and Hu, Chunhua},
    month = {6},
    pages = {2568--2573},
    volume = {15},
    publisher = {American Chemical Society},
    url = {https://pubs.acs.org/doi/full/10.1021/acs.cgd.5b00187},
    issn = {15287505}
}

@article{Tumanova2018OpeningCocrystals,
    title = {{Opening Pandora's Box: Chirality, Polymorphism, and Stoichiometric Diversity in Flurbiprofen/Proline Cocrystals}},
    year = {2018},
    journal = {Crystal Growth and Design},
    author = {Tumanova, Natalia and Tumanov, Nikolay and Robeyns, Koen and Fischer, Franziska and Fusaro, Luca and Morelle, Fabrice and Ban, Voraksmy and Hautier, Geoffroy and Filinchuk, Yaroslav and Wouters, Johan and Leyssens, Tom and Emmerling, Franziska},
    month = {2},
    pages = {954--961},
    volume = {18},
    publisher = {American Chemical Society},
    url = {/doi/pdf/10.1021/acs.cgd.7b01436?ref=article_openPDF},
    issn = {15287505}
}

@article{Chen2018PrevalentAcids,
    title = {{Prevalent intrinsic emission from nonaromatic amino acids and poly (amino acids)}},
    year = {2018},
    journal = {Science China Chemistry},
    author = {Chen, Xiaohong and Luo, Weijian and Ma, Huili and Peng, Qian and Zhang Yuan, Wang and Zhang, Yongming and Chen, X and Luo, W and Yuan, W Z and Zhang, Y and Ma, H and Peng, Q},
    month = {3},
    pages = {351--359},
    volume = {61},
    publisher = {Science in China Press},
    url = {https://link.springer.com/article/10.1007/s11426-017-9114-4},
    issn = {18691870}
}

@article{Waddell2024RODIN:Demonstration,
    title = {{RODIN: Raw Diffraction Data for Teaching, Training, and Demonstration}},
    year = {2024},
    journal = {Journal of Chemical Education},
    author = {Waddell, Paul G. and Probert, Michael R. and Johnson, Natalie T.},
    month = {10},
    pages = {4276--4281},
    volume = {101},
    publisher = {American Chemical Society},
    url = {https://pubs.acs.org/doi/abs/10.1021/acs.jchemed.4c00797},
    issn = {19381328}
}

@article{Gajda2006StructureState,
    title = {{Structure and dynamics of L-selenomethionine in the solid state}},
    year = {2006},
    journal = {Journal of Physical Chemistry B},
    author = {Gajda, Jaroslaw and Pacholczyk, Justyna and Bujacz, Anna and Bartoszak-Adamska, Elzbieta and Bujacz, Grzegorz and Ciesielski, Wlodzimierz and Potrzebowski, Marek J.},
    month = {12},
    pages = {25692--25701},
    volume = {110},
    publisher = {American Chemical Society},
    url = {https://pubs.acs.org/doi/full/10.1021/jp063332k},
    issn = {15206106}
}

@article{Courvoisier2012TheL-arginine,
    title = {{The crystal structure of L-arginine}},
    year = {2012},
    journal = {Chemical Communications},
    author = {Courvoisier, Emilie and Williams, P. Andrew and Lim, Gin Keat and Hughes, Colan E. and Harris, Kenneth D.M.},
    month = {2},
    pages = {2761--2763},
    volume = {48},
    publisher = {The Royal Society of Chemistry},
    url = {https://pubs.rsc.org/en/content/articlehtml/2012/cc/c2cc17203h https://pubs.rsc.org/en/content/articlelanding/2012/cc/c2cc17203h},
    issn = {1364-548X},
    pmid = {22297609}
}

@article{Bendeif2007TheAcid,
    title = {{The experimental library multipolar atom model refinement of L-aspartic acid}},
    year = {2007},
    journal = {Crystal Structure Communications},
    author = {Bendeif, E.E. and Jelsch, Christian},
    pages = {o361-o364},
    volume = {63},
    url = {https://journals.iucr.org/paper?S0108270107021671}
}

@article{Ihlefeldt2014TheL-Phenylalanine,
    title = {{The Polymorphs of L-Phenylalanine}},
    year = {2014},
    journal = {Angewandte Chemie International Edition},
    author = {Ihlefeldt, Franziska Stefanie and Pettersen, Fredrik Bjarte and Von Bonin, Aidan and Zawadzka, Malgorzata and Rbitz, Carl Henrik},
    month = {12},
    pages = {13600--13604},
    volume = {53},
    publisher = {John Wiley {\&} Sons, Ltd},
    url = {https://onlinelibrary.wiley.com/doi/full/10.1002/anie.201406886 https://onlinelibrary.wiley.com/doi/abs/10.1002/anie.201406886 https://onlinelibrary.wiley.com/doi/10.1002/anie.201406886},
    issn = {1521-3773},
    pmid = {25336255}
}

@article{Parsons2013UseRefinement,
    title = {{Use of intensity quotients and differences in absolute structure refinement}},
    year = {2013},
    journal = {Structural Science},
    author = {Parsons, Simon and Flack, Howard D. and Wagner, Trixie},
    month = {6},
    pages = {249--259},
    volume = {69},
    publisher = {International Union of Crystallography},
    url = {//journals.iucr.org/paper?gp5062},
    issn = {20525192},
    pmid = {23719469}
}

@article{Momma2011VESTA3Data,
    title = {{VESTA 3 for three-dimensional visualization of crystal, volumetric and morphology data}},
    year = {2011},
    journal = {Applied Crystallography},
    author = {Momma, Koichi and Izumi, Fujio},
    month = {10},
    pages = {1272--1276},
    volume = {44},
    publisher = {International Union of Crystallography},
    url = {https://journals.iucr.org/paper?db5098 https://journals.iucr.org/j/issues/2011/06/00/db5098/},
    issn = {0021-8898}
}

@article{Regoutz2021AAcids,
    title = {{A combined density functional theory and x-ray photoelectron spectroscopy study of the aromatic amino acids}},
    year = {2021},
    journal = {Electronic Structure},
    author = {Regoutz, Anna and Swolinska, Marta and Fernando, Nathalie K. and Ratcliff, Laura E.},
    number = {},
    month = {1},
    pages = {044005},
    volume = {2},
    publisher = {IOP Publishing},
    url = {https://iopscience.iop.org/article/10.1088/2516-1075/abd63c https://iopscience.iop.org/article/10.1088/2516-1075/abd63c/meta},
    issn = {2516-1075},
    arxivId = {2010.16220}
}

@article{Engh1991AccurateRefinement,
    title = {{Accurate bond and angle parameters for X-ray protein structure refinement}},
    year = {1991},
    journal = {Foundations of Crystallography},
    author = {Engh, Richard A and Huber, Robert},
    pages = {392--400},
    volume = {47},
    url = {https://journals.iucr.org/paper?cnor=li0061&buy=yes}
}

@article{Cavigliasso1999AccurateApproach,
    title = {{Accurate density-functional calculation of core-electron binding energies by a total-energy difference approach}},
    year = {1999},
    journal = {The Journal of Chemical Physics},
    author = {Cavigliasso, Germán and Chong, Delano P.},
    month = {12},
    pages = {9485--9492},
    volume = {111},
    publisher = {AIP Publishing},
    url = {/aip/jcp/article/111/21/9485/530733/Accurate-density-functional-calculation-of-core},
    issn = {0021-9606}
}

@article{Chandel2021AminoMetabolism,
    title = {{Amino acid metabolism}},
    year = {2021},
    journal = {Cold Spring Harbor Perspectives in Biology},
    author = {Chandel, Navdeep S.},
    number = {4},
    pages = {a040584},
    volume = {13},
    publisher = {Cold Spring Harbor Laboratory Press},
    issn = {19430264},
    pmid = {33795250}
}

@incollection{Dietzen2018AminoProteins,
    title = {{Amino acids, peptides, and proteins}},
    year = {2018},
    booktitle = {Principles and Applications of Molecular Diagnostics},
    author = {Dietzen, Dennis J.},
    month = {1},
    pages = {345--380},
    publisher = {Elsevier},
    isbn = {9780128160619}
}

@article{Akram2011AminoArticle,
    title = {{Amino acids: A review article}},
    year = {2011},
    journal = {Journal of Medicinal Plants Research},
    author = {Akram, M and Asif, H M and Uzair, M and Akhtar, Naveed and Madni, Asadullah and Shah, S M Ali and Ul Hasan, Zahoor and Ullah, Asmat},
    pages = {3997--4000},
    volume = {5},
    url = {http://www.academicjournals.org/JMPR},
    issn = {1996-0875}
}

@article{Kumar2000AmphoterizationMolecules,
    title = {{Amphoterization of colloidal gold particles by capping with valine molecules and their phase transfer from water to toluene by electrostatic coordination with fatty amine molecules}},
    year = {2000},
    journal = {The Journal of Physical Chemistry B},
    author = {Kumar, Ashavani and Mukherjee, Priyabrata and Guha, Ayon and Adyantaya, S. D. and Mandale, A. B. and Kumar, Rajiv and Sastry, Murali},
    month = {12},
    pages = {9775--9783},
    volume = {16},
    publisher = {ACS},
    url = {/doi/pdf/10.1021/la000886k?ref=article_openPDF},
    issn = {07437463}
}

@article{Clark1976AnPolypeptides,
    title = {{An experimental and theoretical investigation of the core level spectra of a series of amino acids, dipeptides and polypeptides}},
    year = {1976},
    journal = {Biochimica et Biophysica Acta (BBA)-Protein Structure},
    author = {Clark, D. T. and Peeling, J. and Colling, L.},
    month = {12},
    pages = {533--545},
    volume = {453},
    publisher = {Elsevier},
    issn = {0005-2795},
    pmid = {999903}
}

@article{Briggs2003AnalysisToFSIMS,
    title = {{Analysis of the surface chemistry of oxidized polyethylene: comparison of XPS and ToF‐SIMS}},
    year = {2003},
    journal = {Surface and Interface Analysis},
    author = {Briggs, D. and Brewis, D. M. and Dahm, R. H. and Fletcher, I. W.},
    month = {2},
    pages = {156--167},
    volume = {35},
    url = {https://analyticalsciencejournals.onlinelibrary.wiley.com/doi/abs/10.1002/sia.1515},
    issn = {01422421}
}

@article{Tolbatov2017BenchmarkingAcids,
    title = {{Benchmarking density functionals and Gaussian basis sets for calculation of core-electron binding energies in amino acids}},
    year = {2017},
    journal = {Theoretical Chemistry Accounts},
    author = {Tolbatov, Iogann and Chipman, Daniel M},
    month = {7},
    pages = {82},
    volume = {136},
    publisher = {Springer New York LLC},
    url = {https://link.springer.com/article/10.1007/s00214-017-2115-x}
}

@article{Feng2007ChemistryStudy,
    title = {{Chemistry of glycine on Pd(111): Temperature-programmed desorption and X-ray photoelectron spectroscopic study}},
    year = {2007},
    journal = {Journal of Physical Chemistry C},
    author = {Feng, Gao and Zhenjun, Li and Yilin, Wang and Burkholder, Luke and Tysoe, W. T.},
    month = {7},
    pages = {9981--9991},
    volume = {111},
    publisher = { American Chemical Society },
    url = {/doi/pdf/10.1021/jp071943m},
    issn = {19327447}
}

@article{Perdew1997CommentEigenvalue,
    title = {{Comment on “Significance of the highest occupied Kohn-Sham eigenvalue”}},
    year = {1997},
    journal = {Physical Review B},
    author = {Perdew, John P. and Levy, Mel},
    pages = {16021--16028},
    volume = {56},
    url = {https://journals.aps.org/prb/abstract/10.1103/PhysRevB.56.16021},
    issn = {1550235X}
}

@article{Tolbatov2014ComparativeGlycine,
    title = {{Comparative study of Gaussian basis sets for calculation of core electron binding energies in first-row hydrides and glycine}},
    year = {2014},
    journal = {Theoretical Chemistry Accounts},
    author = {Tolbatov, Iogann and Chipman, Daniel M.},
    month = {9},
    pages = {1560-},
    volume = {133},
    publisher = {Springer Science and Business Media Deutschland GmbH},
    url = {https://link.springer.com/article/10.1007/s00214-014-1560-z},
    issn = {14322234}
}

@article{Muller1998ConnectionsAlloys,
    title = {{Connections between the electron-energy-loss spectra, the local electronic structure, and the physical properties of a material: A study of nickel aluminum alloys}},
    year = {1998},
    journal = {Physical Review B},
    author = {Muller, David A. and Singh, David J.},
    number = {14},
    month = {4},
    pages = {8181},
    volume = {57},
    publisher = {American Physical Society},
    url = {https://journals.aps.org/prb/abstract/10.1103/PhysRevB.57.8181},
    issn = {1550235X}
}

@article{Feyer2008CoreThreonine,
    title = {{Core level study of alanine and threonine}},
    year = {2008},
    journal = {The Journal of Physical Chemistry A},
    author = {Feyer, Vitaliy and Plekan, Oksana and Richter, Robert and Coreno, Marcello and Prince, Kevin C and Carravetta, Vincenzo},
    month = {8},
    pages = {7806--7815},
    volume = {112},
    url = {https://pubs.acs.org/doi/abs/10.1021/jp803017y}
}

@article{Chatterjee2008Core-levelHartreeFock,
    title = {{Core-level electronic structure of solid-phase glycine, glycyl-glycine, diglycyl-glycine, and polyglycine: X-ray photoemission analysis and Hartree–Fock}},
    year = {2008},
    journal = {The Journal of chemical physics},
    author = {Chatterjee, Avisek and Zhao, Liyan and Zhang, Lei and Pradhan, Debabrata and Zhou, Xiaojing and Leung, KT},
    pages = {105104},
    volume = {129},
    url = {https://pubs.aip.org/aip/jcp/article/129/10/105104/919165}
}

@article{Slaughter1988Core-photoelectronAcidity,
    title = {{Core-photoelectron binding energies of gaseous glycine: correlation with its proton affinity and gas-phase acidity}},
    year = {1988},
    journal = {The Journal of Physical Chemistry},
    author = {Slaughter, AR and Banna, MS},
    pages = {2165--2167},
    volume = {92},
    url = {https://pubs.acs.org/doi/pdf/10.1021/j100319a017},
    issn = {00223654}
}

@article{Brizzolara1996CysteineSpectroscopy,
    title = {{Cysteine by X-Ray Photoelectron Spectroscopy}},
    year = {1996},
    journal = {Surface Science Spectra},
    author = {Brizzolara, Robert A.},
    month = {1},
    pages = {102--107},
    volume = {4},
    publisher = {AIP Publishing},
    url = {/avs/sss/article/4/1/102/367060/Cysteine-by-X-Ray-Photoelectron-Spectroscopy},
    isbn = {202512:48:48},
    issn = {1055-5269}
}

@article{Hait2018DelocalizationNumber,
    title = {{Delocalization Errors in Density Functional Theory Are Essentially Quadratic in Fractional Occupation Number}},
    year = {2018},
    journal = {The Journal of Physical Chemistry Letters},
    author = {Hait, Diptarka and Head-Gordon, Martin},
    month = {11},
    pages = {6280--6288},
    volume = {9},
    publisher = {American Chemical Society},
    url = {/doi/pdf/10.1021/acs.jpclett.8b02417?ref=article_openPDF},
    issn = {19487185}
}

@article{Sham1983Density-functionalGap,
    title = {{Density-functional theory of the energy gap}},
    year = {1983},
    journal = {Physical review letters},
    author = {Sham, LJ and letters, M Schlüter},
    number = {20},
    pages = {1888--1891},
    volume = {51},
    url = {https://journals.aps.org/prl/abstract/10.1103/PhysRevLett.51.1888},
    issn = {00319007}
}

@article{Willett2005DifferentialSurfaces,
    title = {{Differential adhesion of amino acids to inorganic surfaces}},
    year = {2005},
    journal = {Proceedings of the National Academy of Sciences},
    author = {Willett, R. L. and Baldwin, K. W. and West, K. W. and Pfeiffer, L. N.},
    month = {5},
    pages = {7817--7822},
    volume = {102},
    url = {https://www.pnas.org/doi/abs/10.1073/pnas.0408565102},
    issn = {00278424},
    pmid = {15901900}
}

@article{Allred1961ElectronegativityData,
    title = {{Electronegativity values from thermochemical data}},
    year = {1961},
    journal = {Journal of Inorganic and Nuclear Chemistry},
    author = {Allred, A. L.},
    month = {6},
    pages = {215--221},
    volume = {17},
    publisher = {Pergamon},
    url = {https://www.sciencedirect.com/science/article/pii/0022190261801425},
    issn = {0022-1902}
}

@article{Dixon1976ElectronicAtoms.,
    title = {{Electronic structure and bonding of the amino acids containing first row atoms.}},
    year = {1976},
    journal = {Journal of Biological Chemistry},
    author = {Dixon, D. A. and Lipscomb, W. N.},
    month = {10},
    pages = {5992--6000},
    volume = {251},
    publisher = {Elsevier},
    url = {https://www.sciencedirect.com/science/article/pii/S0021925817330491},
    issn = {0021-9258},
    pmid = {972149}
}

@article{Sathisaran2018EngineeringMedium,
    title = {{Engineering cocrystals of poorly water-soluble drugs to enhance dissolution in aqueous medium}},
    year = {2018},
    journal = {Pharmaceutics},
    author = {Sathisaran, Indumathi and Dalvi, Sameer Vishvanath},
    month = {9},
    pages = {108},
    volume = {10},
    publisher = {MDPI AG},
    url = {https://pubmed.ncbi.nlm.nih.gov/30065221/},
    issn = {19994923}
}

@article{Peeling1976EvaluationMeals,
    title = {{Evaluation of the ESCA technique as a screening method for the estimation of protein content and quality in seed meals}},
    year = {1976},
    journal = {Journal of the Science of Food and Agriculture},
    author = {Peeling, James and Clark, David T. and Evans, I. Marta and Boulter, Donald},
    pages = {331--340},
    volume = {27},
    issn = {10970010},
    pmid = {1263459}
}

@article{Dawson2024ExploratoryCalculations,
    title = {{Exploratory data science on supercomputers for quantum mechanical calculations}},
    year = {2024},
    journal = {Electronic Structure},
    author = {Dawson, William and Beal, Louis and Ratcliff, Laura E. and Stella, Martina and Nakajima, Takahito and Genovese, Luigi},
    month = {6},
    pages = {027003},
    volume = {6},
    publisher = {IOP Publishing},
    url = {https://iopscience.iop.org/article/10.1088/2516-1075/ad4b80 https://iopscience.iop.org/article/10.1088/2516-1075/ad4b80/meta},
    issn = {2516-1075}
}

@article{Clark2005FirstCASTEP,
    title = {{First principles methods using CASTEP}},
    year = {2005},
    journal = {Crystalline Materials},
    author = {Clark, Stewart J. and Segall, Matthew D. and Pickard, Chris J. and Hasnip, Phil J. and Probert, Matt I.J. and Refson, Keith and Payne, Mike C.},
    pages = {567--570},
    volume = {220},
    url = {https://www.degruyterbrill.com/document/doi/10.1524/zkri.220.5.567.65075/html}
}

@article{Segall2002First-principlesCode,
    title = {{First-principles simulation: ideas, illustrations and the CASTEP code}},
    year = {2002},
    journal = {Journal of Physics},
    author = {Segall, M D and Lindan, Philip J D and Probert, M J and Pickard, C J and Hasnip, P J and Clark, S J and Payne, M C},
    pages = {2717--2744},
    volume = {14}
}

@article{Li2012First-principlesPhases,
    title = {{First-principles study on core-level spectroscopy of arginine in gas and solid phases}},
    year = {2012},
    journal = {The Journal of Physical Chemistry B},
    author = {Li, Hongbao and Hua, Weijie and Lin, Zijing and Luo, Yi},
    month = {10},
    pages = {12641--12650},
    volume = {116},
    publisher = {American Chemical Society},
    url = {https://pubs.acs.org/doi/abs/10.1021/jp302309u}
}

@article{Bachtold2001FullyInfrared,
    title = {{Fully conjugated porphyrin tapes with electronic absorption bands that reach into infrared}},
    year = {2001},
    journal = {Science},
    author = {Bachtold, ; A and Fisher, M P A and Balents, L and Hadley, P and Thorwart, M and Braggio, A and Nazarov, Yu V and Tsuda, Akihiko and Osuka, Atsuhiro},
    month = {7},
    pages = {79--82},
    volume = {293},
    publisher = {Plenum},
    url = {https://www.science.org/doi/abs/10.1126/science.1059552},
    issn = {00368075},
    pmid = {11441176}
}

@article{Jackson2018Galore:Spectroscopy,
    title = {{Galore: Broadening and weighting for simulation of photoelectron spectroscopy}},
    year = {2018},
    journal = {Journal of Open Source Software},
    author = {Jackson, Adam J and Ganose, Alex M and Regoutz, Anna and Egdell, Russell G and Scanlon, David O},
    pages = {773},
    volume = {3},
    url = {https://discovery.ucl.ac.uk/id/eprint/10083764/}
}

@article{Perdew1996GeneralizedSimple,
    title = {{Generalized Gradient Approximation Made Simple}},
    year = {1996},
    journal = {Physical Review Letters},
    author = {Perdew, John P and Burke, Kieron and Ernzerhof, Matthias},
    pages = {073005},
    volume = {102}
}

@techreport{Ihs1990InfraredFl-Alanine,
    title = {{Infrared and Photoelectron Spectroscopy of Amino Acids on Copper: Glycine, L-Alanine and fl-Alanine}},
    year = {1990},
    author = {Ihs, A and Liedberg, B and Uvdal, K and Tornkvist, C and Bod(j, P and Lundstr, I},
    isbn = {00219797/90}
}

@article{Hohenberg1964InhomogeneousGas,
    title = {{Inhomogeneous electron gas}},
    year = {1964},
    journal = {Physical Review},
    author = {Hohenberg, P. and Kohn, W.},
    pages = {B864-B871},
    volume = {136},
    issn = {0031899X}
}

@article{Alexander2001InteractionPseudoboehmite,
    title = {{Interaction of carboxylic acids with the oxyhydroxide surface of aluminium: poly(acrylic acid), acetic acid and propionic acid on pseudoboehmite}},
    year = {2001},
    journal = {Journal of Electron Spectroscopy and Related Phenomena},
    author = {Alexander, M. R. and Beamson, G. and Blomfield, C. J. and Leggett, G. and Duc, T. M.},
    month = {12},
    pages = {19--32},
    volume = {121},
    publisher = {Elsevier},
    url = {https://www.sciencedirect.com/science/article/pii/S0368204801003243},
    issn = {0368-2048}
}

@article{Plekan2007InvestigationSpectroscopy,
    title = {{Investigation of the amino acids glycine, proline, and methionine by photoemission spectroscopy}},
    year = {2007},
    journal = {The Journal of Physical Chemistry A},
    author = {Plekan, Oksana and Feyer, Vitaliy and Richter, Robert and Coreno, Marcello and De Simone, Monica and Prince, Kevin C and Carravetta, Vincenzo},
    month = {11},
    pages = {10998--11005},
    volume = {111},
    publisher = {American Chemical Society},
    url = {https://pubs.acs.org/doi/abs/10.1021/jp075384v}
}

@article{Powis2003InvestigationSpectroscopy,
    title = {{Investigation of the gas-phase amino acid alanine by synchrotron radiation photoelectron spectroscopy}},
    year = {2003},
    journal = {Journal of Physical Chemistry A},
    author = {Powis, Ivan and Rennie, Emma E. and Hergenhahn, Uwe and Kugeler, Oliver and Bussy-Socrate, Reagan},
    month = {1},
    pages = {25--34},
    volume = {107},
    publisher = { American Chemical Society },
    url = {https://pubs.acs.org/doi/full/10.1021/jp0266345},
    issn = {10895639}
}

@techreport{Uvdal1992L-CysteineStudy,
    title = {{L-Cysteine Adsorbed on Gold and Copper: An X-Ray Photoelectron Spectroscopy Study}},
    year = {1992},
    author = {Uvdal, K and Bodo, ~ P and Liedberg, B}
}

@article{Morris2014OptaDOS:Codes,
    title = {{OptaDOS: A tool for obtaining density of states, core-level and optical spectra from electronic structure codes}},
    year = {2014},
    journal = {Computer Physics Communications},
    author = {Morris, Andrew J. and Nicholls, Rebecca J. and Pickard, Chris J. and Yates, Jonathan R.},
    pages = {1477--1485},
    volume = {185},
    publisher = {North-Holland},
    url = {https://www.sciencedirect.com/science/article/pii/S0010465514000460}
}

@article{Flores2008OpticalCrystals,
    title = {{Optical absorption and electronic band structure first-principles calculations of {$\alpha$}-glycine crystals}},
    year = {2008},
    journal = {Physical Review B},
    author = {Flores, M. Z.S. and Freire, V. N. and Dos Santos, R. P. and Farias, G. A. and Caetano, E. W.S. and De Oliveira, M. C.F. and Fernandez, J. R.L. and Scolfaro, L. M.R. and Bezerra, M. J.B. and Oliveira, T. M. and Bezerra, G. A. and Cavada, B. S. and Leite Alves, H. W.},
    month = {3},
    pages = {115104},
    volume = {77},
    issn = {10980121}
}

@article{Tauc1968OpticalSi,
    title = {{Optical properties and electronic structure of amorphous Ge and Si}},
    year = {1968},
    journal = {Materials Research Bulletin},
    author = {Tauc, J.},
    month = {1},
    pages = {37--46},
    volume = {3},
    publisher = {Pergamon},
    url = {https://www.sciencedirect.com/science/article/pii/0025540868900238},
    issn = {0025-5408}
}

@article{Powis2000PhotoelectronL-Alanine,
    title = {{Photoelectron spectroscopy and circular dichroism in chiral biomolecules: L-Alanine}},
    year = {2000},
    journal = {Journal of Physical Chemistry A},
    author = {Powis, Ivan},
    number = {5},
    month = {2},
    pages = {878--882},
    volume = {104},
    publisher = {American Chemical Society},
    doi = {10.1021/jp9933119},
    issn = {10895639}
}

@article{Winter2006PhotoemissionSolutions,
    title = {{Photoemission from liquid aqueous solutions}},
    year = {2006},
    journal = {Chemical Reviews},
    author = {Winter, Bernd and Faubel, Manfred},
    month = {4},
    pages = {1176--1211},
    volume = {106},
    publisher = { American Chemical Society },
    url = {/doi/pdf/10.1021/cr040381p?ref=article_openPDF},
    issn = {00092665},
    pmid = {16608177}
}

@article{Perdew1983PhysicalDiscontinuities,
    title = {{Physical content of the exact Kohn-Sham orbital energies: band gaps and derivative discontinuities}},
    year = {1983},
    journal = {Physical Review Letters},
    author = {Perdew, John P and Levy, Mel},
    pages = {1884--1887},
    volume = {51},
    url = {https://journals.aps.org/prl/abstract/10.1103/PhysRevLett.51.1884},
    issn = {00319007}
}

@article{Segall1996PopulationMaterials,
    title = {{Population analysis of plane-wave electronic structure calculations of bulk materials}},
    year = {1996},
    journal = {Physical Review B},
    author = {Segall, M. and Shah, R. and Pickard, C. and Payne, M.},
    pages = {16317--16320},
    volume = {54},
    url = {https://journals.aps.org/prb/abstract/10.1103/PhysRevB.54.16317},
    issn = {1550235X},
    pmid = {9985733}
}

@article{Pi2020PredictingTheory,
    title = {{Predicting Core Level Photoelectron Spectra of Amino Acids Using Density Functional Theory}},
    year = {2020},
    journal = {The Journal of Physical Chemistry Letters},
    author = {Pi, Jo M. and Stella, Martina and Fernando, Nathalie K. and Lam, Aaron Y. and Regoutz, Anna and Ratcliff, Laura E.},
    number = {},
    month = {3},
    pages = {2256--2262},
    volume = {11},
    publisher = {American Chemical Society},
    url = {/doi/pdf/10.1021/acs.jpclett.0c00333?ref=article_openPDF},
    issn = {19487185},
    pmid = {32125160}
}

@article{Essien2023PredictionElectra,
    title = {{Prediction of protein ion--ligand binding sites with electra}},
    year = {2023},
    journal = {Molecules},
    author = {Essien, Clement and Jiang, Lei and Wang, Duolin and Xu, Dong},
    month = {10},
    pages = {6793},
    volume = {28},
    publisher = {Multidisciplinary Digital Publishing Institute (MDPI)},
    url = {https://pmc.ncbi.nlm.nih.gov/articles/PMC10574437/}
}

@article{Gao2009Probing0,
    title = {{Probing the interaction of the amino acid alanine with the surface of ZnO(1 0 over(1, ̄) 0)}},
    year = {2009},
    journal = {Journal of Colloid and Interface Science},
    author = {Gao, Y. K. and Traeger, F. and Shekhah, O. and Idriss, H. and W{\"{o}}ll, C.},
    month = {10},
    pages = {16--21},
    volume = {338},
    issn = {00219797},
    pmid = {19596338}
}

@article{Wolstenholme2020ProcedureFrequently,
    title = {{Procedure which allows the performance and calibration of an XPS instrument to be checked rapidly and frequently}},
    year = {2020},
    journal = {Journal of Vacuum Science {\&} Technology A: Vacuum, Surfaces, and Films},
    author = {Wolstenholme, John},
    month = {7},
    pages = {43206},
    volume = {38},
    publisher = {American Vacuum Society},
    url = {/avs/jva/article/38/4/043206/246868/Procedure-which-allows-the-performance-and},
    issn = {0734-2101}
}

@article{Stevens2013QuantitativeXPS,
    title = {{Quantitative analysis of complex amino acids and RGD peptides by X-ray photoelectron spectroscopy (XPS)}},
    year = {2013},
    journal = {Surface and Interface Analysis},
    author = {Stevens, Joanna S. and De Luca, Alba C. and Pelendritis, Michalis and Terenghi, Giorgio and Downes, Sandra and Schroeder, Sven L.M.},
    month = {8},
    pages = {1238--1246},
    volume = {45},
    publisher = {John Wiley {\&} Sons, Ltd},
    url = {https://onlinelibrary.wiley.com/doi/full/10.1002/sia.5261 https://onlinelibrary.wiley.com/doi/abs/10.1002/sia.5261 https://analyticalsciencejournals.onlinelibrary.wiley.com/doi/10.1002/sia.5261},
    issn = {1096-9918}
}

@article{Artemenko2021ReferenceAcids,
    title = {{Reference XPS spectra of amino acids}},
    year = {2021},
    journal = {IOP Conference Series: Materials Science and Engineering},
    author = {Artemenko, A. and Shchukarev, A. and {\v{S}}tenclov{\'{a}}, P. and Wagberg, T. and Segervald, J. and Jia, X. and Kromka, A.},
    month = {1},
    pages = {012001},
    volume = {1050},
    publisher = {IOP Publishing},
    url = {https://iopscience.iop.org/article/10.1088/1757-899X/1050/1/012001 https://iopscience.iop.org/article/10.1088/1757-899X/1050/1/012001/meta},
    issn = {1757-899X}
}

@article{Kohn1965Self-consistentEffects,
    title = {{Self-consistent equations including exchange and correlation effects}},
    year = {1965},
    journal = {Physical Review},
    author = {Kohn, W. and Sham, L. J.},
    pages = {A1133-A1138},
    volume = {140},
    issn = {0031899X}
}

@article{Grimme2006SemiempiricalCorrection,
    title = {{Semiempirical GGA-type density functional constructed with a long-range dispersion correction}},
    year = {2006},
    journal = {Journal of Computational Chemistry},
    author = {Grimme, Stefan},
    number = {15},
    month = {11},
    pages = {1787--1799},
    volume = {27},
    publisher = {John Wiley {\&} Sons, Ltd},
    url = {/doi/pdf/10.1002/jcc.20495 https://onlinelibrary.wiley.com/doi/abs/10.1002/jcc.20495 https://onlinelibrary.wiley.com/doi/10.1002/jcc.20495},
    issn = {01928651}
}

@article{Lin2017ShortProtons,
    title = {{Short carboxylic acid–carboxylate hydrogen bonds can have fully localized protons}},
    year = {2017},
    journal = {Biochemistry},
    author = {Lin, Jiusheng and Pozharski, Edwin and Wilson, Mark A.},
    month = {1},
    pages = {391--402},
    volume = {56},
    publisher = {American Chemical Society},
    url = {https://pubs.acs.org/doi/abs/10.1021/acs.biochem.6b00906?casa_token=fdld8lo8RSMAAAAA:PQJ-74WOVgH9OSBO0RHpO0Uv83J1f10Rpnwx7z1cTc5L-QPMLgo76SReE1JK6oAm4pNL6Y3FlS5xFQ},
    doi = {10.1021/ACS.BIOCHEM.6B00906},
    issn = {15204995},
    pmid = {27989121}
}

@article{Zubavichus2004SoftStudy,
    title = {{Soft X-ray induced decomposition of phenylalanine and tyrosine: a comparative study}},
    year = {2004},
    journal = {The Journal of Physical Chemistry A},
    author = {Zubavichus, Yan and Zharnikov, Michael and Shaporenko, Andrey and Fuchs, Oliver and Weinhardt, Lothar and Heske, Clemens and Umbach, Eberhard and Denlinger, Jonathan D and Grunze, Michael},
    month = {5},
    pages = {4557--4565},
    volume = {108},
    url = {https://pubs.acs.org/doi/abs/10.1021/jp049376f}
}

@article{Zubavichus2004SoftStudyb,
    title = {{Soft X-ray-induced decomposition of amino acids: an XPS, mass spectrometry, and NEXAFS study}},
    year = {2004},
    journal = {Radiation Research},
    author = {Zubavichus, Yan and Fuchs, Oliver and Weinhardt, Lothar and Heske, Clemens and Umbach, Eberhard and Denlinger, Jonathan D and Grunze, Michael},
    pages = {346--358},
    volume = {161},
    url = {https://meridian.allenpress.com/radiation-research/article-abstract/161/3/346/205768}
}

@article{Mocellin2017SurfaceXPS,
    title = {{Surface Propensity of Atmospherically Relevant Amino Acids Studied by XPS}},
    year = {2017},
    journal = {Journal of Physical Chemistry B},
    author = {Mocellin, Alexandra and Gomes, Anderson Herbert De Abreu and Ara{\'{u}}jo, Oscar Cardoso and De Brito, Arnaldo Naves and Bj{\"{o}}rneholm, Olle},
    month = {4},
    pages = {4220--4225},
    volume = {121},
    publisher = {American Chemical Society},
    issn = {15205207},
    pmid = {28358197}
}

@article{Eralp2011TheCu110,
    title = {{The adsorption geometry and chemical state of lysine on Cu{\{}110{\}}}},
    year = {2011},
    journal = {Surface Science},
    author = {Eralp, Tugce and Shavorskiy, Andrey and Held, Georg},
    month = {2},
    pages = {468--472},
    volume = {605},
    issn = {00396028}
}

@article{Brunner2018TheCCOcis,
    title = {{The chirality chain in valine: How the configuration at the C{$\alpha$} position through the OcisC′ C{$\alpha$}N torsional system leads to distortion of the planar group C{$\alpha$}C′(Ocis)}},
    year = {2018},
    journal = {ChemistryOpen},
    author = {Brunner, Henri and Tsuno, Takashi},
    month = {9},
    pages = {696--700},
    volume = {7},
    publisher = {Wiley-VCH Verlag},
    url = {https://chemistry-europe.onlinelibrary.wiley.com/doi/abs/10.1002/open.201800137},
    isbn = {107.69110.13},
    issn = {21911363}
}

@article{KazuoTorii1971TheL-isoleucine,
    title = {{The crystal structure of l-isoleucine}},
    year = {1971},
    journal = {Structural Science},
    author = {Kazuo Torii, BY and Iitaka, Yoichi},
    month = {11},
    pages = {2237--2246},
    volume = {27},
    publisher = {International Union of Crystallography},
    url = {https://journals.iucr.org/paper?a08400 https://journals.iucr.org/b/issues/1971/11/00/a08400/},
    issn = {0567-7408}
}

@article{Cannington1979TheSurvey,
    title = {{The photoelectron spectra of amino-acids : A survey}},
    year = {1979},
    journal = {Journal of Electron Spectroscopy and Related Phenomena},
    author = {Cannington, P. H. and Ham, Norman S.},
    number = {1},
    month = {1},
    pages = {79--82},
    volume = {15},
    publisher = {Elsevier},
    url = {https://www.sciencedirect.com/science/article/pii/0368204879870152},
    issn = {0368-2048}
}

@article{Farahani2014Valence-bandFilms,
    title = {{Valence-band density of states and surface electron accumulation in epitaxial SnO 2 films}},
    year = {2014},
    journal = {Physical Review B},
    author = {Farahani, S K Vasheghani and Veal, T D and Mudd, J J and Scanlon, D O and Watson, G W and Bierwagen, O and White, M E and Speck, J S and Mcconville, C F},
    pages = {155413},
    volume = {90}
}

@article{Tzvetkov2010X-rayStudy,
    title = {{X-ray induced irradiation effects in glycine thin films: A time-dependent XPS and TPD study}},
    year = {2010},
    journal = {Journal of Electron Spectroscopy and Related Phenomena},
    author = {Tzvetkov, George and Netzer, Falko P.},
    month = {11},
    pages = {41--46},
    volume = {182},
    publisher = {Elsevier},
    url = {https://www.sciencedirect.com/science/article/pii/S0368204810001301},
    issn = {0368-2048}
}

@article{Baer2020XPSSamples,
    title = {{XPS guide: Charge neutralization and binding energy referencing for insulating samples}},
    year = {2020},
    journal = {Journal of Vacuum Science {\&} Technology A},
    author = {Baer, Donald R. and Artyushkova, Kateryna and Cohen, Hagai and Easton, Christopher D. and Engelhard, Mark and Gengenbach, Thomas R. and Greczynski, Grzegorz and Mack, Paul and Morgan, David J. and Roberts, Adam},
    pages = {031204},
    volume = {38},
    publisher = {American Vacuum Society},
    url = {/avs/jva/article/38/3/031204/1063946/XPS-guide-Charge-neutralization-and-binding-energy}
}

\section*{TOC Graphic}

\begin{figure}[htp]
    \centering
    \includegraphics[width=0.5\textwidth]{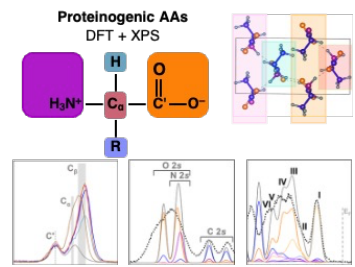}
\end{figure}

\end{document}


\maketitle
\cleardoublepage
\section{Molecular and Crystal Structures}

\begin{figure}[htp]
    \centering
    \includegraphics[width=1.0\textwidth]{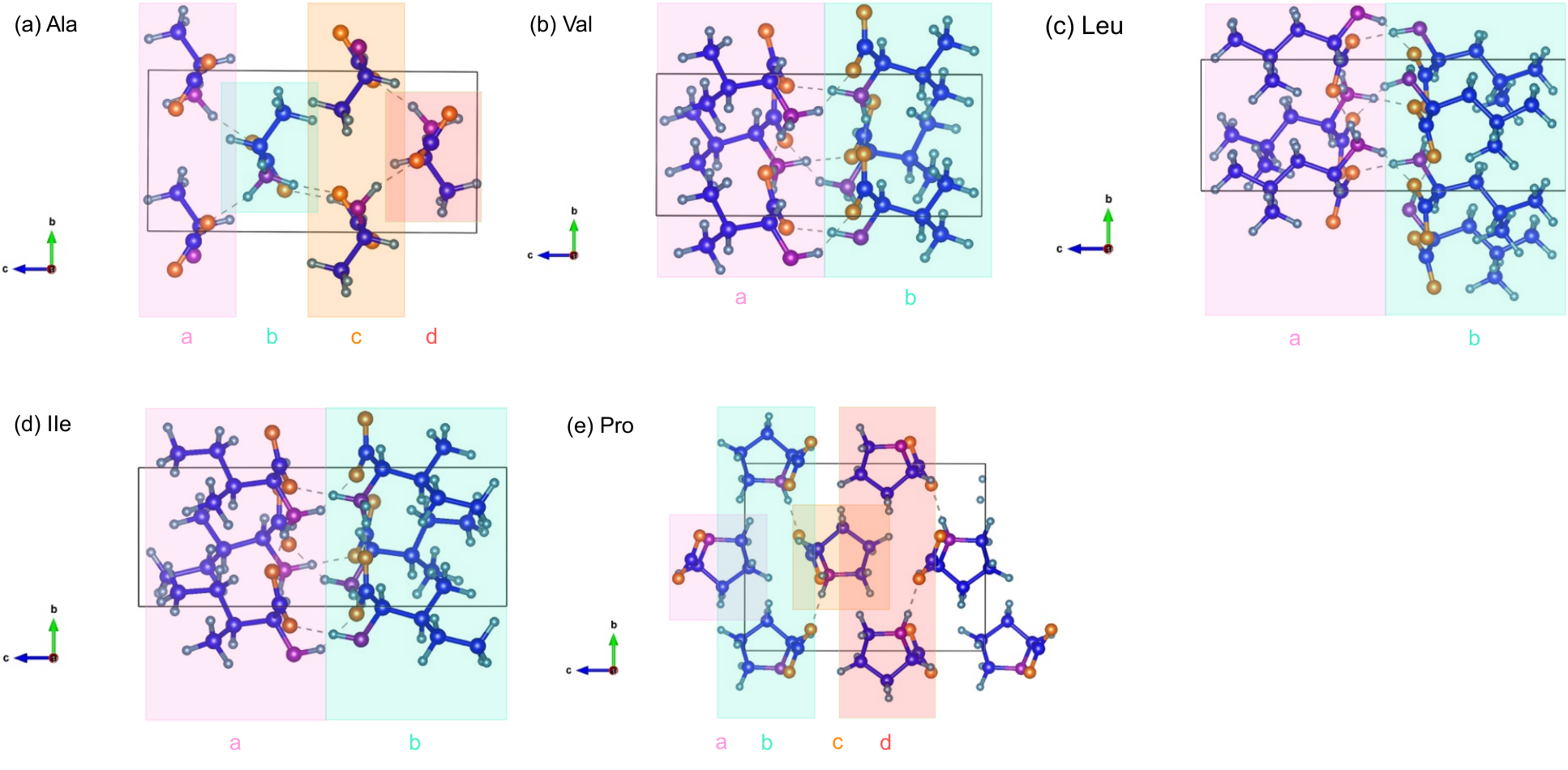}
    \caption{Visualisation of the unit cell crystal structures of the representative AAs with different columns and orientations highlighted by different colours for (a) Ala, (b) Val, (c) Leu, (d) Ile, and (e) Pro. The first and second columns are coloured by pink and cyan for all aliphatic AAs. The third and fourth columns are inked orange and red for Ala and Pro, respectively. All H, C, N, and O atoms are inked in steel blue, deep blue, purple, and orange, respectively. All structures were prepared in the VESTA software package.~\cite{Momma2011VESTA3Data}}
    \label{Ali_crystal structure}
\end{figure} 

\begin{figure}[htp]
    \centering
    \includegraphics[width=1.0\textwidth]{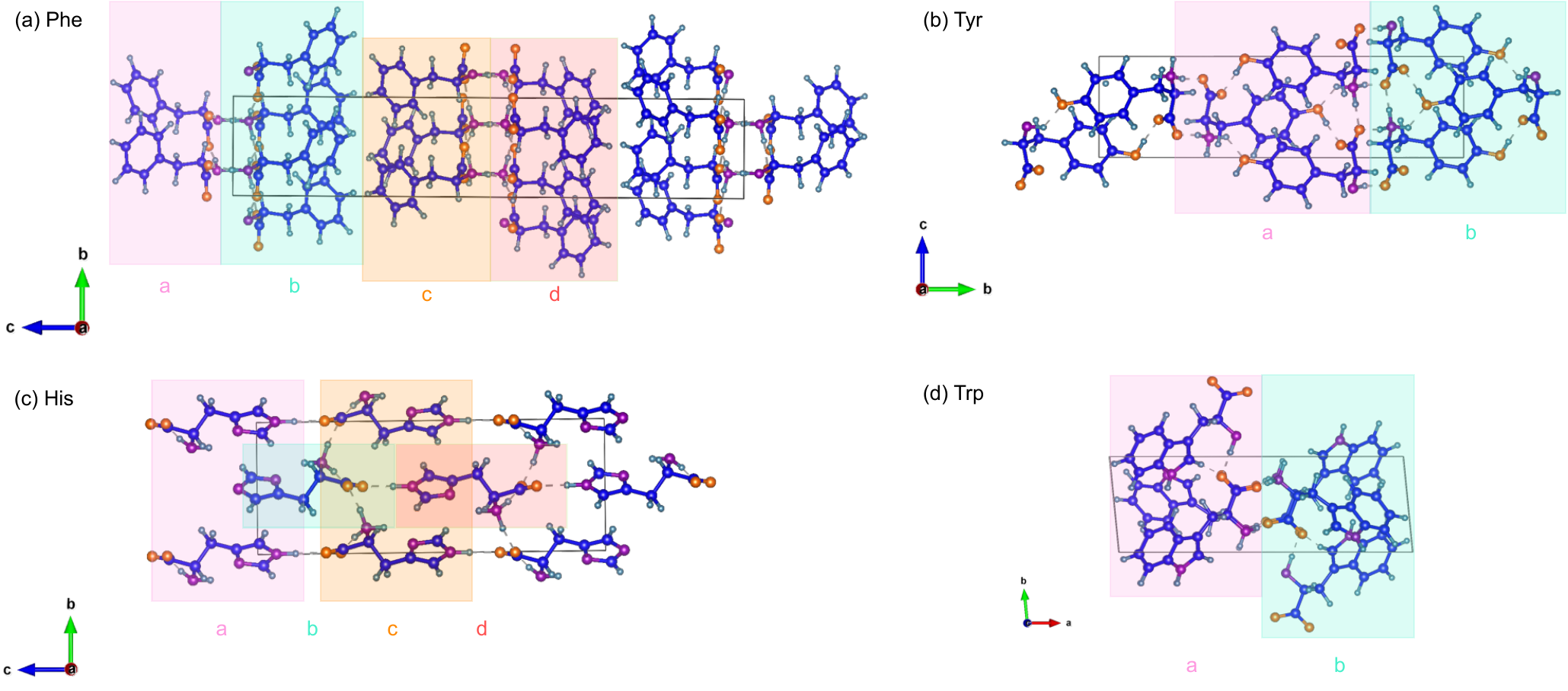}
    \caption{Visualisation of the unit cell crystal structures of the aromatic AAs with different columns and orientations highlighted by different colours for (a) Phe, (b) Tyr, (c) His, and (d) Trp. The first and second columns and orientations are coloured by pink and cyan green for all aromatic AAs. The third and fourth columns and orientations are inked by orange and vibrant red for Phe, His, and Trp respectively. The colour scheme of atoms used is the same as the one used in Figure~\ref{Ali_crystal structure}. All structures were prepared in the VESTA software package.~\cite{Momma2011VESTA3Data}}
    \label{Aro_crystal structure}
\end{figure} 
\cleardoublepage

\begin{figure}[ht]
    \centering
    \includegraphics[width=0.9\textwidth]{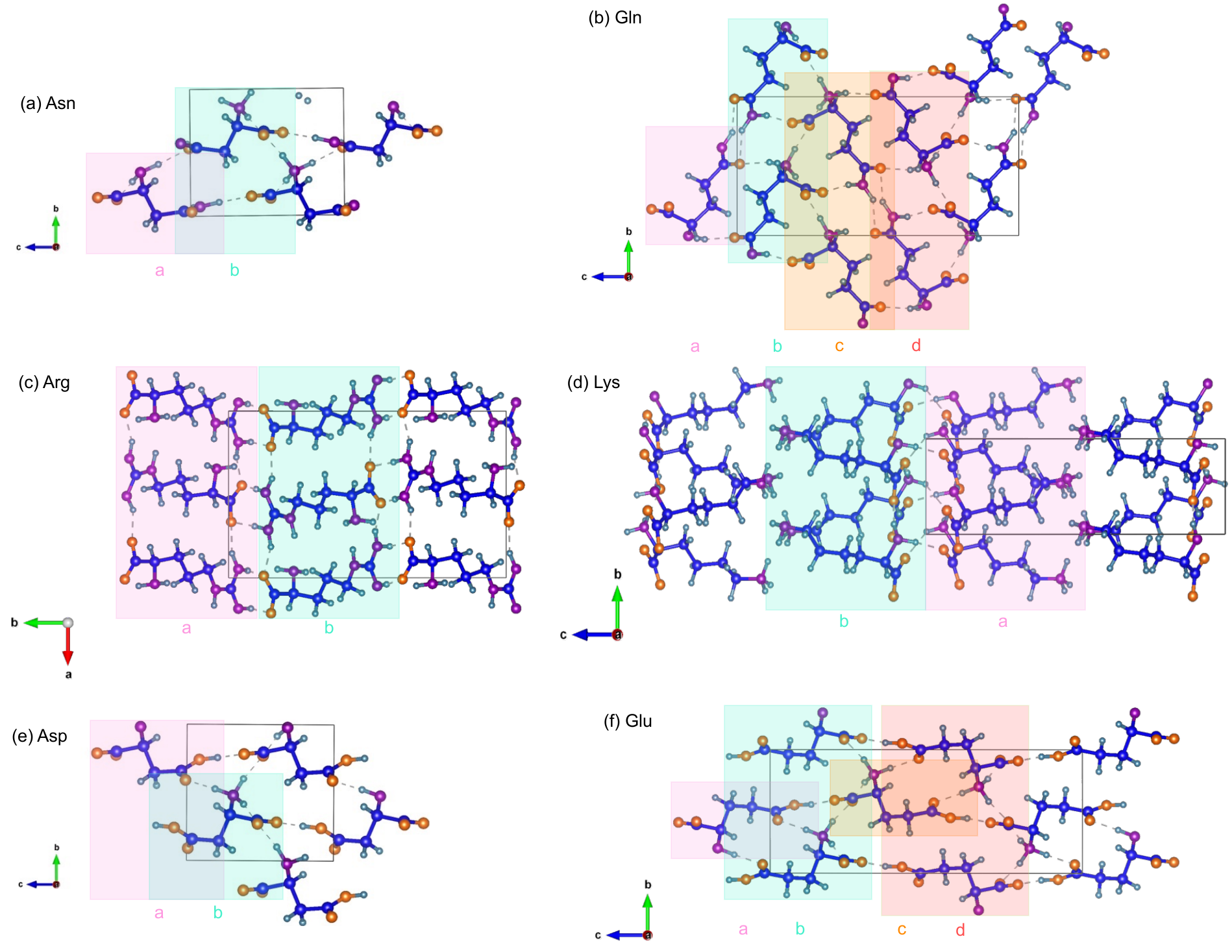}
    \caption{Visualisation of the unit cell crystal structures of the polar side chain-containing AAs with different columns and orientations highlighted by different colours for (a) Asn, (b) Gln, (c) Arg, (d) Lys, (e) Asp, and (f) Glu. The first and second columns and orientations are coloured by pink and cyan green for all AAs in the group. The third and fourth columns and orientations are inked orange and vibrant red for Gln and Glu, respectively. The colour scheme of atoms used is the same as the one used in Figure~\ref{Ali_crystal structure}. All structures were prepared in the VESTA software package.~\cite{Momma2011VESTA3Data}}
    \label{other_crystal structure}
\end{figure} 

\begin{figure}[ht]
    \centering
    \includegraphics[width=0.9\textwidth]{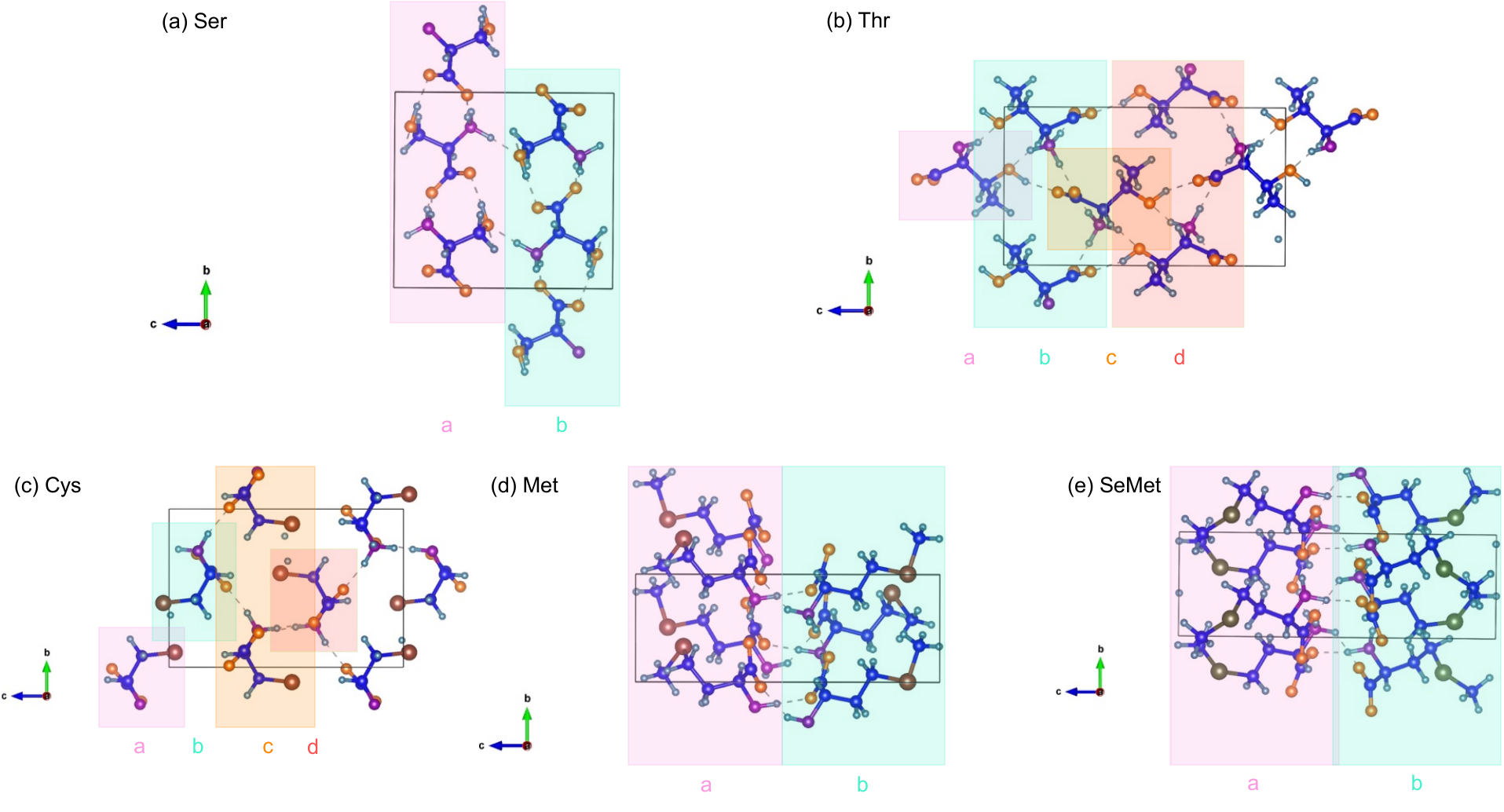}
    \caption{Visualisation of the unit cell crystal structures of the S/Se-containing AAs with different columns and orientations highlighted by different colours for (a) Ser, (b) Thr, (c) Cys, (d) Met, and (e) SeMet. The first and second columns are coloured by pink and cyan for all AAs in the group. The third and fourth columns are inked orange and red for Thr and Cys, respectively. The colour scheme of H, C, N, and O atoms used is the same as the one used in Figure~\ref{Ali_crystal structure}. S and Se are inked in brown and green, respectively. All structures were prepared in the VESTA software package.~\cite{Momma2011VESTA3Data}}
    \label{S_Se_crystal structure}
\end{figure} 

\begin{figure}[htp]
    \centering
    \includegraphics[width=1.0\textwidth]{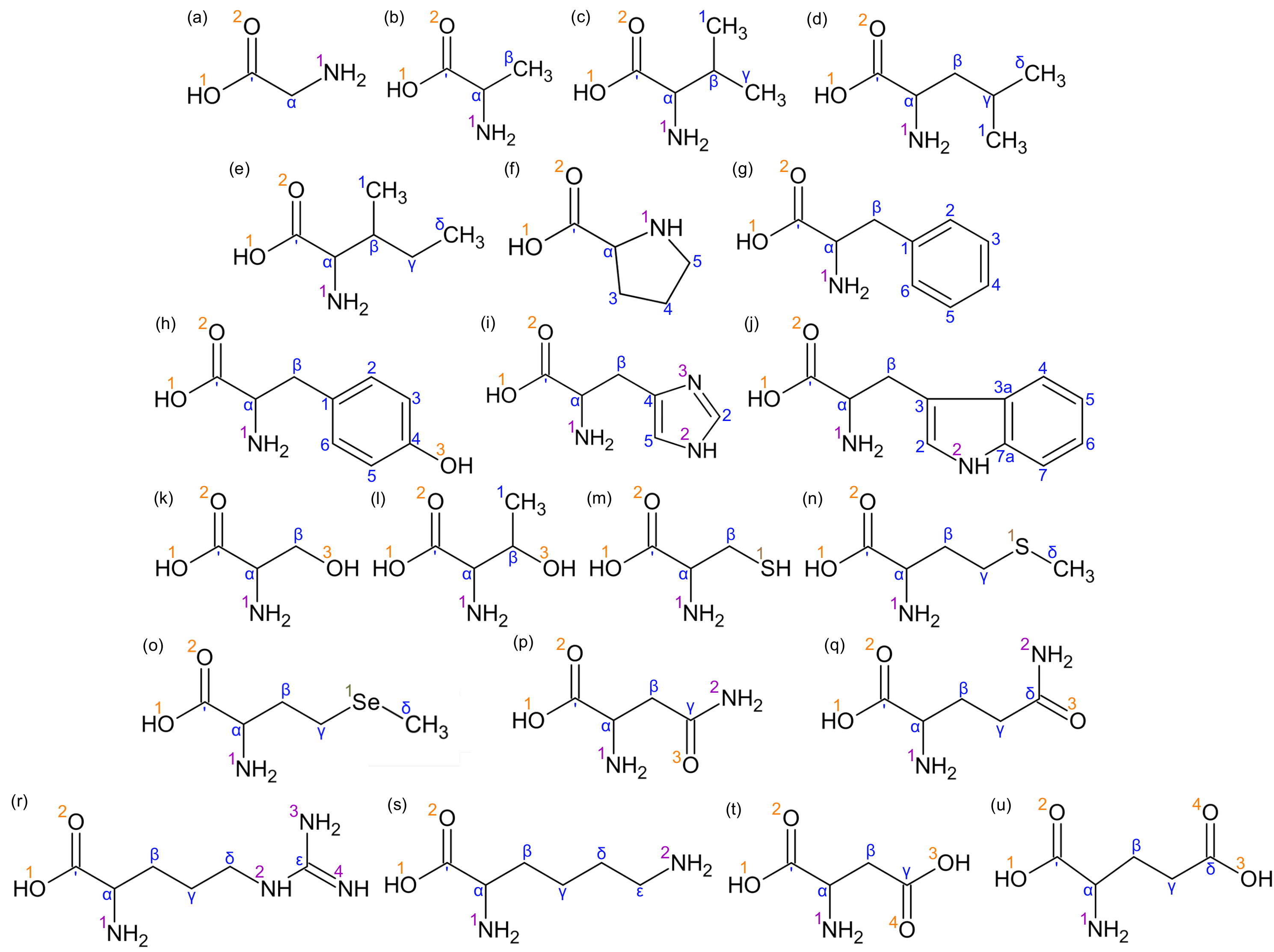}
    \caption{Schematic of (a) Gly, (b) Ala, (c) Val, (d) Leu, (e) Ile, (f) Pro, (g) Phe, (h) Tyr, (i) His, (j) Trp, (k) Ser, (l) Thr, (m) Cys, (n) Met, (o) SeMet, (p) Asn, (q) Gln, (r) Arg, (s) Lys, (t) Asp, and (u) Glu showing the atomic structures and atom labels, which are used throughout the manuscript. C atoms in the main chain are labelled using Greek letters. All C in the side chain and aromatic rings, N, and O atoms are labelled using numbers. N attached to the \ce{C_{$\alpha$}} is labelled as 1, hydroxyl O is labelled as 1 and carbonyl O as 2. The rest N and O atoms are labelled with ascending numbers and follow the IUPAC numbering rules. S and Se atoms in Cys, Met, and SeMet are all labelled as 1. All C, N, O, S, and Se atoms are inked in blue, purple, orange, brown, and dark green, respectively. This figure was made using ChemSketch.}
    \label{2D structure}
\end{figure}

\clearpage
\newpage
\section{Additional Computational Details}

Initial crystal structures of AAs were selected from the Cambridge Structural Database (CSD). Some AAs, including Gly, Trp, His, Ser, and Glu, show polymorphism, and multiple structures were selected to perform valence state calculations to probe the differences. 

\begin{figure}[ht!]
    \centering
    \includegraphics[keepaspectratio, width=0.42\textwidth]{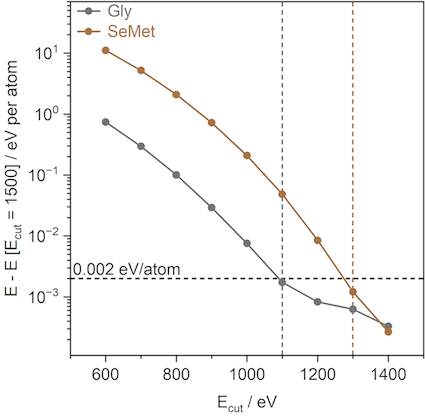}
    \caption{The cut-off energy convergence plot for Gly (representative of AAs with the same cut-off energy required), and for SeMet with a higher cut-off energy required. Note that a log arithmetic scale is used for the y-axis.}
    \label{cutoff}
\end{figure}

\begin{table}[!hb]
    \caption{Summary of the converged ground state (gs) and spectral $k$-points used for the PDOS calculations for AAs. The corresponding type of polymorph is indicated as a subscript for those AAs where multiple structures were calculated, using Greek letters.}
    \centering
    \begin{tabular}{c c c}
    
    \hline
    AAs &  Converged gs $k$-point & Converged spectral $k$-point \\
    \hline
    Gly & 2 $\times$ 1 $\times$ 2 & 8 $\times$ 8 $\times$ 8 \\
    Ala & 2 $\times$ 2 $\times$ 1 & 7 $\times$ 7 $\times$ 7 \\
    Val & 2 $\times$ 2 $\times$ 1 & 5 $\times$ 5 $\times$ 5 \\
    Leu & 2 $\times$ 2 $\times$ 1 & 5 $\times$ 5 $\times$ 5 \\
    Ile & 2 $\times$ 2 $\times$ 1 & 6 $\times$ 6 $\times$ 6 \\
    Pro & 2 $\times$ 2 $\times$ 1 & 6 $\times$ 6 $\times$ 6 \\
    Phe & 2 $\times$ 2 $\times$ 1 & 7 $\times$ 7 $\times$ 7 \\
    Tyr & 2 $\times$ 1 $\times$ 2 & 8 $\times$ 8 $\times$ 8 \\
    His & 2 $\times$ 2 $\times$ 1 & 7 $\times$ 7 $\times$ 7 \\
    Trp & 1 $\times$ 2 $\times$ 2 & 8 $\times$ 8 $\times$ 8 \\
    Arg & 2 $\times$ 1 $\times$ 2 & 5 $\times$ 5 $\times$ 5 \\
    Lys & 2 $\times$ 2 $\times$ 1 & 8 $\times$ 8 $\times$ 8 \\
    Asn & 2 $\times$ 2 $\times$ 2 & 6 $\times$ 6 $\times$ 6 \\
    Gln & 2 $\times$ 2 $\times$ 1 & 8 $\times$ 8 $\times$ 8 \\
    Asp & 2 $\times$ 2 $\times$ 2 & 6 $\times$ 6 $\times$ 6 \\
    Glu & 2 $\times$ 2 $\times$ 1 & 7 $\times$ 7 $\times$ 7 \\
    \ce{Ser_\alpha} & 2 $\times$ 2 $\times$ 1 & 7 $\times$ 7 $\times$ 7 \\
    \ce{Ser_\gamma} & 1 $\times$ 2 $\times$ 2 & 7 $\times$ 7 $\times$ 7 \\
    Cys & 2 $\times$ 2 $\times$ 1 & 6 $\times$ 6 $\times$ 6 \\
    Met & 2 $\times$ 2 $\times$ 1 & 7 $\times$ 7 $\times$ 7 \\
    SeMet & 2 $\times$ 2 $\times$ 1 & 8 $\times$ 8 $\times$ 8 \\
    \hline
    \end{tabular}
    \label{spectral kpoints}
\end{table}

\cleardoublepage
\begin{table}[ht!]
    \caption{Experimental (Exp.) and theoretical (Theor.) unit cell parameters for the PBE-relaxed crystals for Gly, aliphatic and aromatic AAs. Lattice parameters, $a$, $b$, and $c$, $\alpha$, $\beta$, and $\gamma$ unit cell angles, and differences (Diff.) are given in \(\text{\AA}\), degree, and $\%$, respectively. The Exp. values were taken from the references listed.}
\begin{center}
\begin{tabular}{c c c c c c c c c c} 

 \hline
 Molecule &  & $a$ & $b$ & $c$ &  & $\alpha$ & $\beta$ & $\gamma$ & Ref. \\ [0.5ex] 
 \hline
 \multirow{3}{*}{Gly} 
  & Exp. & 5.10 & 11.97 & 5.46 &  & 90.00 & 111.7 & 90.00 & ~\cite{Jiang2015Non-topotactic-glycine} \\
  & Theor. & 5.05 & 11.78 & 5.46 &  & 90.00 & 112.6 & 90.00 & ~ \\
  \rowcolor{lightgray}
\cellcolor{white}  & Diff. & 0.27 & 0.55 & -0.12 &  & 0.00 & -0.21 & 0.00 & ~ \\
 \multirow{3}{*}{Ala} 
  & Exp. & 5.79 & 5.96 & 12.29 &  & 90.00 & 90.00 & 90.00 & ~\cite{Waddell2024RODIN:Demonstration} \\
  & Theor. & 5.82 & 5.90 & 12.08 &  & 90.00 & 90.00 & 90.00 & ~ \\ 
  \rowcolor{lightgray}
\cellcolor{white}  & Diff. & -0.53 & 1.09 & 1.71 &  & 0.00 & 0.00 & 0.00 & ~ \\
 \multirow{3}{*}{Val} 
  & Exp. & 9.66 & 5.25 & 11.94 &  & 90.00 & 90.59 & 90.00 \\
  & Theor. & 9.64 & 5.14 & 11.79 &  & 90.00 & 90.11 & 90.00 \\ 
  \rowcolor{lightgray}
\cellcolor{white}  & Diff. & 0.26 & 2.01 & 1.26 &  & 0.00 & 0.53 & 0.00 & ~ \\
 \multirow{3}{*}{Leu} 
  & Exp. & 9.56 & 5.30 & 14.52 &  & 90.00 & 94.20 & 90.00 & ~\cite{Binns2016AccurateL-leucine} \\
  & Theor. & 9.53 & 5.21 & 14.48 &  & 90.00 & 94.29 & 90.00 & ~ \\ 
  \rowcolor{lightgray}
\cellcolor{white}  & Diff. & 0.31 & 1.84 & 0.28 &  & 0.00 & -0.09 & 0.00 & ~ \\
 \multirow{3}{*}{Ile} 
  & Exp. & 9.67 & 5.28 & 13.94 &  & 90.00 & 95.74 & 90.00 & ~\cite{Chen2018PrevalentAcids} \\
  & Theor. & 9.65 & 5.19 & 13.82 &  & 90.00 & 96.13 & 90.00 & ~ \\
  \rowcolor{lightgray}
\cellcolor{white}  & Diff. & 0.23 & 1.78 & 0.85 &  & 0.00 & -0.41 & 0.00 & ~ \\
 \multirow{3}{*}{Pro} 
  & Exp. & 5.28 & 8.88 & 11.53 &  & 90.00 & 90.00 & 90.00 & ~\cite{Tumanova2018OpeningCocrystals} \\
  & Theor. & 5.29 & 8.77 & 11.27 &  & 90.00 & 90.00 & 90.00 & ~ \\ 
  \rowcolor{lightgray}
\cellcolor{white}  & Diff. & -0.12 & 1.29 & 2.29 &  & 0.00 & 0.00 & 0.00 & ~ \\
 \multirow{3}{*}{Phe} 
  & Exp. & 8.78 & 5.60 & 31.02 &  & 90.00 & 96.95 & 90.00 & ~\cite{Ihlefeldt2014TheL-Phenylalanine} \\
  & Theor. & 8.72 & 5.93 & 30.75 &  & 90.00 & 96.95 & 90.00 & ~ \\
  \rowcolor{lightgray}
\cellcolor{white}  & Diff. & 0.71 & 1.13 & 0.87 &  & 0.00 & 0.01 & 0.00 & ~ \\
 \multirow{3}{*}{Tyr} 
  & Exp. & 6.91 & 21.12 & 5.83 &  & 90.00 & 90.00 & 90.00 \\
  & Theor. & 6.70 & 21.01 & 5.82 &  & 90.00 & 90.00 & 90.00 \\ 
  \rowcolor{lightgray}
\cellcolor{white}  & Diff. & 3.16 & 0.49 & 0.24 &  & 0.00 & 0.00 & 0.00 & ~ \\
 \multirow{3}{*}{His} 
  & Exp. & 5.15 & 7.22 & 18.84 &  & 90.00 & 90.00 & 90.00 & ~\cite{Novelli2021AccurateK} \\
  & Theor. & 5.08 & 7.14 & 18.83 &  & 90.00 & 90.00 & 90.00 & ~ \\
  \rowcolor{lightgray}
\cellcolor{white}  & Diff. & 1.48 & 1.21 & 0.08 &  & 0.00 & 0.00 & 0.00 & ~ \\
  \multirow{3}{*}{Trp} 
  & Exp. & 18.32 & 5.77 & 9.94 &  & 89.02 & 104.27 & 95.53 & ~\cite{Sosa-Rivadeneyra2023CrystallizationPolytypism} \\
  & Theor. & 18.26 & 5.78 & 9.82 &  & 89.01 & 104.08 & 95.65 & ~ \\
  \rowcolor{lightgray}
\cellcolor{white}  & Diff. & 0.31 & -0.08 & 1.24 &  & 0.00 & 0.18 & -0.13 & ~ \\
 \hline
\end{tabular}
\end{center}
\label{structural table 1}
\end{table}

\begin{table}[ht!]
    \caption{Experimental (Exp.) and theoretical (Theor.) unit cell parameters for the PBE-relaxed crystals for other, and S/Se-containing AAs. Lattice parameters, $a$, $b$, and $c$, $\alpha$, $\beta$, and $\gamma$ unit cell angles, and differences (Diff.) are given in \(\text{\AA}\), degree, and $\%$, respectively. The Exp. values were taken from the references listed.}
\begin{center}
\begin{tabular}{c c c c c c c c c c} 

 \hline
 Molecule &  & $a$ & $b$ & $c$ &  & $\alpha$ & $\beta$ & $\gamma$ & Ref. \\ [0.5ex] 
 \hline
 \multirow{3}{*}{Asn} 
  & Exp. & 5.06 & 6.70 & 8.05 &  & 90.00 & 91.71 & 90.00 & ~\cite{Yamada2007L-Asparagine} \\
  & Theor. & 5.05 & 6.62 & 8.04 &  & 90.00 & 91.88 & 90.00 & ~ \\
  \rowcolor{lightgray}
\cellcolor{white}  & Diff. & 0.32 & 1.18 & 0.23 &  & 0.00 & -0.19 & 0.00 & ~ \\
 \multirow{3}{*}{Gln} 
  & Exp. & 5.09 & 7.75 & 15.94 &  & 90.00 & 90.00 & 90.00 & ~\cite{Parsons2013UseRefinement} \\
  & Theor. & 5.03 & 7.76 & 15.77 &  & 90.00 & 90.00 & 90.00 & ~ \\ 
  \rowcolor{lightgray}
\cellcolor{white}  & Diff. & 1.21 & -0.16 & 1.08 &  & 0.00 & 0.00 & 0.00 & ~ \\
 \multirow{3}{*}{Arg} 
  & Exp. & 9.76 & 16.02 & 5.58 &  & 90.00 & 98.06 & 90.00 & ~\cite{Courvoisier2012TheL-arginine} \\
  & Theor. & 9.83 & 15.73 & 5.64 &  & 90.00 & 98.44 & 90.00 & ~ \\ 
  \rowcolor{lightgray}
\cellcolor{white}  & Diff. & -0.75 & 1.90 & -0.99 &  & 0.00 & 0.39 & 0.00 & ~ \\
 \multirow{3}{*}{Lys} 
  & Exp. & 9.51 & 5.13 & 16.99 &  & 90.00 & 97.70 & 90.00 & ~\cite{Williams2015L-Lysine:Acids} \\
  & Theor. & 9.36 & 5.20 & 17.87 &  & 90.00 & 100.01 & 90.00 & ~ \\
  \rowcolor{lightgray}
\cellcolor{white}  & Diff. & 1.52 & -1.27 & -5.18 &  & 0.00 & -2.31 & 0.00 & ~ \\
 \multirow{3}{*}{Asp} 
  & Exp. & 5.11 & 6.90 & 7.59 &  & 90.00 & 100.66 & 90.00 & ~\cite{Bendeif2007TheAcid} \\
  & Theor. & 5.08 & 6.91 & 7.59 &  & 90.00 & 100.33 & 90.00 & ~ \\ 
  \rowcolor{lightgray}
\cellcolor{white}  & Diff. & 0.62 & -0.06 & 0.01 &  & 0.00 & 0.33 & 0.00 & ~ \\
 \multirow{3}{*}{Glu} 
  & Exp. & 5.14 & 6.88 & 17.25 &  & 90.00 & 90.00 & 90.00 & ~\cite{Ruggiero2016ExaminationSpectroscopy} \\
  & Theor. & 5.09 & 6.84 & 17.30 &  & 90.00 & 90.00 & 90.00 & ~ \\ 
  \rowcolor{lightgray}
\cellcolor{white}  & Diff. & 1.08 & 0.60 & -0.32 &  & 0.00 & 0.00 & 0.00 & ~ \\
 \multirow{3}{*}{Ser} 
  & Exp. & 4.79 & 9.18 & 10.45 &  & 90.00 & 99.25 & 90.00 & ~\cite{Boldyreva2006AA} \\
  & Theor. & 4.70 & 9.25 & 10.45 &  & 90.00 & 99.08 & 90.00 & ~ \\
  \rowcolor{lightgray}
\cellcolor{white}  & Diff. & 1.93 & -0.67 & -0.02 &  & 0.00 & 0.17 & 0.00 & ~ \\
 \multirow{3}{*}{Thr} 
  & Exp. & 5.15 & 7.74 & 13.62 &  & 90.00 & 90.00 & 90.00 \\
  & Theor. & 5.02 & 7.61 & 13.66 &  & 90.00 & 90.00 & 90.00 \\
  \rowcolor{lightgray}
\cellcolor{white}  & Diff. & 2.68 & 0.91 & -0.28 &  & 0.00 & 0.00 & 0.00 & ~ \\
 \multirow{3}{*}{Cys} 
  & Exp. & 5.41 & 8.13 & 12.05 &  & 90.00 & 90.00 & 90.00 & ~\cite{Waddell2024RODIN:Demonstration} \\
  & Theor. & 5.43 & 8.09 & 12.00 &  & 90.00 & 90.00 & 90.00 & ~ \\
  \rowcolor{lightgray}
\cellcolor{white}  & Diff. & -0.34 & 0.54 & 0.39 &  & 0.00 & 0.00 & 0.00 & ~ \\
 \multirow{3}{*}{Met} 
  & Exp. & 9.51 & 5.21 & 14.86 &  & 90.00 & 99.70 & 90.00 \\
  & Theor. & 9.49 & 5.12 & 14.75 &  & 90.00 & 100.64 & 90.00 \\ 
  \rowcolor{lightgray}
\cellcolor{white}  & Diff. & 0.21 & 1.73 & 0.72 &  & 0.00 & -0.93 & 0.00 & ~ \\
 \multirow{3}{*}{SeMet} 
  & Exp. & 9.50 & 5.10 & 15.66 &  & 90.00 & 100.54 & 90.00 & ~\cite{Gajda2006StructureState} \\
  & Theor. & 9.49 & 5.03 & 15.56 &  & 90.00 & 100.27 & 90.00 & ~ \\
  \rowcolor{lightgray}
\cellcolor{white}  & Diff. & 0.16 & 1.41 & 0.65 &  & 0.00 & 0.27 & 0.00 & ~ \\
 \hline
\end{tabular}
\end{center}
\label{structural table 2}
\end{table}

The following parameters were used to generate the ground-state PSP for each element presented in AAs:

\begin{center}
H 1|0.8|14|16|19|10N(qc = 8) \\
C 1|1.2|17|20|23|20N:21L(qc = 8) \\
N 1|1.1|23|26|31|20N:21L(qc = 9) \\
O 1|1.2|23|26|31|20NN:21NN(qc = 9) \\
\end{center}

\clearpage
\newpage
\section{Survey Spectra}

\begin{figure}[htp]
\centering
    \includegraphics[keepaspectratio, width = 0.8\linewidth]{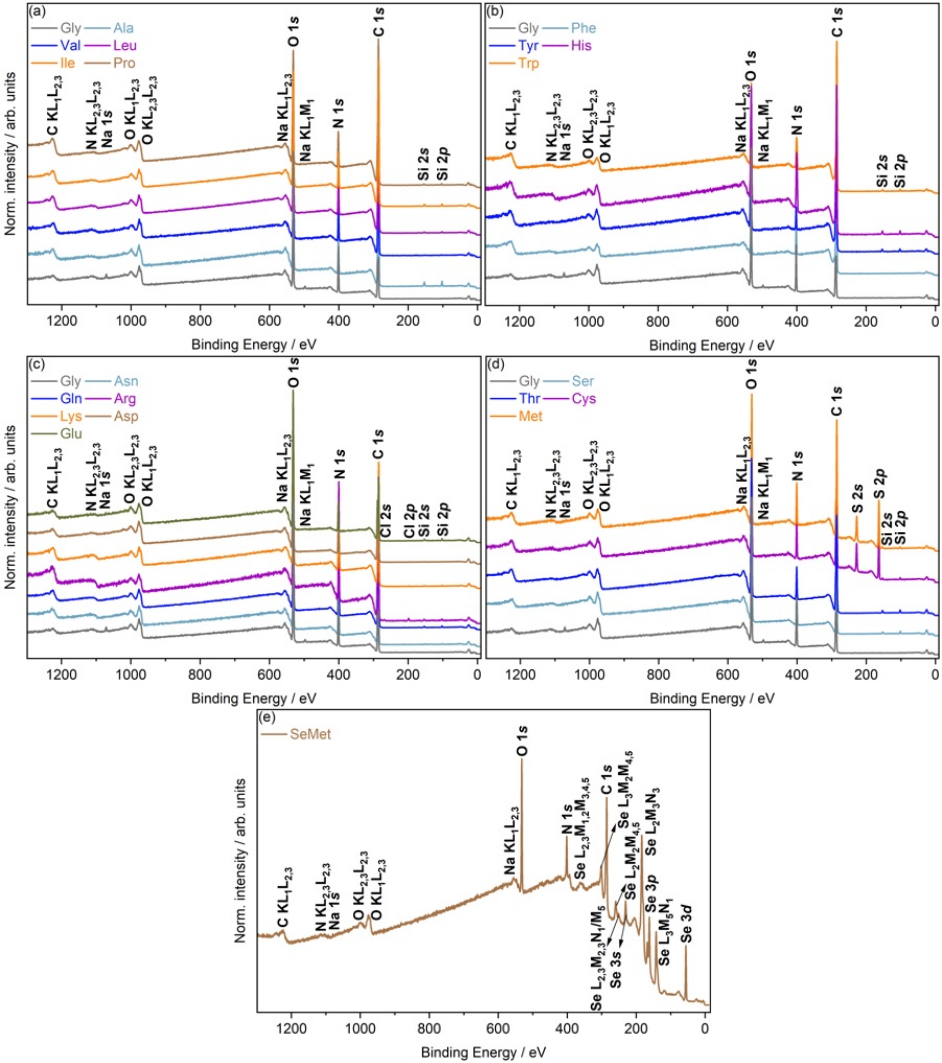}
    \caption{Survey spectra of (a) aliphatic AAs, (b) aromatic AAs, (c) polar side chain-containing AAs, and (d) S/Se-containing AAs with Gly as reference in each group. The survey spectrum of SeMet is plotted separately, presented in (e). All observed core and Auger-Meitner features are labelled.}
    \label{AAs_Survey}
\end{figure}

Figure~\ref{AAs_Survey} presents the survey spectra of the 21 studied AAs, subdivided into groups as classified in the main manuscript. 

\cleardoublepage

\section{Core Level Spectra}
\subsection{Aliphatic Group}

\begin{figure}[htp]
\centering
    \includegraphics[keepaspectratio, width = 0.8\linewidth]{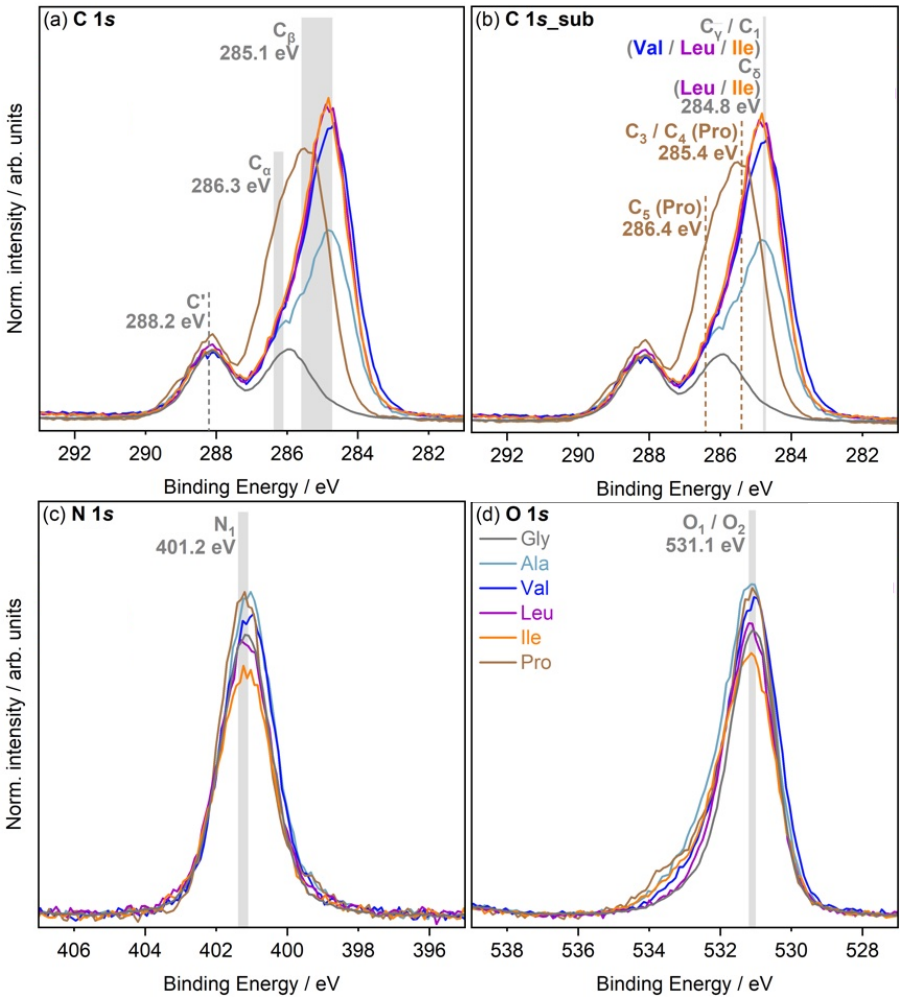}
    \caption{Combined CL spectra for the aliphatic group of (a) and (b) C~1\textit{s} with (a) common features and (b) features to specific AAs highlighted, (c) N~1\textit{s} spectra, and (d) O~1\textit{s} spectra with Gly included as reference in all Subfigures. Common features are labelled with grey solid lines of varying widths, representing the BE differences after alignment across chemical states for each CL, and the annotated values reflect the average BE values. Features specific to individual AAs are labelled using coloured, dashed lines, where the colour of the dashed lines is the same as the label colour of the corresponding AA. Spectra are aligned to the average value of the BEs of C~1\textit{s} features, assigned to the \ce{C^$'$} atom, of each AA in the group.}
    \label{Ali_CLs}
\end{figure}

\begin{figure}[htp]
\centering
    \includegraphics[keepaspectratio, width = 0.65\linewidth]{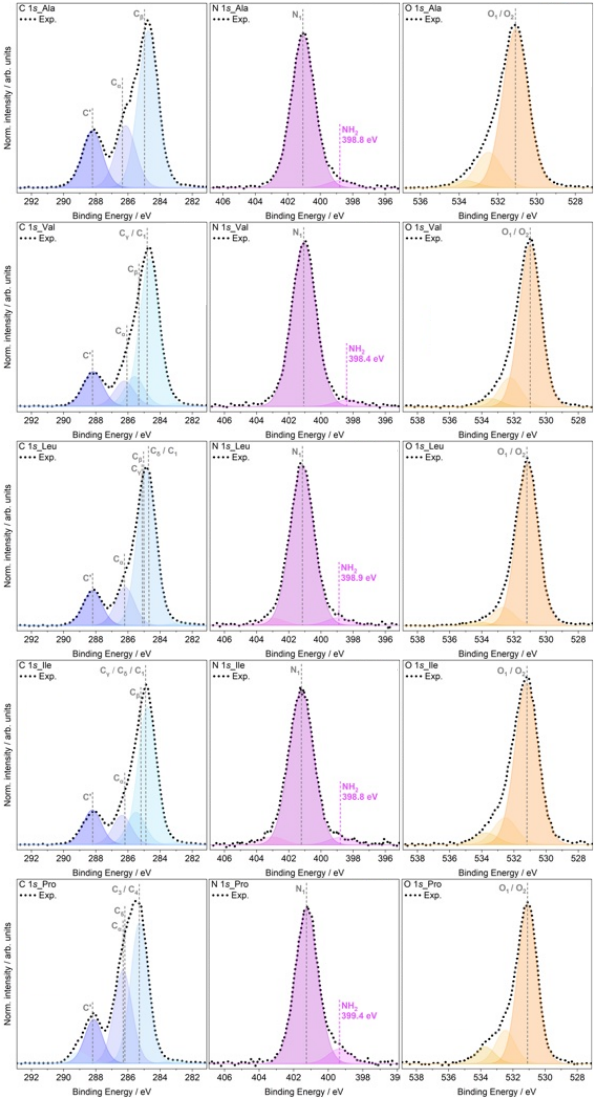}
    \caption{Complete set of the CL spectra for the aliphatic group, with experimental spectra depicted as black dots, experimental peak fits denoted as colour-shaded peaks, and calculated BEs shown as grey dash lines. Calculated BEs were performed using the PBE functional throughout all AAs, and have been aligned to the average value of the BE of C~1\textit{s} feature, assigned to the \ce{C^$'$} atom of each AA in the group for C~1\textit{s}, the experimental peak contributed to the \ce{N1} atom for N~1\textit{s}, and the experimental peak with the lowest BE, taking the average calculated BE of \ce{O1} and \ce{O2} for O~1\textit{s}, respectively. A Shirley-type background has been subtracted from all CL spectra to aid comparison with theory.}
    \label{Ali_CL_peakfitting}
\end{figure}

\begin{table}[!ht]
    \caption{Experimental BEs (Exp.) extracted from peak fit analysis, as well as the relative experimental (Exp.\ Rel.) and theoretical BEs (PBE) for each C and O atom to the \ce{C$'$} and \ce{O1} atoms, respectively, for the aliphatic group. Relative theoretical BEs are obtained from solid-state $\Delta$SCF calculations using the PBE functional.}
 \setlength{\tabcolsep}{4pt}
 \begin{center}
 \begin{tabular}{c c c c c c c c c c c c c} 
 \hline
 ~ & ~ & \ce{C^$'$} & \ce{C_\alpha} & \ce{C_\beta} & \ce{C_\gamma} & \ce{C_\delta} & \ce{C1} & \ce{C3} & \ce{C4} & \ce{C5} & \ce{O1} & \ce{O2} \\ [0.5ex] 
 \hline
 \multirow{3}{*}{Ala} & Exp. & 288.2 & 286.1 & 284.7 & - & - & - & - & - & - & 531.1 & 531.1 \\
 ~ & Exp.\ Rel. & 0.0 & -2.1 & -3.5 & - & - & - & - & - & - & 0.0 & 0.0 \\
 ~ & PBE & 0.0 & -1.9 & -3.2 & - & - & - & - & - & - & 0.0 & -0.17 \\
 \multirow{3}{*}{Val} & Exp. & 288.2 & 286.2 & 285.6 & 284.7 & - & 284.7 & - & - & - & 531.0 & 531.0 \\
 ~ & Exp.\ Rel. & 0.0 & -2.0 & -2.6 & -3.5 & - & -3.5 & - & - & - & 0.0 & 0.0 \\
 ~ & PBE & 0.0 & -2.0 & -2.8 & -3.3 & - & -3.3 & - & - & - & 0.0 & -0.03 \\
 \multirow{3}{*}{Leu} & Exp. & 288.2 & 286.2 & 284.8 & 284.8 & 284.8 & 284.8 & - & - & - & 531.2 & 531.2 \\
 ~ & Exp.\ Rel. & 0.0 & -2.0 & -3.4 & -3.4 & -3.4 & -3.4 & - & - & - & 0.0 & 0.0 \\
 ~ & PBE & 0.0 & -1.9 & -3.1 & -3.0 & -3.4 & -3.4 & - & - & - & 0.0 & 0.07 \\
 \multirow{3}{*}{Ile} & Exp. & 288.2 & 286.4 & 285.5 & 284.8 & 284.8 & 284.8 & - & - & - & 531.2 & 531.2 \\
 ~ & Exp.\ Rel. & 0.0 & -1.8 & -2.7 & -3.4 & -3.4 & -3.4 & - & - & - & 0.0 & 0.0 \\
 ~ & PBE & 0.0 & -2.0 & -2.9 & -3.3 & -3.3 & -3.3 & - & - & - & 0.0 & 0.08 \\
 \multirow{3}{*}{Pro} & Exp. & 288.2 & 286.4 & - & - & - & - & 285.4 & 285.4 & 286.4 & 531.2 & 531.2 \\
 ~ & Exp.\ Rel. & 0.0 & -1.8 & - & - & - & - & -2.8 & -2.8 & -1.8 & 0.0 & 0.0 \\
 ~ & PBE & 0.0 & -1.9 & - & - & - & - & -2.9 & -2.9 & -2.0 & 0.0 & -0.13 \\
 \hline
\end{tabular}
\end{center}
\label{Ali_CL_comp}
\end{table}

\clearpage
\newpage
\subsection{Aromatic Group}

\begin{figure}[htp]
\centering
    \includegraphics[keepaspectratio, width = 0.8\linewidth]{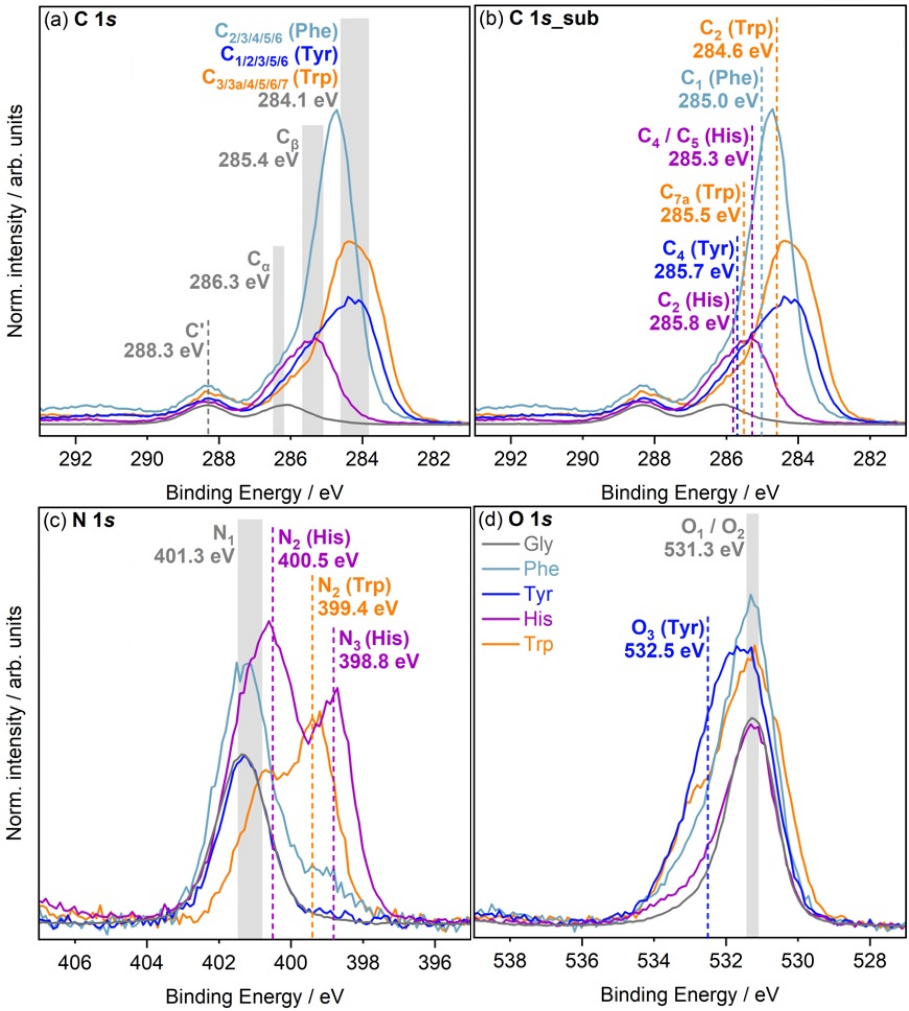}
    \caption{Combined CL spectra for the aromatic group of (a) and (b) C~1\textit{s} with (a) common features and (b) features to specific AAs highlighted, (c) N~1\textit{s} spectra, and (d) O~1\textit{s} spectra with Gly included as reference in all Subfigures. Common features are labelled with grey solid lines of varying widths, representing the BE differences after alignment across chemical states for each CL, and the annotated values reflect the average BE values. Features specific to individual AAs are labelled using coloured, dashed lines, where the colour of the dashed lines is the same as the label colour of the corresponding AA. Spectra are aligned to the average value of the BEs of C~1\textit{s} features, assigned to the \ce{C^$'$} atom, of each AA in the group.}
    \label{Aro_CLs}
\end{figure}

\begin{figure}[htp]
\centering
    \includegraphics[keepaspectratio, width = 0.75\linewidth]{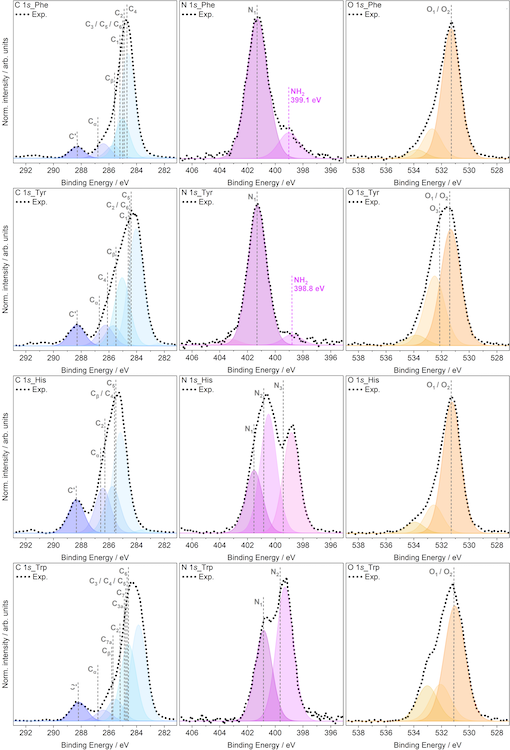}
    \caption{Complete set of the CL spectra for the aromatic group, with experimental spectra depicted as black dots, experimental peak fits denoted as colour-shaded peaks, and calculated BEs shown as grey dash lines. Calculated BEs were performed using the PBE functional throughout all AAs, and have been aligned to the average value of the BE of C~1\textit{s} feature, assigned to the \ce{C^$'$} atom of each AA in the group for C~1\textit{s}, the experimental peak contributed to the \ce{N1} atom for N~1\textit{s}, and the experimental peak with the lowest BE, taking the average calculated BE of \ce{O1} and \ce{O2} for O~1\textit{s}, respectively. A Shirley-type background has been subtracted from all CL spectra to aid comparison with theory.}
    \label{Aro_CL_peakfitting}
\end{figure}

\begin{landscape}
\begin{table}
    \caption{Experimental BEs (Exp.) extracted from peak fit analysis, as well as the relative experimental (Exp.\ Rel.) and theoretical BEs (PBE) for each C, N, and O atom to the \ce{C$'$}, \ce{N1}, and \ce{O1} atoms, respectively, for the aromatic group. Relative theoretical BEs are obtained from solid-state $\Delta$SCF calculations using the PBE functional.}
 \setlength{\tabcolsep}{5pt}
 \begin{center}
 \begin{tabular}{c c c c c c c c c c c c c c c c c c c c} 
 \hline
 ~ & ~ & \ce{C^$'$} & \ce{C_\alpha} & \ce{C_\beta} & \ce{C1} & \ce{C2} & \ce{C3} & \ce{C_{3a}} & \ce{C4} & \ce{C5} & \ce{C6} & \ce{C7} & \ce{C_{7a}} & \ce{N1} & \ce{N2} & \ce{N3} & \ce{O1} & \ce{O2} & \ce{O3} \\ [0.5ex] 
 \hline
 \multirow{3}{*}{Phe} & Exp. & 288.3 & 286.4 & 285.7 & 285.0 & 284.6 & 284.6 & - & 284.6 & 284.6 & 284.6 & - & - & 401.3 & - & - & 531.3 & 531.3 & - \\
 ~ & Exp.\ Rel. & 0.0 & -1.9 & -2.6 & -3.3 & -3.7 & -3.7 & - & -3.7 & -3.7 & -3.7 & - & - & 0.0 & - & - & 0.0 & 0.0 & - \\
 ~ & PBE & 0.0 & -1.5 & -2.7 & -3.1 & -3.4 & -3.3 & - & -3.6 & -3.3 & -3.3 & - & - & 0.0 & - & - & 0.0 & -0.03 & - \\
 \multirow{3}{*}{Tyr} & Exp. & 288.3 & 286.2 & 285.1 & 284.0 & 284.0 & 284.0 & - & 285.7 & - & 284.0 & 284.0 & 401.3 & - & - & 531.4 & 531.4 & 532.5 \\
 ~ & Exp.\ Rel. & 0.0 & -2.1 & -3.2 & -4.3 & -4.3 & -4.3 & - & -4.3 & -4.3 & -4.3 & 0.0 & - & - & 0.0 & 0.0 & 1.1 \\
 ~ & PBE & 0.0 & -1.6 & -2.8 & -3.7 & -3.8 & -3.9 & - & -2.2 & -4.0 & -3.8 & 0.0 & - & - & 0.0 & 0.14 & 0.67 \\
 \multirow{3}{*}{His} & Exp. & 288.3 & 286.5 & 285.3 & - & 285.8 & - & - & 285.3 & 285.3 & - & - & - & 401.5 & 400.5 & 398.8 & 531.3 & 531.3 & - \\
 ~ & Exp.\ Rel. & 0.0 & -1.8 & -3.0 & - & -2.5 & - & - & -3.0 & -3.0 & - & - & - & 0.0 & -1.0 & -2.7 & 0.0 & 0.0 & - \\
 ~ & PBE & 0.0 & -1.7 & -2.7 & - & -2.0 & - & - & -2.7 & -2.8 & 0.0 & -0.7 & -2.1 & 0.0 & -0.08 & - \\
 \multirow{3}{*}{Trp} & Exp. & 288.3 & 286.2 & 285.5 & - & 284.6 & 283.8 & 283.8 & 283.8 & 283.8 & 283.8 & 283.8 & 285.5 & 400.8 & 399.4 & - & 531.1 & 531.1 & - \\
 ~ & Exp.\ Rel. & 0.0 & -2.1 & -2.8 & - & -3.7 & -4.5 & -4.5 & -4.5 & -4.5 & -4.5 & -4.5 & -2.8 & -1.4 & - & 0.0 & 0.0 & - \\
 ~ & PBE & 0.0 & -1.5 & -2.5 & - & -3.1 & -3.6 & -3.4 & -3.6 & -3.6 & -3.7 & -3.5 & -2.6 & 0.0 & -1.2 & - & 0.0 & 0.05 & - \\
 \hline
\end{tabular}
\end{center}
\label{Aro_CL_comp}
\end{table}
\end{landscape}

\clearpage
\newpage
\subsection{Polar Side Chain-Containing Group}

\begin{figure}[ht]
\centering
    \includegraphics[keepaspectratio, width = 0.7\linewidth]{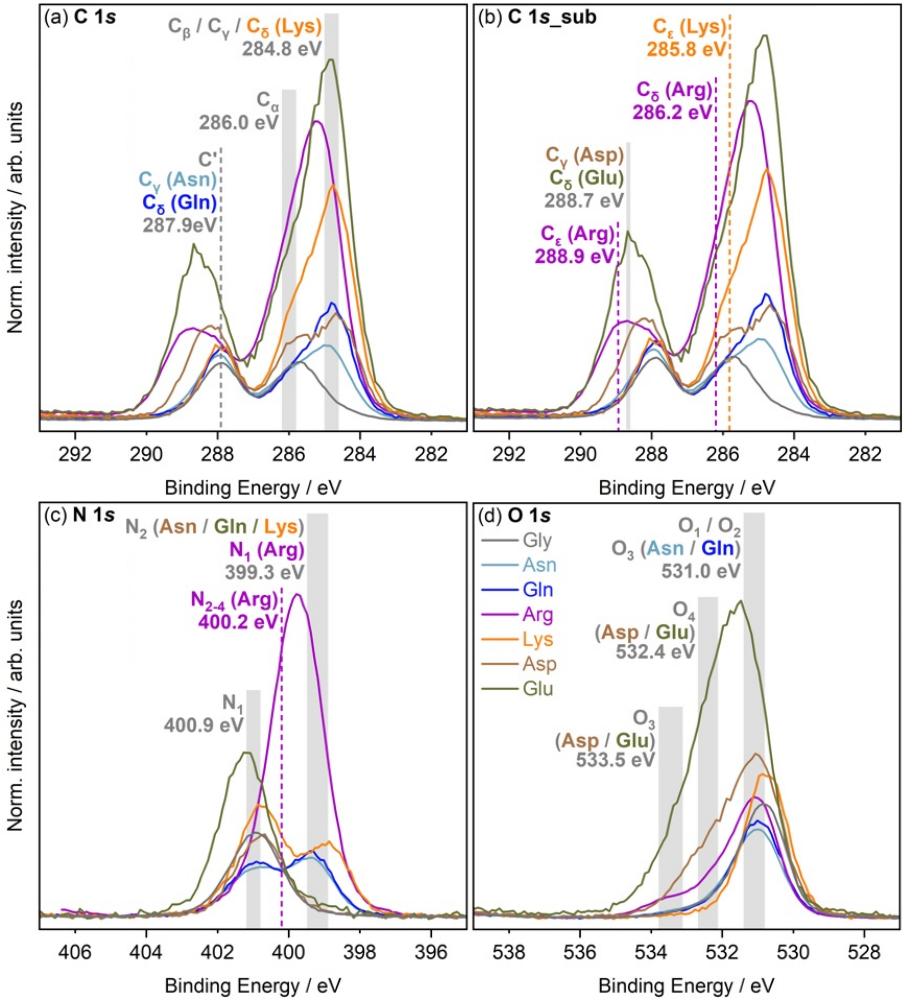}
    \caption{Combined CL spectra for the polar side chain-containing group of (a) and (b) C~1\textit{s} with (a) common features and (b) features to specific AAs highlighted, (c) N~1\textit{s} spectra, and (d) O~1\textit{s} spectra with Gly included as reference in all Subfigures. Common features are labelled with grey solid lines of varying widths, representing the BE differences after alignment across chemical states for each CL, and the annotated values reflect the average BE values. Features specific to individual AAs are labelled using coloured, dashed lines, where the colour of the dashed lines is the same as the label colour of the corresponding AA. Spectra are aligned to the average value of the BEs of C~1\textit{s} features, assigned to the \ce{C^$'$} atom, of each AA in the group.}
    \label{other_CLs}
\end{figure}

\begin{figure}[htp]
\centering
    \includegraphics[keepaspectratio, width = 0.5\linewidth]{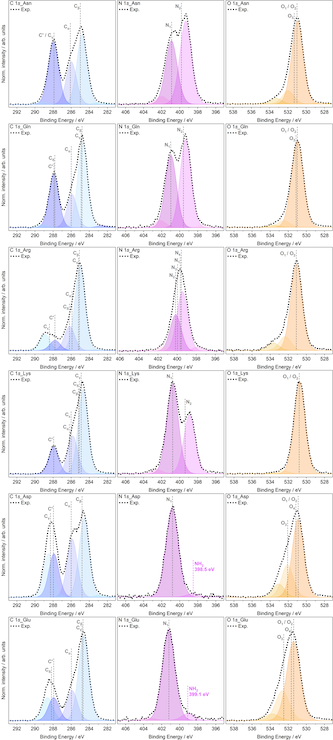}
    \caption{Complete set of the CL spectra for the polar side chain-containing group, with experimental spectra depicted as black dots, experimental peak fits denoted as colour-shaded peaks, and calculated BEs shown as grey dash lines. Calculated BEs were performed using the PBE functional throughout all AAs, and have been aligned to the average value of the BE of C~1\textit{s} feature, assigned to the \ce{C^$'$} atom of each AA in the group for C~1\textit{s}, the experimental peak contributed to the \ce{N1} atom for N~1\textit{s}, and the experimental peak with the lowest BE, taking the average calculated BE of \ce{O1} and \ce{O2} for O~1\textit{s}, respectively. A Shirley-type background has been subtracted from all CL spectra to aid comparison with theory.}
    \label{other_CL_peakfitting}
\end{figure}

\begin{landscape}
\begin{table}
    \caption{Experimental BEs (Exp.) extracted from peak fit analysis, as well as the relative experimental (Exp.\ Rel.) and theoretical BEs (PBE) for each C, N, and O atom relative to the \ce{C$'$}, \ce{N1}, and \ce{O1} atoms, respectively, for the polar side chain-containing group. Theoretical BEs are obtained from solid-state $\Delta$SCF calculations using the PBE functional.}
\begin{center}
\begin{tabular}{c c c c c c c c c c c c c c c c} 
 \hline
 ~ & ~ & \ce{C^$'$} & \ce{C_\alpha} & \ce{C_\beta} & \ce{C_\gamma} & \ce{C_\delta} & \ce{C_\epsilon} & \ce{N1} & \ce{N2} & \ce{N3} & \ce{N4} & \ce{O1} & \ce{O2} & \ce{O3} & \ce{O4} \\ [0.5ex] 
 \hline
 \multirow{3}{*}{Asn} & Exp. & 287.9 & 286.0 & 284.8 & 287.9 & - & - & 400.9 & 399.3 & - & - & 531.0 & 531.0 & 532.0 & - \\
 ~ & Exp.\ Rel. & 0.0 & -1.9 & -3.1 & 0.0 & - & - & 0.0 & -1.6 & - & - & 0.0 & 0.0 & 0.0 & - \\
 ~ & PBE & 0.0 & -1.8 & -2.9 & 0.01 & - & - & 0.0 & -1.09 & - & - & 0.0 & -0.08 & 0.3 & - \\
 \multirow{3}{*}{Gln} & Exp. & 287.9 & 285.9 & 284.8 & 284.8 & 287.9 & - & 401.0 & 399.3 & - & - & 531.0 & 531.0 & 532.4 & - \\
 ~ & Exp.\ Rel. & 0.0 & -2.0 & -3.1 & -3.1 & 0.0 & - & 0.0 & -1.7 & - & - & 0.0 & 0.0 & 0.0 & - \\
 ~ & PBE & 0.0 & -1.9 & -3.1 & -3.0 & -0.1 & - & 0.0 & -1.4 & - & - & 0.0 & -0.07 & 0.18 & - \\
 \multirow{3}{*}{Arg} & Exp. & 287.9 & 286.2 & 285.0 & 285.0 & 286.2 & 288.9 & 399.5 & 400.1 & 400.1 & 400.1 & 531.1 & 531.1 & - & - \\
 ~ & Exp.\ Rel. & 0.0 & -1.7 & -2.9 & -2.9 & -1.7 & 1.0 & 0.0 & 0.7 & 0.7 & 0.7 & 0.0 & 0.0 & - & - \\
 ~ & PBE & 0.0 & -2.0 & -2.8 & -2.7 & -1.7 & 0.6 & 0.0 & 0.5 & 0.3 & 0.1 & 0.0 & -0.01 & - & - \\
 \multirow{3}{*}{Lys} & Exp. & 287.9 & 285.8 & 284.7 & 284.7 & 284.7 & 285.8 & 400.8 & 398.9 & - & - & 530.8 & 530.8 & - & - \\
 ~ & Exp.\ Rel. & 0.0 & -2.1 & -3.2 & -3.2 & -3.2 & -2.1 & 0.0 & -1.9 & - & - & 0.0 & 0.0 & - & - \\
 ~ & PBE & 0.0 & -1.7 & -2.7 & -3.0 & -2.8 & -1.9 & 0.0 & -1.4 & - & - & 0.0 & -0.13 & - & - \\
 \multirow{3}{*}{Asp} & Exp. & 287.9 & 285.9 & 284.6 & 288.5 & - & - & 400.8 & - & - & - & 530.9 & 530.9 & 533.1 & 532.1 \\
 ~ & Exp.\ Rel. & 0.0 & -2.0 & -3.3 & 0.6 & - & - & 0.0 & - & - & - & 0.0 & 0.0 & 2.2 & 1.2 \\
 ~ & PBE & 0.0 & -1.9 & -3.1 & 0.4 & - & - & 0.0 & - & - & - & 0.0 & -0.06 & 1.17 & 0.26 \\
 \multirow{3}{*}{Glu} & Exp. & 287.9 & 286.1 & 284.8 & 284.8 & 288.7 & - & 401.2 & - & - & - & 531.4 & 531.4 & 533.8 & 532.7 \\
 ~ & Exp.\ Rel. & 0.0 & -1.8 & -3.1 & -3.1 & 0.8 & - & 0.0 & - & - & - & 0.0 & 0.0 & 2.4 & 1.3 \\
 ~ & PBE & 0.0 & -1.9 & -3.2 & -3.1 & 0.3 & - & 0.0 & - & - & - & 0.0 & -0.01 & 1.06 & 0.26 \\
 \hline
\end{tabular}
\end{center}
\label{other_CL_comp}
\end{table}
\end{landscape}

\cleardoublepage
\subsection{S/Se-containing Group}

\begin{figure}[ht]
\centering
    \includegraphics[keepaspectratio, width = 0.7\linewidth]{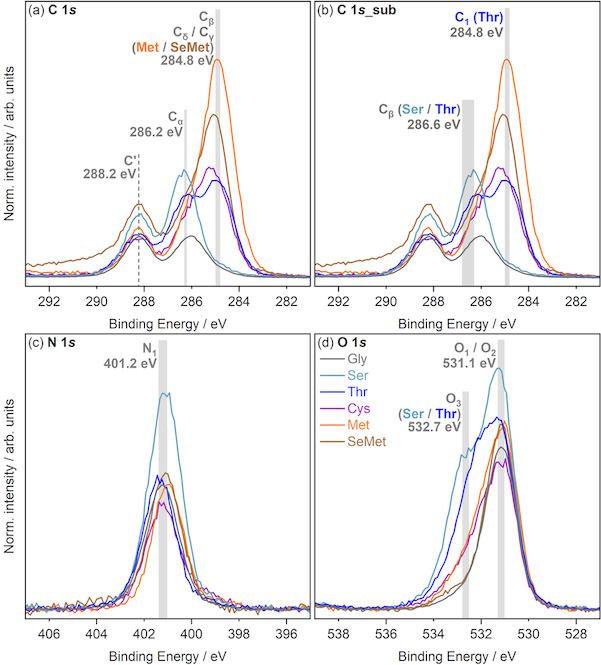}
    \caption{Combined CL spectra for the S/Se-containing group of (a) and (b) C~1\textit{s} with (a) common features and (b) features to specific AAs highlighted, (c) N~1\textit{s} spectra, and (d) O~1\textit{s} spectra with Gly included as reference in all Subfigures. Common features are labelled with grey solid lines of varying widths, representing the BE differences after alignment across chemical states for each CL, and the annotated values reflect the average BE values. Features specific to individual AAs are labelled using coloured, dashed lines, where the colour of the dashed lines is the same as the label colour of the corresponding AA. Spectra are aligned to the average value of the BEs of C~1\textit{s} features, assigned to the \ce{C^$'$} atom, of each AA in the group.}
    \label{SHOH_CLs}
\end{figure}

\begin{figure}[ht]
\centering
    \includegraphics[keepaspectratio, width = 0.7\linewidth]{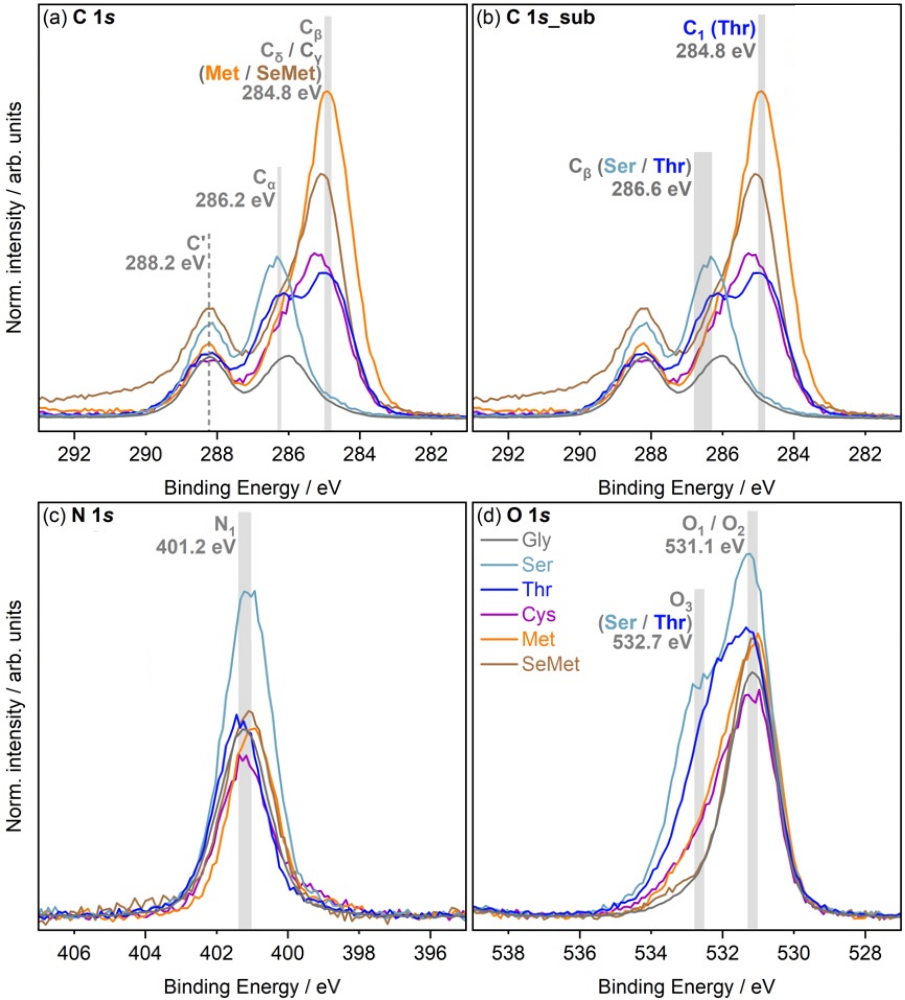}
    \caption{S~2\textit{p} spectra of (a) Cys and Met, and (b) Se~3\textit{d} spectrum of SeMet. S~2\textit{p} features are labelled using grey solid lines, representing the BE range after alignment for Cys and Met. Se~3\textit{d} feature is labelled using brown dash lines, where the colour of the dash lines is the same as the colour of SeMet. Spectra are aligned to the average value of the BEs of C~1\textit{s} features, assigned to the \ce{C^'} atom, of each AA for each group.}
    \label{S_Se_CLs}
\end{figure}

\begin{figure}[htp]
\centering
    \includegraphics[keepaspectratio, width = 0.65\linewidth]{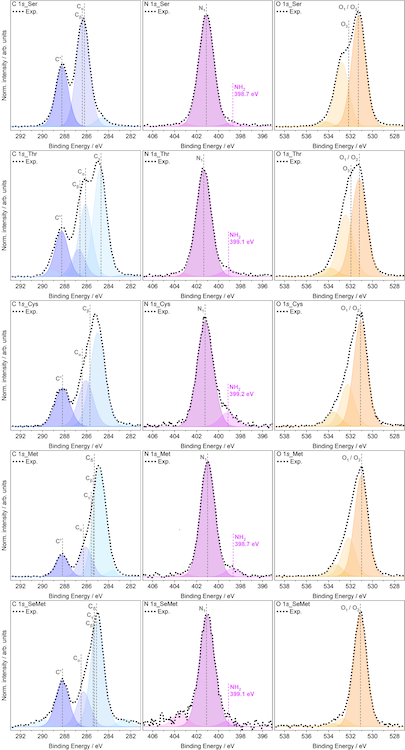}
    \caption{Complete set of the CL spectra for the S/Se-containing group, with experimental spectra depicted as black dots, experimental peak fits denoted as colour-shaded peaks, and calculated BEs shown as grey dash lines. Calculated BEs were performed using the PBE functional throughout all AAs, and have been aligned to the average value of the BE of C~1\textit{s} feature, assigned to the \ce{C^$'$} atom of each AA in the group for C~1\textit{s}, the experimental peak contributed to the \ce{N1} atom for N~1\textit{s}, and the experimental peak with the lowest BE, taking the average calculated BE of \ce{O1} and \ce{O2} for O~1\textit{s}, respectively. A Shirley-type background has been subtracted from all CL spectra to aid comparison with theory.}
    \label{S_Se_CL_peakfitting}
\end{figure}

\cleardoublepage
\begin{table}[htp]
    \caption{Experimental BEs (Exp.) extracted from peak fit analysis, as well as the relative experimental (Exp.\ Rel.) and theoretical BEs (PBE) for each C and O atom to the \ce{C$'$} and \ce{O1} atoms, respectively, for the S/Se-containing group. Relative theoretical BEs are obtained from solid-state $\Delta$SCF calculations using the PBE functional.}
\begin{center}
\begin{tabular}{c c c c c c c c c c} 
 \hline
 ~ & ~ & \ce{C^$'$} & \ce{C_\alpha} & \ce{C_\beta} & \ce{C_\gamma} & \ce{C_\delta} & \ce{O1} & \ce{O2} & \ce{O3} \\ [0.5ex] 
 \hline
 \multirow{3}{*}{Ser} & Exp. & 288.2 & 286.3 & 286.3 & - & - & 531.3 & 531.3 & 532.8 \\
 ~ & Exp.\ Rel. & 0.0 & -1.9 & -1.9 & - & - & 0.0 & 0.0 & 1.5 \\
 ~ & PBE & 0.0 & -2.0 & -1.8 & - & - & 0.0 & 0.0 & 0.87 \\
 \multirow{3}{*}{Thr} & Exp. & 288.2 & 286.2 & 286.8 & 284.8 & - & 531.2 & 531.2 & 532.5 \\
 ~ & Exp.\ Rel. & 0.0 & -2.0 & -1.4 & -3.4 & - & 0.0 & 0.0 & 1.3 \\
 ~ & PBE & 0.0 & -2.1 & -1.6 & -3.5 & - & 0.0 & -0.04 & 0.76 \\
 \multirow{3}{*}{Cys} & Exp. & 288.2 & 286.2 & 285.0 & - & - & 531.1 & 531.1 & - \\
 ~ & Exp.\ Rel. & 0.0 & -2.0 & -3.2 & - & 0.0 & 0.0 & - \\
 ~ & PBE & 0.0 & -1.8 & -2.5 & - & 0.0 & -0.13 & - \\
 \multirow{3}{*}{Met} & Exp. & 288.2 & 286.2 & 284.8 & 284.8 & 284.8 & 531.0 & 531.0 & - \\
 ~ & Exp.\ Rel. & 0.0 & -2.0 & -3.4 & -3.4 & -3.4 & 0.0 & 0.0 & - \\
 ~ & PBE & 0.0 & -1.9 & -2.8 & -2.6 & -2.9 & 0 & -0.05 & - \\
 \multirow{3}{*}{SeMet} & Exp. & 288.2 & 286.3 & 285.0 & 285.0 & 285.0 & 531.1 & 531.1 & - \\
 ~ & Exp.\ Rel. & 0.0 & -1.9 & -3.2 & -3.2 & -3.2 & 0.0 & 0.0 & - \\
 ~ & PBE & 0.0 & -1.7 & -2.8 & -2.9 & -3.1 & 0.0 & -0.1 & - \\
 \hline
\end{tabular}
\end{center}
\label{S_Se_CL_comp}
\end{table}

\cleardoublepage
\section{Comparison of PDOS to XPS Semi-Core States}

\subsection{Aliphatic Group}

\begin{figure}[htp]
\centering
    \includegraphics[keepaspectratio, width = 1.0\linewidth]{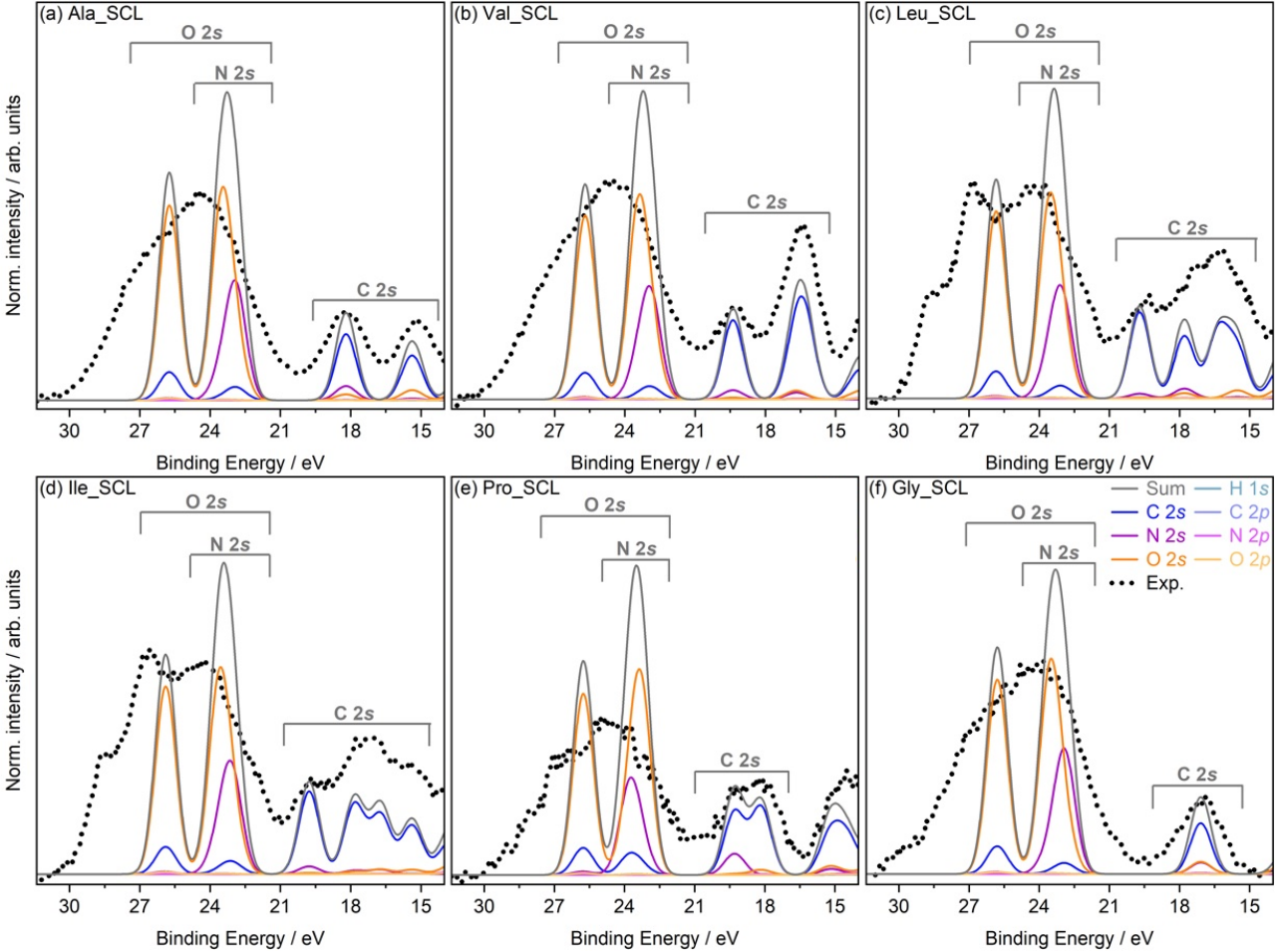}
    \caption{SCL spectra of the aliphatic AAs, including (a) Ala, (b) Val, (c) Leu, (d) Ile, (e) Pro, and (f) Gly as reference. All plots display the one-electron photoionisation cross-section weighted PDOS, as well as the sum of all PDOS, from PBE-based DFT calculations and the experimental XP spectra. The labels in dark grey indicate the majority orbital contribution to the spectral features determined from DFT. The weighted PDOS have been aligned and normalised to the experimental BE peak of the highest theoretical C~2\textit{s} feature. The legend shown in (f) applies to all subfigures.}
    \label{Ali_SCL}
\end{figure}

\begin{figure}[htp]
\centering
    \includegraphics[keepaspectratio, width = 1.0\linewidth]{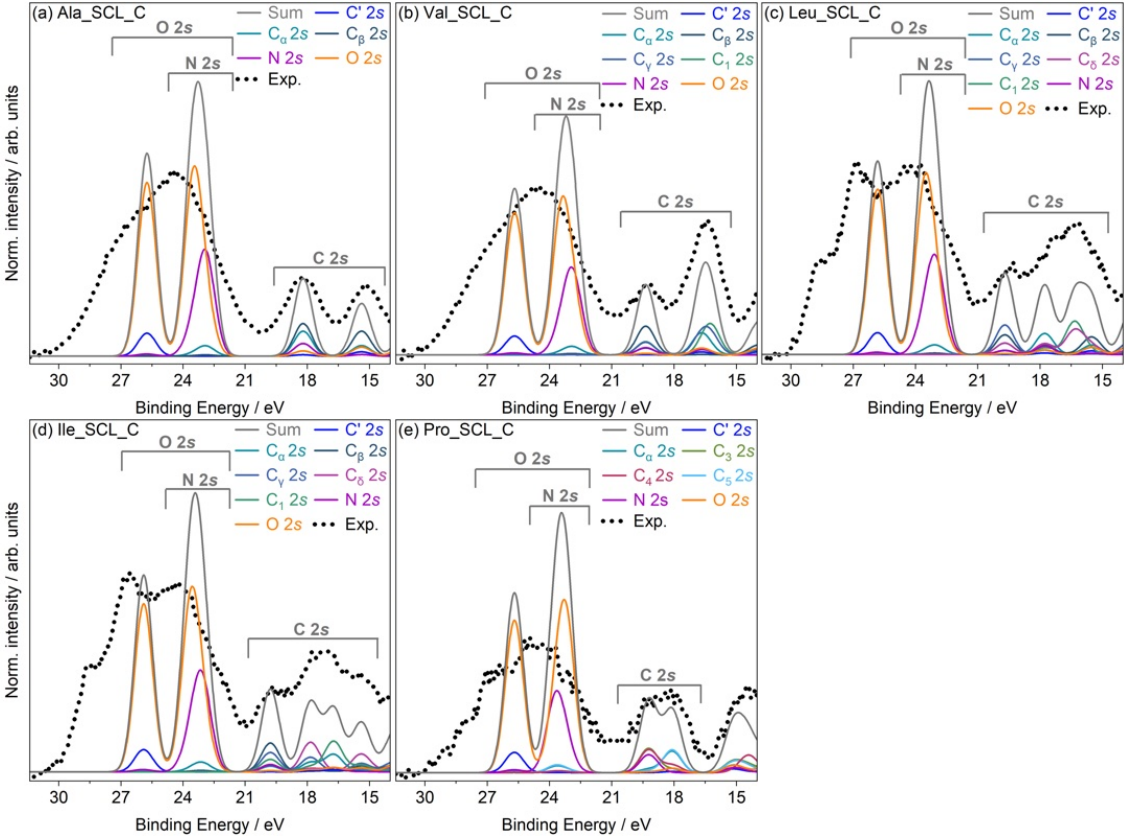}
    \caption{SCL spectra of the aliphatic AAs with contributions projected onto specific C atoms, including (a) Ala, (b) Val, (c) Leu, (d) Ile, and (e) Pro. All plots display the one-electron photoionisation cross-section weighted PDOS, as well as the sum of all PDOS, from PBE-based DFT calculations and the experimental XP spectra. To better visualise the contributions from specific C atoms, only 2\textit{s} states are shown as 2\textit{p} states do not present contributions in the SCL region. The labels in dark grey indicate the majority orbital contribution to the spectral features determined from DFT. The weighted PDOS have been aligned and normalised to the experimental BE peak of the highest theoretical C~2\textit{s} feature.}
    \label{Ali_SCL_C}
\end{figure}

\begin{figure}[htp]
\centering
    \includegraphics[keepaspectratio, width = 1.0\linewidth]{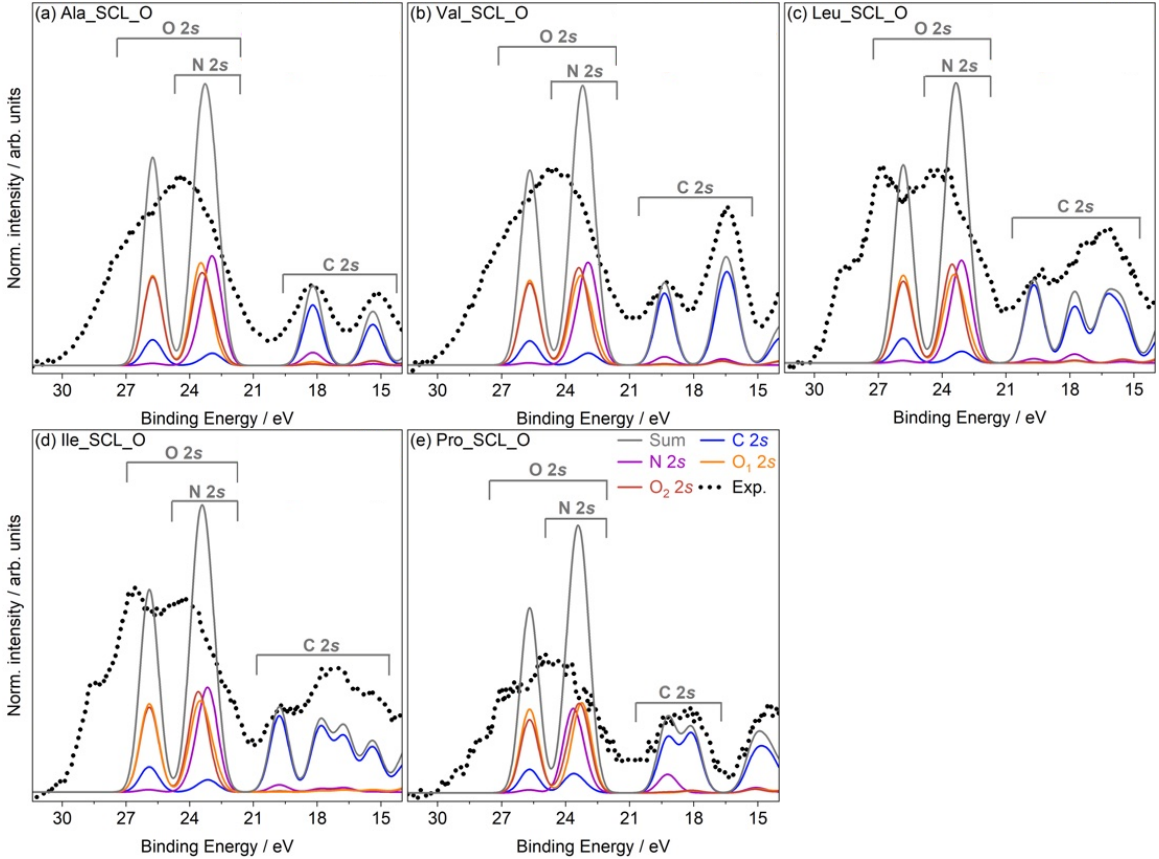}
    \caption{SCL spectra of the aliphatic AAs with contributions projected onto specific O atoms, including (a) Ala, (b) Val, (c) Leu, (d) Ile, and (e) Pro. All plots display the one-electron photoionisation cross-section weighted PDOS, as well as the sum of all PDOS, from PBE-based DFT calculations and the experimental XP spectra. To better visualise the contributions from specific O atoms, only 2\textit{s} states are shown as 2\textit{p} states do not present contributions in the SCL region. The labels in dark grey indicate the majority orbital contribution to the spectral features determined from DFT. The weighted PDOS have been aligned and normalised to the experimental BE peak of the highest theoretical C~2\textit{s} feature.}
    \label{Ali_SCL_O}
\end{figure}

\textbf{\begin{table}
    \caption{Four 2\textit{s} peak area ratios, including \ce{C$'$}:\ce{O1}; \ce{C$'$}:\ce{O2}; \ce{N1}:O; and \ce{C_$\alpha$}:\ce{N1}, based on PDOS calculations projected onto specific C atoms, which were determined using the integration function in the Origin software package for aliphatic AAs, as well as their corresponding bond lengths, $r$, determined by visualising the relaxed crystal structure in the VESTA software package.~\cite{Momma2011VESTA3Data}}
\begin{center}
\begin{tabular}{c c c c c c} 
 \hline
 ~ & Ala & Val & Leu & Ile & Pro \\ [0.5ex] 
 \hline
\rowcolor{lightgray} \ce{C$'$}:\ce{O1} & 1:3.48 & 1:3.50 & 1:3.61 & 1:3.57 & 1:3.74 \\
 \multirow{2}{*}{$r_{C'-O_1}$ / \(\text{\AA}\)} & \multirow{2}{*}{1.27631} & 1.26998 & 1.26993 & 1.26665 & \multirow{2}{*}{1.27040} \\ 
 ~ & ~ & 1.27657 & 1.27951 & 1.27694 & ~ \\
\rowcolor{lightgray} \ce{C$'$}:\ce{O2} & 1:3.45 & 1:3.34 & 1:3.30 & 1:3.38 & 1:3.08 \\ 
 \multirow{2}{*}{$r_{C'-O_2}$ / \(\text{\AA}\)} & \multirow{2}{*}{1.25753} & 1.25846 & 1.25803 & 1.25763 & \multirow{2}{*}{1.26115} \\ 
 ~ & ~ & 1.26415 & 1.26606 & 1.26721 & ~ \\
\rowcolor{lightgray} \ce{N1}:O & 1:1.33 & 1:1.32 & 1:1.32 & 1:3.34 & 1:1.53 \\ 
\rowcolor{lightgray} \ce{C_$\alpha$}:\ce{N1} & 1:10.61 & 1:10.74 & 1:10.78 & 1:10.71 & 1:12.11 \\
 \multirow{2}{*}{$r_{C_\alpha-N_1}$ / \(\text{\AA}\)} & \multirow{2}{*}{1.49339} & 1.49377 & 1.49398 & 1.49609 & \multirow{2}{*}{1.50516} \\
 ~ & ~ & 1.49949 & 1.49718 & 1.49616 & ~ \\
 \hline
\end{tabular}
\end{center}
\label{area ratio table_Ali}
\end{table}} 

\begin{figure}[!ht]
    \centering
    \includegraphics[width=1.0\textwidth]{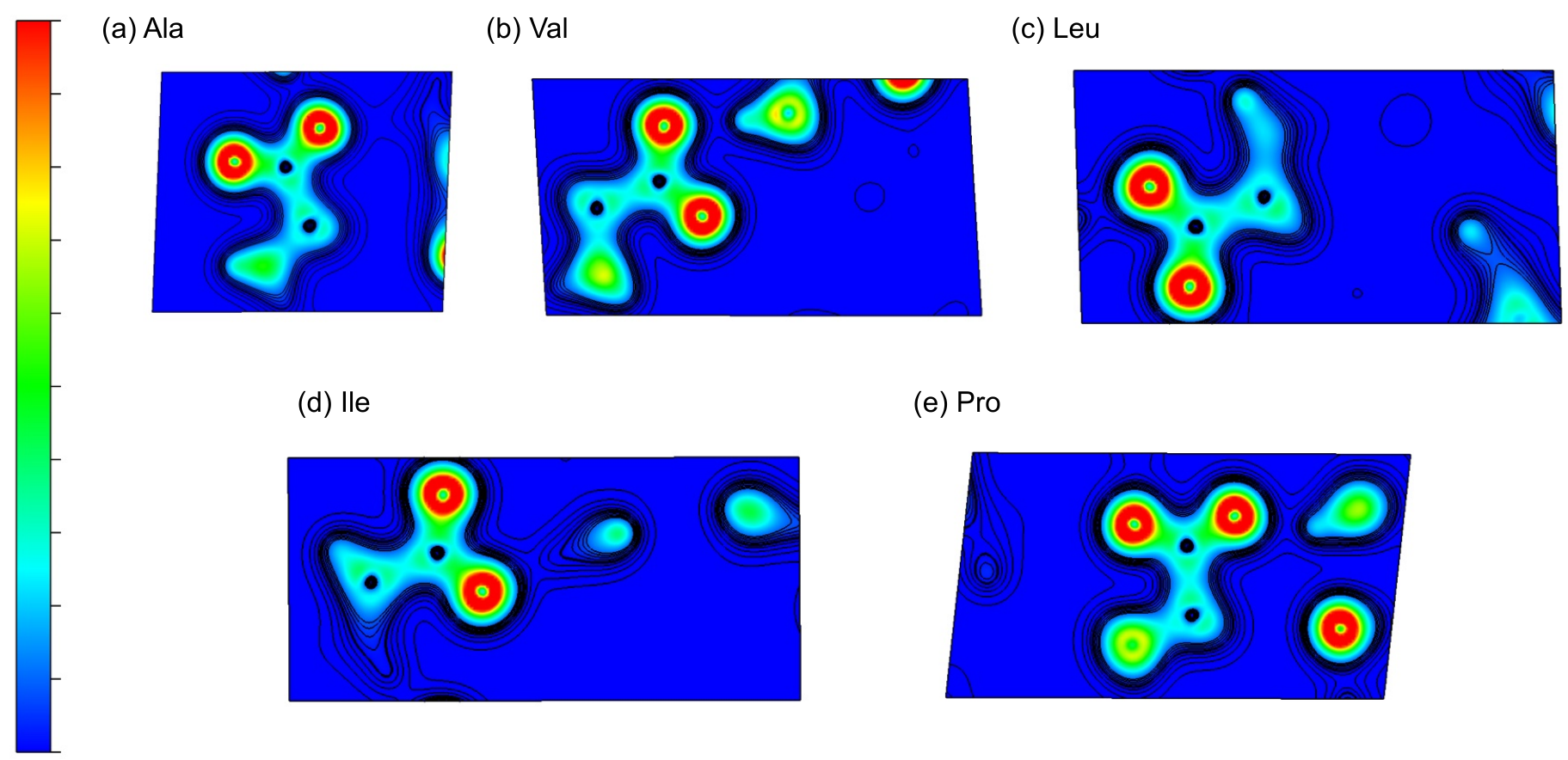}
    \caption{Visualisation of the electron density distribution around the \ce{COO^-} group of the aliphatic AAs for (a) Ala, (b) Val, (c) Leu, (d) Ile, and (e) Pro, with contour lines present. The colour gradient on the left presents how electron density changes, where blue represents the least electron density, and red represents the highest electron density. All electron density distribution figures were prepared in the VESTA software package.~\cite{Momma2011VESTA3Data}}
    \label{Ali_electron density}
\end{figure}

\textbf{\begin{table}
    \caption{Summary of the calculated Mulliken bond population, Mulliken charges, and the bond length of $r_{C'-O_1}$ and $r_{C'-O_1}$ for the aliphatic group, including different conformations for Val, Leu, and Ile. Different conformations observed in the crystal structure for Val, Leu, and Ile are distinguished by the different torsion angles, denoted as \ce{\Psi_x} in the bracket. n(\ce{C$'$}\ce{O1}) and n(\ce{C$'$}\ce{O2}) represent the Mulliken bond population between the \ce{C$'$} and \ce{O1}, and between \ce{C$'$} and \ce{O2}, respectively.}
\begin{center}
\begin{tabular}{c c c c c c c} 
 \hline
 AAs & \ce{O1} / e & n(\ce{C$'$}\ce{O1}) & $r_{C'-O_1}$ / \(\text{\AA}\) & \ce{O2} / e & n(\ce{C$'$}\ce{O2}) & $r_{C'-O_2}$ / \(\text{\AA}\) \\ [0.5ex] 
 \hline
 Ala & -0.674 & 0.87 & 1.27631 & -0.679 & 0.94 & 1.25752 \\
 Val(\ce{\Psi_1}) & -0.687 & 0.89 & 1.26998 & -0.660 & 0.92 & 1.26415 \\ 
 Val(\ce{\Psi_2}) & -0.686 & 0.87 & 1.27657 & -0.667 & 0.93 & 1.25846 \\
 Leu(\ce{\Psi_1}) & -0.695 & 0.90 & 1.26993 & -0.669 & 0.91 & 1.26606 \\ 
 Ile(\ce{\Psi_2}) & -0.682 & 0.86 & 1.27951 & -0.663 & 0.93 & 1.25803 \\ 
 Ile(\ce{\Psi_1}) & -0.690 & 0.91 & 1.26721 & -0.662 & 0.90 & 1.26665 \\
 Ile(\ce{\Psi_2}) & -0.683 & 0.87 & 1.27694 & -0.663 & 0.93 & 1.25763 \\ 
 Pro & -0.672 & 0.89 & 1.27040 & -0.681 & 0.92 & 1.26115 \\
 \hline
\end{tabular}
\end{center}
\label{bond populaion_Ali}
\end{table}} 

\begin{figure}[htp]
    \centering
    \includegraphics[width=0.6\textwidth]{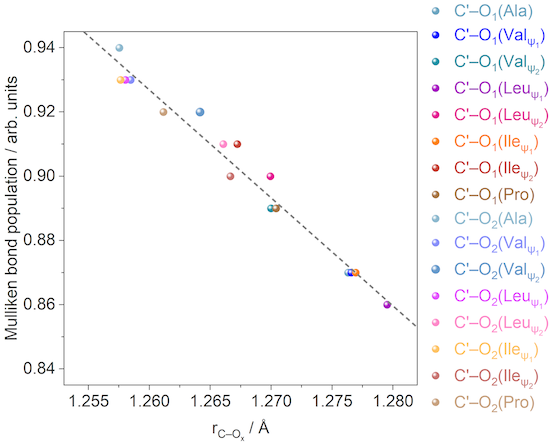}
    \caption{Correlation between the Mulliken bond population between the \ce{C$'$} and \ce{O1}, and between \ce{C$'$} and \ce{O2}, respectively, and the corresponding bond lengths, $r_{C'-O_1}$, and $r_{C'-O_2}$ for the aliphatic group, including different conformations for Val, Leu, and Ile.}
    \label{Ali_population correlation}
\end{figure}

\begin{table}
    \caption{Various bond lengths of $r_{C'-O_1}$ and $r_{C'-O_2}$ and the corresponding bond distances of $d_{O_1-H}$, $d_{O_2-H}$, and $r_{H-N}$ of the aliphatic AAs, including different conformations of Val, Leu, and Ile distinguished by the torsion angle, $\psi$. Bond lengths were determined by visualising the relaxed crystal structure in the VESTA software package.~\cite{Momma2011VESTA3Data} The letters and torsion angles in the bracket denote the specific columns and orientations of molecules involved in the formation of intermolecular hydrogen bonds.}
\begin{center}
\begin{tabular}{c c c c c c c c} 
 \hline
 AAs & ~ & \multicolumn{3}{c}{~} & \multicolumn{3}{c}{~} \\ [0.5ex] 
 \hline
 \multirow{5}{*}{\textbf{Ala}} & ~ & \multicolumn{4}{c}{\ce{O1}} & \multicolumn{2}{c}{\ce{O2}} \\
 ~ & $r_{C'-O}$ / \(\text{\AA}\) & \multicolumn{4}{c}{1.27631} & \multicolumn{2}{c}{1.25752} \\ 
 ~ & \multirow{2}{*}{$d_{O-H}$ / \(\text{\AA}\)} & \multicolumn{2}{c}{1.71128} & \multicolumn{2}{c}{1.76454} & \multicolumn{2}{c}{1.78438} \\ 
 ~ & ~ & \multicolumn{2}{c}{(ab)} & \multicolumn{2}{c}{(aa)} & \multicolumn{2}{c}{(ad)} \\
 ~ & $r_{H-N}$ / \(\text{\AA}\) & \multicolumn{2}{c}{1.06054} & \multicolumn{2}{c}{1.04710} & \multicolumn{2}{c}{1.04335} \\ 
 \multirow{6}{*}{\textbf{Val}} & ~ & \multicolumn{3}{c}{$\psi_1=17.544^\circ$} & \multicolumn{3}{c}{$\psi_2=41.430^\circ$} \\
 ~ & ~ & \multicolumn{2}{c}{\ce{O1}} & \ce{O2} & \multicolumn{2}{c}{\ce{O1}} & \ce{O2} \\
 ~ & $r_{C'-O}$ / \(\text{\AA}\) & \multicolumn{2}{c}{1.26998} & 1.26451 & \multicolumn{2}{c}{1.27657} & 1.25846 \\ 
 ~ & \multirow{2}{*}{$d_{O-H}$ / \(\text{\AA}\)} & 1.77366 & 1.79073 & 1.69430 & 1.65453 & 1.95721 & 1.81105 \\ 
 ~ & ~ & (aa $\&$ $\psi_1\psi_1$) & (aa $\&$ $\psi_1\psi_2$) & (ab $\&$ $\psi_1\psi_2$) & (ab $\&$ $\psi_1\psi_1$) & (ab $\&$ $\psi_1\psi_2$) & (aa $\&$ $\psi_1\psi_1$) \\
 ~ & $r_{H-N}$ / \(\text{\AA}\) & 1.04847 & 1.05051 & 1.05114 & 1.06494 & 1.03630 & 1.04904 \\ 
 \multirow{6}{*}{\textbf{Leu}} & ~ & \multicolumn{3}{c}{$\psi_1=26.953^\circ$} & \multicolumn{3}{c}{$\psi_2=32.462^\circ$} \\ 
 ~ & ~ & \ce{O1} & \multicolumn{2}{c}{\ce{O2}} & \multicolumn{2}{c}{\ce{O1}} & \ce{O2} \\
 ~ & $r_{C'-O}$ / \(\text{\AA}\) & 1.26993 & \multicolumn{2}{c}{1.26606} & \multicolumn{2}{c}{1.27951} & 1.25803 \\ 
 ~ & \multirow{2}{*}{$d_{O-H}$ / \(\text{\AA}\)} & 1.67455 & 1.90674 & 1.81687 & 1.64021 & 1.82202 & 1.78016 \\ 
 ~ & ~ & (ab $\&$ $\psi_1\psi_2$) & (aa $\&$ $\psi_1\psi_2$) & (aa $\&$ $\psi_1\psi_1$) & (aa $\&$ $\psi_1\psi_2$) & (ab $\&$ $\psi_1\psi_2$) & (aa $\&$ $\psi_1\psi_1$) \\
 ~ & $r_{H-N}$ / \(\text{\AA}\) & 1.05162 & 1.04292 & 1.04459 & 1.06628 & 1.04115 & 1.05080 \\
 \multirow{6}{*}{\textbf{Ile}} & ~ & \multicolumn{3}{c}{$\psi_1=19.271^\circ$} & \multicolumn{3}{c}{$\psi_2=42.138^\circ$} \\ 
 ~ & ~ & \ce{O1} & \multicolumn{2}{c}{\ce{O2}} & \multicolumn{2}{c}{\ce{O1}} & \ce{O2} \\
 ~ & $r_{C'-O}$ / \(\text{\AA}\) & 1.26721 & \multicolumn{2}{c}{1.26665} & \multicolumn{2}{c}{1.27694} & 1.25763 \\ 
 ~ & \multirow{2}{*}{$r_{O-H}$ / \(\text{\AA}\)} & 1.68015 & 1.77221 & 1.87328 & 1.65827 & 1.89113 & 1.77935 \\ 
 ~ & ~ & (ab $\&$ $\psi_1\psi_2$) & (aa $\&$ $\psi_1\psi_2$) & (aa $\&$ $\psi_1\psi_1$) & (aa $\&$ $\psi_1\psi_2$) & (ab $\&$ $\psi_1\psi_2$) & (aa $\&$ $\psi_1\psi_1$) \\
 ~ & $r_{H-N}$ / \(\text{\AA}\) & 1.05113 & 1.05008 & 1.04547 & 1.06509 & 1.03754 & 1.05213 \\
 \multirow{5}{*}{\textbf{Pro}} & ~ & \multicolumn{4}{c}{\ce{O1}} & \multicolumn{2}{c}{\ce{O2}} \\
 ~ & $r_{C'-O}$ / \(\text{\AA}\) & \multicolumn{4}{c}{1.27040} & \multicolumn{2}{c}{1.26115} \\ 
 ~ & \multirow{2}{*}{$d_{O-H}$ / \(\text{\AA}\)} & \multicolumn{4}{c}{1.63724} & \multicolumn{2}{c}{1.75326} \\ 
 ~ & ~ & \multicolumn{4}{c}{(aa)} & \multicolumn{2}{c}{(ad)} \\
 ~ & $r_{H-N}$ / \(\text{\AA}\) & \multicolumn{4}{c}{1.06888} & \multicolumn{2}{c}{1.05007} \\ 
 \hline
\end{tabular}
\end{center}
\label{bond length study_Ali}
\end{table}

\clearpage
\newpage
\subsection{Aromatic Group}

\begin{figure}[htp]
\centering
    \includegraphics[keepaspectratio, width = 1.0\linewidth]{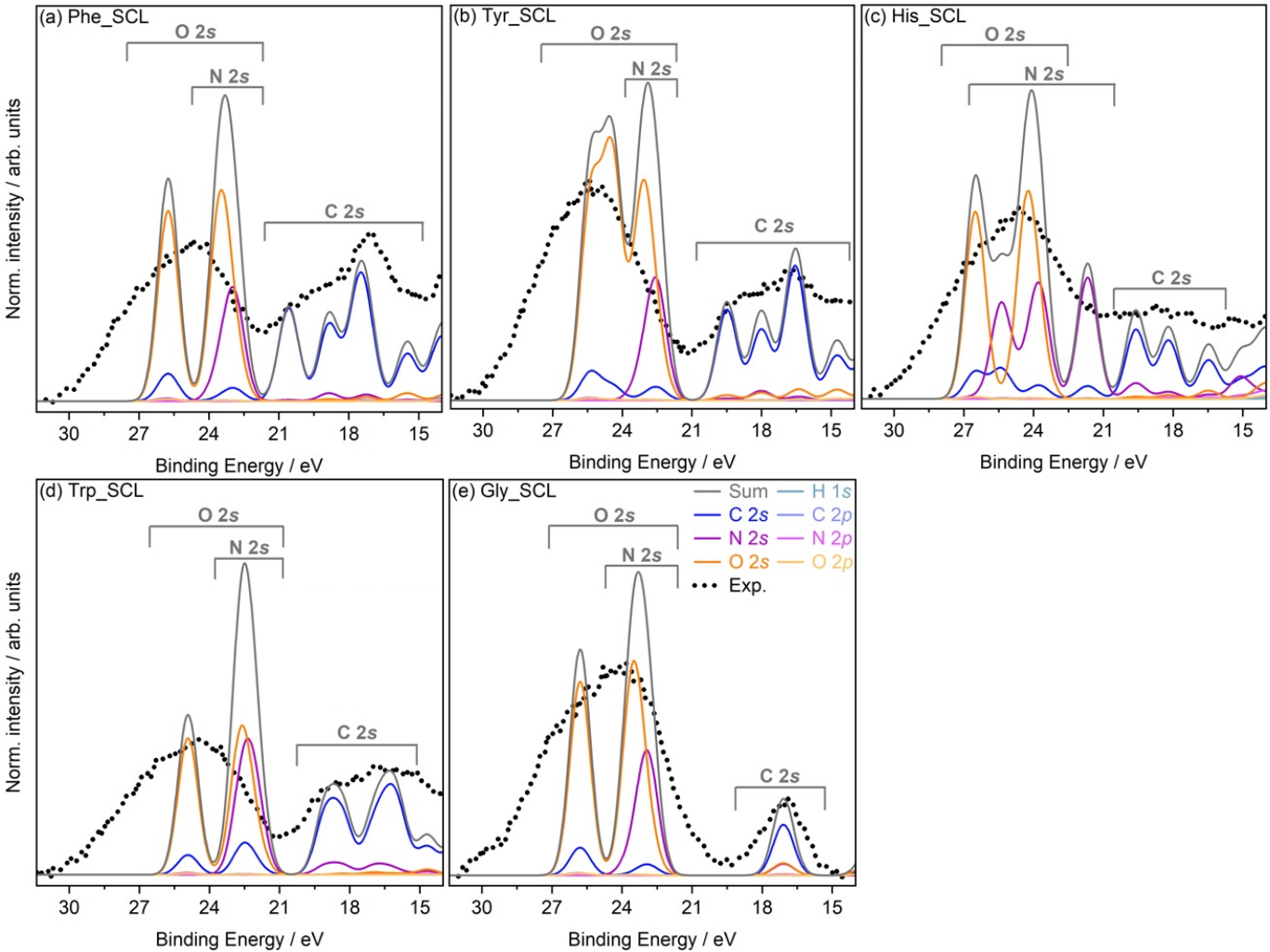}
    \caption{SCL spectra of the aromatic AAs, including (a) Phe, (b) Tyr, (c) His, (d) Trp, and (e) Gly as reference. All plots display the one-electron photoionisation cross-section weighted PDOS, as well as the sum of all PDOS, from PBE-based DFT calculations and the experimental XP spectra. The labels in dark grey indicate the majority orbital contribution to the spectral features determined from DFT. The weighted PDOS have been aligned and normalised to the experimental BE peak of the highest theoretical C~2\textit{s} feature. The legend shown in (e) applies to all subfigures.}
    \label{Aro_SCL}
\end{figure}

\begin{figure}[htp]
\centering
    \includegraphics[keepaspectratio, width = 0.75\linewidth]{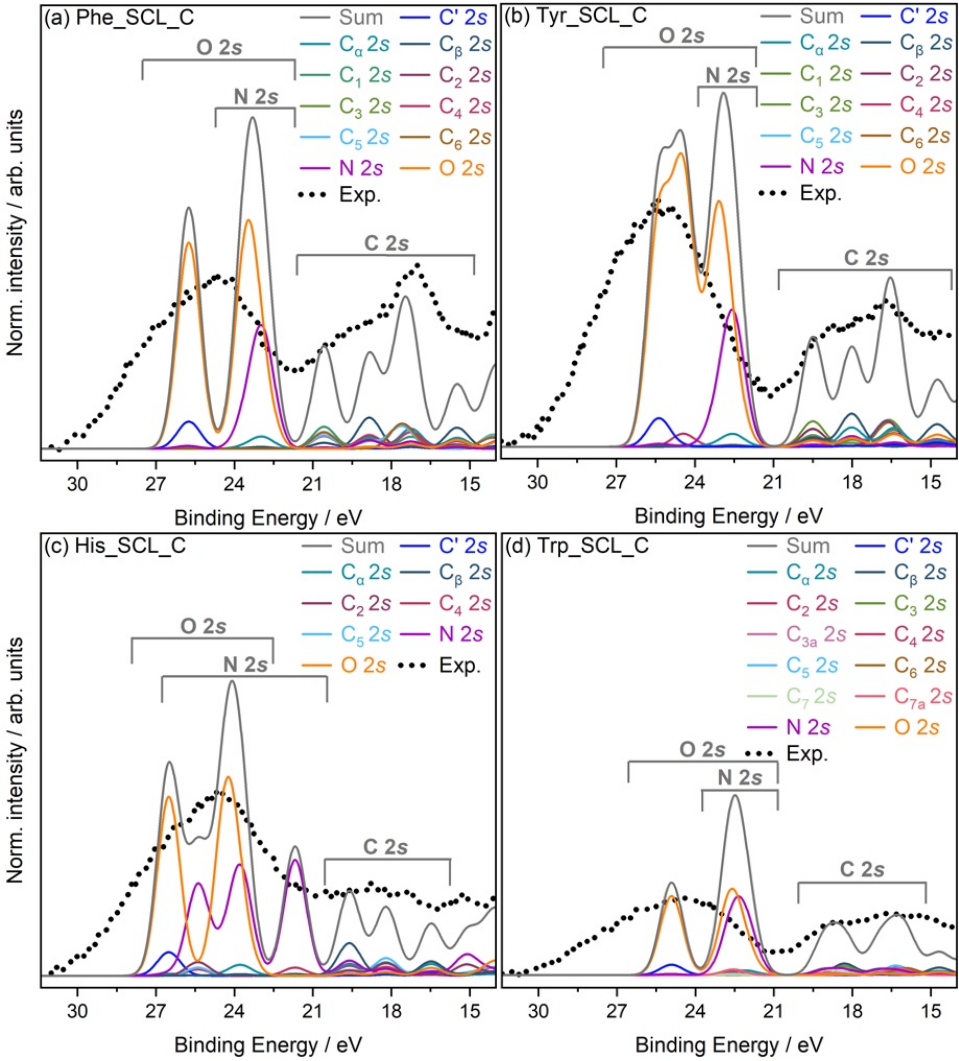}
    \caption{SCL spectra of the aromatic AAs, including (a) Phe, (b) Tyr, (c) His, and (d) Trp with contributions projected onto specific C atoms. All plots display the one-electron photoionisation cross-section weighted PDOS, as well as the sum of all PDOS, from PBE-based DFT calculations and the experimental XP spectra. To better visualise the contributions from specific C atoms, only 2\textit{s} states are shown as 2\textit{p} states do not present contributions in the SCL region. The labels in dark grey indicate the majority orbital contribution to the spectral features determined from DFT. The weighted PDOS have been aligned and normalised to the experimental BE peak of the highest theoretical C~2\textit{s} feature.}
    \label{Aro_SCL_C}
\end{figure}

\begin{figure}[htp]
\centering
    \includegraphics[keepaspectratio, width = 0.75\linewidth]{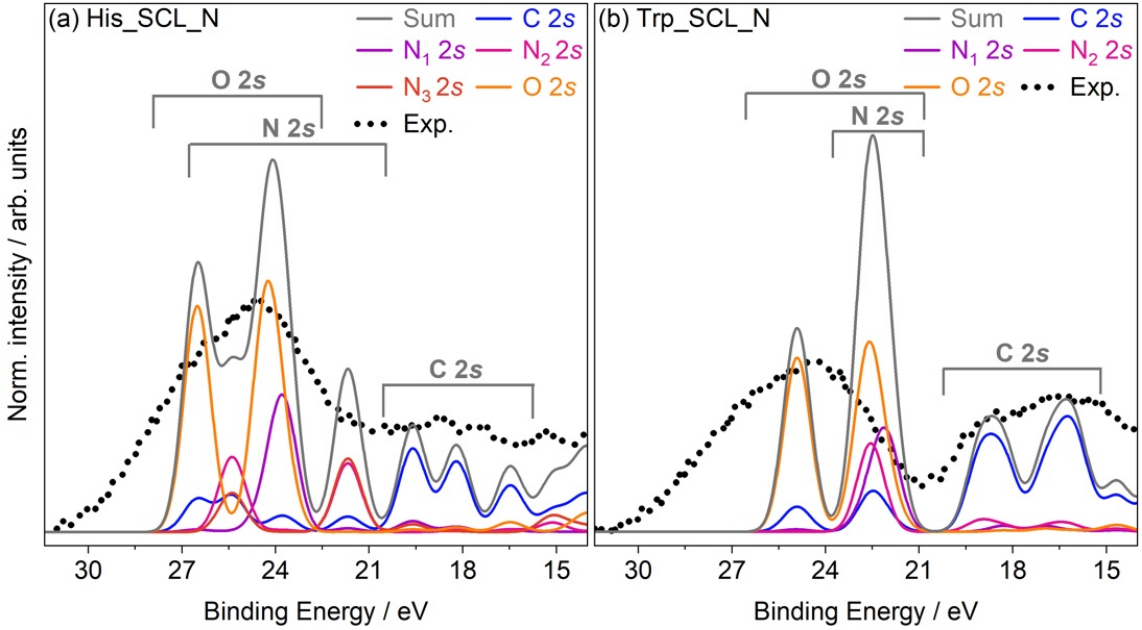}
    \caption{SCL spectra of the aromatic AAs, including (a) His, and (b) Trp with contributions projected onto specific N atoms. All plots display the one-electron photoionisation cross-section weighted PDOS, as well as the sum of all PDOS, from PBE-based DFT calculations and the experimental XP spectra. To better visualise the contributions from specific N atoms, only 2\textit{s} states are shown as 2\textit{p} states do not present contributions in the SCL region. The labels in dark grey indicate the majority orbital contribution to the spectral features determined from DFT. The weighted PDOS have been aligned and normalised to the experimental BE peak of the highest theoretical C~2\textit{s} feature.}
    \label{Aro_SCL_N}
\end{figure}

\begin{figure}[htp]
\centering
    \includegraphics[keepaspectratio, width = 0.75\linewidth]{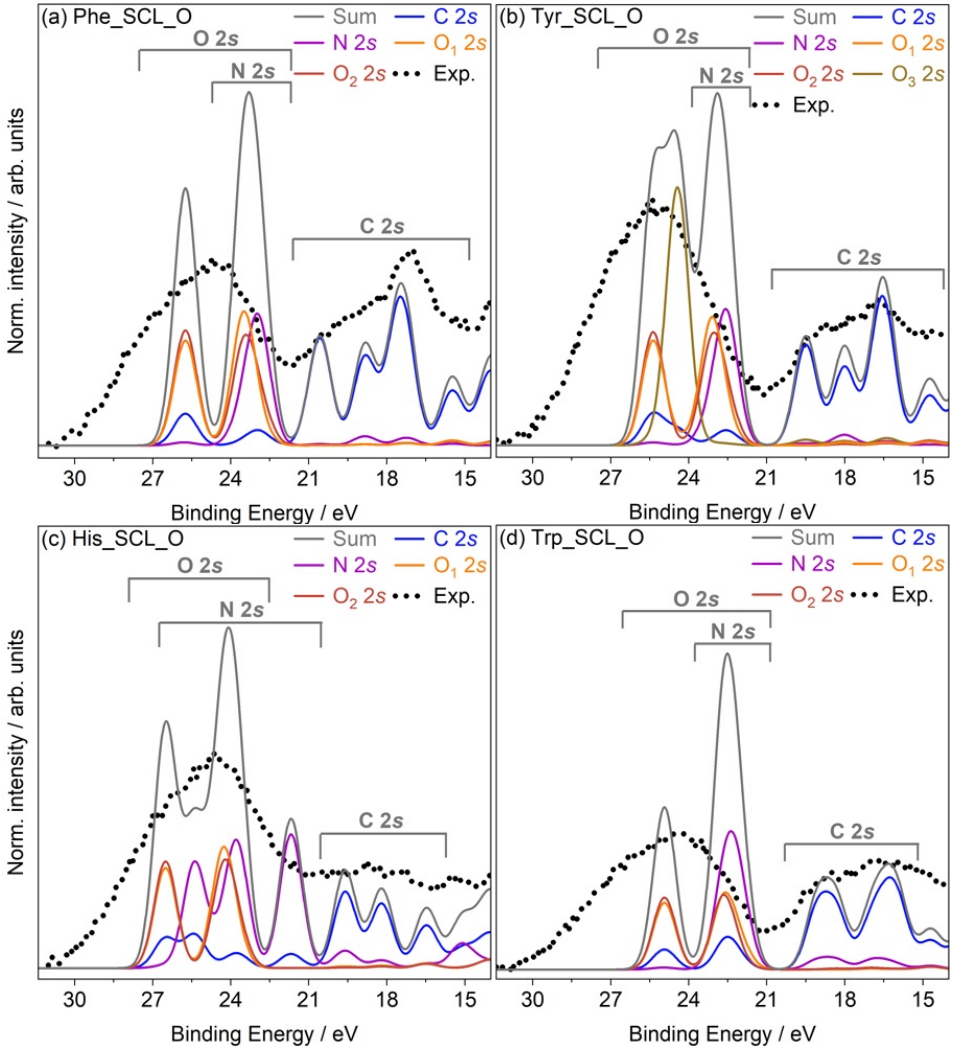}
    \caption{SCL spectra of the aromatic AAs, including (a) Phe, (b) Tyr, (c) His, and (d) Trp with contributions projected onto specific O atoms. All plots display the one-electron photoionisation cross-section weighted PDOS, as well as the sum of all PDOS, from PBE-based DFT calculations and the experimental XP spectra. To better visualise the contributions from specific O atoms, only 2\textit{s} states are shown as 2\textit{p} states do not present contributions in the SCL region. The labels in dark grey indicate the majority orbital contribution to the spectral features determined from DFT. The weighted PDOS have been aligned and normalised to the experimental BE peak of the highest theoretical C~2\textit{s} feature.}
    \label{Aro_SCL_O}
\end{figure}

\begin{table}[htp]
    \caption{Four common 2\textit{s} peak area ratios, including \ce{C$'$}:\ce{O1}; \ce{C$'$}:\ce{O2}; \ce{N1}:O; and \ce{C_$\alpha$}:\ce{N1} for all AAs in the group; and multiple additional 2\textit{s} peak area ratios, containing \ce{C4}:\ce{O3} for Tyr, \ce{C2}:\ce{N2}, and \ce{C_{7a}}:\ce{N2} for Trp, based on PDOS calculations projected onto specific C, N and O atoms, which were determined using the integration function in the Origin software package for the aromatic AAs, as well as their corresponding bond lengths, $r$, determined by visualising the relaxed crystal structure in the VESTA software package.~\cite{Momma2011VESTA3Data} N.A. in the table represents not applicable to the specific AAs as these environments do not exist.}
\begin{center}
\begin{tabular}{c c c c c} 
 \hline
 ~ & Phe & Tyr & His & Trp \\ [0.5ex] 
 \hline
 \rowcolor{lightgray} \ce{C$'$}:\ce{O1} & 1:3.27 & 1:3.28 & 1:3.26 & 1:3.28 \\
 \multirow{4}{*}{$r_{C'-O_1}$ / \(\text{\AA}\)} & 1.26834 & \multirow{4}{*}{1.27441} & \multirow{4}{*}{1.26828} & 1.27509 \\
 ~ & 1.26848 & ~ & ~ & 1.27744 \\
 ~ & 1.26798 & ~ & ~ & 1.27698 \\
 ~ & 1.26797 & ~ & ~ & 1.27479 \\
 \rowcolor{lightgray} \ce{C$'$}:\ce{O2} & 1:3.65 & 1:3.63 & 1:3.61& 1:3.53 \\ 
 \multirow{4}{*}{$r_{C'-O_2}$ / \(\text{\AA}\)} & 1.26501 & \multirow{4}{*}{1.26006} & \multirow{4}{*}{1.26063} & 1.25988 \\ 
 ~ & 1.26565 & ~ & ~& 1.25846 \\
 ~ & 1.26461 & ~ & ~ & 1.25774 \\
 ~ & 1.26660 & ~ & ~& 1.25874 \\
 \rowcolor{lightgray} \ce{N1}:O & 1:1.35 & 1:1.31 & 1:1.40 & 1:1.31\\ 
 \rowcolor{lightgray} \ce{C_$\alpha$}:\ce{N1} & 1:10.55 & 1:11.13 & 1:10.32 & 1:10.41 \\
 \multirow{4}{*}{$r_{C_\alpha-N_1}$ / \(\text{\AA}\)} & 1.49793 & \multirow{4}{*}{1.49015} & \multirow{4}{*}{1.49332} & 1.48834 \\
 ~ & 1.49880 & ~ & ~ & 1.49101 \\
 ~ & 1.49835 & ~ & ~ & 1.49044 \\
 ~ & 1.49808 & ~ & ~ & 1.48854 \\
 \rowcolor{lightgray} \ce{C4}:\ce{O3} & N.A. & 1:20.12 & N.A. & N.A. \\
 $r_{C_4-O_3}$ / \(\text{\AA}\) & N.A. & 1.36800 & N.A. & N.A. \\
 \rowcolor{lightgray} \ce{C2}:\ce{N2} & N.A. & N.A. & N.A. & 1:8.35 \\
 \multirow{4}{*}{$r_{C_2-N_2}$ / \(\text{\AA}\)} & \multirow{4}{*}{N.A.} & \multirow{4}{*}{N.A.} & \multirow{4}{*}{N.A.} & 1.38464 \\
 ~ & ~ & ~ & ~ & 1.38313 \\
 ~ & ~ & ~ & ~ & 1.38087 \\
 ~ & ~ & ~ & ~ & 1.38252 \\
 \rowcolor{lightgray} \ce{C_{7a}}:\ce{N2} & N.A. & N.A. & N.A. & 1:7.16 \\
 \multirow{4}{*}{$r_{C_{7a}-N_2}$ / \(\text{\AA}\)} & \multirow{4}{*}{N.A.} & \multirow{4}{*}{N.A.} & \multirow{4}{*}{N.A.} & 1.37932 \\
 ~ & ~ & ~ & ~ & 1.38101 \\
 ~ & ~ & ~ & ~ & 1.37903 \\
 ~ & ~ & ~ & ~ & 1.37724 \\
 \hline
\end{tabular}
\end{center}
\label{area ratio table_Aro}
\end{table} 

\textbf{\begin{table}[htp]
    \caption{Various bond lengths of $r_{C'-O_1}$, and $r_{C'-O_2}$, and the corresponding bond distances of $d_{O_1-H}$, $d_{O_2-H}$, and $r_{H-N_x}$ for Tyr and His. Additional bond length $r_{C_4-O_3}$ and the corresponding bond distances $r_{O_3-H}$, $d_{H-O_2}$ and $r_{H-N_1}$ for Tyr are also included. Letters in the brackets indicate the intermolecular hydrogen bonds are formed between which two columns and atoms. Bond lengths were determined by visualising the relaxed crystal structure in the VESTA software package.~\cite{Momma2011VESTA3Data} The letters in the brackets denote the specific columns of molecules involved in the formation of intermolecular hydrogen bonds. N.A. in the table represents not applicable.}
\begin{center}
\begin{tabular}{c c c c c c c} 
 \hline
 AAs & ~ & \multicolumn{2}{c}{\ce{O1}} & \ce{O2} & \multicolumn{2}{c}{\ce{O3}} \\ [0.5ex] 
 \hline
 \multirow{5}{*}{\textbf{Tyr}} & $r_{C'-O_1}$ / \(\text{\AA}\) & \multicolumn{2}{c}{1.27441} & 1.26006 & \multicolumn{2}{c}{1.36800} \\
 ~ & \multirow{2}{*}{$d_{O-H}$ / \(\text{\AA}\)} & 1.71070 & 1.75196 & 1.61800 & 1.00254 & 2.00474 \\ 
 ~ & ~ & (ab) & (aa) & (aa) & (aa) & (aa) \\
 ~ & \multirow{2}{*}{$r_{H-N}$ / \(\text{\AA}\)} & 1.06023 & 1.055110 & N.A. & N.A. & 1.03603 \\ 
 ~ &  ~ & (\ce{N1}) & (\ce{N1}) & N.A. & N.A. & (\ce{N1}) \\
 \multirow{5}{*}{\textbf{His}} & $r_{C'-O_1}$ / \(\text{\AA}\) & \multicolumn{2}{c}{1.26828} & 1.26063 & \multicolumn{2}{c}{N.A.} \\
 ~ & \multirow{2}{*}{$d_{O-H}$ / \(\text{\AA}\)} & 1.72605 & 1.74713 & 1.69362 & \multicolumn{2}{c}{N.A.} \\ 
 ~ & ~ & (aa) & (ad) & (ac) & \multicolumn{2}{c}{N.A.} \\
 ~ & \multirow{2}{*}{$r_{H-N}$ / \(\text{\AA}\)} & 1.05173 & 1.04409 & 1.04980 & \multicolumn{2}{c}{N.A.} \\ 
 ~ &  ~ & (\ce{N1}) & (\ce{N1}) & (\ce{N2}) & \multicolumn{2}{c}{N.A.} \\
 \hline
\end{tabular}
\end{center}
\label{bond length study1_aro}
\end{table}}

\textbf{\begin{table}[htp]
    \caption{Various bond lengths of $r_{C'-O_1}$ and $r_{C'-O_2}$ and the corresponding bond distances of $d_{O_1-H}$, $d_{O_2-H}$, and $r_{H-N}$ of different conformations of Phe and Trp, distinguished by the torsion angle, $\psi$. Bond lengths were determined by visualising the relaxed crystal structure in the VESTA software package.~\cite{Momma2011VESTA3Data} The letters and torsion angles in the brackets denote the specific columns and orientations of molecules involved in the formation of intermolecular hydrogen bonds.}
\begin{center}
\begin{tabular}{c c c c c c c c}     
\hline
 \textbf{AAs} & ~ & \multicolumn{3}{c}{$\psi_x$} & \multicolumn{3}{c}{$\psi_x$} \\
 \hline
 \multirow{12}{*}{\textbf{Phe}} & ~ & \multicolumn{3}{c}{$\psi_1=131.488^\circ$} & \multicolumn{3}{c}{$\psi_2=137.237^\circ$} \\
 ~ & ~ & \multicolumn{2}{c}{\ce{O1}} & \ce{O2} & \multicolumn{2}{c}{\ce{O1}} & \ce{O2} \\
 ~ & $r_{C'-O}$ / \(\text{\AA}\) & \multicolumn{2}{c}{1.26848} & 1.26565 & \multicolumn{2}{c}{1.26834} & 1.26501 \\ 
 ~ & \multirow{2}{*}{$d_{O-H}$ / \(\text{\AA}\)} & 1.81651 & 1.84954 & 1.59974 & 1.79392 & 1.79797 & 1.59122 \\ 
 ~ & ~ & (aa $\&$ $\psi_1\psi_2$) & (ab $\&$ $\psi_1\psi_2$) & (aa $\&$ $\psi_1\psi_2$) & (aa $\&$ $\psi_2\psi_1$) & (ab $\&$ $\psi_2\psi_1$) & (aa $\&$ $\psi_2\psi_1$) \\
 ~ & $r_{H-N}$ / \(\text{\AA}\) & 1.04448 & 1.04414 & 1.07495 & 1.04444 & 1.04765 & 1.07491 \\ 
 ~ & ~ & \multicolumn{3}{c}{$\psi_3=133.331^\circ$} & \multicolumn{3}{c}{$\psi_4=134.593^\circ$} \\ 
  ~ & ~ & \multicolumn{2}{c}{\ce{O1}} & \ce{O2} & \multicolumn{2}{c}{\ce{O1}} & \ce{O2} \\
 ~ & $r_{C'-O}$ / \(\text{\AA}\) & \multicolumn{2}{c}{1.26797} & 1.26660 & \multicolumn{2}{c}{1.26798} & 1.26461 \\ 
 ~ & \multirow{2}{*}{$d_{O-H}$ / \(\text{\AA}\)} & 1.78730 & 1.87892 & 1.58334 & 1.78118 & 1.81719 & 1.59265 \\ 
 ~ & ~ & (cc $\&$ $\psi_3\psi_4$) & (cd $\&$ $\psi_3\psi_4$) & (cc $\&$ $\psi_3\psi_4$) & (cc $\&$ $\psi_4\psi_3$) & (cd $\&$ $\psi_4\psi_3$) & (cc $\&$ $\psi_4\psi_3$) \\
 ~ & $r_{H-N}$ / \(\text{\AA}\) & 1.04556 & 1.04343 & 1.07875 & 1.04948 & 1.04265 & 1.07477 \\
 \multirow{12}{*}{\textbf{Trp}} & ~ & \multicolumn{3}{c}{$\psi_1=15.000^\circ$} & \multicolumn{3}{c}{$\psi_2=16.310^\circ$} \\
 ~ & ~ & \multicolumn{2}{c}{\ce{O1}} & \ce{O2} & \multicolumn{2}{c}{\ce{O1}} & \ce{O2} \\
 ~ & $r_{C'-O}$ / \(\text{\AA}\) & \multicolumn{2}{c}{1.27509} & 1.25988 & \multicolumn{2}{c}{1.27744} & 1.25846 \\ 
 ~ & \multirow{2}{*}{$d_{O-H}$ / \(\text{\AA}\)} & 1.78494 & 1.79631 & 1.61077 & 1.76718 & 1.80583 & 2.08586 \\ 
 ~ & ~ & (ab $\&$ $\psi_1\psi_3$) & (ab $\&$ $\psi_1\psi_4$) & (aa $\&$ $\psi_1\psi_1$) & (ab $\&$ $\psi_2\psi_4$) & (ab $\&$ $\psi_2\psi_3$) & (ab $\&$ $\psi_2\psi_1$) \\
 ~ & \multirow{2}{*}{$r_{H-N}$ / \(\text{\AA}\)} & 1.04548 & 1.04352 & 1.06927 & 1.04412 & 1.04499 & 1.08619 \\ 
 ~ & ~ & (\ce{N1}) & (\ce{N1}) & (\ce{N1}) & (\ce{N1}) & (\ce{N1}) & (\ce{C2}) \\
 ~ & \multicolumn{3}{c}{$\psi_3=14.582^\circ$} & \multicolumn{3}{c}{$\psi_4=17.198^\circ$} \\
 ~ & ~ & \multicolumn{2}{c}{\ce{O1}} & \ce{O2} & \multicolumn{2}{c}{\ce{O1}} & \ce{O2} \\
 ~ & $r_{C'-O}$ / \(\text{\AA}\) & \multicolumn{2}{c}{1.27698} & 1.25774 & \multicolumn{2}{c}{1.27479} & 1.25874 \\ 
 ~ & \multirow{2}{*}{$d_{O-H}$ / \(\text{\AA}\)} & 1.78409 & 1.80432 & 2.06777 & 1.76649 & 1.81023 & 1.62083 \\ 
 ~ & ~ & (ba $\&$ $\psi_3\psi_1$) & (ba $\&$ $\psi_3\psi_2$) & (bb $\&$ $\psi_3\psi_4$) & (ba $\&$ $\psi_4\psi_1$) & (ba $\&$ $\psi_4\psi_2$) & (bb $\&$ $\psi_4\psi_4$) \\
 ~ & \multirow{2}{*}{$r_{H-N}$ / \(\text{\AA}\)} & 1.04380 & 1.04497 & 1.08547 & 1.04340 & 1.04375 & 1.06876 \\ 
 ~ & ~ & (\ce{N1}) & (\ce{N1}) & (\ce{C2}) & (\ce{N1}) & (\ce{N1}) & (\ce{N1}) \\
 \hline
\end{tabular}
\end{center}
\label{bond length study2_aro}
\end{table}}

\textbf{\begin{table}[htp]
    \caption{Summary of the calculated Mulliken bond population, Mulliken charges, and the bond length of $r_{C'-O_1}$ and $r_{C'-O_1}$ for the aromatic group, including different conformations for Phe, Tyr, and Trp. Different conformations observed in the crystal structure for Phe, Tyr, and Trp are distinguished by the different torsion angles, denoted as \ce{\Psi_x} in the bracket. n(\ce{C$'$}\ce{O1}) and n(\ce{C$'$}\ce{O2}) represent the Mulliken bond population between the \ce{C$'$} and \ce{O1}, and between \ce{C$'$} and \ce{O2}, respectively.}
\begin{center}
\begin{tabular}{c c c c c c c} 
 \hline
 AAs & \ce{O1} / e & n(\ce{C$'$}\ce{O1}) & $r_{C'-O_1}$ / \(\text{\AA}\) & \ce{O2} / e & n(\ce{C$'$}\ce{O2}) & $r_{C'-O_2}$ / \(\text{\AA}\) \\ [0.5ex] 
 \hline
 Phe(\ce{\Psi_1}) & -0.675 & 0.90 & 1.26848 & -0.657 & 0.91 & 1.26565 \\
 Phe(\ce{\Psi_2}) & -0.671 & 0.89 & 1.26834 & -0.657 & 0.91 & 1.26501 \\ 
 Phe(\ce{\Psi_3}) & -0.675 & 0.90 & 1.26797 & -0.659 & 0.91 & 1.26660 \\ 
 Phe(\ce{\Psi_4}) & -0.671 & 0.89 & 1.26798 & -0.656 & 0.92 & 1.26461 \\
 Tyr & -0.663 & 0.88 & 1.27441 & -0.653 & 0.93 & 1.26006 \\ 
 His & -0.676 & 0.89 & 1.26828 & -0.671 & 0.92 & 1.26063 \\
 Trp(\ce{\Psi_1}) & -0.707 & 0.87 & 1.27509 & -0.636 & 0.93 & 1.25988 \\
 Trp(\ce{\Psi_2}) & -0.708 & 0.87 & 1.27744 & -0.635 & 0.93 & 1.25846 \\ 
 Trp(\ce{\Psi_3}) & -0.708 & 0.87 & 1.27698 & -0.636 & 0.93 & 1.25774 \\ 
 Trp(\ce{\Psi_4}) & -0.706 & 0.88 & 1.27479 & -0.640 & 0.93 & 1.25874 \\
 \hline
\end{tabular}
\end{center}
\label{bond populaion_aro}
\end{table}} 

\begin{figure}[htp]
    \centering
    \includegraphics[width=0.6\textwidth]{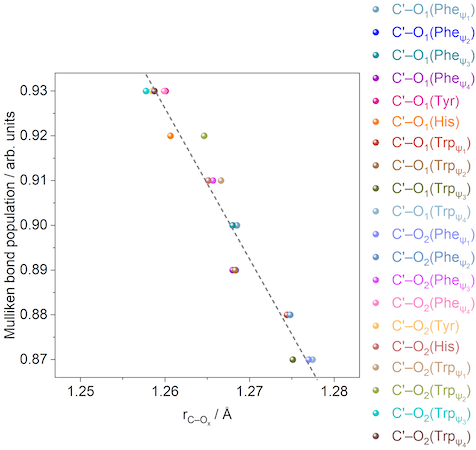}
    \caption{Correlation between the Mulliken bond population between the \ce{C$'$} and \ce{O1}, and between \ce{C$'$} and \ce{O2}, respectively, and the corresponding bond lengths, $r_{C'-O_1}$, and $r_{C'-O_2}$ for the aromatic group, including different conformations for Phe and Trp.}
    \label{Aro_population correlation}
\end{figure}

\clearpage
\newpage
\subsection{Polar Side Chain-Containing Group}

\begin{figure}[!ht]
\centering
    \includegraphics[keepaspectratio, width = 0.9\linewidth]{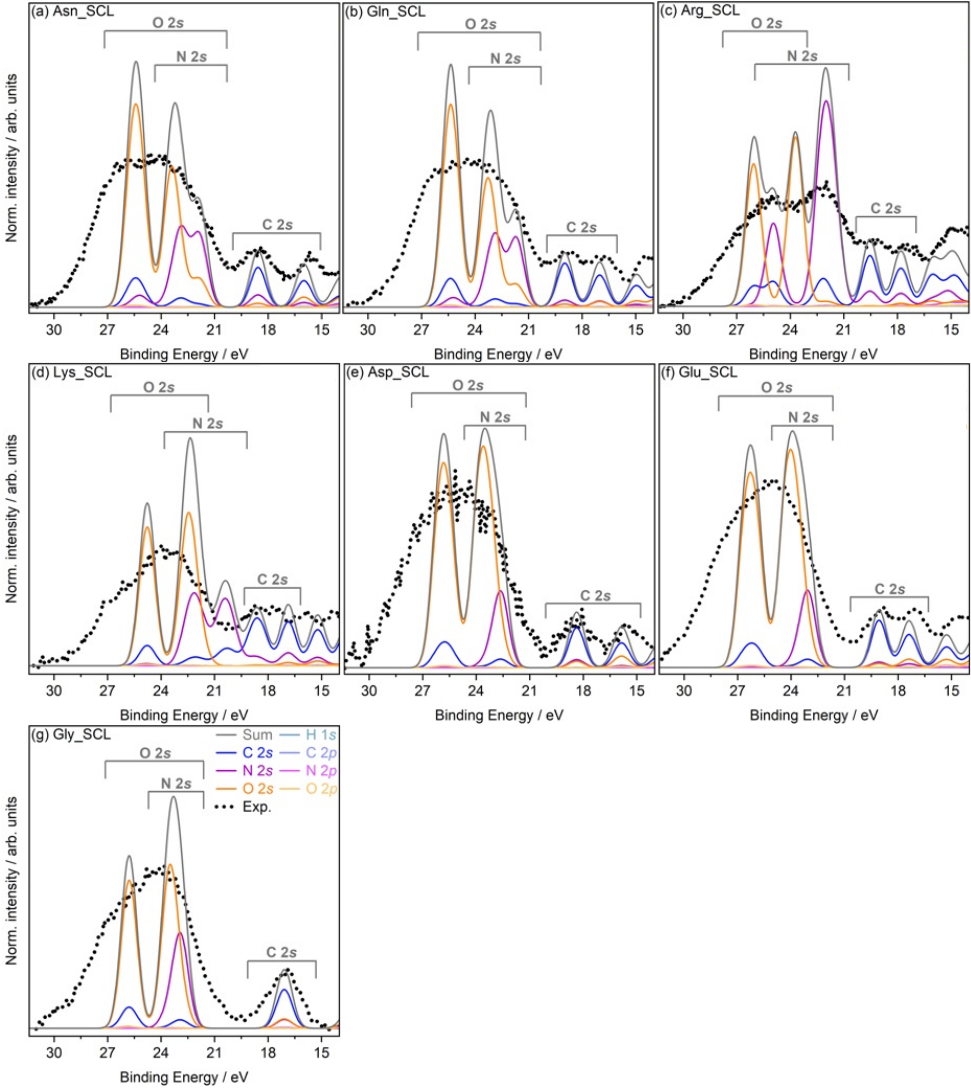}
    \caption{SCL spectra of the polar side chain-containing AAs, including (a) Asn, (b) Gln, (c) Arg, (d) Lys, (e) Asp, (f) Glu, and (g) Gly as reference. All plots display the one-electron photoionisation cross-section weighted PDOS, as well as the sum of all PDOS, from PBE-based DFT calculations and the experimental XP spectra. The labels in dark grey indicate the majority orbital contribution to the spectral features determined from DFT. The weighted PDOS have been aligned and normalised to the experimental BE peak of the highest theoretical C~2\textit{s} feature. The legend shown in (g) applies to all subfigures.}
    \label{other_SCL}
\end{figure}

\clearpage
\begin{figure}[htp]
\centering
    \includegraphics[keepaspectratio, width = 1.0\linewidth]{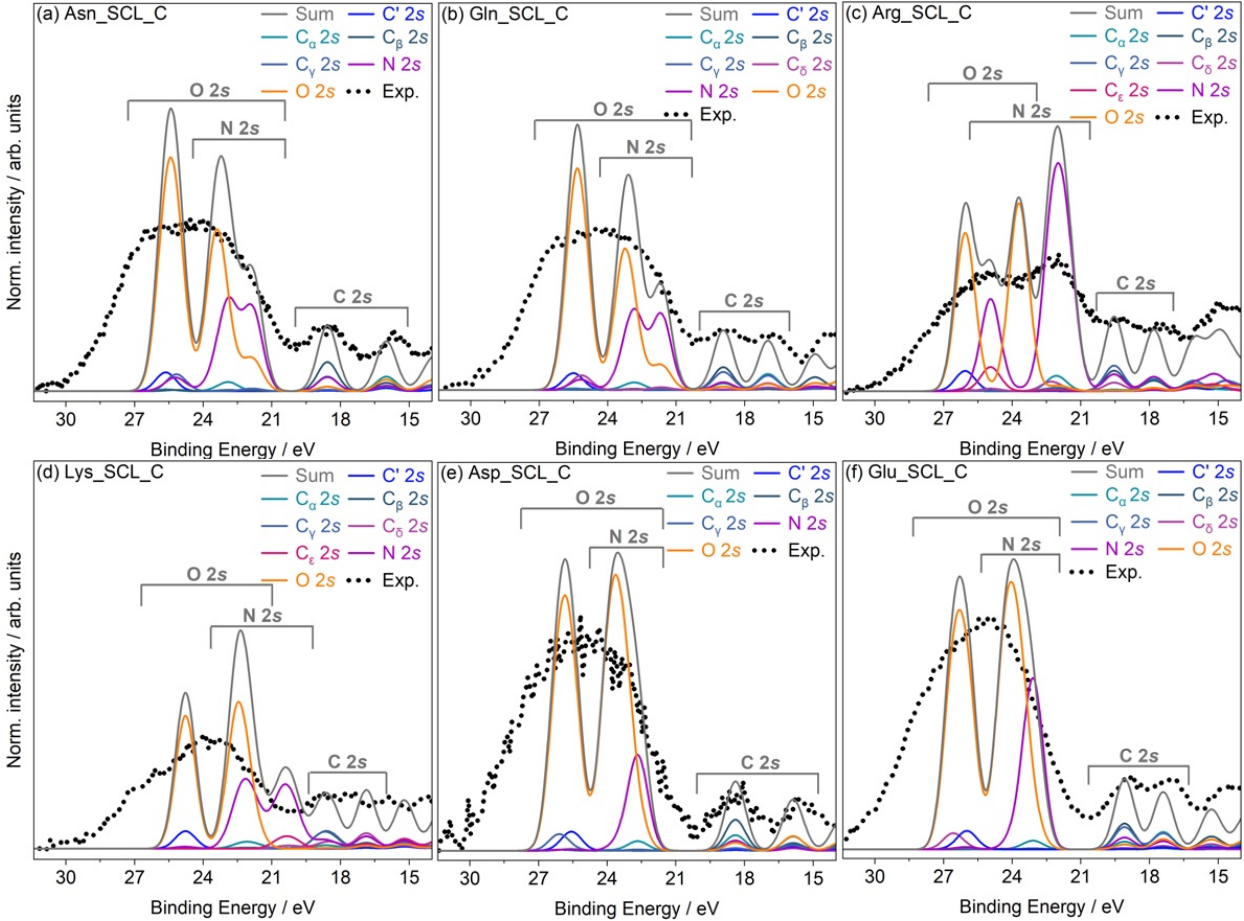}
    \caption{SCL spectra of the polar side chain-containing AAs, including (a) Asn, (b) Gln, (c) Arg, (d) Lys, (e) Asp, and (f) Glu with contributions projected onto specific C atoms. All plots display the one-electron photoionisation cross-section weighted PDOS, as well as the sum of all PDOS, from PBE-based DFT calculations and the experimental XP spectra. To better visualise the contributions from specific C atoms, only 2\textit{s} states are shown as 2\textit{p} states do not present contributions in the SCL region. The labels in dark grey indicate the majority orbital contribution to the spectral features determined from DFT. The weighted PDOS have been aligned and normalised to the experimental BE peak of the highest theoretical C~2\textit{s} feature.}
    \label{other_SCL_C}
\end{figure}

\begin{figure}[htp]
\centering
    \includegraphics[keepaspectratio, width = 0.75\linewidth]{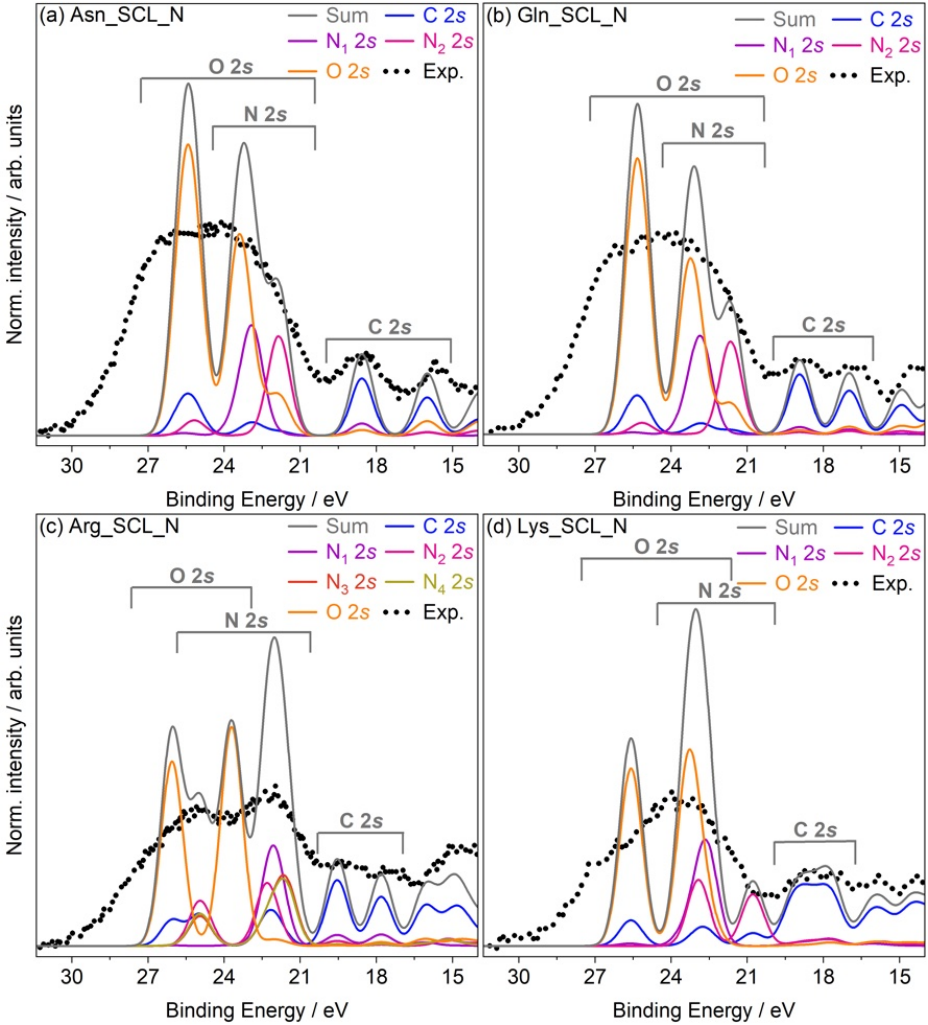}
    \caption{SCL spectra of the polar side chain-containing AAs, including (a) Asn, (b) Gln, (c) Arg, and (d) Lys with contributions projected onto specific N atoms. All plots display the one-electron photoionisation cross-section weighted PDOS, as well as the sum of all PDOS, from PBE-based DFT calculations and the experimental XP spectra. To better visualise the contributions from specific N atoms, only 2\textit{s} states are shown as 2\textit{p} states do not present contributions in the SCL region. The labels in dark grey indicate the majority orbital contribution to the spectral features determined from DFT. The weighted PDOS have been aligned and normalised to the experimental BE peak of the highest theoretical C~2\textit{s} feature.}
    \label{other_SCL_N}
\end{figure}

\begin{figure}[htp]
\centering
    \includegraphics[keepaspectratio, width = 1.0\linewidth]{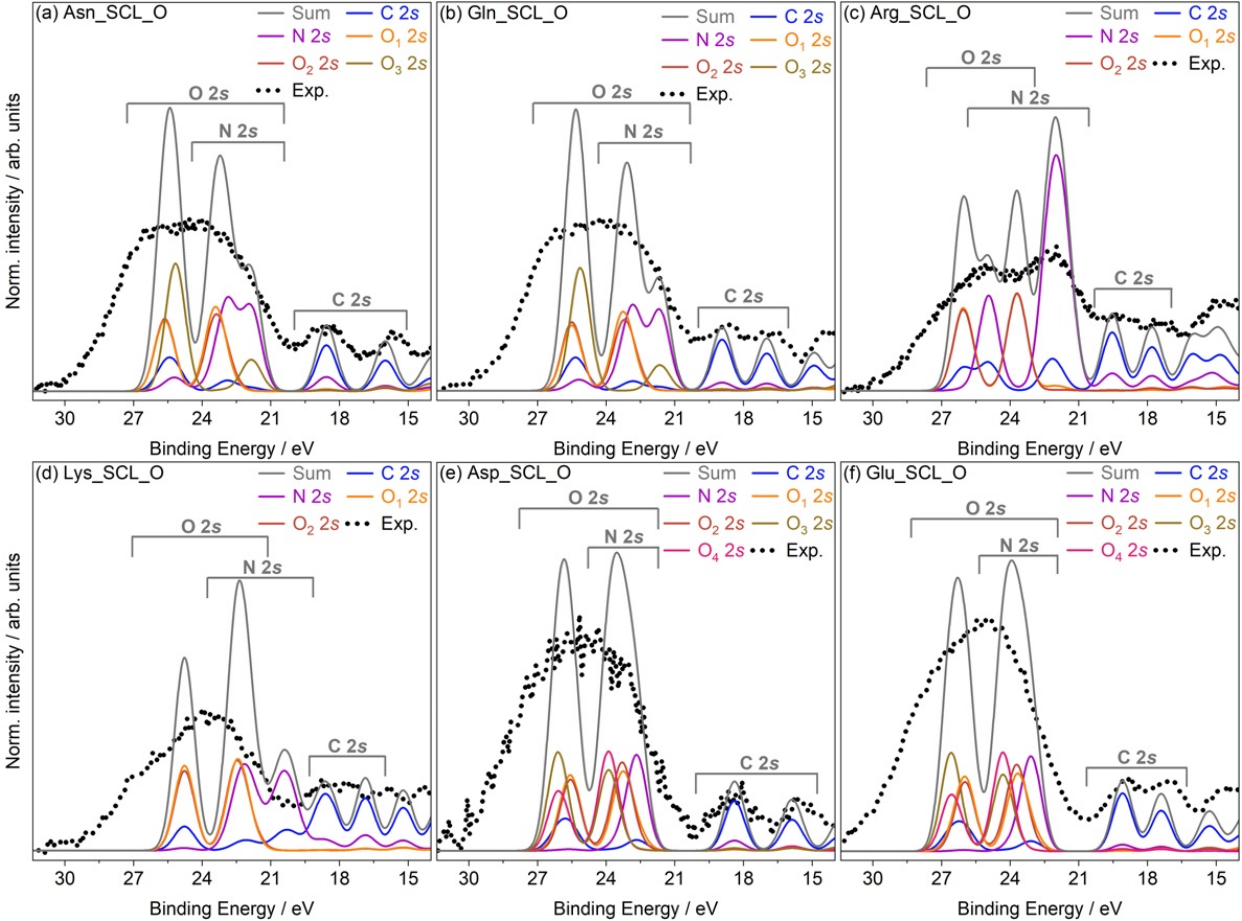}
    \caption{SCL spectra of the polar side chain-containing AAs, including (a) Asn, (b) Gln, (c) Arg, (d) Lys, (e) Asp, and (f) Glu with contributions projected onto specific O atoms. All plots display the one-electron photoionisation cross-section weighted PDOS, as well as the sum of all PDOS, from PBE-based DFT calculations and the experimental XP spectra. To better visualise the contributions from specific O atoms, only 2\textit{s} states are shown as 2\textit{p} states do not present contributions in the SCL region. The labels in dark grey indicate the majority orbital contribution to the spectral features determined from DFT. The weighted PDOS have been aligned and normalised to the experimental BE peak of the highest theoretical C~2\textit{s} feature.}
    \label{other_SCL_O}
\end{figure}
\clearpage

\begin{table}[htp]
    \caption{Eight 2\textit{s} peak area ratios, including \ce{C$'$}:\ce{O1}; \ce{C$'$}:\ce{O2}; \ce{N1}:\ce{O}; \ce{C_\alpha}:\ce{N1}; \ce{C_x}:\ce{O3}; \ce{N2}:\ce{O2}; \ce{N2}:\ce{O3}; and \ce{C_x}:\ce{N2}, where \ce{C_x} represents the C atom attached with the carbamido group, based on theoretical calculations, which were determined using the integration function in the Origin software package for Asn and Gln, as well as their corresponding bond lengths, $r$, determined by visualising the relaxed crystal structure in the VESTA software package.~\cite{Momma2011VESTA3Data}}
\begin{center}
\begin{tabular}{c c c} 
 \hline
 ~ & Asn & Gln \\ [0.5ex] 
 \hline
 \rowcolor{lightgray} \ce{C$'$}:\ce{O1} & 1:3.29 & 1:3.27 \\
 $r_{C'-O_1}$ / \(\text{\AA}\) & 1.27351 & 1.27893 \\ 
 \rowcolor{lightgray} \ce{C$'$}:\ce{O2} & 1:3.41 & 1:3.40 \\ 
 $r_{C'-O_2}$ / \(\text{\AA}\) & 1.25470 & 1.25220 \\
 \rowcolor{lightgray} \ce{N1}:O & 1:1.32 & 1:1.33 \\ 
 \rowcolor{lightgray} \ce{C_$\alpha$}:\ce{N1} & 1:10.09 & 1:11.00 \\
 $r_{C_\alpha-N_1}$ / \(\text{\AA}\) & 1.49077 & 1.49663 \\
 \multirow{2}{*}{\ce{C_x}:\ce{O3}} & 1:7.48 & 1:7.96 \\
 ~ & (\ce{C_\gamma}:\ce{O3}) & (\ce{C_\delta}:\ce{O3}) \\
 \multirow{2}{*}{$r_{C_x-O_3}$ / \(\text{\AA}\)} & 1.25658 & 1.25308 \\
 ~ & ($r_{C_\gamma-O_3}$) & ($r_{C_\delta-O_3}$) \\
 \rowcolor{lightgray} \ce{N2}:\ce{O2} & 1:5.37 & 1:6.48 \\
 \rowcolor{lightgray} \ce{N2}:\ce{O3} & 1:10.48 & 1:12.87 \\
 \multirow{2}{*}{\ce{C_x}:\ce{N2}} & 1:0.71 & 1:0.62 \\
 ~ & (\ce{C_\gamma}:\ce{N2}) & (\ce{C_\delta}:\ce{N2}) \\
 \multirow{2}{*}{$r_{C_x-N_2}$ / \(\text{\AA}\)} & 1.33956 & 1.34076 \\
 ~ & ($r_{C_\gamma-N_2}$) & ($r_{C_\delta-N_2}$) \\
 \hline
\end{tabular}
\end{center}
\label{area ratio table1_other}
\end{table}

\textbf{\begin{table}[htp]
    \caption{Various bond lengths of $r_{C'-O_1}$, $r_{C'-O_2}$, and $r_{C_x-O_3}$ and the corresponding bond distances of $d_{O_1-H}$, $d_{O_2-H}$, $d_{O_3-H}$ and $r_{H-N_x}$ of Asn and Gln, where \ce{Cx} represents different C atoms attached to the carbamido group, and \ce{Nx} represents different N atoms, respectively. Letters in the brackets indicate the intermolecular hydrogen bonds are formed between which two columns and atoms. Bond lengths were determined by visualising the relaxed crystal structure in the VESTA software package.~\cite{Momma2011VESTA3Data} The letters in the bracket denote the specific columns of molecules involved in the formation of intermolecular hydrogen bonds.}
\begin{center}
\begin{tabular}{c c c c c c c} 
 \hline
 AAs & ~ & \multicolumn{2}{c}{\ce{O1}} & \ce{O2} & \multicolumn{2}{c}{\ce{O3}} \\ [0.5ex] 
 \hline
 \multirow{6}{*}{\textbf{Asn}} & \multirow{2}{*}{$r_{C_x-O}$ / \(\text{\AA}\)} & \multicolumn{2}{c}{1.27351} & 1.25470 & \multicolumn{2}{c}{1.25658} \\
 ~ & ~ & \multicolumn{2}{c}{($r_{C'-O_1}$)} & ($r_{C'-O_2}$) & \multicolumn{2}{c}{($r_{C_\gamma-O_3}$)} \\
 ~ & \multirow{2}{*}{$d_{O-H}$ / \(\text{\AA}\)} & 1.68014 & 1.70752 & 1.77754 & 1.80318 & 1.88175 \\ 
 ~ & ~ & (ab) & (aa) & (ab) & (ab) & (aa) \\
 ~ & \multirow{2}{*}{$r_{H-N}$ / \(\text{\AA}\)} & 1.05687 & 1.05364 & 1.03995 & 1.02438 & 1.04515 \\ 
 ~ &  ~ & (\ce{N1}) & (\ce{N1}) & (\ce{N2}) & (\ce{N2}) & (\ce{N1}) \\
 \multirow{6}{*}{\textbf{Gln}} & \multirow{2}{*}{$r_{C_x-O}$ / \(\text{\AA}\)} & \multicolumn{2}{c}{1.27893} & 1.25220 & \multicolumn{2}{c}{1.25308} \\
 ~ & ~ & \multicolumn{2}{c}{($r_{C'-O_1}$)} & ($r_{C'-O_2}$) & \multicolumn{2}{c}{($r_{C_\delta-O_3}$)} \\
 ~ & \multirow{2}{*}{$d_{O-H}$ / \(\text{\AA}\)} & 1.67917 & 1.75257 & 1.84881 & 1.82481 & 2.00614 \\ 
 ~ & ~ & (ad) & (aa) & (ad) & (ab) & (ab) \\
 ~ & \multirow{2}{*}{$r_{H-N}$ / \(\text{\AA}\)} & 1.05687 & 1.05364 & 1.03995 & 1.02438 & 1.04515 \\ 
 ~ & ~ & (\ce{N1}) & (\ce{N1}) & (\ce{N2}) & (\ce{N2}) & (\ce{N1}) \\
 \hline
\end{tabular}
\end{center}
\label{bond length study1_other}
\end{table}} 

\begin{table}[htp]
    \caption{Eight 2\textit{s} peak area ratios, including \ce{C$'$}:\ce{O1}; \ce{C$'$}:\ce{O2}; \ce{N1}:\ce{O2}; \ce{N1}:\ce{O4}; \ce{C_\alpha}:\ce{N1}; \ce{C_x}:\ce{O3}; \ce{C_x}:\ce{O4}; and \ce{O1}:\ce{O3}, where \ce{C_x} represents different C atoms attached to the additional --COOH group, based on theoretical calculations, which were determined using the integration function in the Origin software package for Asp and Glu, as well as their corresponding bond lengths, $r$, determined by visualising the relaxed crystal structure in the VESTA software package.~\cite{Momma2011VESTA3Data}}
\begin{center}
\begin{tabular}{c c c} 
 \hline
 ~ & Asp & Glu \\ [0.5ex] 
 \hline
 \rowcolor{lightgray} \ce{C$'$}:\ce{O1} & 1:3.64 & 1:3.68 \\
 $r_{C'-O_1}$ / \(\text{\AA}\) & 1.26779 & 1.27107 \\ 
 \rowcolor{lightgray} \ce{C$'$}:\ce{O2} & 1:3.25 & 1:3.20 \\ 
 $r_{C'-O_2}$ / \(\text{\AA}\) & 1.26404 & 1.26327 \\
 \rowcolor{lightgray} \ce{N1}:\ce{O2} & 1:0.62 & 1:0.60 \\ 
 \rowcolor{lightgray} \ce{N1}:\ce{O4} & 1:0.51 & 1:0.49 \\
 \rowcolor{lightgray} \ce{C_$\alpha$}:\ce{N1} & 1:10.41 & 1:11.10 \\
 $r_{C_\alpha-N_1}$ / \(\text{\AA}\) & 1.49336 & 1.49665 \\
 \multirow{2}{*}{\ce{C_x}:\ce{O3}} & 1:5.41 & 1:5.54 \\
 ~ & (\ce{C_\gamma}:\ce{O3}) & (\ce{C_\delta}:\ce{O3}) \\
 \multirow{2}{*}{$r_{C_x-O_3}$ / \(\text{\AA}\)} & 1.32148 & 1.32363 \\
 ~ & ($r_{C_\gamma-O_3}$) & ($r_{C_\delta-O_3}$) \\
 \multirow{2}{*}{\ce{C_x}:\ce{O4}} & 1:3.17 & 1:3.06 \\
 ~ & (\ce{C_\gamma}:\ce{O4}) & (\ce{C_\delta}:\ce{O4}) \\
 \multirow{2}{*}{$r_{C_x-O_4}$ / \(\text{\AA}\)} & 1.23458 & 1.23838 \\
 ~ & ($r_{C_\gamma-O_4}$) & ($r_{C_\delta-O_4}$) \\
 \hline
\end{tabular}
\end{center}
\label{area ratio table2_other}
\end{table}

\textbf{\begin{table}[htp]
    \caption{Various bond lengths of $r_{C'-O_1}$, $r_{C'-O_2}$, $r_{C_x-O_3}$, and $r_{C_x-O_4}$ and the corresponding bond distances of $d_{O_1-H}$, $d_{O_2-H}$, $r_{O_3-H}$, $d_{O_4-H}$ and $r_{H-N_1}$ for Asp, and Glu, where \ce{Cx} represents different C atoms attached to the --COOH group. Letters in the brackets indicate the intermolecular hydrogen bonds are formed between which two columns and atoms. Bond lengths were determined by visualising the relaxed crystal structure in the VESTA software package.~\cite{Momma2011VESTA3Data} The letters in the bracket denote the specific columns of molecules involved in the formation of intermolecular hydrogen bonds.}
\begin{center}
\begin{tabular}{c c c c c c c c} 
 \hline
 AAs & ~ & \multicolumn{2}{c}{\ce{O1}} & \multicolumn{2}{c}{\ce{O2}} & \ce{O3} & \ce{O4} \\ [0.5ex] 
 \hline
 \multirow{6}{*}{\textbf{Asp}} & \multirow{2}{*}{$r_{C_x-O}$ / \(\text{\AA}\)} & \multicolumn{2}{c}{1.26779} & \multicolumn{2}{c}{1.26404} & 1.32148 & 1.23458 \\
 ~ & ~ & \multicolumn{2}{c}{($r_{C'-O_1}$)} & \multicolumn{2}{c}{($r_{C'-O_3}$)} & ($r_{C_\gamma-O_3}$) & ($r_{C_\gamma-O_4}$) \\
 ~ & \multirow{2}{*}{$d_{O-H}$ / \(\text{\AA}\)} & \multicolumn{2}{c}{1.49217} & 1.73029 & 1.77029 & 1.05056 & 1.73207 \\ 
 ~ & ~ & \multicolumn{2}{c}{(ab)} & (aa) & (ab) & (aa) & (ab) \\
 ~ & \multirow{2}{*}{$r_{H-N}$ / \(\text{\AA}\)} & \multicolumn{2}{c}{1.05056} & 1.05117 & 1.04677 & 1.32148 & 1.05129 \\ 
 ~ &  ~ & \multicolumn{2}{c}{(\ce{O3})} & (\ce{N1}) & (\ce{N1}) & (\ce{O1}) & (\ce{N1}) \\
 \multirow{6}{*}{\textbf{Glu}} & \multirow{2}{*}{$r_{C_x-O}$ / \(\text{\AA}\)} & \multicolumn{2}{c}{1.27107} & \multicolumn{2}{c}{1.26327} & 1.32363 & 1.23838 \\
 ~ & ~ & \multicolumn{2}{c}{($r_{C'-O_1}$)} & \multicolumn{2}{c}{($r_{C'-O_2}$)} & ($r_{C_\delta-O_3}$) & ($r_{C_\delta-O_4}$) \\
 ~ & \multirow{2}{*}{$d_{O-H}$ / \(\text{\AA}\)} & \multicolumn{2}{c}{1.46042} & 1.76861 & 1.77537 & 1.05640 & 1.83094 \\ 
 ~ & ~ & \multicolumn{2}{c}{(ac)} & (ad) & (aa) & (ac) & (ab) \\
 ~ & \multirow{2}{*}{$r_{H-N}$ / \(\text{\AA}\)} & \multicolumn{2}{c}{1.05640} & 1.05088 & 1.04650 & 1.46042 & 1.04619 \\ 
 ~ & ~ & \multicolumn{2}{c}{(\ce{O3})} & (\ce{N1}) & (\ce{N1}) & (\ce{O1}) & (\ce{N1}) \\
 \hline
\end{tabular}
\end{center}
\label{bond length study2_other}
\end{table}} 

\begin{table}[htp]
    \caption{Various 2\textit{s} peak area ratios, including \ce{C$'$}:\ce{O1}; \ce{C$'$}:\ce{O2}; \ce{N1}:\ce{O}; and \ce{C_\alpha}:\ce{N1} for Arg and Lys; \ce{N3}:\ce{O}; \ce{N4}:\ce{O}; \ce{C_\epsilon}:\ce{N2}; \ce{C_\epsilon}:\ce{N3}; and \ce{C_\epsilon}:\ce{N4} for Arg; and \ce{C_\epsilon}:\ce{N2} for Lys, based on theoretical calculations, which were determined using the integration function in the Origin software package, as well as their corresponding bond lengths, $r$, determined by visualising the relaxed crystal structure in the VESTA software package.~\cite{Momma2011VESTA3Data} N.A. in the table represents not applicable.}
\begin{center}
\begin{tabular}{c c c} 
 \hline
 ~ & Arg & Lys \\ [0.5ex] 
 \hline
 \rowcolor{lightgray} \ce{C$'$}:\ce{O1} & 1:3.59 & 1:3.55 \\
 \multirow{2}{*}{$r_{C'-O_1}$ / \(\text{\AA}\)} & 1.26699 & 1.27281 \\ 
 ~ & 1.26906 & 1.27697 \\
 \rowcolor{lightgray} \ce{C$'$}:\ce{O2} & 1:3.50 & 1:3.38 \\ 
 \multirow{2}{*}{$r_{C'-O_2}$ / \(\text{\AA}\)} & 1.26500 & 1.26360 \\
 ~ & 1.26398 & 1.25838 \\
 \rowcolor{lightgray} \ce{N1}:\ce{O} & 1:1.65 & 1:1.34 \\ 
 \rowcolor{lightgray} \ce{C_$\alpha$}:\ce{N1} & 1:5.63 & 1:10.01 \\
 \multirow{2}{*}{$r_{C_\alpha-N_1}$ / \(\text{\AA}\)} & 1.48237 & 1.49059 \\
 ~ & 1.48397 & 1.48956 \\
 \rowcolor{lightgray} \ce{N3}:\ce{O} & 1:5.46 & N.A. \\
 \rowcolor{lightgray} \ce{N4}:\ce{O} & 1:5.89 & N.A. \\
 \rowcolor{lightgray} \ce{C_\epsilon}:\ce{N2} & 1:1.55 & 1:10.64 \\
 \multirow{2}{*}{$r_{C_\epsilon-N_2}$ / \(\text{\AA}\)} & 1.34449 & 1.50789 \\
 ~ & 1.34551 & 1.47909 \\
 \rowcolor{lightgray} \ce{C_\epsilon}:\ce{N3} & 1:1.03 & N.A. \\
 \multirow{2}{*}{$r_{C_\epsilon-N_3}$ / \(\text{\AA}\)} & 1.34187 & \multirow{2}{*}{N.A.} \\
 ~ & 1.34157 & ~ \\
 \rowcolor{lightgray} \ce{C_\epsilon}:\ce{N4} & 1:1.12 & N.A. \\
 \multirow{2}{*}{$r_{C_\epsilon-N_4}$ / \(\text{\AA}\)} & 1.33704 & \multirow{2}{*}{N.A.} \\
 ~ & 1.33731 & ~ \\
 \hline
\end{tabular}
\end{center}
\label{area ratio table3_other}
\end{table}

\begin{landscape}
\begin{table}
    \caption{Various bond lengths of $r_{C'-O_1}$ and $r_{C'-O_2}$ and the corresponding bond distances of $d_{O_1-H}$, $d_{O_2-H}$, and $r_{H-N}$ of different conformations of Arg and Lys distinguished by the torsion angle, $\psi$. Bond lengths were determined by visualising the relaxed crystal structure in the VESTA software package.~\cite{Momma2011VESTA3Data} The letters and torsion angles in the bracket denote the specific columns and orientations of molecules involved in the formation of intermolecular hydrogen bonds.}
\begin{center}
\begin{tabular}{c c c c c c c c c c} 
 \hline
 AAs & ~ & \multicolumn{4}{c}{$\psi_1$} & \multicolumn{4}{c}{$\psi_2$} \\ [0.5ex] 
 \hline
 \multirow{7}{*}{\textbf{Arg}} & ~ & \multicolumn{4}{c}{$9.680^\circ$} & \multicolumn{4}{c}{$13.254^\circ$} \\
 ~ & ~ & \multicolumn{2}{c}{\ce{O1}} & \multicolumn{2}{c}{\ce{O2}} & \multicolumn{2}{c}{\ce{O1}} & \multicolumn{2}{c}{\ce{O2}} \\
 ~ & $r_{C'-O}$ / \(\text{\AA}\) & \multicolumn{2}{c}{1.26699} & \multicolumn{2}{c}{1.26500} & \multicolumn{2}{c}{1.26906} & \multicolumn{2}{c}{1.26360} \\ 
 ~ & \multirow{2}{*}{$d_{O-H}$ / \(\text{\AA}\)} & 1.73465 & 1.74325 & 1.78226 & 1.87230 & 1.68294 & 1.71732 & 1.87579 & 1.88285 \\ 
 ~ & ~ & (ab $\&$ $\psi_1\psi_1$) & (aa $\&$ $\psi_1\psi_2$) & (ab $\&$ $\psi_1\psi_1$) & (aa $\&$ $\psi_1\psi_2$) & (ab $\&$ $\psi_2\psi_2$) & (aa $\&$ $\psi_2\psi_1$) & (ab $\&$ $\psi_2\psi_2$) & (aa $\&$ $\psi_2\psi_1$) \\
 ~ & \multirow{2}{*}{$r_{H-N}$ / \(\text{\AA}\)} & 1.04096 & 1.03308 & 1.04315 & 1.01743 & 1.04317 & 1.03347 & 1.03522 & 1.02243 \\ 
 ~ & ~ & (\ce{N3}) & (\ce{N3}) & (\ce{N4}) & (\ce{N4}) & (\ce{N4}) & (\ce{N4}) & (\ce{N3}) & (\ce{N3}) \\
 \multirow{7}{*}{\textbf{Lys}} & ~ & \multicolumn{4}{c}{$30.295^\circ$} & \multicolumn{4}{c}{$165.749^\circ$} \\
 ~ & ~ & \multicolumn{2}{c}{\ce{O1}} & \multicolumn{2}{c}{\ce{O2}} & \multicolumn{2}{c}{\ce{O1}} & \multicolumn{2}{c}{\ce{O2}} \\
 ~ & $r_{C'-O}$ / \(\text{\AA}\) & \multicolumn{2}{c}{1.27697} & \multicolumn{2}{c}{1.25838} & \multicolumn{2}{c}{1.27281} & \multicolumn{2}{c}{1.26398} \\
 ~ & \multirow{2}{*}{$d_{O-H}$ / \(\text{\AA}\)} & 1.60872 & 1.91580 & \multicolumn{2}{c}{1.78197} & 1.73759 & 1.76292 & \multicolumn{2}{c}{1.70458} \\ 
 ~ & ~ & (aa $\&$ $\psi_1\psi_2$) & (ab $\&$ $\psi_1\psi_2$) & \multicolumn{2}{c}{(aa $\&$ $\psi_1\psi_1$)} & (aa $\&$ $\psi_2\psi_2$) & (aa $\&$ $\psi_2\psi_1$) & \multicolumn{2}{c}{(ab $\&$ $\psi_2\psi_1$)} \\
 ~ & \multirow{2}{*}{$r_{H-N}$ / \(\text{\AA}\)} & 1.07118 & 1.03821 & \multicolumn{2}{c}{1.05214} & 1.04817 & 1.04829 & \multicolumn{2}{c}{1.05227} \\  
 ~ & ~ & (\ce{N1}) & (\ce{N1}) & \multicolumn{2}{c}{(\ce{N1})} & (\ce{N1}) & (\ce{N1}) & \multicolumn{2}{c}{(\ce{N1})} \\
 \hline
\end{tabular}
\end{center}
\label{bond length study3_other}
\end{table}
\end{landscape}

\textbf{\begin{table}[htp]
    \caption{Summary of the calculated Mulliken bond population, Mulliken charges, and the bond length of $r_{C'-O_1}$ and $r_{C'-O_1}$ for the polar side chain-containing group, including different conformations for Arg, and Lys. Different conformations observed in the crystal structure for Arg, and Lys are distinguished by the different torsion angles, denoted as \ce{\Psi_x} in the bracket. n(\ce{C$'$}\ce{O1}) and n(\ce{C$'$}\ce{O2}) represent the Mulliken bond population between the \ce{C$'$} and \ce{O1}, and between \ce{C$'$} and \ce{O2}, respectively.}
\begin{center}
\begin{tabular}{c c c c c c c} 
 \hline
 AAs & \ce{O1} / e & n(\ce{C$'$}\ce{O1}) & $r_{C'-O_1}$ / \(\text{\AA}\) & \ce{O2} / e & n(\ce{C$'$}\ce{O2}) & $r_{C'-O_2}$ / \(\text{\AA}\) \\ [0.5ex] 
 \hline
 Asn & -0.663 & 0.87 & 1.27351 & -0.662 & 0.94 & 1.25470 \\
 Gln & -0.675 & 0.86 & 1.27893 & -0.675 & 0.95 & 1.25220 \\ 
 Arg(\ce{\Psi_1}) & -0.667 & 0.88 & 1.26699 & -0.677 & 0.90 & 1.26500 \\
 Arg(\ce{\Psi_2}) & -0.666 & 0.88 & 1.26906 & -0.682 & 0.92 & 1.26360 \\ 
 Lys(\ce{\Psi_1}) & -0.675 & 0.87 & 1.27697 & -0.666 & 0.92 & 1.25838 \\ 
 Lys(\ce{\Psi_2}) & -0.670 & 0.88 & 1.27281 & -0.692 & 0.93 & 1.26398 \\
 Asp & -0.650 & 0.89 & 1.26779 & -0.657 & 0.91 & 1.26404 \\ 
 Glu & -0.655 & 0.89 & 1.27107 & -0.655 & 0.91 & 1.26327 \\
 \hline
\end{tabular}
\end{center}
\label{bond populaion_other}
\end{table}} 

\begin{figure}[hb]
    \centering
    \includegraphics[width=0.6\textwidth]{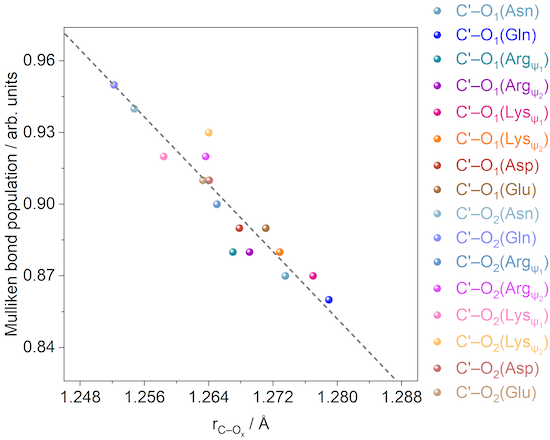}
    \caption{Correlation between the Mulliken bond population between the \ce{C$'$} and \ce{O1}, and between \ce{C$'$} and \ce{O2}, respectively, and the corresponding bond lengths, $r_{C'-O_1}$, and $r_{C'-O_2}$ for the polar side chain-containing group, including different conformations for Arg, and Lys.}
    \label{other_population correlation}
\end{figure}

\cleardoublepage
\newpage
\subsection{S/Se-containing Group}

\begin{figure}[htp]
\centering
    \includegraphics[keepaspectratio, width = 1.0\linewidth]{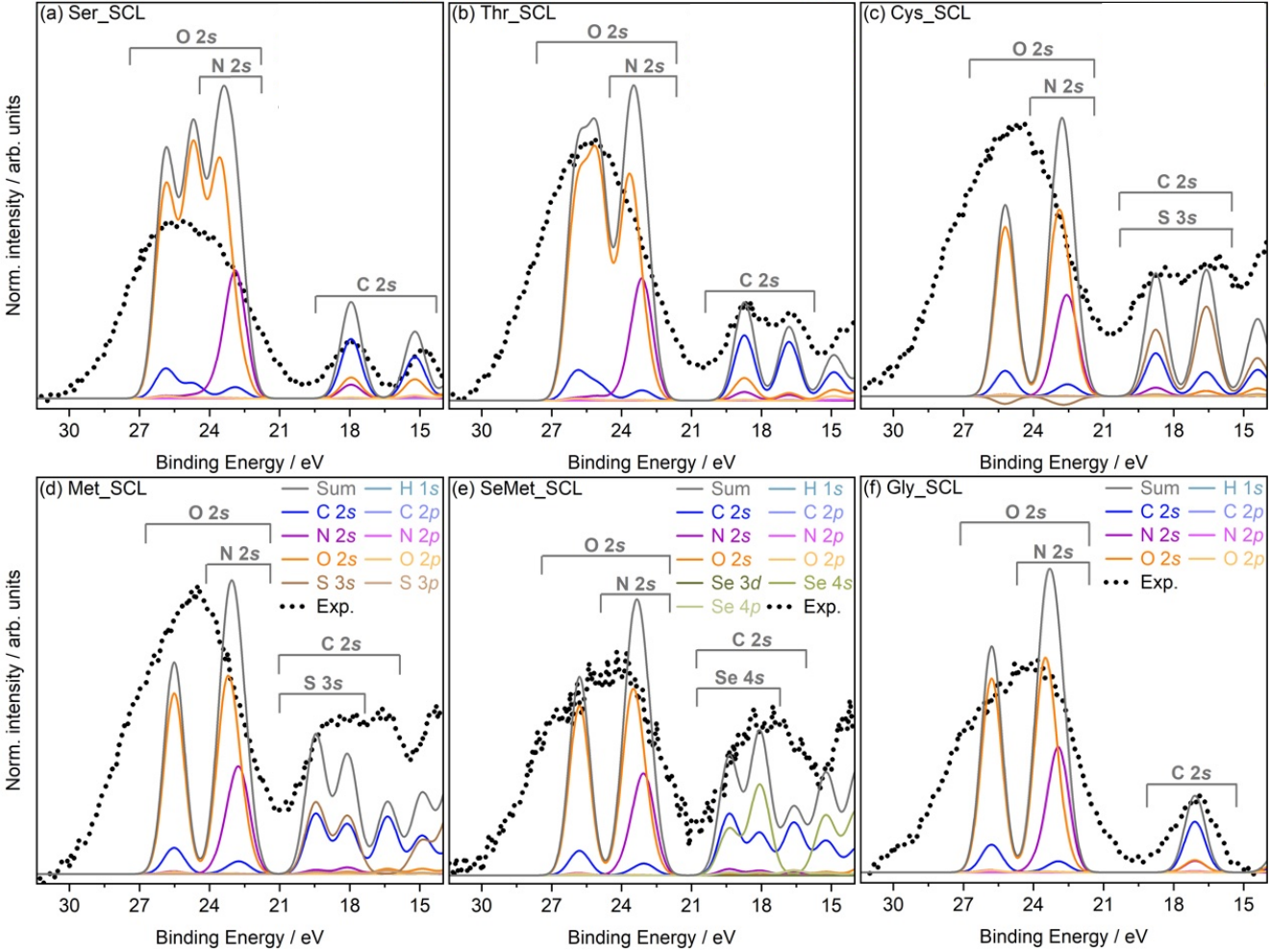}
    \caption{SCL spectra of the S/Se-containing AAs, including (a) Ser, (b) Thr, (c) Cys, (d) Met, (e) SeMet, and (f) Gly as reference. All plots display the one-electron photoionisation cross-section weighted PDOS, as well as the sum of all PDOS, from PBE-based DFT calculations and the experimental XP spectra. The labels in dark grey indicate the majority orbital contribution to the spectral features determined from DFT. The weighted PDOS have been aligned and normalised to the experimental BE peak of the highest theoretical C~2\textit{s} feature. The legend shown in (f) is applied to subfigures (a) and (b); the legend present in (d) is applied to subfigure (c); and the legend in (e) is only applied to itself.}
    \label{S_Se_SCL}
\end{figure}

\begin{figure}[htp]
\centering
    \includegraphics[keepaspectratio, width = 1.0\linewidth]{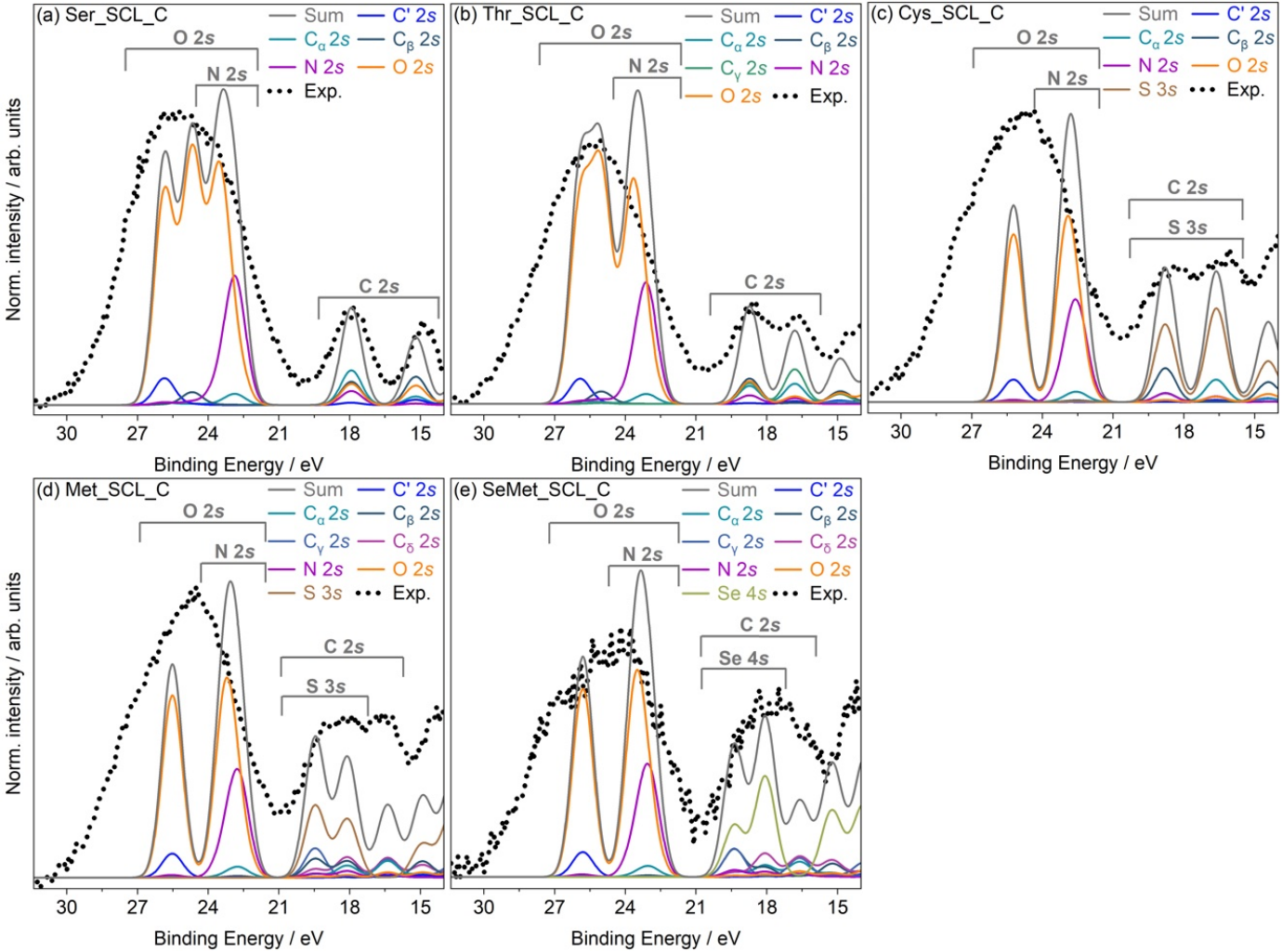}
    \caption{SCL spectra of the S/Se-containing AAs, including (a) Ser, (b) Thr, (c) Cys, (d) Met, and (e) SeMet with contributions projected onto specific C atoms. All plots display the one-electron photoionisation cross section weighted PDOS, as well as the sum of all PDOS, from PBE-based DFT calculations and the experimental XP spectra. To better visualise the contributions from specific C atoms, only 2\textit{s} states are shown as 2\textit{p} states do not present contributions in the SCL region. The labels in dark grey indicate the majority orbital contribution to the spectral features determined from DFT. The weighted PDOS have been aligned and normalised to the experimental BE peak of the highest theoretical C~2\textit{s} feature.}
    \label{S_Se_SCL_C}
\end{figure}

\begin{figure}[htp]
\centering
    \includegraphics[keepaspectratio, width = 1.0\linewidth]{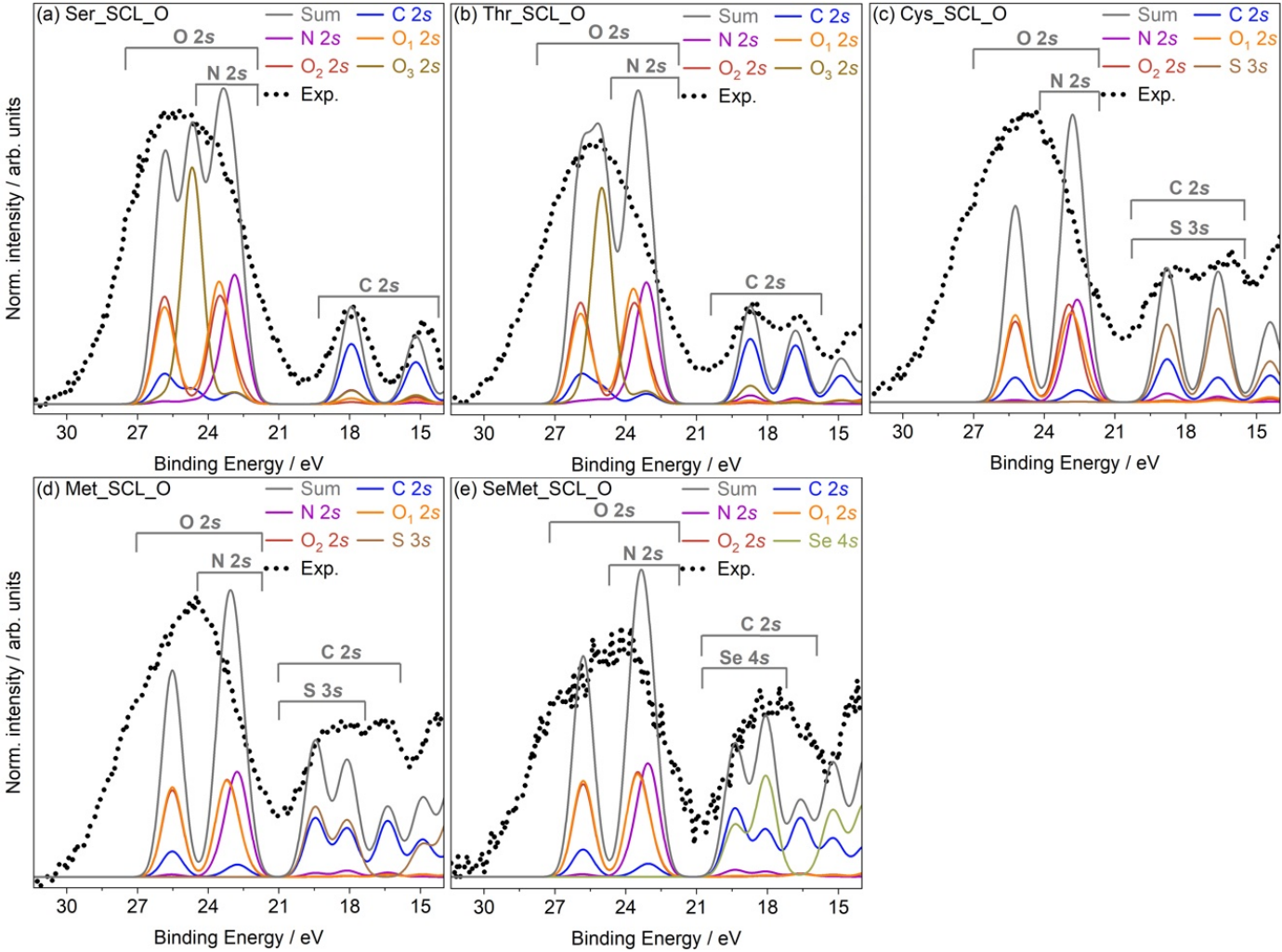}
    \caption{SCL spectra of (a) Ser, (b) Thr, (c) Cys, (d) Met, and (e) SeMet with contributions projected onto specific O atoms. All plots display the one-electron photoionisation cross-section weighted PDOS, as well as the sum of all PDOS, from PBE-based DFT calculations and the experimental XP spectra. To better visualise the contributions from specific O atoms, only 2\textit{s} states are shown as 2\textit{p} states do not present contributions in the SCL region. The labels in dark grey indicate the majority orbital contribution to the spectral features determined from DFT. The weighted PDOS have been aligned and normalised to the experimental BE peak of the highest theoretical C~2\textit{s} feature.}
    \label{S_Se_SCL_O}
\end{figure}

\begin{table}[htp]
    \caption{Four common 2\textit{s} peak area ratios, including \ce{C$'$}:\ce{O1}; \ce{C$'$}:\ce{O2}; \ce{N1}:O; and \ce{C_$\alpha$}:\ce{N1} for all AAs in the group; and one additional 2\textit{s} peak area ratio, \ce{C_$\beta$}:\ce{O3}, for Ser and Thr, based on PDOS calculations projected onto specific C and O atoms, which were determined using the integration function in the Origin software package for the S/Se-containing AAs, as well as their corresponding bond lengths, $r$, determined by visualising the relaxed crystal structure in the VESTA software package.~\cite{Momma2011VESTA3Data} N.A. in the table represents not applicable.}
\begin{center}
\begin{tabular}{c c c c c c} 
 \hline
 ~ & Ser & Thr & Cys & Met & SeMet \\ [0.5ex] 
 \hline
\rowcolor{lightgray} \ce{C$'$}:\ce{O1} & 1:3.03 & 1:2.99 & 1:3.59 & 1:3.50 & 1:3.48 \\
 \multirow{2}{*}{$r_{C'-O_1}$ / \(\text{\AA}\)} & \multirow{2}{*}{1.27249} & \multirow{2}{*}{1.26969} & \multirow{2}{*}{1.27310} & 1.27186 & 1.27180 \\ 
 ~ & ~ & ~ & ~ & 1.27769 & 1.27543 \\
\rowcolor{lightgray} \ce{C$'$}:\ce{O2} & 1:3.55 & 1:3.53 & 1:3.27 & 1:3.39 & 1:3.36 \\ 
 \multirow{2}{*}{$r_{C'-O_2}$ / \(\text{\AA}\)} & \multirow{2}{*}{1.26365} & \multirow{2}{*}{1.25956} & \multirow{2}{*}{1.25738} & 1.26266 & 1.25960 \\ 
 ~ & ~ & ~ & ~ & 1.25943 & 1.25969 \\
\rowcolor{lightgray} \ce{N1}:O & 1:1.28 & 1:1.36 & 1:1.33 & 1:1.33 & 1:1.32 \\ 
\rowcolor{lightgray} \ce{C_$\alpha$}:\ce{N1} & 1:13.52 & 1:13.52 & 1:10.40 & 1:10.42 & 1:10.68 \\
 \multirow{2}{*}{$r_{C_\alpha-N_1}$ / \(\text{\AA}\)} & \multirow{3}{*}{1.49408} & \multirow{3}{*}{1.49413} & \multirow{3}{*}{1.49100} & 1.49065 & \multirow{2}{*}{1.49138}\\
 ~ & ~ & ~ & ~ & 1.49161 & 1.49666 \\
 ~ & ~ & ~ & ~ & 1.49162 & ~ \\
\rowcolor{lightgray} \ce{C_\beta}:\ce{O3} & 1:16.46 & 1:15.62 & N.A. & N.A. & N.A. \\
 $r_{C_\beta-O_3}$ / \(\text{\AA}\) & 1.42639 & 1.43320 & N.A. & N.A. & N.A. \\
 \hline
\end{tabular}
\end{center}
\label{area ratio table_S_Se}
\end{table} 

\textbf{\begin{table}[htp]
    \caption{Various bond lengths of $r_{C'-O_1}$, $r_{C'-O_2}$ and $r_{C_\beta-O_3}$, and the corresponding bond distances of $d_{O_1-H}$, $d_{O_2-H}$, and $r_{H-N}$ for the S/Se-containing AAs. Different conformations of Met and SeMet are distinguished by the torsion angle, $\psi$. Bond lengths were determined by visualising the relaxed crystal structure in the VESTA software package.~\cite{Momma2011VESTA3Data} The letters and torsion angles in the bracket denote the specific columns and orientations of molecules involved in the formation of intermolecular hydrogen bonds.} 
\begin{center}    
\begin{tabular}{c c c c c c c} 
\hline
 \textbf{Ser} & \multicolumn{2}{c}{\ce{O1}} & \multicolumn{2}{c}{\ce{O2}} & \multicolumn{2}{c}{\ce{O3}} \\
 \hline
 \multirow{2}{*}{$r_{C-O}$ / \(\text{\AA}\)} & \multicolumn{2}{c}{1.27249} & \multicolumn{2}{c}{1.26365} & \multicolumn{2}{c}{1.42639} \\
 ~ & \multicolumn{2}{c}{(\ce{C^$'$-O})} & \multicolumn{2}{c}{(\ce{C^$'$-O})} & \multicolumn{2}{c}{(\ce{C_\beta-O})} \\
 \multirow{2}{*}{$d_{O-H}$ / \(\text{\AA}\)} & 1.73162 & 1.80408 & \multicolumn{2}{c}{1.61884} & \multicolumn{2}{c}{1.71200} \\
 ~ & (aa) & (aa) & \multicolumn{2}{c}{(aa)} & \multicolumn{2}{c}{(ab)} \\
 \multirow{2}{*}{$r_{H-N}/r_{O-H}$ / \(\text{\AA}\)} & 1.05050 & 1.05191 & \multicolumn{2}{c}{1.00284} & \multicolumn{2}{c}{1.05253} \\
 ~ & (\ce{N1}) & (\ce{N1}) & \multicolumn{2}{c}{(\ce{O3})} & \multicolumn{2}{c}{(\ce{N1})} \\
 \textbf{Thr} & \multicolumn{2}{c}{\ce{O1}} & \multicolumn{2}{c}{\ce{O2}} & \multicolumn{2}{c}{\ce{O3}} \\
 \multirow{2}{*}{$r_{C-O}$ / \(\text{\AA}\)} & \multicolumn{2}{c}{1.26969} & \multicolumn{2}{c}{1.25956} & \multicolumn{2}{c}{1.43320} \\
 ~ & \multicolumn{2}{c}{(\ce{C^$'$-O})} & \multicolumn{2}{c}{(\ce{C^$'$-O})} & \multicolumn{2}{c}{(\ce{C_\beta-O})} \\
 \multirow{2}{*}{$d_{O-H}$ / \(\text{\AA}\)} & 1.73524 & 1.76629 & \multicolumn{2}{c}{1.65490} & \multicolumn{2}{c}{1.99303} \\
 ~ & (ad) & (aa) & \multicolumn{2}{c}{(ac)} & \multicolumn{2}{c}{(ab)} \\
 \multirow{2}{*}{$r_{H-N}/r_{O-H}$ / \(\text{\AA}\)} & 1.05262 & 1.05353 & \multicolumn{2}{c}{1.00007} & \multicolumn{2}{c}{1.03997} \\
 ~ & (\ce{N1}) & (\ce{N1}) & \multicolumn{2}{c}{(\ce{O3})} & \multicolumn{2}{c}{(\ce{N1})} \\
 \textbf{Cys} & \multicolumn{4}{c}{\ce{O1}} & \multicolumn{2}{c}{\ce{O2}} \\
 $r_{C'-O}$ / \(\text{\AA}\) & \multicolumn{2}{c}{1.27310} & \multicolumn{2}{c}{1.25738} \\
 \multirow{2}{*}{$d_{O-H}$ / \(\text{\AA}\)} & \multicolumn{2}{c}{1.67479} &\multicolumn{2}{c}{1.93696} & \multicolumn{2}{c}{1.73506} \\ 
 ~ & \multicolumn{2}{c}{(aa)} & \multicolumn{2}{c}{(ab)} & \multicolumn{2}{c}{(ad)} \\
 $r_{H-N}$ / \(\text{\AA}\) & \multicolumn{2}{c}{1.06673} &\multicolumn{2}{c}{1.03536} & \multicolumn{2}{c}{1.04851} \\ 
 \textbf{Met} & \multicolumn{3}{c}{$\psi_1=31.759^\circ$} & \multicolumn{3}{c}{$\psi_1=165.632^\circ$} \\ [0.5ex] 
 ~ & \multicolumn{2}{c}{\ce{O1}} & \ce{O2} & \multicolumn{2}{c}{\ce{O1}} & \ce{O2} \\
 $r_{C'-O}$ / \(\text{\AA}\) & \multicolumn{2}{c}{1.27769} & 1.25943 & \multicolumn{2}{c}{1.27186} & 1.26266 \\ 
 \multirow{2}{*}{$d_{O-H}$ / \(\text{\AA}\)} & 1.63004 & 1.93237 & 1.79782 & 1.76513 & 1.77168 & 1.67535 \\ 
 ~ & (aa $\&$ $\psi_1\psi_2$) & (ab $\&$ $\psi_1\psi_2$) & (aa $\&$ $\psi_1\psi_1$) & (aa $\&$ $\psi_2\psi_2$) & (aa $\&$ $\psi_2\psi_1$) & (ab $\&$ $\psi_2\psi_1$) \\
 $r_{H-N}$ / \(\text{\AA}\) & 1.06773 & 1.03724 & 1.04922 & 1.04937 & 1.04936 & 1.04819 \\ 
 \textbf{SeMet} & \multicolumn{3}{c}{$\psi_1=39.364^\circ$} & \multicolumn{3}{c}{$\psi_1=170.347^\circ$} \\ [0.5ex] 
 ~ & \ce{O1} & \multicolumn{2}{c}{\ce{O2}} & \multicolumn{2}{c}{\ce{O1}} & \ce{O2} \\
 $r_{C'-O}$ / \(\text{\AA}\) & 1.27180 & \multicolumn{2}{c}{1.25960} & \multicolumn{2}{c}{1.27281} & 1.25969 \\ 
 \multirow{2}{*}{$d_{O-H}$ / \(\text{\AA}\)} & 1.63504 & 2.06421 & 1.76909 & 1.72041 & 1.75650 & 1.74373 \\ 
 ~ & (aa $\&$ $\psi_1\psi_2$) & (ab $\&$ $\psi_1\psi_2$) & (aa $\&$ $\psi_1\psi_1$) & (aa $\&$ $\psi_2\psi_1$) & (aa $\&$ $\psi_2\psi_2$) & (ab $\&$ $\psi_2\psi_1$) \\
 $r_{H-N}$ / \(\text{\AA}\) & 1.06437 & 1.03493 & 1.05093 & 1.04897 & 1.05665 & 1.04587 \\
 \hline
\end{tabular}
\end{center}
\label{bond length study_S_Se}
\end{table}}

\begin{figure}[!hb]
    \centering
    \includegraphics[width=1.0\textwidth]{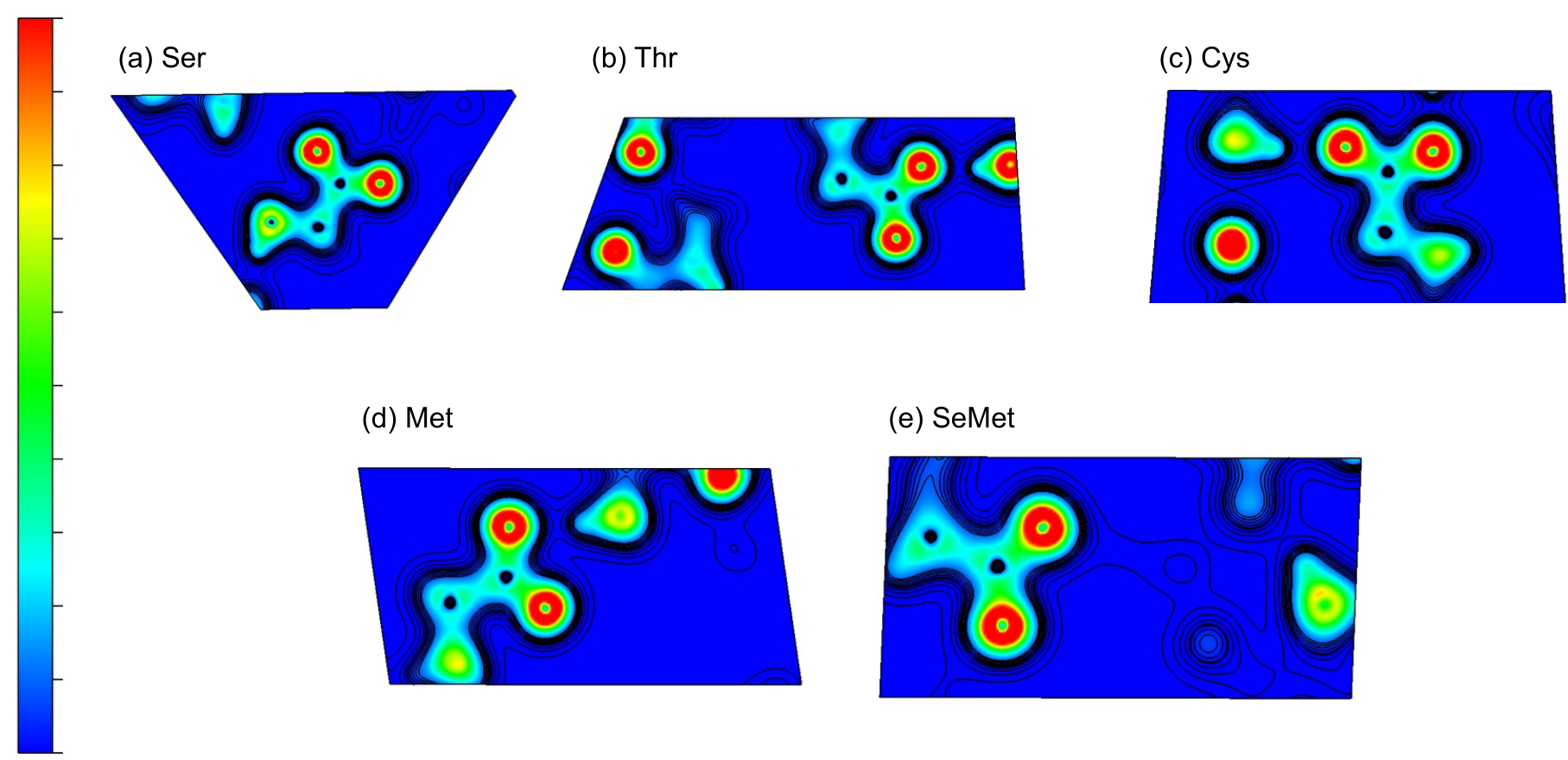}
    \caption{Visualisation of the electron density distribution around the \ce{COO^-} group of the S/Se-containing AAs for (a) Ser, (b) Thr, (c) Cys, (d) Met, and (e) SeMet, with contour lines present. The colour gradient on the left presents how electron density changes, where blue represents the least electron density, and red represents the highest electron density. All electron density distribution figures were prepared in the VESTA software package.~\cite{Momma2011VESTA3Data}}
    \label{S_Se_electron density}
\end{figure}

\textbf{\begin{table}[htp]
    \caption{Summary of the calculated Mulliken bond population, Mulliken charges, and the bond length of $r_{C'-O_1}$ and $r_{C'-O_1}$ for the S/Se-containing group, including different conformations for Met and SeMet. Different conformations observed in the crystal structure for Met and SeMet are distinguished by the different torsion angles, denoted as \ce{\Psi_x} in the bracket. n(\ce{C$'$}\ce{O1}) and n(\ce{C$'$}\ce{O2}) represent the Mulliken bond population between the \ce{C$'$} and \ce{O1}, and between \ce{C$'$} and \ce{O2}, respectively.}
\begin{center}
\begin{tabular}{c c c c c c c} 
 \hline
 AAs & \ce{O1} / e & n(\ce{C$'$}\ce{O1}) & $r_{C'-O_1}$ / \(\text{\AA}\) & \ce{O2} / e & n(\ce{C$'$}\ce{O2}) & $r_{C'-O_2}$ / \(\text{\AA}\) \\ [0.5ex] 
 \hline
 Ser & -0.673 & 0.89 & 1.27249 & -0.660 & 0.91 & 1.26365 \\
 Thr & -0.665 & 0.89 & 1.26969 & -0.667 & 0.91 & 1.25956 \\
 Cys & -0.670 & 0.89 & 1.27310 & -0.669 & 0.94 & 1.25738 \\ 
 Met(\ce{\Psi_1}) & -0.685 & 0.87 & 1.27769 & -0.669 & 0.93 & 1.25943 \\ 
 Met(\ce{\Psi_2}) & -0.664 & 0.89 & 1.27186 & -0.680 & 0.92 & 1.26266 \\
 SeMet(\ce{\Psi_1}) & -0.688 & 0.88 & 1.27180 & -0.677 & 0.92 & 1.25960 \\ 
 SeMet(\ce{\Psi_2}) & -0.665 & 0.87 & 1.27543 & -0.684 & 0.93 & 1.25969 \\
 \hline
\end{tabular}
\end{center}
\label{bond populaion_S_Se}
\end{table}} 

\begin{figure}[htp]
    \centering
    \includegraphics[width=0.6\textwidth]{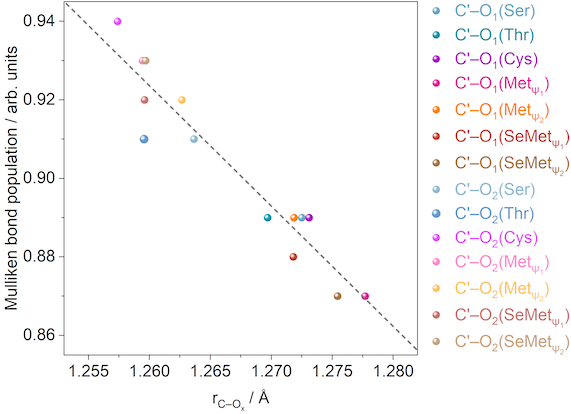}
    \caption{Correlation between the Mulliken bond population between the \ce{C$'$} and \ce{O1}, and between \ce{C$'$} and \ce{O2}, respectively, and the corresponding bond lengths, $r_{C'-O_1}$, and $r_{C'-O_2}$ for the S/Se-containing group, including different conformations for Met, and SeMet.}
    \label{S_Se_population correlation}
\end{figure}

\cleardoublepage

\section{Comparison of PDOS to XPS Valence Band}

\subsection{Aliphatic Group}

\begin{figure}[htp]
\centering
    \includegraphics[keepaspectratio, width = 1.0\linewidth]{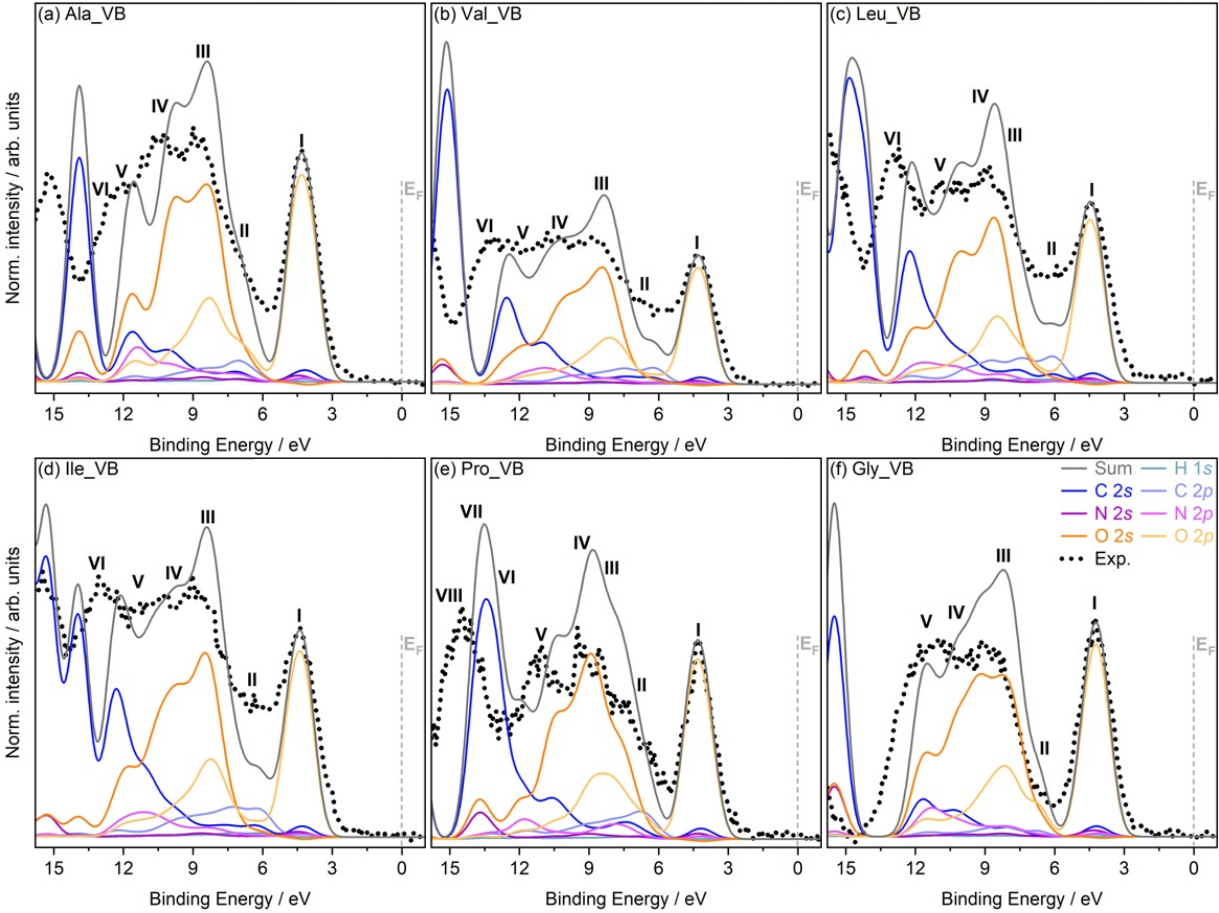}
    \caption{VB spectra of the aliphatic AAs, including (a) Ala, (b) Val, (c) Leu, (d) Ile, (e) Pro, and (f) Gly as reference. All plots display the one-electron photoionisation cross-section weighted PDOS, as well as the sum of all PDOS, from PBE-based DFT calculations and the experimental XP spectra. The Greek letters in black indicate the main spectral features with the Fermi level labelled as \ce{E_F} and marked by a light grey dashed line. The weighted PDOS have been aligned and normalised to the lowest BE feature in the experiment. The legend shown in (f) applies to all subfigures.}
    \label{Ali_VB}
\end{figure}

\begin{figure}[htp]
\centering
    \includegraphics[keepaspectratio, width = 1.0\linewidth]{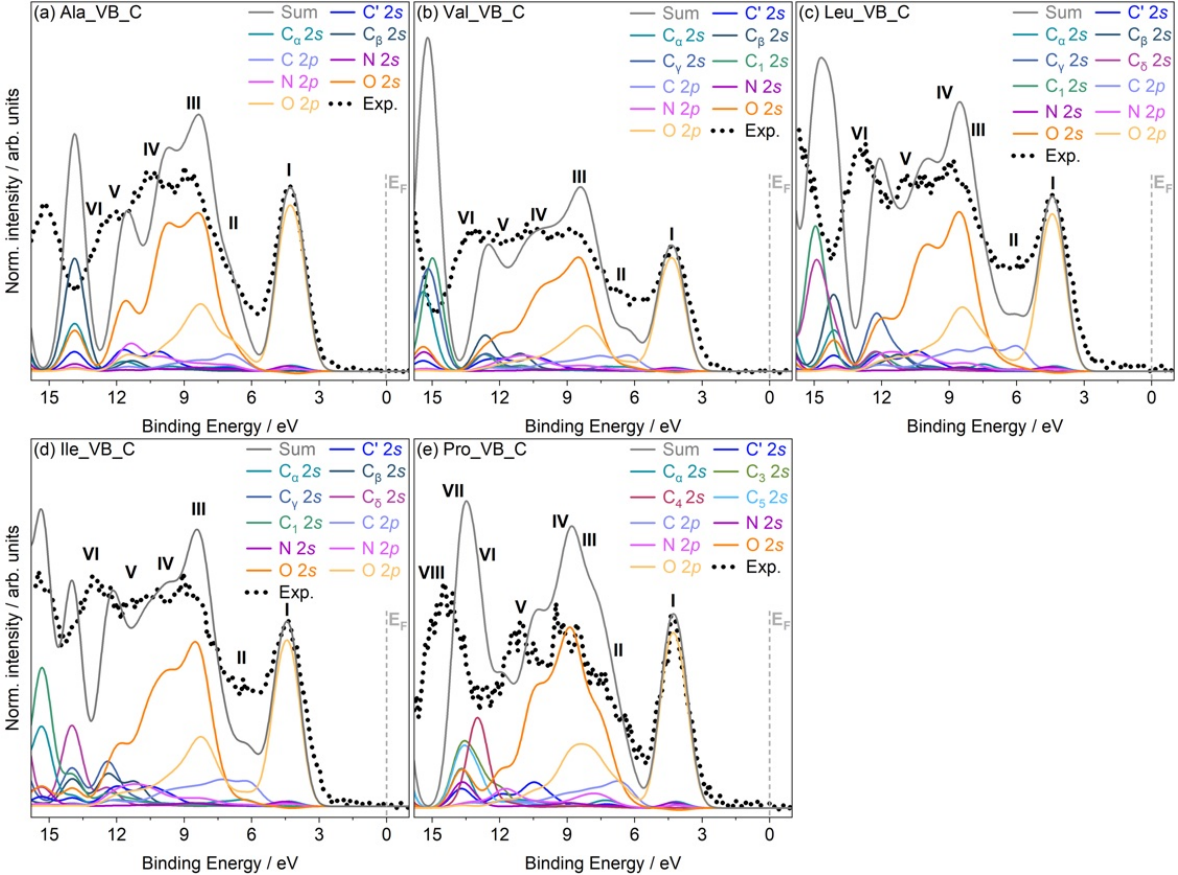}
    \caption{VB spectra of the aliphatic AAs, including (a) Ala, (b) Val, (c) Leu, (d) Ile, and (e) Pro with contributions projected onto specific C atoms. All plots display the one-electron photoionisation cross-section weighted PDOS, as well as the sum of all PDOS, from PBE-based DFT calculations and the experimental XP spectra. To better visualise the contributions from specific C atoms, only C~2\textit{s} states from specific C atoms are shown. C~2\textit{p} state is the sum for all C atoms. The Greek letters in black indicate the main spectral features with the Fermi level labelled as \ce{E_F} and marked by a light grey dashed line. The weighted PDOS have been aligned and normalised to the lowest BE feature in the experiment.}
    \label{Ali_VB_C}
\end{figure}

\clearpage
\newpage
\subsection{Aromatic Group}

\begin{figure}[htp]
\centering
    \includegraphics[keepaspectratio, width = 1.0\linewidth]{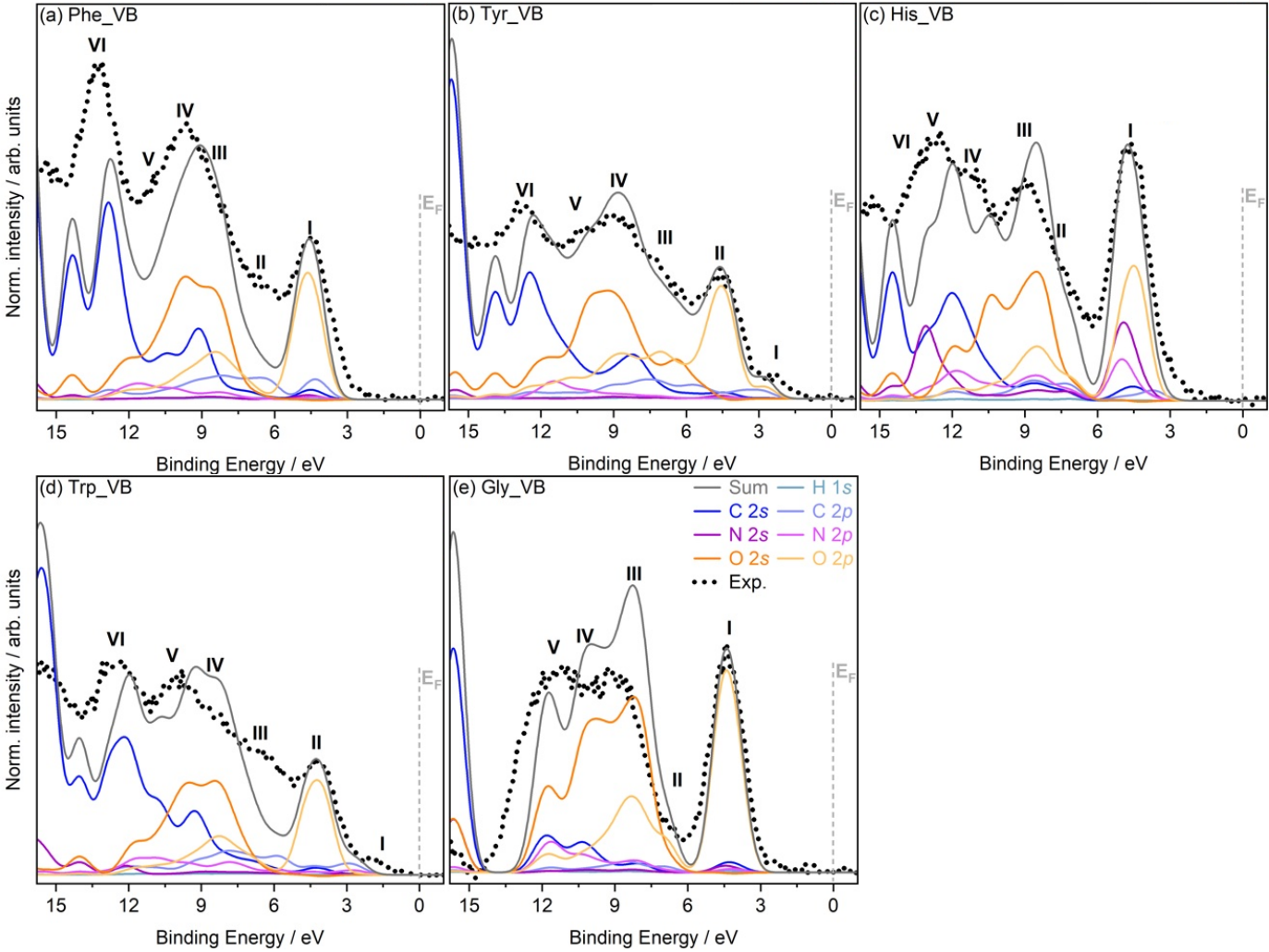}
    \caption{VB spectra of the aromatic AAs, including (a) Phe, (b) Tyr, (c) His, (d) Trp, and (e) Gly as reference. All plots display the one-electron photoionisation cross-section weighted PDOS, as well as the sum of all PDOS, from PBE-based DFT calculations and the experimental XP spectra. The labels in dark grey indicate the majority orbital contribution to the spectral features determined from DFT. The weighted PDOS have been aligned and normalised to the lowest BE feature in the experiment. The legend shown in (e) applies to all subfigures.}
    \label{Aro_VB}
\end{figure}

\begin{figure}[htp]
\centering
    \includegraphics[keepaspectratio, width = 0.75\linewidth]{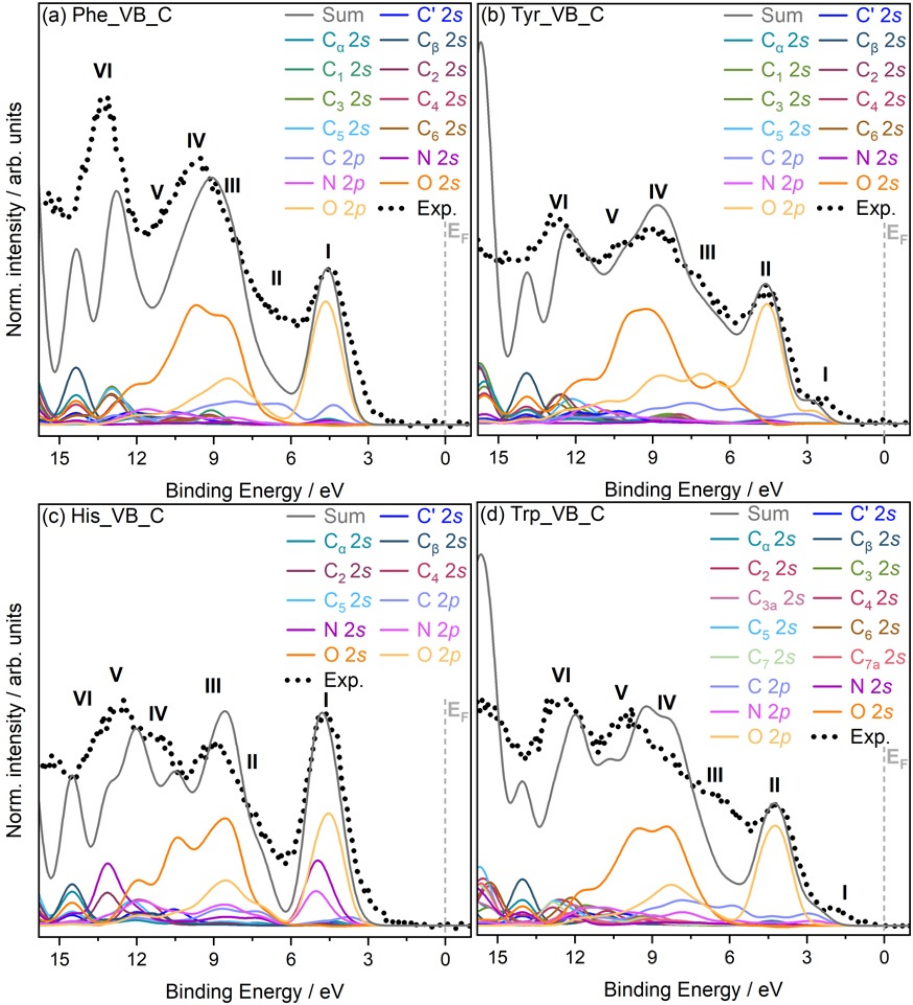}
    \caption{VB spectra of the aromatic AAs, including (a) Phe, (b) Tyr, (c) His, and (d) Trp with contributions projected onto specific C atoms. All plots display the one-electron photoionisation cross-section weighted PDOS, as well as the sum of all PDOS, from PBE-based DFT calculations and the experimental XP spectra. To better visualise the contributions from specific C atoms, only C~2\textit{s} states from specific C atoms are shown. C~2\textit{p} state is the sum for all C atoms. The Greek letters in black indicate the main spectral features with the Fermi level labelled as \ce{E_F} and marked by a light grey dashed line. The weighted PDOS have been aligned and normalised to the lowest BE feature in the experiment.}
    \label{Aro_VB_C}
\end{figure}

\begin{figure}[htp]
\centering
    \includegraphics[keepaspectratio, width = 0.75\linewidth]{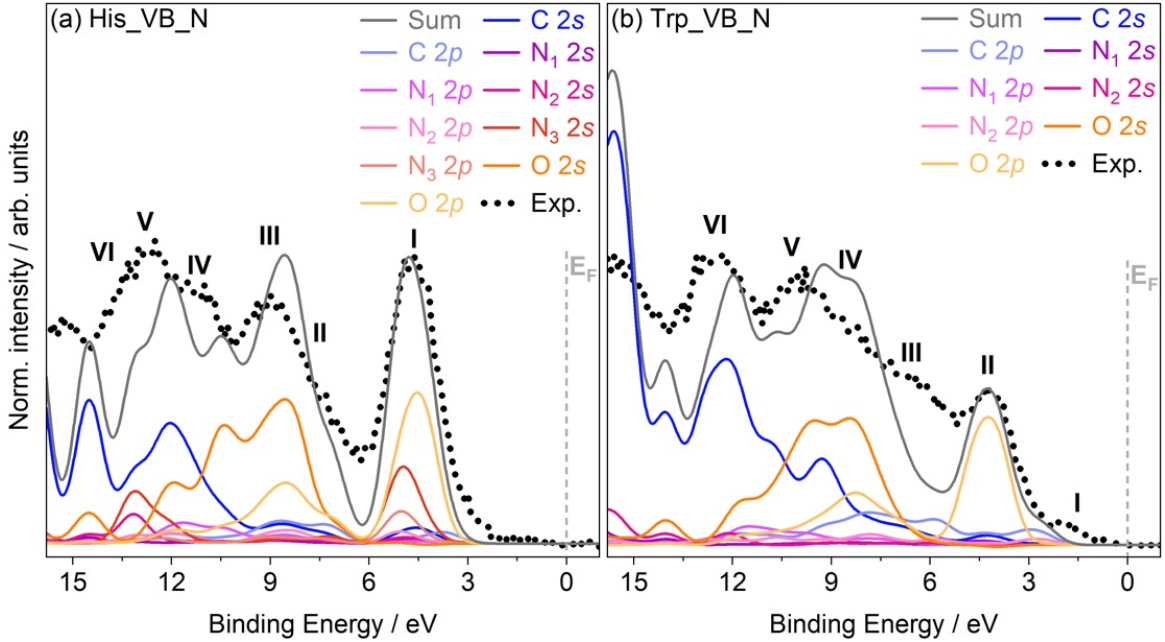}
    \caption{VB spectra of the aromatic AAs, including (a) His, and (b) Trp with contributions projected onto specific N atoms. All plots display the one-electron photoionisation cross-section weighted PDOS, as well as the sum of all PDOS, from PBE-based DFT calculations and the experimental XP spectra. The Greek letters in black indicate the main spectral features with the Fermi level labelled as \ce{E_F} and marked by a light grey dashed line. The weighted PDOS have been aligned and normalised to the lowest BE feature in the experiment.}
    \label{Aro_VB_N}
\end{figure}

\begin{figure}[htp]
\centering
    \includegraphics[keepaspectratio, width = 0.75\linewidth]{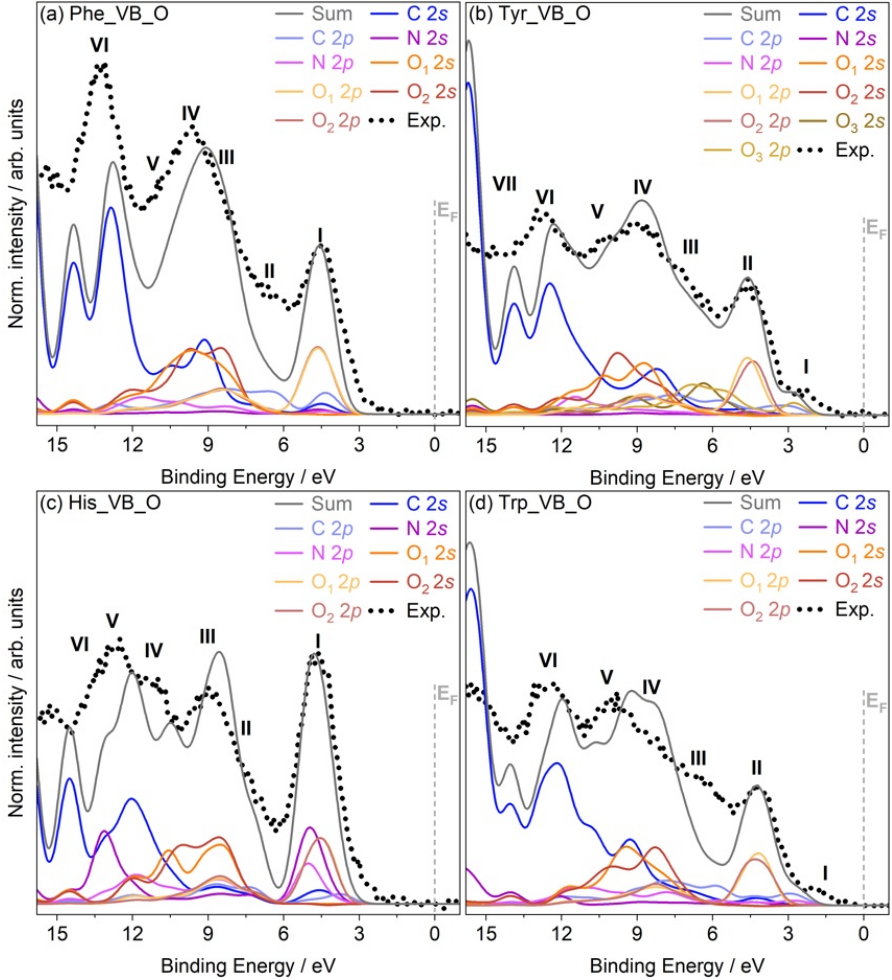}
    \caption{VB spectra of the aromatic AAs, including (a) Phe, (b) Tyr, (c) His, and (d) Trp with contributions projected onto specific O atoms. All plots display the one-electron photoionisation cross-section weighted PDOS, as well as the sum of all PDOS, from PBE-based DFT calculations and the experimental XP spectra. The Greek letters in black indicate the main spectral features with the Fermi level labelled as \ce{E_F} and marked by a light grey dashed line. The weighted PDOS have been aligned and normalised to the lowest BE feature in the experiment.}
    \label{Aro_VB_O}
\end{figure}

\clearpage
\newpage
\subsection{Polar Side Chain-Containing Group}

\begin{figure}[htp]
\centering
    \includegraphics[keepaspectratio, width = 0.8\linewidth]{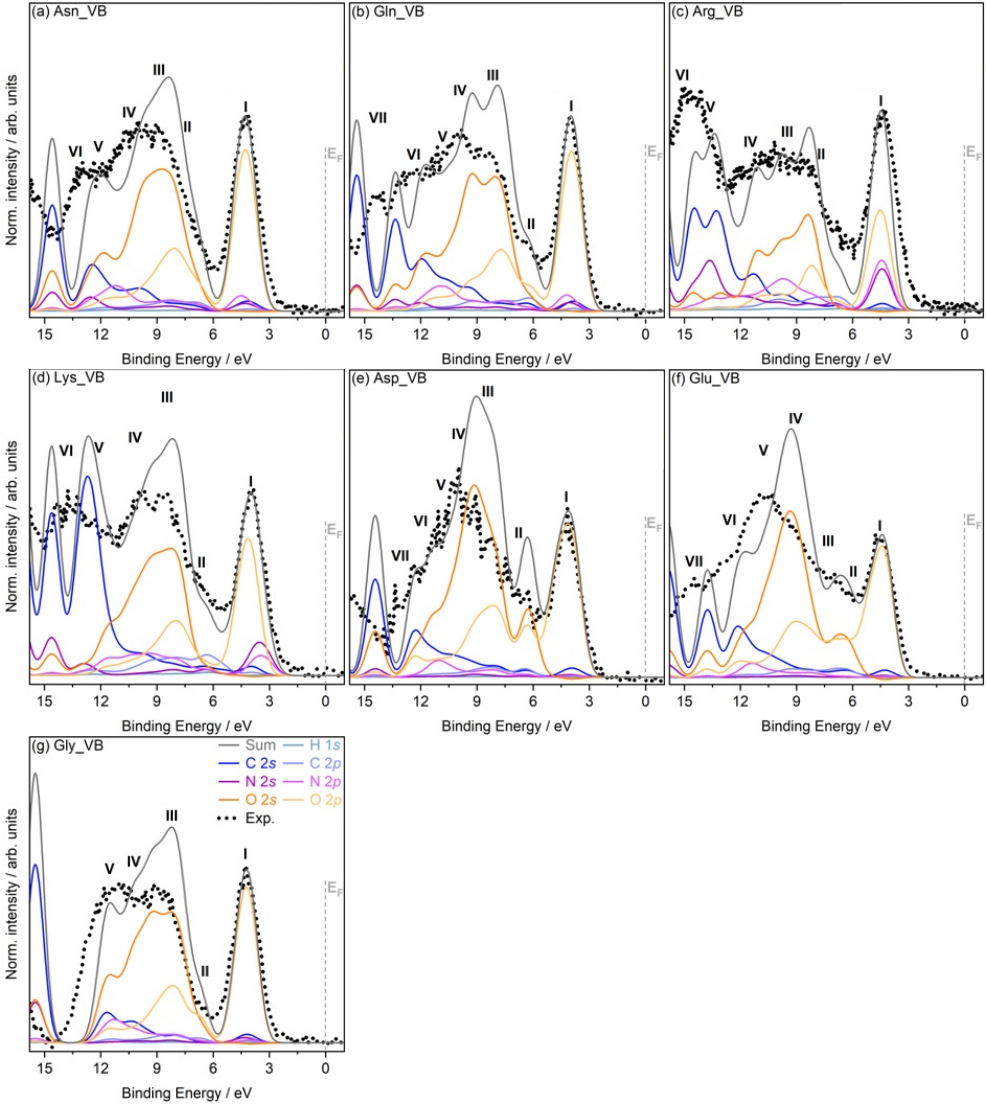}
    \caption{VB spectra of the polar side chain-containing AAs, including (a) Asn, (b) Gln, (c) Arg, (d) Lys, (e) Asp, (f) Glu, and (g) Gly as reference. All plots display the one-electron photoionisation cross-section weighted PDOS, as well as the sum of all PDOS, from PBE-based DFT calculations and the experimental XP spectra. The labels in dark grey indicate the majority orbital contribution to the spectral features determined from DFT. The weighted PDOS have been aligned and normalised to the lowest BE feature in the experiment. The legend shown in (g) applies to all subfigures.}
    \label{other_VB}
\end{figure}

\clearpage
\begin{figure}[htp]
\centering
    \includegraphics[keepaspectratio, width = 1.0\linewidth]{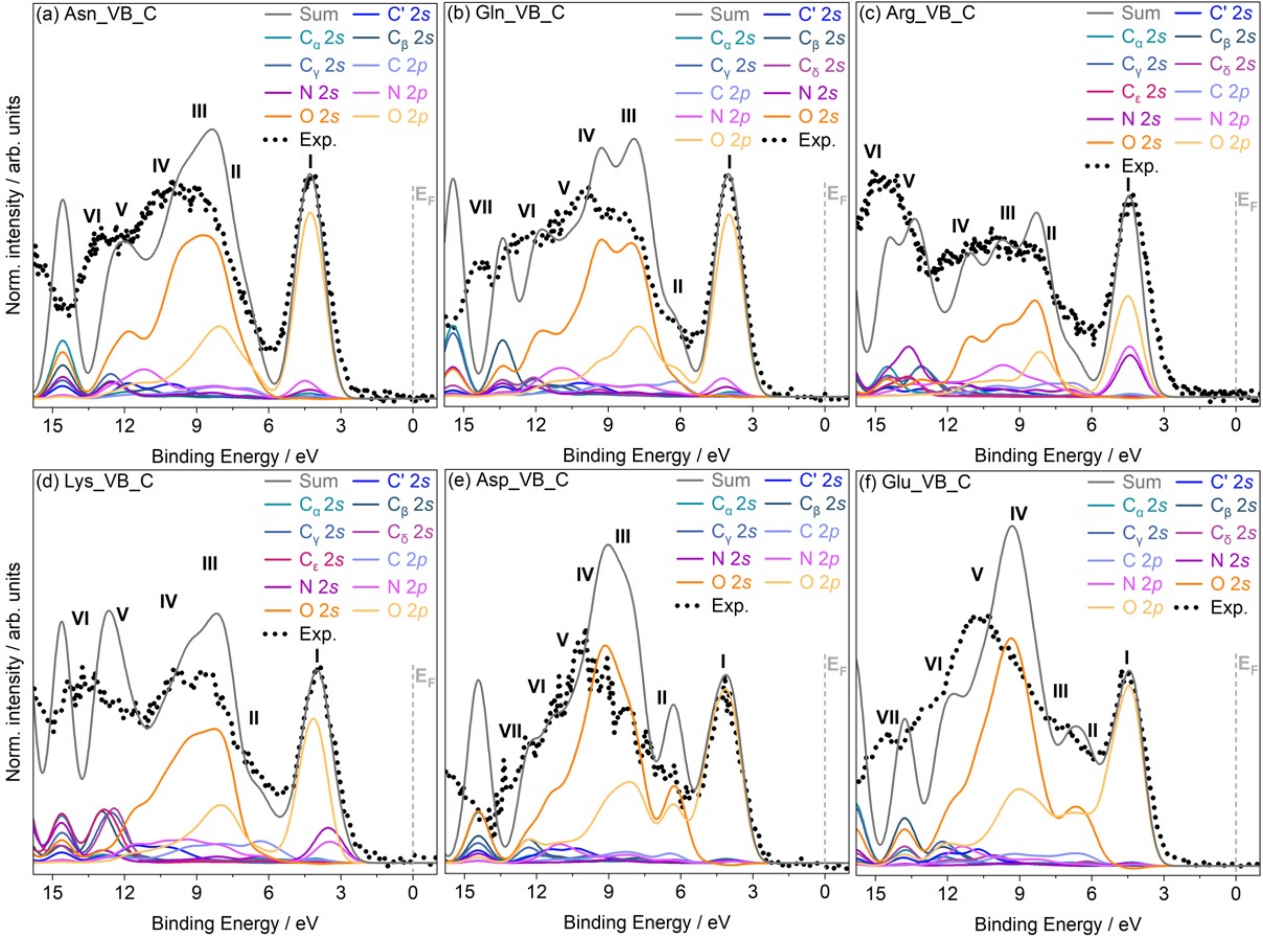}
    \caption{VB spectra of the polar side chain-containing AAs, including (a) Asn, (b) Gln, (c) Arg, (d) Lys, (e) Asp, and (f) Glu with contributions from specific C atoms. All plots display the one-electron photoionisation cross-section weighted PDOS, as well as the sum of all PDOS, from PBE-based DFT calculations and the experimental XP spectra. To better visualise the contributions from specific C atoms, only C~2\textit{s} states from specific C atoms are shown. C~2\textit{p} state is the sum for all C atoms. The Greek letters in black indicate the main spectral features with the Fermi level labelled as \ce{E_F} and marked by a light grey dashed line. The weighted PDOS have been aligned and normalised to the lowest BE feature in the experiment.}
    \label{other_VB_C}
\end{figure}

\begin{figure}[htp]
\centering
    \includegraphics[keepaspectratio, width = 0.75\linewidth]{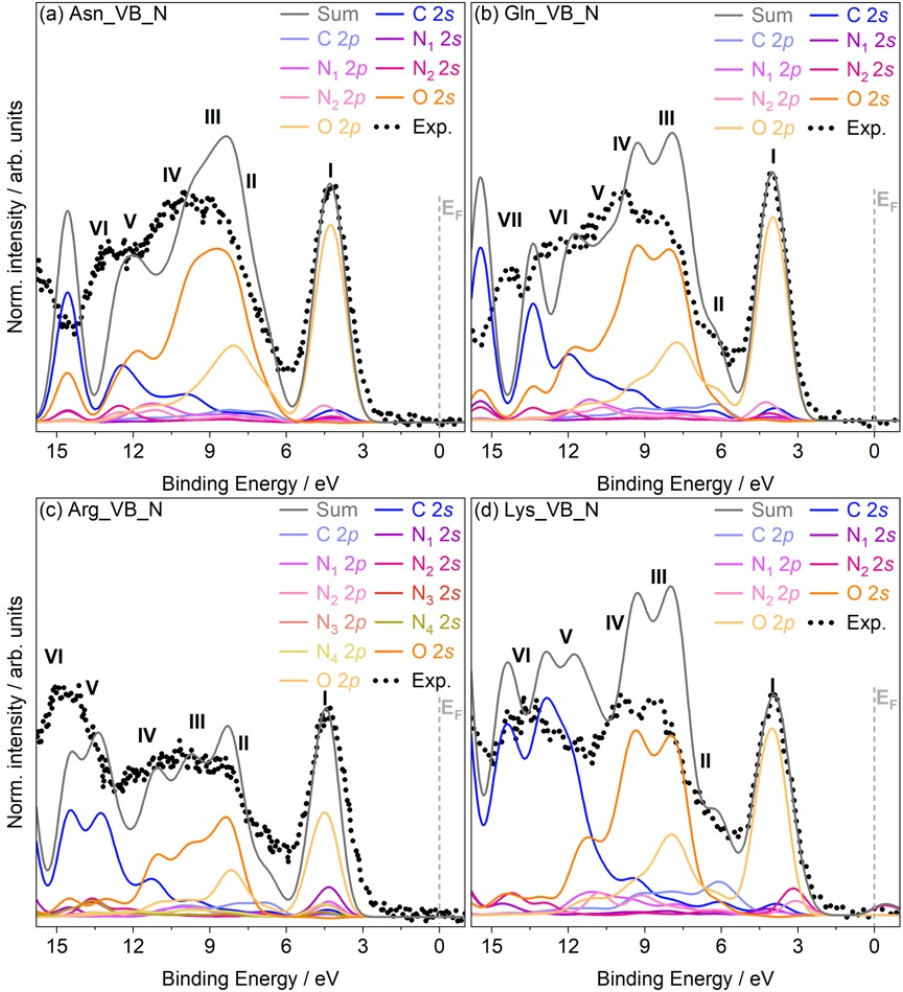}
    \caption{VB spectra of the polar side chain-containing AAs, including (a) Asn, (b) Gln, (c) Arg, and (d) Lys with contributions from specific N atoms. All plots display the one-electron photoionisation cross-section weighted PDOS, as well as the sum of all PDOS, from PBE-based DFT calculations and the experimental XP spectra. The Greek letters in black indicate the main spectral features with the Fermi level labelled as \ce{E_F} and marked by a light grey dashed line. The weighted PDOS have been aligned and normalised to the lowest BE feature in the experiment.}
    \label{other_VB_N}
\end{figure}

\begin{figure}[htp]
\centering
    \includegraphics[keepaspectratio, width = 1.0\linewidth]{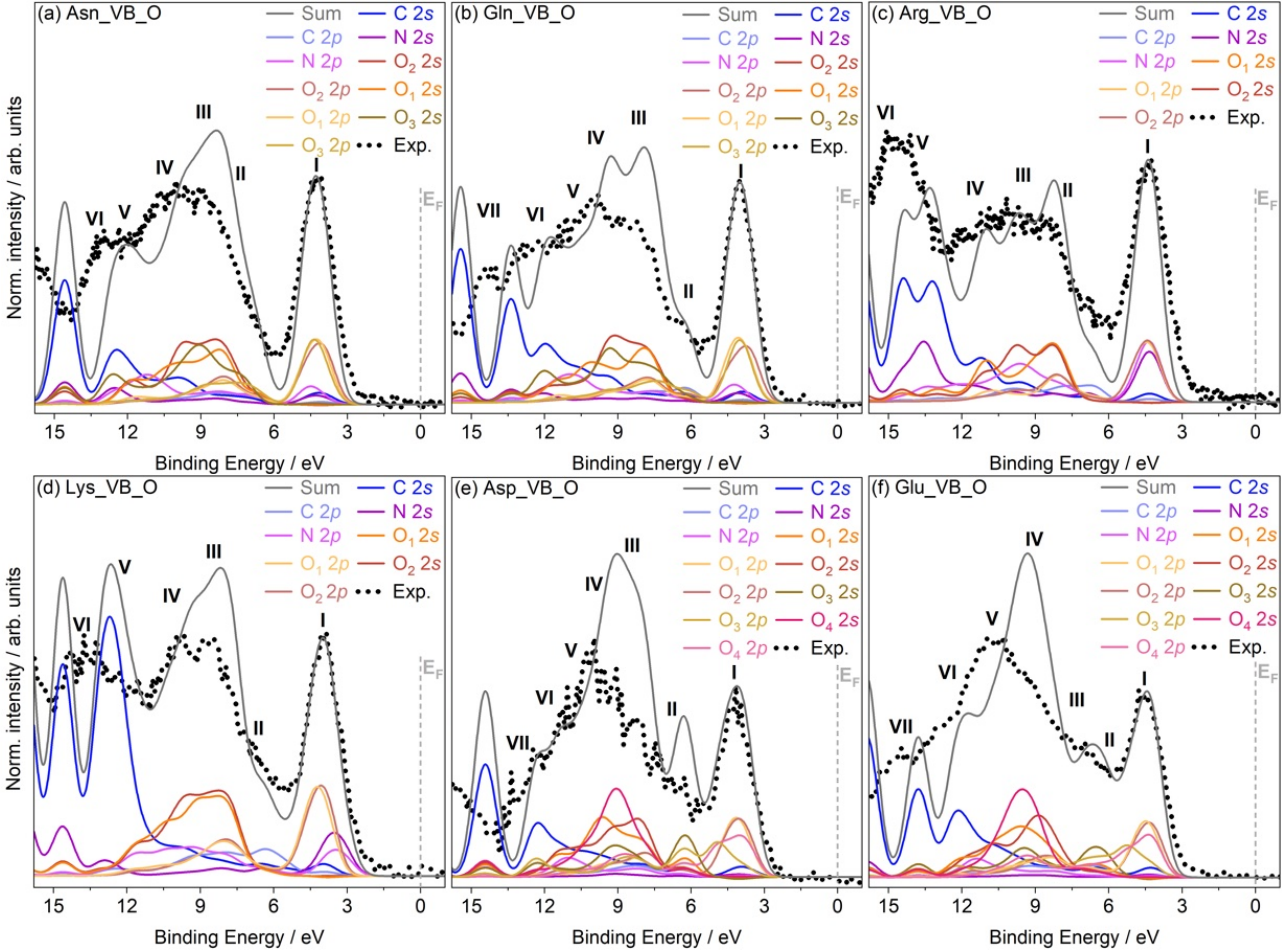}
    \caption{VB spectra of the polar side chain-containing AAs, including (a) Asn, (b) Gln, (c) Arg, (d) Lys, (e) Asp, and (f) Glu with contributions from specific O atoms. All plots display the one-electron photoionisation cross-section weighted PDOS, as well as the sum of all PDOS, from PBE-based DFT calculations and the experimental XP spectra. The Greek letters in black indicate the main spectral features with the Fermi level labelled as \ce{E_F} and marked by a light grey dashed line. The weighted PDOS have been aligned and normalised to the lowest BE feature in the experiment.}
    \label{other_VB_O}
\end{figure}

\clearpage
\newpage
\subsection{S/Se-containing Group}

\begin{figure}[htp]
\centering
    \includegraphics[keepaspectratio, width = 1.0\linewidth]{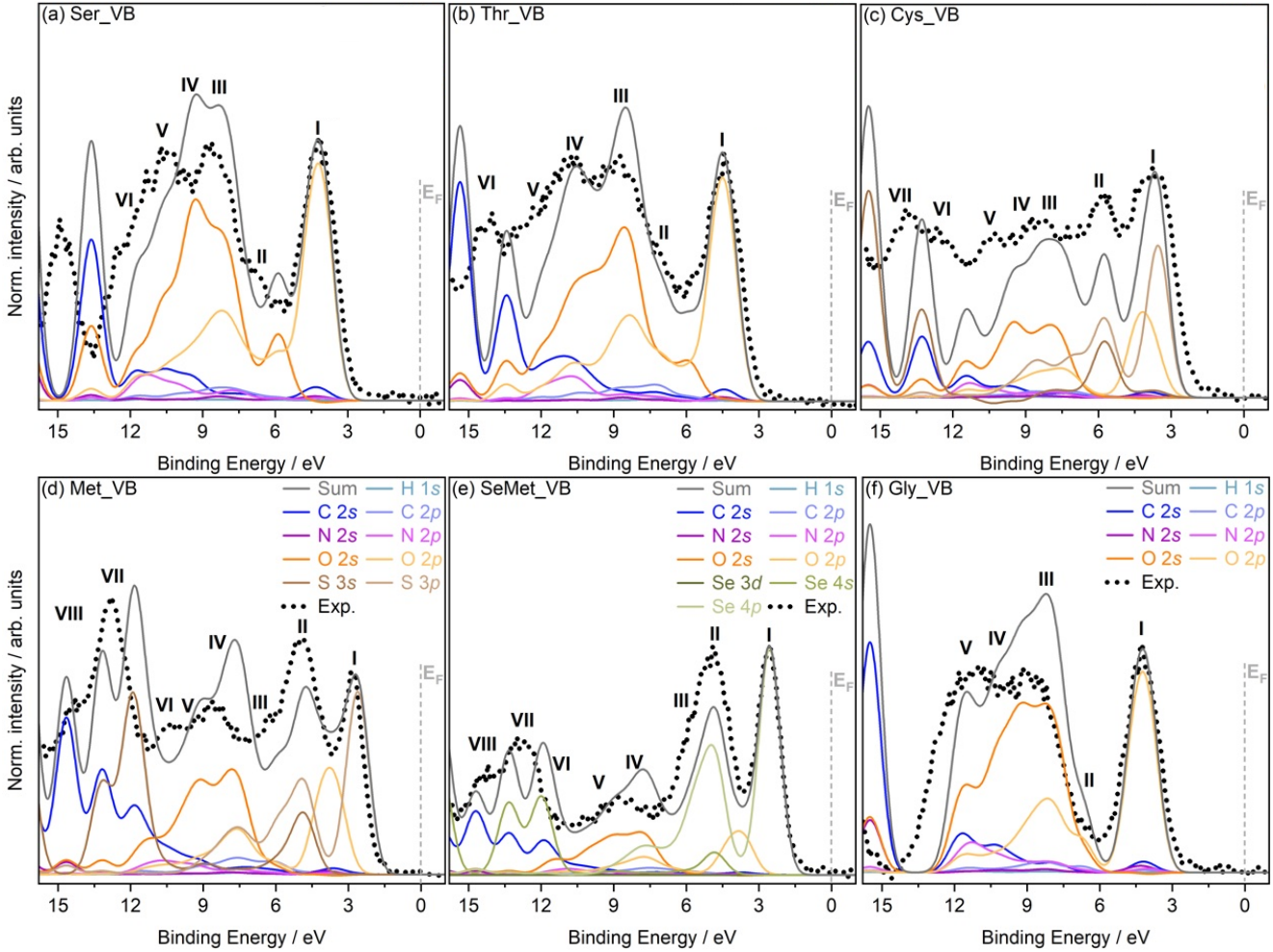}
    \caption{VB spectra of the S/Se-containing AAs, including (a) Ser, (b) Thr, (c) Cys, (d) Met, (e) SeMet, and (f) Gly as reference. All plots display the one-electron photoionisation cross-section weighted PDOS, as well as the sum of all PDOS, from PBE-based DFT calculations and the experimental XP spectra. The Greek letters in black indicate the main spectral features with the Fermi level labelled as \ce{E_F} and marked by a light grey dashed line. The weighted PDOS have been aligned and normalised to the lowest BE feature in the experiment. The legend shown in (f) applies to subfigures (a) and (b); the legend present in (d) applies to subfigure (c); and the legend in (e) only applies to itself.}
    \label{S_Se_VB}
\end{figure}

\begin{figure}[htp]
\centering
    \includegraphics[keepaspectratio, width = 1.0\linewidth]{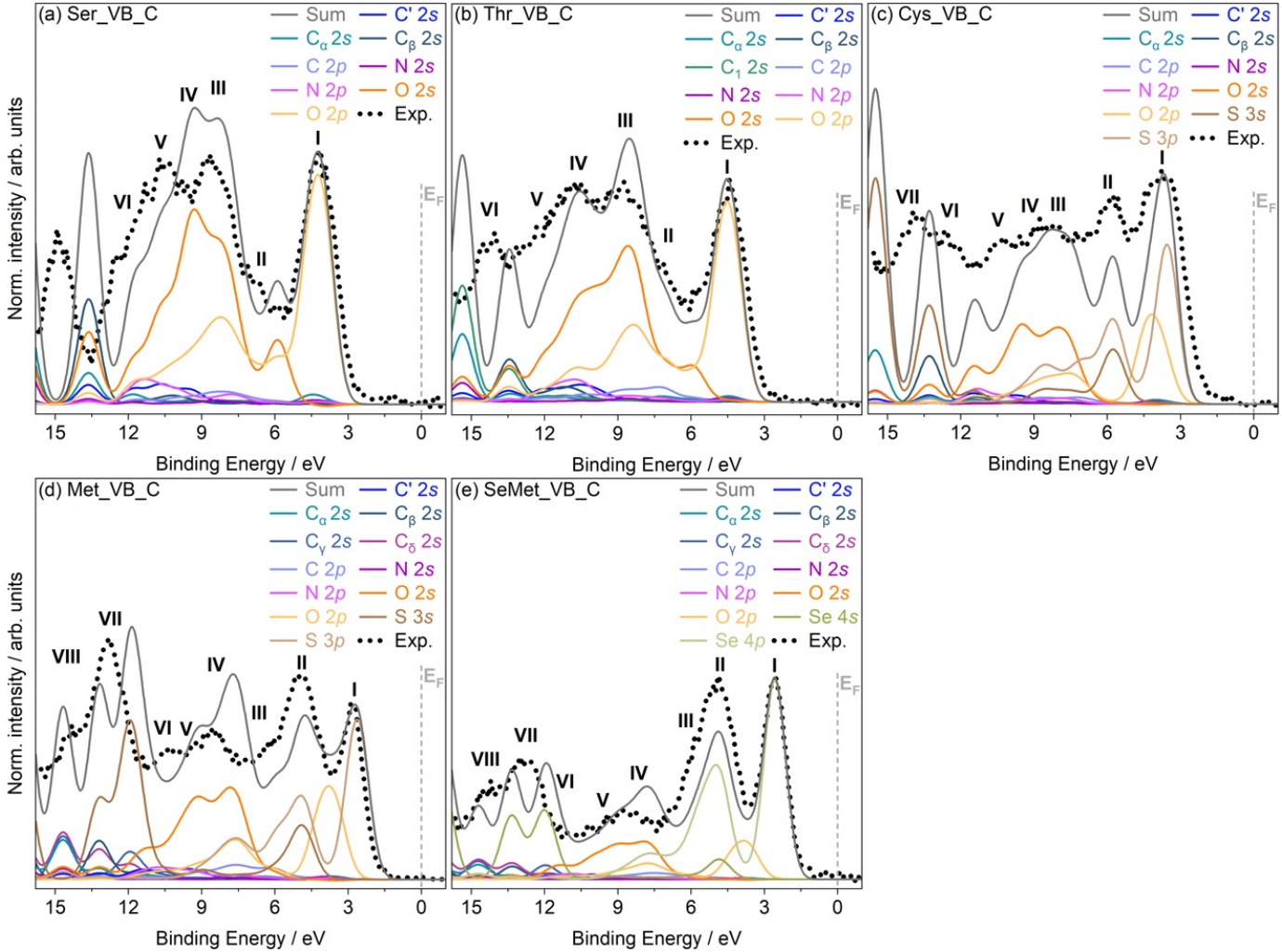}
    \caption{VB spectra of the S/Se-containing AAs, including (a) Ser, (b) Thr, (c) Cys, (d) Met, and (e) SeMet with contributions from specific C atoms. All spectra show the one-electron photoionisation cross-section weighted PDOS from PBE-based DFT calculations and the experimental XP spectra. To better visualise the contributions from C~2\textit{s} states, the sum of C~2\textit{p} states from specific C atoms is present. The Greek letters in black indicate the main spectral features with the Fermi level labelled as \ce{E_F} and marked by a light grey dashed line. The weighted PDOS have been aligned and normalised to the lowest BE feature in the experiment.}
    \label{S_Se_VB_C}
\end{figure}

\begin{figure}[htp]
\centering
    \includegraphics[keepaspectratio, width = 1.0\linewidth]{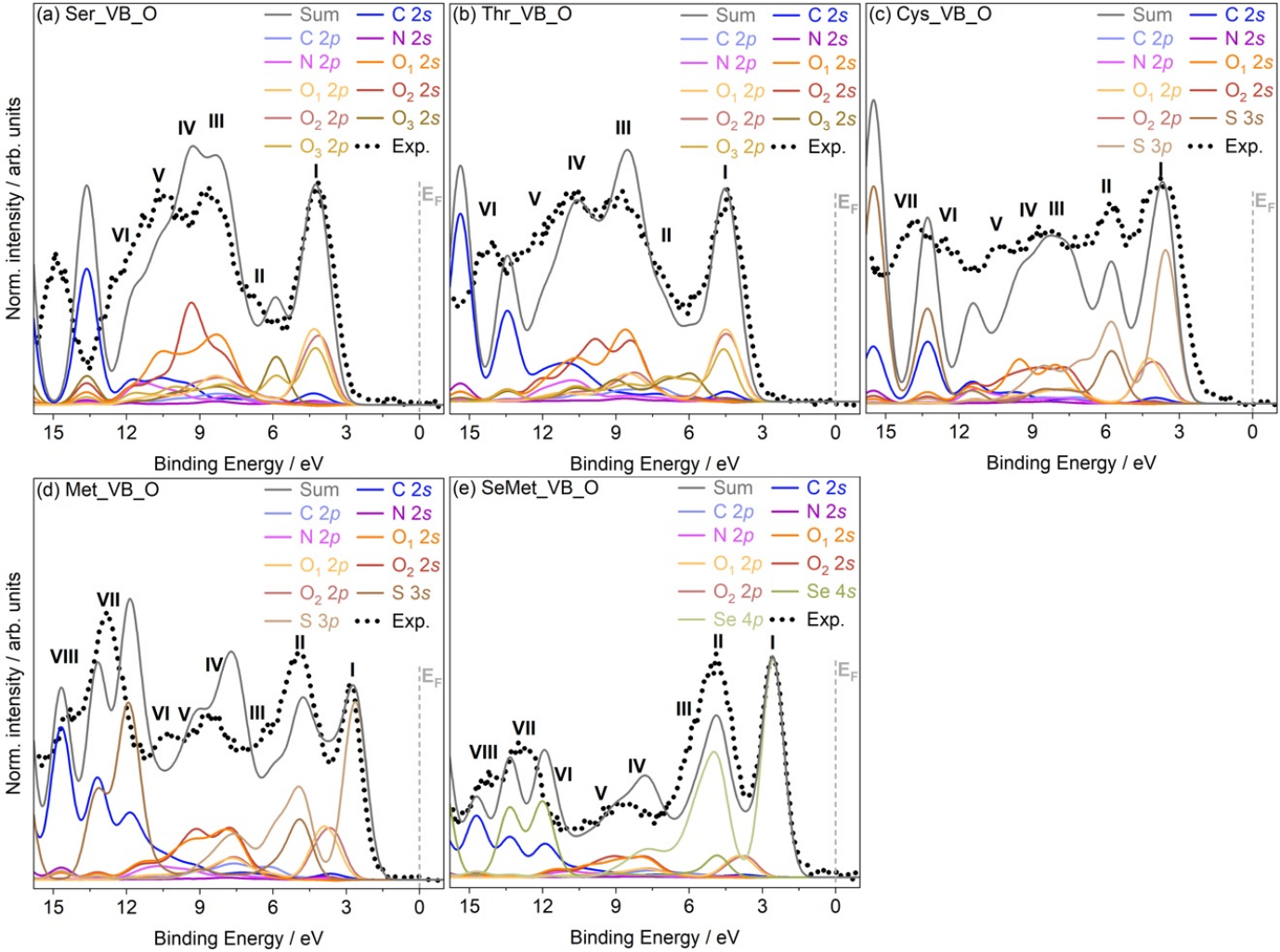}
    \caption{VB spectra of the S/Se-containing AAs, including (a) Ser, (b) Thr, (c) Cys, (d) Met, and (e) SeMet with contributions from specific O atoms. All plots display the one-electron photoionisation cross-section weighted PDOS, as well as the sum of all PDOS, from PBE-based DFT calculations and the experimental XP spectra. The Greek letters in black indicate the main spectral features with the Fermi level labelled as \ce{E_F} and marked by a light grey dashed line. The weighted PDOS have been aligned and normalised to the lowest BE feature in the experiment.}
    \label{S_Se_VB_O}
\end{figure}

\clearpage
\newpage
\subsection{Correlation of Valence Band Maximum and Optical Band Gap}

\begin{figure}[htp]
    \centering
    \includegraphics[keepaspectratio, width=0.8\textwidth]{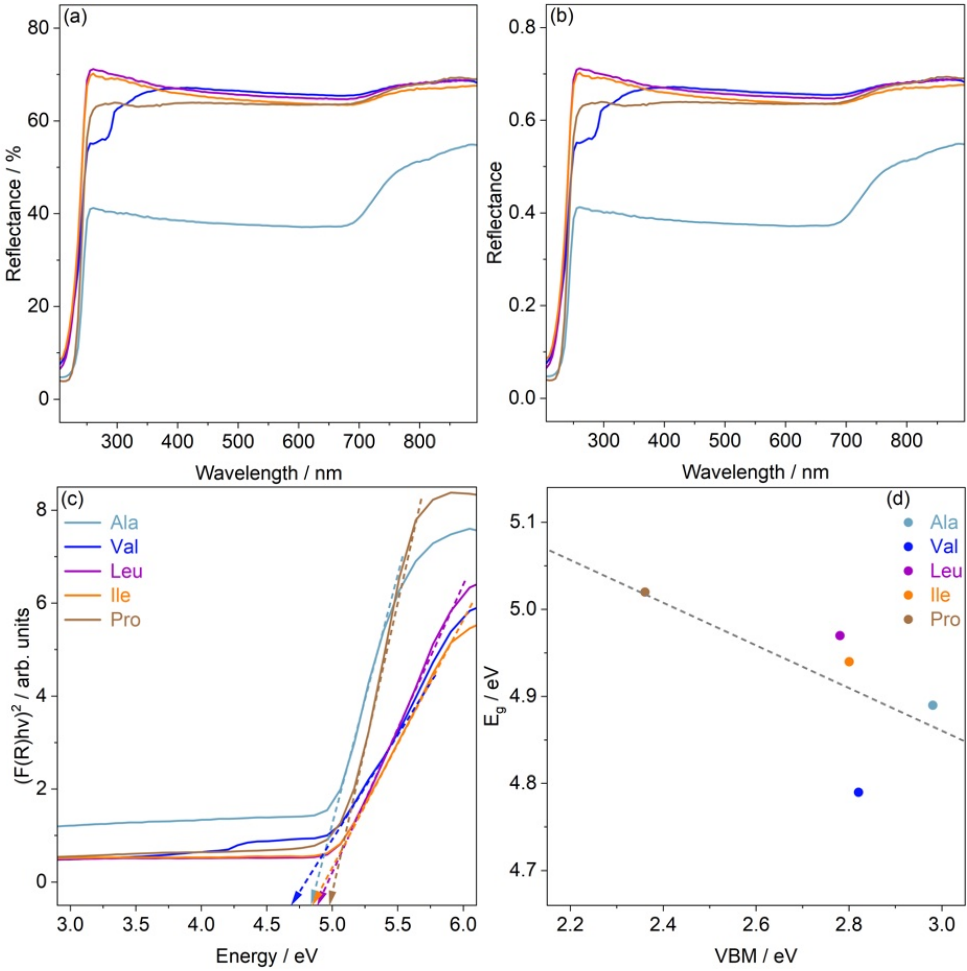}
    \caption{The UV-Vis reflectance data for the aliphatic group in (a) R($\%$) vs.$\lambda$, (b) R vs.$\lambda$, (c) Tauc plot, and (d) correlation plot of VBM and optical \ce{E^{solid}_g} values. The legend in (c) is also applied to (a) and (b).}
    \label{Ali_UV}
\end{figure} 

\textbf{\begin{table}
    \caption{Summary of VBM values which are determined from the as-collected VB spectra as the intersection of the linear fits to the final drop in intensity and the background lines and their corresponding solid-state band gap values, \ce{E^{solid}_g}, which are dictated as the intersection of the linear fits to the Tauc plots made for the reflectance data and the x-axis, for the aliphatic group.}
\begin{center}
\begin{tabular}{c c c } 
 \hline
 AAs & \ce{VBM_{exp.}} / eV & \ce{E^{solid}_g} / eV \\ [0.5ex] 
 \hline
 Ala & 2.98 & 4.89 \\
 Val & 2.82 & 4.79 \\
 Leu & 2.78 & 4.97 \\
 Ile & 2.80 & 4.94 \\ 
 Pro & 2.36 & 5.02 \\
 \hline
\end{tabular}
\end{center}
\label{Ali_VBM}
\end{table}}

\begin{figure}[htp]
    \centering
    \includegraphics[keepaspectratio, width=1.0\textwidth]{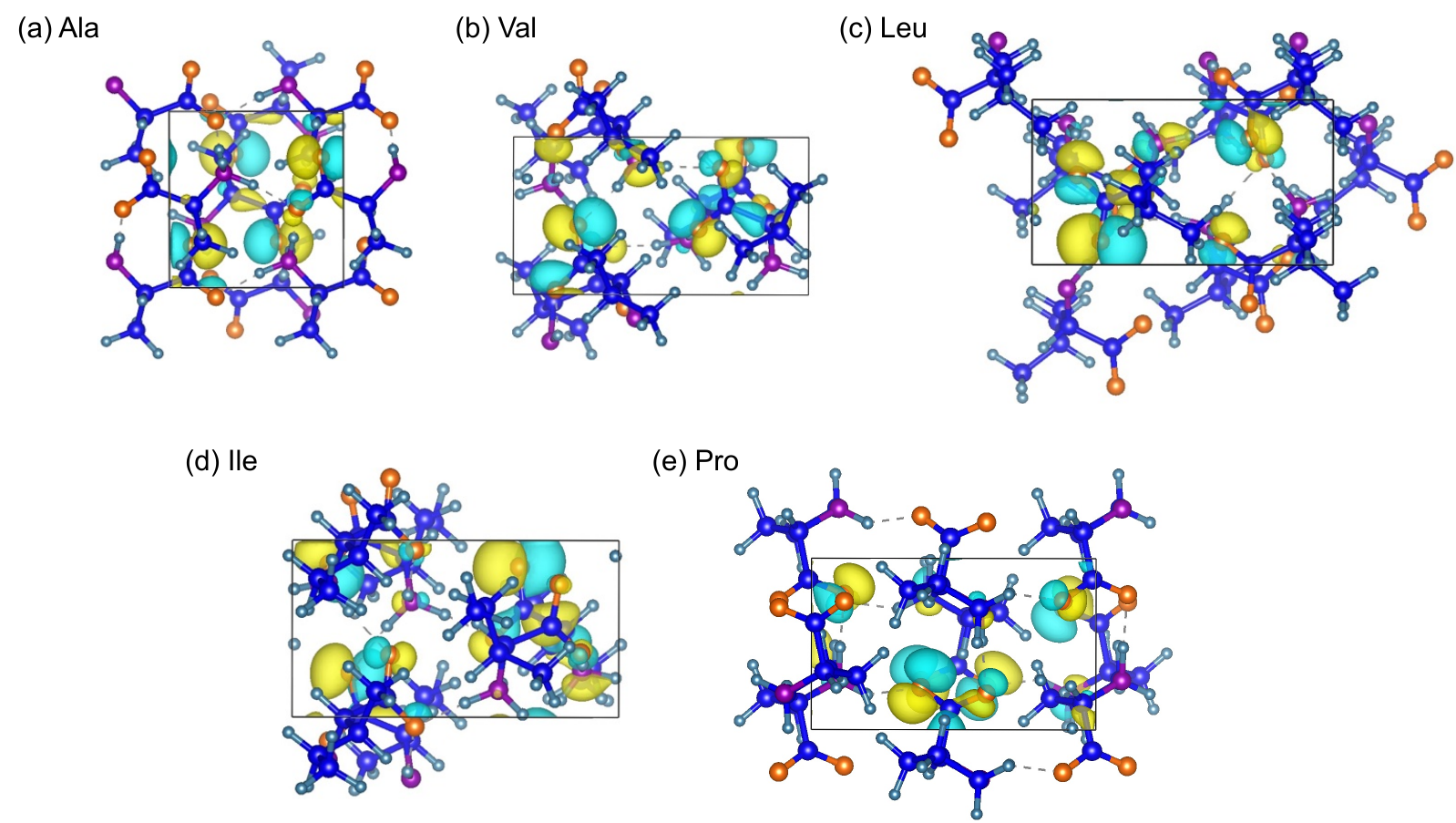}
    \caption{Visualisation of the highest occupied molecular orbitals (HOMOs) of the aliphatic AAs for (a) Ala, (b) Val, (c) Leu, (d) Ile, and (e) Pro using the VESTA software package.~\cite{Momma2011VESTA3Data} All H, C, N, and O atoms are inked in steel blue, deep blue, purple, and orange, respectively.}
    \label{Ali_HOMOs}
\end{figure} 

\begin{figure}[htp]
    \centering
    \includegraphics[keepaspectratio, width=0.8\textwidth]{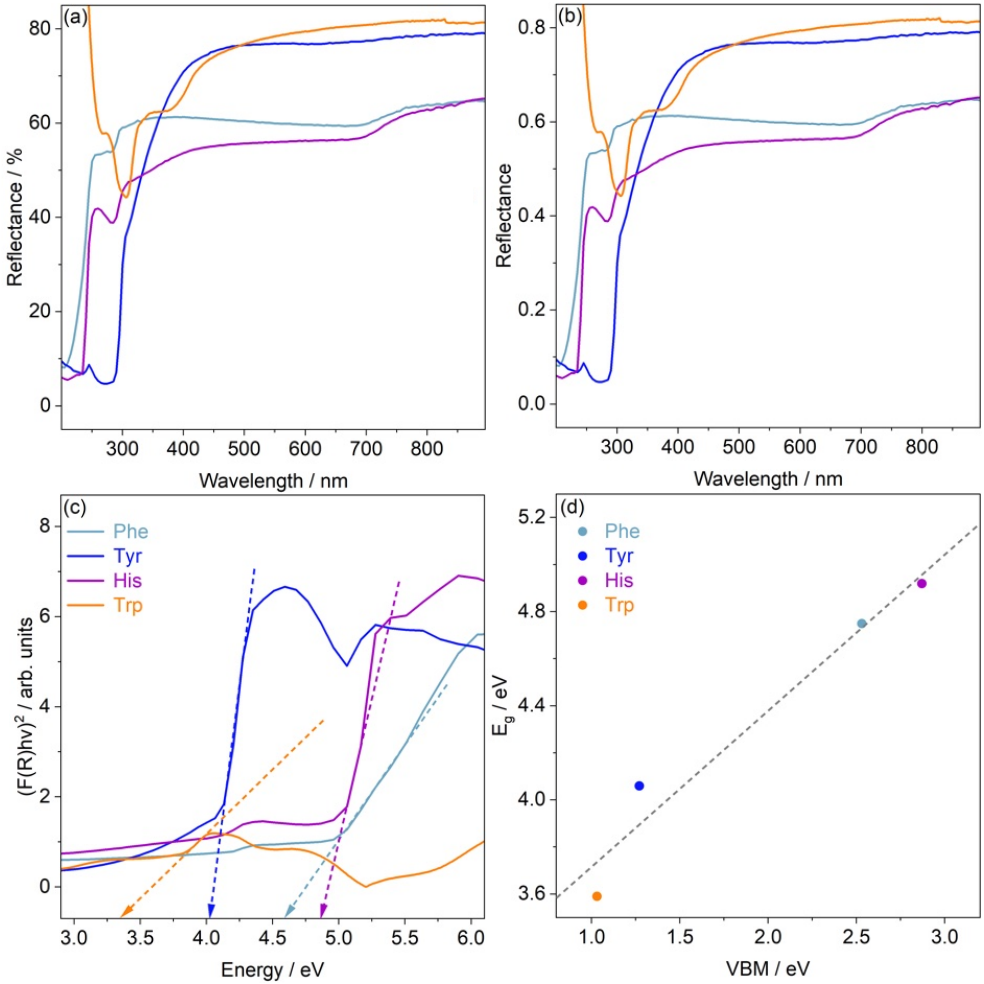}
    \caption{The UV-Vis reflectance data for the aromatic group in (a) R($\%$) vs.$\lambda$, (b) R vs.$\lambda$, (c) Tauc plot, and (d) correlation plot of VBM and optical \ce{E^{solid}_g} values. The legend in (c) is also applied to (a) and (b).}
    \label{Aro_UV}
\end{figure} 

\textbf{\begin{table}
    \caption{Summary of VBM values which are determined from the as-collected VB spectra as the intersection of the linear fits to the final drop in intensity and the background lines and their corresponding solid-state band gap values, \ce{E^{solid}_g}, which are dictated as the intersection of the linear fits to the Tauc plots made for the reflectance data and the x-axis, for the aromatic group.}
\begin{center}
\begin{tabular}{c c c} 
 \hline
 AAs & \ce{VBM_{exp.}} / eV & \ce{E^{solid}_g} / eV \\ [0.5ex] 
 \hline
 Phe & 2.53 & 4.75 \\
 Tyr & 1.27 & 4.06 \\ 
 His & 2.87 & 4.92 \\
 Trp & 1.03 & 3.59 \\ 
 \hline
\end{tabular}
\end{center}
\label{Aro_VBM}
\end{table}}

\begin{figure}[htp]
    \centering
    \includegraphics[keepaspectratio, width=1.0\textwidth]{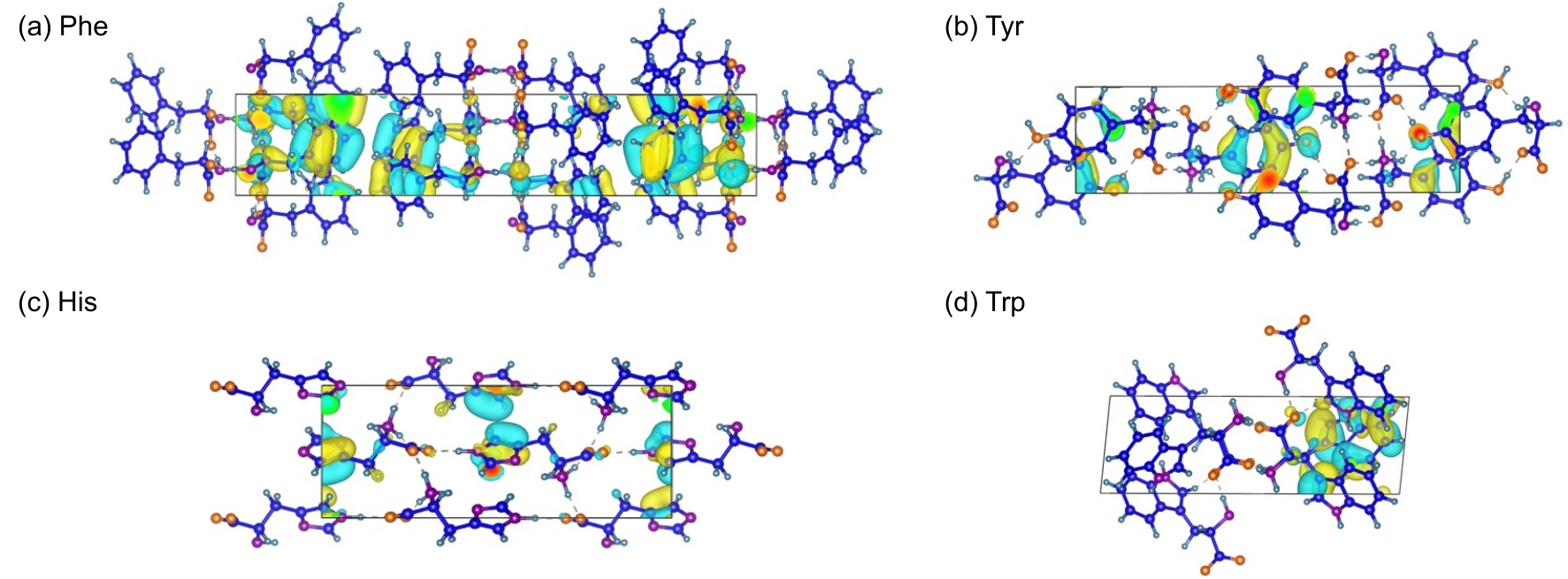}
    \caption{Visualisation of the highest occupied molecular orbitals (HOMOs) of the aromatic AAs for (a) Phe, (b) Tyr, (c) His, and (d) Trp using the VESTA software package.~\cite{Momma2011VESTA3Data} All H, C, N, and O atoms are inked in steel blue, deep blue, purple, and orange, respectively.}
    \label{Aro_HOMOs}
\end{figure} 

\begin{figure}[htp]
    \centering
    \includegraphics[keepaspectratio, width=0.8\textwidth]{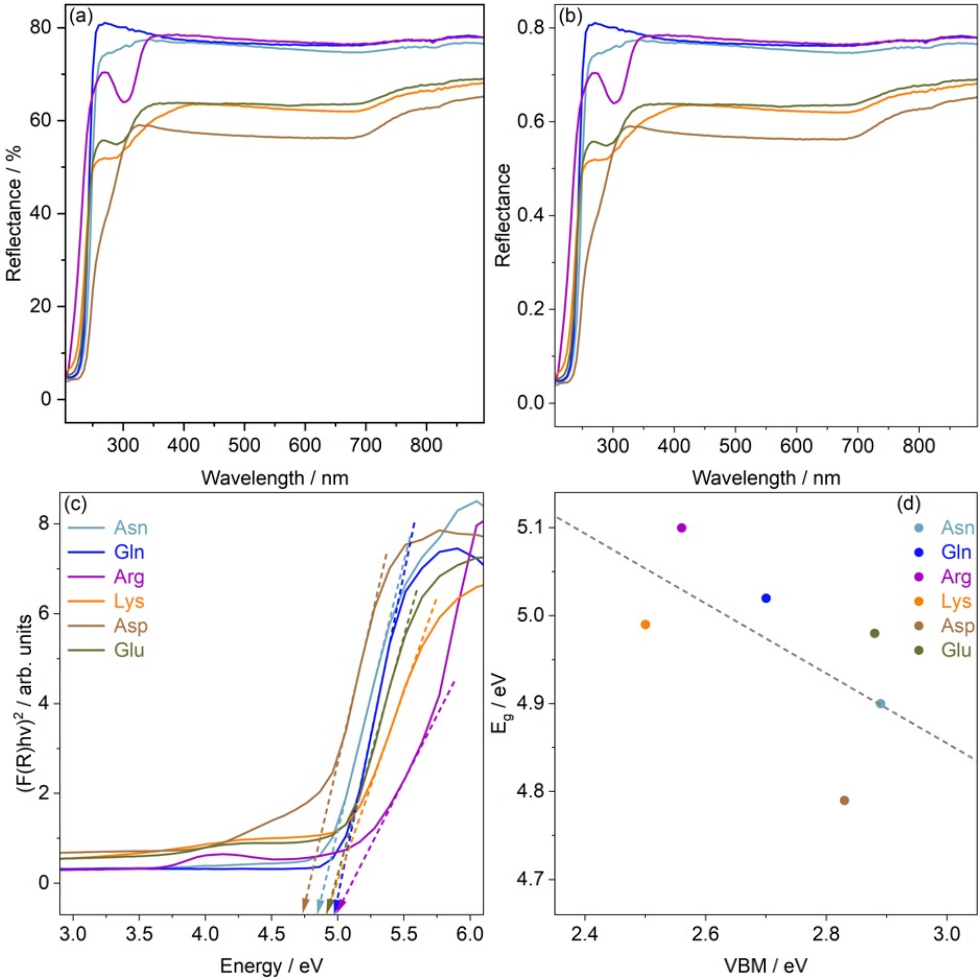}
    \caption{The UV-Vis reflectance data for the polar side chain-containing group in (a) R($\%$) vs.$\lambda$, (b) R vs.$\lambda$, (c) Tauc plot, and (d) correlation plot of VBM and optical \ce{E^{solid}_g} values. The legend in (c) is also applied to (a) and (b).}
    \label{other_UV}
\end{figure} 

\textbf{\begin{table}[htp]
    \caption{Summary of VBM values which are determined from the as-collected VB spectra as the intersection of the linear fits to the final drop in intensity and the background lines and their corresponding solid-state band gap values, \ce{E^{solid}_g}, which are dictated as the intersection of the linear fits to the Tauc plots made for the reflectance data and the x-axis, for the polar side chain-containing group.}
\begin{center}
\begin{tabular}{c c c} 
 \hline
 AAs & \ce{VBM_{exp.}} / eV & \ce{E^{solid}_g} / eV \\ [0.5ex] 
 \hline
 Asn & 2.89 & 4.90 \\
 Gln & 2.70 & 5.02 \\ 
 Arg & 2.56 & 5.10 \\
 Lys & 2.50 & 4.99 \\ 
 Asp & 2.83 & 4.79 \\
 Glu & 2.88 & 4.98 \\
 \hline
\end{tabular}
\end{center}
\label{other_VBM}
\end{table}} 

\begin{figure}[htp]
    \centering
    \includegraphics[keepaspectratio, width=1.0\textwidth]{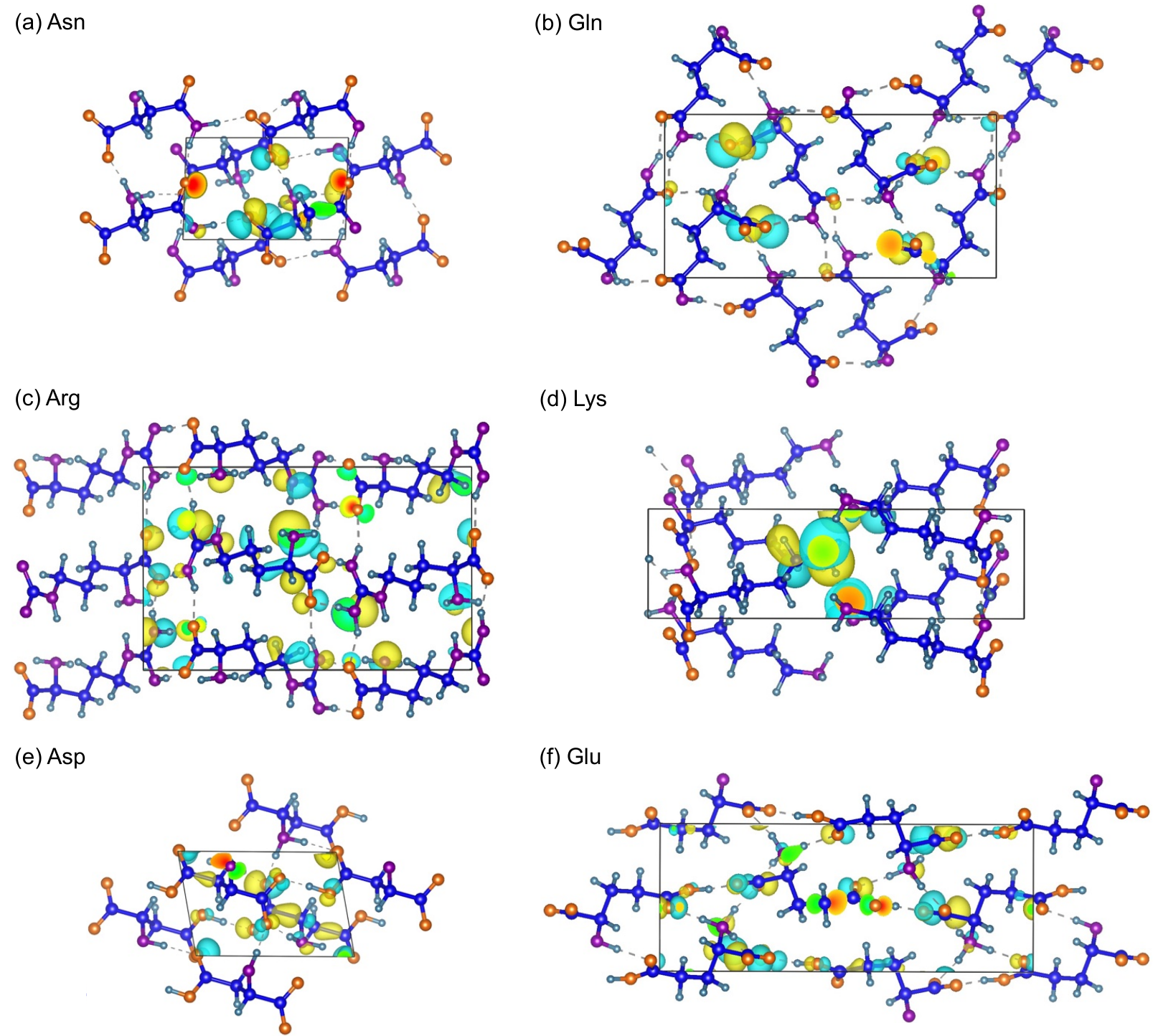}
    \caption{Visualisation of the highest occupied molecular orbitals (HOMOs) of the polar side chain-containing AAs for (a) Asn, (b) Gln, (c) Arg, (d) Lys, (e) Asp, and (f) Glu using the VESTA software package.~\cite{Momma2011VESTA3Data} All H, C, N, and O atoms are inked in steel blue, deep blue, purple, and orange, respectively.}
    \label{other_HOMOs}
\end{figure} 

\begin{figure}[htp]
    \centering
    \includegraphics[width=0.8\textwidth]{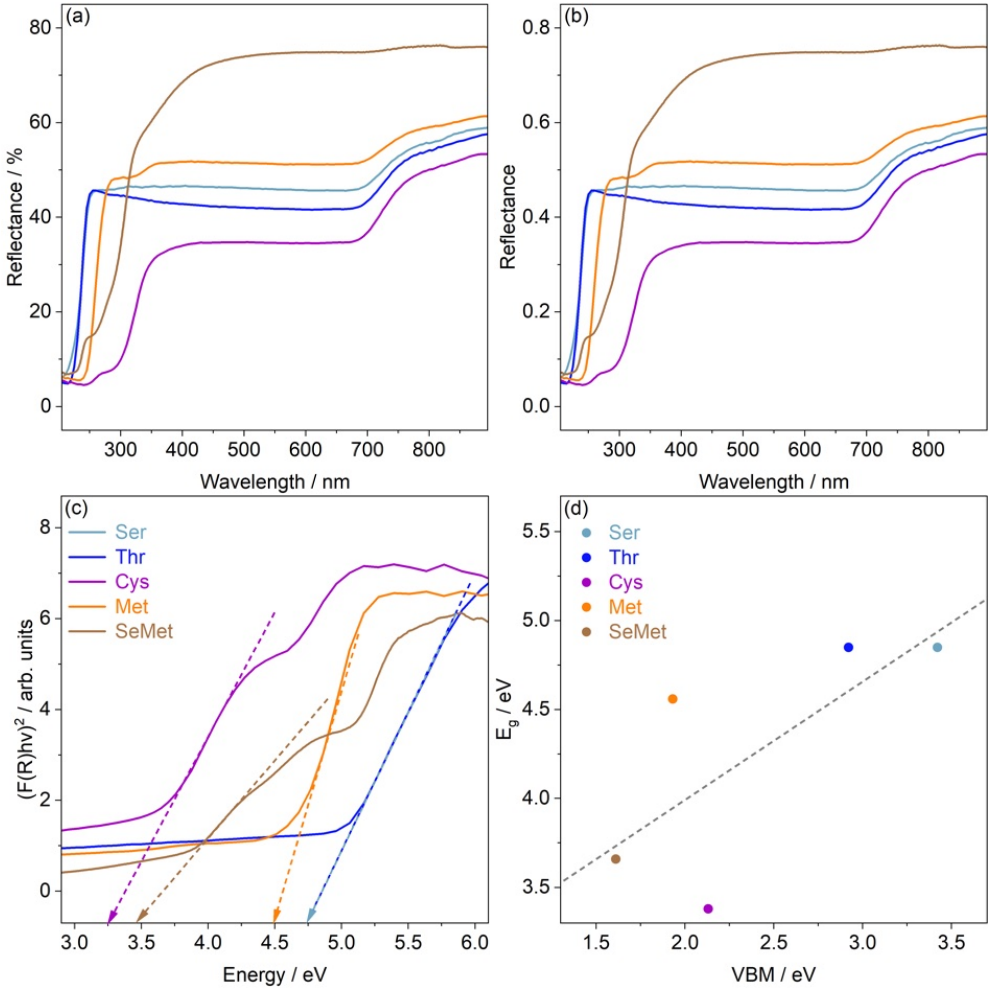}
    \caption{The UV-Vis reflectance data for the S/Se-containing AAs in (a) R($\%$) vs.$\lambda$, (b) R vs.$\lambda$, (c) Tauc plot, and (d) correlation plot of VBM and optical \ce{E^{solid}_g} values. The legend in (c) is also applied to (a) and (b).}
    \label{S_Se_UV}
\end{figure} 

\textbf{\begin{table}[htp]
    \caption{Summary of VBM values which are determined from the as-collected VB spectra as the intersection of the linear fits to the final drop in intensity and the background lines and their corresponding solid-state band gap values, \ce{E^{solid}_g}, which are dictated as the intersection of the linear fits to the Tauc plots made for the reflectance data and the x-axis, for the S/Se-containing AAs.}
\begin{center}
\begin{tabular}{c c c} 
 \hline
 AAs & \ce{VBM_{exp.}} / eV & \ce{E^{solid}_g} / eV \\ [0.5ex] 
 \hline
 Ser & 3.42 & 4.85 \\
 Thr & 2.92 & 4.85 \\ 
 Cys & 2.13 & 3.38 \\
 Met & 1.93 & 4.56 \\ 
 SeMet & 1.61 & 3.66 \\
 \hline
\end{tabular}
\end{center}
\label{S_Se_VBM}
\end{table}}

\begin{figure}[htp]
    \centering
    \includegraphics[keepaspectratio, width=1.0\textwidth]{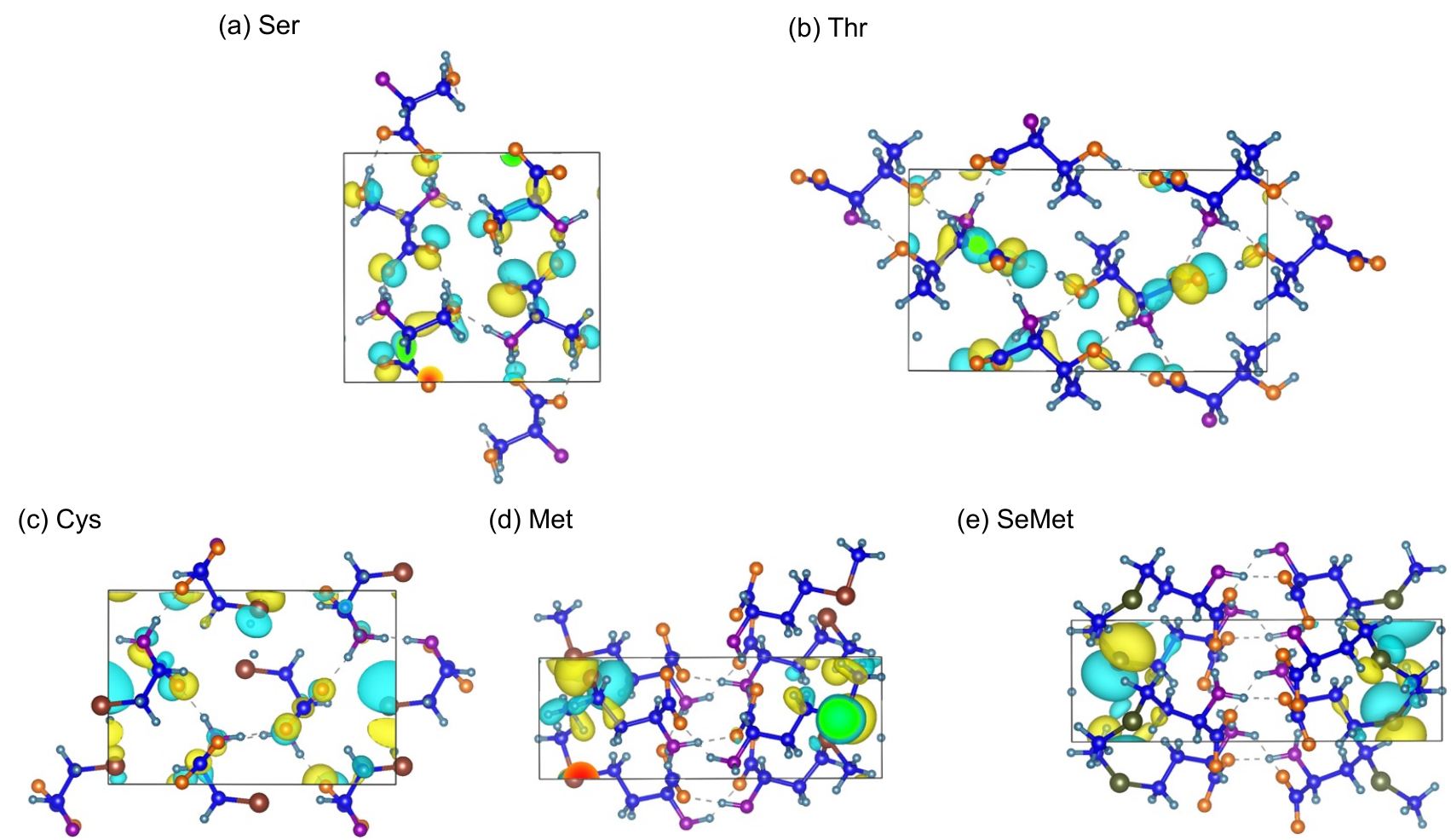}
    \caption{Visualisation of the highest occupied molecular orbitals (HOMOs) of the S/Se-containing AAs for (a) Ser, (b) Thr, (c) Cys, (d) Met, and (e) SeMet using the VESTA software package.~\cite{Momma2011VESTA3Data} All H, C, N, and O atoms are inked in steel blue, deep blue, purple, and orange, respectively.}
    \label{S_Se_HOMOs}
\end{figure}


\newpage
\printbibliography